\documentclass[twoside,titlepage,openright,12pt,a4paper]{report}
\pdfoutput=1

\usepackage{amsmath}
\usepackage{amssymb}

\usepackage{mathtools}

\usepackage[pdftex]{graphicx}
\graphicspath{{./Plots/}{./}}

\usepackage[T1]{fontenc}
\usepackage{lmodern}

\usepackage{bm}
\usepackage{enumerate}

\usepackage{tikz}
\usepackage{tikz-3dplot}
\usetikzlibrary{decorations.pathmorphing}

\usepackage[version=4]{mhchem}

\usepackage[numbers,sort&compress]{natbib}

\usepackage[nottoc,notlof]{tocbibind}

\usepackage{commath}

\usepackage{bbold}

\usepackage{color}

\usepackage{pdfpages}

\usepackage[labelfont=bf,font=small]{caption}

\usepackage{floatrow}
\usepackage{afterpage}

\usepackage{lastpage}

\usepackage[a4paper,width=150mm,top=25mm,bottom=25mm]{geometry}

\usepackage{fancyhdr}

\fancypagestyle{fancystyle}{%
	\fancyhf{}
	\fancyhead[CO]{\textsl{\leftmark}\vspace{0.2cm}}
	\fancyhead[RO]{\textbf{\thepage}\vspace{0.2cm}}
	\fancyhead[CE]{\S$\;$\thesection$\quad$\textsl{\rightmark}\vspace{0.2cm}}
	\fancyhead[LE]{\textbf{\thepage}\vspace{0.2cm}}
	
}
\fancypagestyle{bibstyle}{%
	\fancyhf{}
	\fancyhead[CO]{\textsl{Bibliography}\vspace{0.2cm}}
	\fancyhead[RO]{\textbf{\thepage}\vspace{0.2cm}}
	\fancyhead[CE]{}
	\fancyhead[LE]{\textbf{\thepage}\vspace{0.2cm}}
	
}
\fancypagestyle{plain}{%
	\fancyhf{} 
	\fancyfoot[C]{\textbf{\thepage}}
}

\usepackage{titlesec}

\renewcommand{\thechapter}{\arabic{chapter}}
\titleformat{\chapter}[display]
  {\bfseries\Huge}
    {\filleft{\large\chaptertitlename}$\;$\large\thechapter}
      {2ex}
	  {\titlerule
	     \vspace{1ex}%
	        \filright}
		    [\vspace{1ex}%
		    \titlerule]

\usepackage{hepparticles}

\usepackage{changepage}

\usepackage{setspace}

\DeclareMathOperator*{\SumInt}{%
\mathchoice%
  {\ooalign{$\displaystyle\sum$\cr\hidewidth$\displaystyle\int$\hidewidth\cr}}
    {\ooalign{\raisebox{.14\height}{\scalebox{.7}{$\textstyle\sum$}}\cr\hidewidth$\textstyle\int$\hidewidth\cr}}
      {\ooalign{\raisebox{.2\height}{\scalebox{.6}{$\scriptstyle\sum$}}\cr$\scriptstyle\int$\cr}}
	  {\ooalign{\raisebox{.2\height}{\scalebox{.6}{$\scriptstyle\sum$}}\cr$\scriptstyle\int$\cr}}
}

\usepackage{float}
\usepackage{flafter}

\usepackage{doi}

\usepackage{notoccite}

\usepackage{epsfig,amsbsy,amssymb}
\usepackage{graphics}
\usepackage{amsfonts}

\newcommand{\isum}{\mathop{\hbox{$\displaystyle\sum\kern-13.2pt\int\kern1.5pt$}}}
\newcommand{\Hm}{H$^-\,$}

\newcommand{\ra}{\rangle}
\newcommand{\la}{\langle}
\newcommand{\nn}{\nonumber}
\newcommand{\be}{\begin{equation}}
\newcommand{\ee}{\end{equation}}
\newcommand{\br}{\begin{eqnarray*}}
\newcommand{\er}{\end{eqnarray*}}
\newcommand{\ba}{\begin{eqnarray}}
\newcommand{\ea}{\end{eqnarray}}
\newcommand{\bp}{\begin{minipage}}
\newcommand{\ep}{\end{minipage}}

\renewcommand{\l}{\lambda}

\usepackage{hyperref}
\hypersetup
	{
	colorlinks=true,
	linkcolor=blue,
	citecolor=blue,
	urlcolor=blue,
	}
\makeatletter
\newcommand\org@hypertarget{}
\let\org@hypertarget\hypertarget
\renewcommand\hypertarget[2]{%
\Hy@raisedlink{\org@hypertarget{#1}{}}#2%
} \makeatother

\usepackage{nameref}
\usepackage{cleveref}

\title{{\large 2016 Honours Thesis}\\\bf{Laser assisted electron-atom collisions}}
\author{Alexander Bray\\
	{\footnotesize u6005108}\\[0.2cm]
	\includegraphics[width=0.30\textwidth]{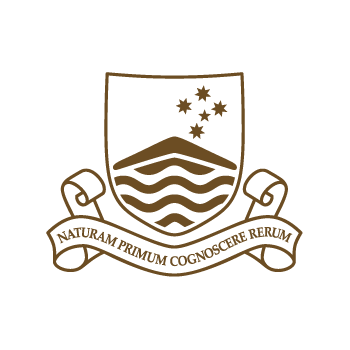}\\[0.2cm]
	{\footnotesize{Research School of Physics and Engineering}}\\
	{\footnotesize{Australian National University}}
	}
\date{\today}

\begin{document}
\pagenumbering{roman}
\onehalfspacing
\pagestyle{plain}
\begin{titlepage}
	\begin{center}
	\vspace*{1cm}
	\Large
	2016 Honours Thesis\\[0.2cm]
	\LARGE
	\textbf{Laser assisted electron dynamics}\\
	\vspace{0.5cm}
	\vspace{1.5cm}
	\Large{Alexander Bray}\\
	\vspace{2.0cm}
	\includegraphics[width=0.4\textwidth]{anu.png}\\[0.2cm]
	\large
	Research School of Physics and Engineering\\
	Australian National University\\
	\vfill
	\Large
	A thesis presented in partial fulfilment for the degree of \\
	Bachelor of Science (Honours) in Physics\\
	\vspace{2.0cm}
	October 27, 2016
	\vspace{1.5cm}
	\end{center}
\end{titlepage}
\afterpage{\null\newpage}
\chapter*{Declaration}
\addcontentsline{toc}{chapter}{Declaration}

This thesis is an account of research undertaken between February 2016 and October 2016 in partial fulfilment of the assessment requirements for the degree of Bachelor of Science with Honours in Physics
at The Australian National University. The pages of this document constitute an original work which has not been submitted in whole or part to any other university. In particular, all figures are my own
unless stated otherwise.
\begin{flushright}
{ Alexander Bray}\\[0.2cm]
October 27, 2016
\end{flushright}

\afterpage{\null\newpage}
\chapter*{Acknowledgements}
\addcontentsline{toc}{chapter}{Acknowledgements}
I must firstly thank my supervisor, Professor Anatoli Kheifets, for without his time, effort, and guidance this project would not be possible.
Thanks is also heartily due to Professor Igor Bray and the members of the Curtin University Institute of Theoretical Physics for access to the convergent close-coupling code base and their continued
support in its use.
I must also thank those who funded and orchestrated the Dunbar scholarship. Without their generosity I would not have had the opportunity to undertake my Honours year studies at the ANU.
I also acknowledge the members of the School Computer Unit for their help in setting up technological access and all of the staff at the Pawsey Centre and the NCI for their continued efforts in providing
the computational infrastructure that this project so heavily relied upon.

Now I would like to that all those of whom their efforts helped to keep me (comparatively) sane throughout this year.
Sam and Edmund for organising the Fenner Hall and ANU based table tennis. 
The many, many games played were a most welcome alternative to the hours spent on a computer and was a great source of challenge and fun.
Brian, Ashley, and Ee-Faye for their efforts in running the Fenner Hall Ensemble.
It was great to be able to crack out the ol' saxophone and be a part of such a fun and talented group.
Yifa, Sihui, Yi, Abhijeet, Satomi, and Kirsty for their friendship and regular conversation over dinner.
Helen, for her continued moral support from afar and hours upon hours worth of Skype and Facebook messages.
Josh, for his lasting assistance throughout the year and in hunting down typos.
Finally, I would like to thank Lunch Lord (or equivalent title) Matt and all my fellow Honours students.
Each other's support made coursework all the more bearable and the weekly lunches with the ridiculous discussions they entailed were always a highlight.
We all got there eventually!

This work was supported by resources provided by the Pawsey Supercomputing Centre and the National Computing Infrastructure.

\afterpage{\null\newpage}
\chapter*{Abstract}
\addcontentsline{toc}{chapter}{Abstract}
We apply the convergent close-coupling (CCC) formalism to analyse the processes of laser assisted electron impact ionisation of He, and the attosecond time delay in the photodetachment of the \Hm 
ion and the photoionisation of He. 
Such time dependent atomic collision processes are of considerable interest as experimental measurements on the relevant timescale (attoseconds $10^{-18}$ s) are now possible 
utilising ultrafast and intense laser pulses. 
These processes in particular are furthermore of interest as they are strongly influenced by many-electron correlations. 
In such cases their theoretical description requires a more comprehensive treatment than that offered by first order perturbation theory. 
We apply such a treatment through the use of the CCC formalism which involves the complete numeric solution of the integral Lippmann-Schwinger equations pertaining to a particular scattering event. 
For laser assisted electron impact ionisation of He such a treatment is of a considerably greater accuracy than the majority of previous theoretical descriptions applied to this problem
which treat the field-free scattering event within the first Born approximation. 
For the photodetachment of \Hm and photoionisation of He, the CCC approach allows for accurate calculation of the attosecond time delay and comparison with the 
companion processes of photoelectron scattering on H and He$^+$, respectively. 

Results of our CCC calculations for laser assisted electron impact ionisation of He are consistent with the previous findings reported in the literature 
[C. H\"ohr et al, \href{http://journals.aps.org/prl/abstract/10.1103/PhysRevLett.94.153201}{\textit{Phys}.\ \textit{Rev}.\ \textit{Lett}.\ \textbf{94}, 153201 (2005)}].
Our results provide further confirmation that the cause of the theoretical discrepancy is in the treatment of the laser field interaction as opposed to the that of the field-free scattering. 
Concurrently, our application of the CCC method to attosecond time delay in the photodetachment of H$^-$ and contrasting processes 
has led to the discovery of the measurable opening time of the inelastic channel
[A.\ Kheifets, A.\ Bray, and I.\ Bray, \href{http://journals.aps.org/prl/abstract/10.1103/PhysRevLett.117.143202}{\textit{Phys}.\ \textit{Rev}.\ \textit{Lett}.\ \textbf{117}, 143202 (2016)}].
Additionally, for calculations across this channel threshold we employ the newly developed numerical treatments of the singularity within the aforementioned integral Lippmann-Schwinger equations
[A.\ Bray et al, \href{http://www.sciencedirect.com/science/article/pii/S001046551500260X}{\textit{Comput}.\ \textit{Phys}.\ \textit{Commun}.\ \textbf{196}, 276-279 (2015)} 
and
\href{http://www.sciencedirect.com/science/article/pii/S0010465516300431}{\textbf{203}, 147-151 (2016)}]
which has been extended for application to charged targets 
as part of this work for the purposes of the He$^+$ calculations 
[A.\ Bray et al, \href{http://www.sciencedirect.com/science/article/pii/S0010465516303149}{\textit{Comput}.\ \textit{Phys}.\ \textit{Commun}. (accepted October 2016)}].

{\hypersetup{linkcolor=black}
\tableofcontents

\listoffigures
}

\afterpage{\null\newpage}
\chapter*{Project Summary}
\addcontentsline{toc}{chapter}{Project Summary}
	The work I have been involved in throughout my Honours year falls into four categories:
	\begin{enumerate}[1)]
		\item Implementation of the soft photon approximation for laser assisted collisions within CCC.\\[0.2cm]
			This task was such that I was able to undertake it largely autonomously with the main result being made clear around the time of mid-year presentations.
			It involved running the CCC code to produce convergent results for the field-free triply differential cross section (TDCS) for electron scattering on atomic helium and then calculating
			the laser field-assisted cross section under the soft photon approximation which is expressed as a sum of field-free cross sections weighted by squared Bessel functions
			(see \Cref{SoftPhotonResult}).
			In doing so we were able to reproduce results of a similar form to that of 
			[C. H\"ohr et al, \href{http://www.sciencedirect.com/science/article/pii/S0368204807000631}
			{\textit{J}.\ \textit{Electron}.\ \textit{Spectrosc}.\ \textit{Relat}.\ \textit{Phenom}.\ \textbf{161}, 172-177 (2007)}] 
			and concluded that the introduction of a more elaborate
			treatment of the field-free scattering was not sufficient to rectify their presented discrepancy with their experiment.
			Further investigation of this discrepancy via this approach was deemed to require considerably more time and likely only lead to minor benefit, and as such we moved to work on other
			problems.
		\item Implementation of the alternative treatment of the singularities occurring in the integral Lippmann-Schwinger equations solved within CCC for charged targets.\\[0.2cm]
			This task was an extension of work I had undergone in 2015.
			It involves modifying the CCC formalism to incorporate an analytic form of an integral involving the Green's function 
			[A.\ Bray et al, \href{http://www.sciencedirect.com/science/article/pii/S001046551500260X}{\textit{Comput}.\ \textit{Phys}.\
			\textit{Commun}.\ \textbf{196}, 276-279 (2015)} and 
			\href{http://www.sciencedirect.com/science/article/pii/S0010465516300431}{\textbf{203}, 147-151 (2016)}].
			Doing so removes the need for a numerical treatment of integration across the point of singularity occurring in open channels. 
			The original formulation can become error prone for energies near threshold in which the singularity occurs close to zero.
			For charged targets the Green's function takes on a form with Coulomb functions as opposed to Riccati-Bessel functions. 
			However, the analytic result of the integral expression is of the same form as the original and such it was relatively simple to extend this method to charged
			targets [A.\ Bray et al, \href{http://www.sciencedirect.com/science/article/pii/S0010465516303149}{\textit{Comput}.\ \textit{Phys}.\ \textit{Commun}. (accepted October 2016)}].
			The alternative treatment is utilised for the e-H scattering across the $n=2$ threshold required for photoemission time delay calculations of \Hm.
			The extension to charged targets allowed application to e-He$^+$ scattering which is necessary to calculate the photoemission time delay for He.
		\item Calculation of Wigner time delay for H$^{-}$ and He.\\[0.2cm]
			This involved using the CCC approach to calculate amplitudes for the photodetachment of the \Hm ion across a large range of photoelectron 
			energies and from which calculate the photoemission time delay.
			Doing so requires the half off-shell $T$-matrix for the associated elastic scattering event, which in this case is the elastic scattering of an electron on H 
			in the dipole singlet channel (see \Cref{Photoemission}).
			This scattering event also has an associated time delay calculated from the phase shift in the $L=1$ partial wave, which we compare to the photoemission delay.
			Of particular interest is the behaviours exhibited across the $n=2$ threshold (at 10.2 eV) where the opening of this channel leads to significant contrast 
			due to the different electron-electron correlations present in the ground states of the targets.
			For comparison with that of \Hm we also consider the photoionisation of atomic He and the associated scattering event of elastic e-He$^+$ scattering, again in the dipole singlet channel.
			The major result of this investigation was the large ($\simeq 40$ as) and potentially measurable photoemission time delay of \Hm above this threshold.
			This was a highly exciting result and led to a publication in Physical Review Letters
			[A.\ Kheifets, A.\ Bray, and I.\ Bray, \href{http://journals.aps.org/prl/abstract/10.1103/PhysRevLett.117.143202}{\textit{Phys}.\ \textit{Rev}.\ \textit{Lett}.\ \textbf{117},
			143202 (2016)}].
		\item Investigation of using a TDSE code for atomic systems interacting with ultrashort laser pulses.\\[0.2cm]
			This involved using the newly published TDSE code of 
			[S.\ Patchkovskii and H.G.\ Muller, \href{http://www.sciencedirect.com/science/article/pii/S001046551500394X}
			{\textit{Comput}.\ \textit{Phys}.\ \textit{Commun}.\ \textbf{199}, 153-169 (2016)}] 
			in attempt to reproduce the photoemission spectra presented in 
			[L.\ Torlina et al, \href{http://www.nature.com/nphys/journal/v11/n6/abs/nphys3340.html}{\textit{Nature} \textit{Phys}.\ \textbf{11}, 503-508 (2015)}].
			Despite initial success in producing photoelectron spectra as a function of energy, angular dependences with momentum proved more challenging.
			Eventually this was also rectified, but we were still unable to produce the angular dependencies of the attoclock paper for calculations taking the better part of a day.
			This remains the case as of the current moment.
			We intend to further investigate the use of a time dependent formalism as part of a PhD project in the coming year (see \Cref{Further Work}).
	\end{enumerate}

	\begin{samepage}
	\noindent \textbf{Resulting publications}:

	\begin{enumerate}[i)]
		\item 
		A.S.\ Kheifets, A.W.\ Bray, and I.\ Bray, ``Attosecond time delay in photoemission and electron scattering near threshold,'' 
		\href{http://journals.aps.org/prl/abstract/10.1103/PhysRevLett.117.143202}{Phys.\ Rev.\ Lett.\ \textbf{117}, 143202 (2016)}
		\item 
		A.W.\ Bray, I.B.\ Abdurakhmanov, A.S.\ Kadyrov, D.V.\ Fursa, and I.\ Bray, ``Solving close-coupling equations in momentum space without singularities for charged targets,'' 
		\href{http://www.sciencedirect.com/science/article/pii/S0010465516303149}{Comput.\ Phys.\ Commun.\
		(accepted October 2016)}.
	\end{enumerate}
	\noindent \textbf{Other publications}:

	\begin{enumerate}[i)]
		\setcounter{enumi}{2}
		\item
		A.W.\ Bray, I.B.\ Abdurakhmanov, A.S.\ Kadyrov, D.V.\ Fursa, and I.\ Bray, 
		``Solving close-coupling equations in momentum space without singularities,''
		\href{http://www.sciencedirect.com/science/article/pii/S001046551500260X}{Comput.\ Phys.\ Commun.\ \textbf{196}, 276-279 (2015)}
		\item 
		A.W.\ Bray, I.B.\ Abdurakhmanov, A.S.\ Kadyrov, D.V.\ Fursa, and I.\ Bray, ``Solving close-coupling equations in momentum space without singularities II,'' 
		\href{http://www.sciencedirect.com/science/article/pii/S0010465516300431}{Comput.\ Phys.\ Commun.\ \textbf{203}, 147-151 (2016)}
		\item
		I.I.\ Fabrikant, A.W.\ Bray, A.S.\ Kadyrov, and I.\ Bray, ``Near-threshold behavior of positronium-antiproton scattering,'' 
		\href{http://journals.aps.org/pra/abstract/10.1103/PhysRevA.94.012701}{Phys.\ Rev.\ A \textbf{94}, 012701 (2016)}
	\end{enumerate}
	\end{samepage}
\chapter{Introduction}
\setlength{\abovedisplayskip}{8pt plus 6pt minus 6pt}
\setlength{\belowdisplayskip}{8pt plus 6pt minus 6pt}
\setlength{\abovedisplayshortskip}{4pt plus 3pt minus 3pt}
\setlength{\belowdisplayshortskip}{4pt plus 3pt minus 3pt}
\setlength{\parskip}{0.0cm plus 0.0cm minus 0.0cm}
\pagestyle{fancystyle}
\pagenumbering{arabic}
	The fundamental drive behind all scientific endeavours is to observe and explain the physical world.
	Two of the most prevalent aspects that make up this world in which we reside are matter and light (dark matter/energy notwithstanding),
	of which our understanding provides some of the greatest insight into our universe.
	Let alone the knowledge to be gained from the examination of each in isolation, the interaction between the two tests our understanding like no other.
	Take for example, the photoelectric effect \cite{einstein1905photoelectric}, 
	in which the interaction between matter and light immensely elucidated the nature of both, and subsequently led to the birth of quantum mechanics \cite{born1924quantenmechanik}.
	Despite the passing of more than a century since this point, there are still innumerable questions
	to be answered in the description of this fundamental interaction.
	The two such questions that we investigate within this work are the laser assisted electron impact ionisation of He, and the 
	attosecond time delay in the photodetachment of the H$^-$ ion and the photoionisation of He.
	\section{Contextual Background}
	The scope of this work encompasses both field-free (no laser) and field-assisted (laser present) atomic scattering and as such an introduction to these two related fields are provided.
	This is a common theme through the work and this structure is repeated in a similar vein in subsequent chapters.
	\subsection{Atomic and Molecular Scattering}
	Scattering events at the atomic scale have always been one of the sources of greatest insight into the nature of our world. 
	The earliest experiments of electron scattering \cite{franck1911zusammenhang} were key in establishing the existence of orbitals of quantised energy as suggested by the Bohr model of the
	atom.
	With the onset of quantum mechanical theory the first measurement of electron-atom total cross sections were conducted by \citet{ramsauer1921wirkungsquerschnitt} and theoretical attempts to
	calculate scattering amplitudes by \citet{massey1931collision}.
	But despite this early progress within the field many fundamental problems remained.

	The theoretical description of electron-atom collisions for all incident energies and scattering angles remained elusive for the better part of the century.
	Such a problem is inherently complicated, requiring the solution of the Schr\"odinger or Dirac (relativistic) equation with three or more bodies.
	Additionally, the nature of atomic targets with a countably infinite number of bound discrete states (negative energy) as well as a uncountably infinite number of free states (positive energy) for
	each electron, all of which are coupled to one another provides a considerably challenge for formal theoretical description.
	Furthermore, in collisions involving ionisation or initially charged targets the $1/r$ Coulomb interaction potential which continues to infinite distance constitutes a further source of difficulty.
	Yet another complication
	encountered specific to electron scattering is the non-uniqueness of solution coming from the indistinguishability of electrons and the possibility of projectile exchange
	with the target electrons.
	Despite these complexities of the underlying problem, one of the greatest successes of early theory is the wide applicability of the so called Born approximation \cite{Born1926}.
	In which, large parts of the problem are omitted in the assumption that the interaction between the projectile and target atom is weak.
	This assumption is most appropriate in the case of large incident energies (compared to the ground state energy of the target).
	However, for processes at energies which are comparable with the ground state of the atomic target an equivalently effective description was beyond the reach of theory.
	The work of \citet{massey1956} provides a good summary of the attempts and difficulties faced by the early theoretical attempts to tackle this realm of the parameter space.

	The electron-hydrogen scattering system is considered the most fundamental physical scattering problem, yet it was the source of considerable discrepancy in the 1980's.
	Experiment had progressed to the 1s-2p excitation of atomic hydrogen \cite{weigold1980large,williams1981electron} of which there was theory available for comparison.
	However, the best theoretical attempts of the time \cite{scholz1991electron,van1986elastic}, though in reasonable agreement between themselves, both failed to explain the results of the two
	experiments for backward scattering angles.
	This challenge for theorists in conjunction with advances in computation power led to the development of a number of non-perturbative numerical treatments in order to reconcile this discrepancy.
	Among these are the $R$-matrix with pseudostates (RMPS) \cite{bartschat1996electron}, exterior complex scaling (ECS) \cite{rescigno1999collisional}, time dependent close-coupling (TDCC) 
	\cite{pindzola1996time} and convergent close-coupling (CCC) \cite{CCC1992} methods.
	Despite these advances, much to the initial dismay of theorists, these new techniques produced results similar to the older theories.
	However with the increasing weight of theoretical support these experiments were conducted once more with more modern techniques \cite{yalim19971s,o1998polarization} which finally resolved this 
	disagreement, demonstrating excellent agreement with the theoretically predicted values in the region of previously greatest discrepancy.

	The next major hurdle within the field were ionisation collisions events (break-up) for the three body system that is electron scattering on hydrogen.
	This remained one of the unsolved fundamental problems within quantum mechanics.
	The mathematical formalism was initially given for this system in the 60's by \citet{peterkop1962wave} and \citet{rudge1965ionization}, however 
	this formulation involves a boundary condition enforced 
	onto the wavefunction such that all three charged particles were interacting up to infinite distance that proved so intractable that no computational method has incorporated it in its entirety.
	The first detailed calculations of the ionisation were given in the late 80's and early 90's \cite{bartschat1987r,bray1993calculation}. 
	However, it was not until the incredible success of the exterior complex scaling work of \citet{rescigno1999collisional} and \citet{PhysRevA.63.022712} that the problem was considered solved, 
	with a flurry of subsequent papers published
	as other methods provided their own contributions \cite{bray2002close,bartschat2002convergent,colgan2006double}.
	These computational methods despite their success, lacked formal grounding in their treatment of the Coulomb boundary condition, and it was only recently in a series of works
	\cite{kadyrov2003integral,kadyrov2004theory,kadyrov2008coulomb}
	culminating in that of \citet{Kadyrov20091516}, that this grounding was provided.

	With the fundamental scattering interactions solved for hydrogen, the next decade saw agreement between theory and experiment for all manner of atomic targets, including helium
	\cite{sawey1990electron,pindzola2000time}, 
	hydrogen-like metals \cite{kwan1991total}, helium-like metals \cite{fursaHelike}, heavy noble gases \cite{dzuba1996many} and ions \cite{bray1993calculation}.
	Largely this was due to the structural problems in describing these various targets being solved far earlier 
	(helium for example famously by \citet{hylleraas1929neue} and subsequently \citet{pekeris1958ground}) than those of scattering and the generality of the developed computational methods.
	Scattering on basic molecular targets such as $\ce{H_2^+}$ \cite{hanssen1996differential}, $\ce{H_2}$ \cite{zammitMol}, 
	and $\ce{H_2O}$ \cite{abdurakhmanovH20} (the latter within a neon-like approximation) have also been theoretically described.
	Additionally, various projectiles have been successfully treated by theory including photons \cite{bartschat1998r} (considered as a half-collision), 
	positrons \cite{kernoghan1996positron}, and heavy projectiles such as protons \cite{abdurakhmanovPro}, 
	anti-protons \cite{abdurakhmanovHeAntiPro}, and ions \cite{abdurakhmanovC6}.
	In the case of positively charged projectiles, multi-centre treatments \cite{fainstein1991two} are often required due to the possibility of electron capture.
	Such was the success of theory within recent years that many consider the field to be solved.
	The modern frontiers of atomic and molecular scattering are in the description of increasingly complex molecular targets \cite{abdurakhmanovH20}, near threshold behaviours \cite{Fabrikant2016}, 
	and in multi-interaction processes such as those involved in stopping power calculations \cite{bailey2015antiproton}.
	\subsection{Laser Assisted Electron Dynamics}
	The development of the laser in the 60's \cite{maiman1960stimulated} (maser in the 50's \cite{maser}) provided physicist with a coherent source of light that was readily controllable, 
	and led to a subsequent 
	surge in the science to describe the interaction of light with atomic targets \cite{voronov1965ionization,tozer1965theory,agostini1968multiphoton}.
	From this point onwards, the study of light interacting with atoms and molecules has largely being driven by the continual improvement and development of laser technologies.
	For a long period of time the intensities of light sources were sufficiently low such that their interaction with atomic targets could be adequately described using first order perturbation theory
	\cite{Faisal1973}.
	As such a push for greater intensities was present in order to observe more complicated phenomena.
	The technique known as mode-locking in which a series of laser frequencies are combined to produce an increasingly short and intense pulse at regular intervals 
	was demonstrated in the early Ruby \cite{DeutschRuby}
	and $\ce{Nd}:\ce{YAG}$ \cite{DiDomenicoNd} lasers, the second of which is still commonly in use today.
	The intensity of a pulse generated in such a manner is inversely related to the spread of frequencies in the original laser source, and as such a large number of `colours' are desirable.
	Typically solid state lasers have the largest frequency bandwidth and are hence favoured for the production of intense pulses.
	Using these techniques laser pulses with peak intensities of the order of $10^{14}$ W/cm$^2$ are able to be generated, firmly in the region dubbed `intense' where the interaction is no longer
	trivially described as a perturbation of the laser free system. 

	In conjunction with the drive for more intense laser sources comes the requirement for increasingly short pulse duration.
	Even if a particularly intense laser source is exposed to an atomic target, if the dynamics of the system all occur within the pulse `wings' rather than across the entire pulse, then regardless of
	the intensity the source cannot be used to observe high intensity effects.
	The development of the $\ce{Ti}\!:$\hspace{0.10cm}Sapphire laser in the 80's \cite{moulton1982ti} was a major revolution, providing a 
	highly tunable solid state laser source with which it was possible to 
	produce pulses with 
	durations of the order of femtoseconds ($10^{-15}$ s) and intensities of the order of $10^{18}$ W/cm$^2$.
	It is in this region, dubbed `super intense', that the electric field of the laser now becomes the dominant influence over the atomic system. 
	For example, in the case of atomic hydrogen at an intensity of $3.5\times 10^{16}$ W/cm$^2$ the influence of the electric field of the laser becomes equal to the force that binds the electron.
	In this super intense region extremely short pulses are particularly necessary as even for low laser frequencies (photon energies) the pulse is able to easily ionise the target before the peak
	intensity is reached.
	Regardless, it is interesting that despite the extreme dominance of the laser in terms of sheer magnitude, that due to the electron inertia and the oscillatory nature of the laser pulse 
	that the effects of the atomic structure still play a significant role.
	Sources of coherent and extremely intense radiation 
	have additionally been produced with free-electron lasers \cite{YuFEL}, of which they are uniquely tunable to produce photons over a massive frequency range (microwaves to X-rays). 
	However, due to their significant expense and size, for the purposes of physicists interested in short and intense coherent pulses they are only used for their upper frequency range as elsewhere
	solid state laser sources are a considerably more convenient and readily available alternative.

	Femtosecond pulses of laser light have been used to great effect most notably leading to the 1999 Nobel prize in Chemistry being awarded to Professor Ahmed H.\ Zewail 
	for his work in resolving in time the motion of molecules breaking apart under exposure of such pulses \cite{zewail1988laser,zewail1988real}.
	This was possible as the resolution of such a short pulse of light is comparable to the timescale of molecular dynamics (the vibration period of $\ce{H_2}$ is $\approx 8$ fs).
	Even more recently with the advent of high harmonic generation processes \cite{LewensteinHHG} pulses of attosecond ($10^{-18}$ s) duration with intensities as high as $10^{18}$ W/cm$^2$ are now
	readily available \cite{AntoineAttosecond}.
	This process involves using an existing femtosecond pulse to ionise an atomic target (such as neon \cite{AntoineAttosecond}),
	accelerate it away from the nucleus as the pulse rises, then accelerate it back toward the nucleus as the crest
	passes and then falls, and upon recombination release the gained energy in the form of a burst of attosecond duration.
	Pulses on this timescale have moved into the regime of electronic motion within atoms (the orbital period of the electron in the Bohr model of atomic hydrogen is $\approx 150$ as) 
	and their use in resolving such
	dynamics was soon formally theorised \cite{itatani2002attosecond}, birthing the field of attosecond science \cite{corkum2007attosecond}. 
	However, this has yet to be fully realised as to do so requires a rigorous understanding of the various time delays involved in atomic interactions with laser pulses of this nature 
	\cite{ivanov2011accurate}.
	As one can imagine, this has become an incredibly active 
	area of research and has led to intense scrutiny toward the application of these attosecond pulses and the field of laser assisted
	electron dynamics as a whole.
	\section{Motivation and Project Goals}
	Here we present a short introduction to the existing literature regarding the problems we investigate within this work. 
	Furthermore, we look to justify how our contribution fits into this framework and why it constitutes a worthwhile addition. 
	\subsection{Soft Photon Approximation Implementation with CCC}
	The first considerations of laser assisted charged particle scattering were given by \citet{Kroll1973}, which still forms a significant basis for comparison with theory and experiment 
	in the current day \cite{ladino2011high}.
	This basis being that under a number of approximations (see \Cref{Range of Validity}) that the field-assisted cross section involving $n$ laser photons can be expressed as 
	the field-free cross section with adjusted kinematics multiplied by a squared Bessel function (see \Cref{SoftPhotonFree}), 
	and has come to be known as the soft photon approximation (alternatively the Kroll and Watson approximation).
	An important consequence of this result is that not only is it possible to solve a fundamentally time dependent problem through a time independent formalism, but that the ratio of field-free to
	field-assisted cross sections becomes independent of the scattering centre
	for free-free scattering (equivalent to elastic scattering but with the possible emission or absorption of laser photons) from an atomic target.
	The same result was concisely 
	rederived by \citet{Rahman1974} within the scope of free-free charged particle scattering from an arbitrary potential in the presence of an intense electromagnetic wave and
	tested extensively in a series of experiments \cite{Wein1977,Wein1979,wallbank1987experimental}, and was found generally to be in qualitative agreement.
	However, in the experiment by \citet{wallbank1993} it was found that for small scattering angle and low incident electron energy that the approximation did begin to break down.
	This motivated the subsequent comprehensive study by \citet{geltman1996laser} which provided further confirmation of its inadequacy for small scattering angles.
	These experiments all were well positioned to test the predictions under the soft photon approximation as due to the independence of the atomic target, experimentally convenient noble gases
	could be used, simply measuring the ratio with and without the laser.
	However, when considering either ionising collisions or theoretical descriptions that include target dressing effects \cite{byron1984electron} of the laser this is no longer the case.

	The same principles behind those of Kroll and Watson were brought to particle-atom ionising collisions by \citet{cavaliere1980particle}
	resulting in an expression with a similar form (see \Cref{SoftPhotonResult}) to the free-free case.
	Due to the sum over multiple cross sections leading to the total field-assisted cross section, no longer may the ratio be considered independent from the scattering system.
	Because of this, the usual divide rears its head between the theorists of which atomic hydrogen becomes the ideal target for consideration and experimentalists for which noble gases are much
	preferred.
	Initial application of the theory to the electron impact ionisation of hydrogen was included in the original paper \cite{cavaliere1980particle} with a number of increasingly comprehensive 
	applications in following years \cite{cavaliere1981effects,banerji1981electron,Joachain1988,Martin1989}.
	The latter two of these incorporate and pay particular attention to the target dressing effects of the laser and find the triply differential cross section (TDCS) 
	to be strongly dependent on such effects.
	An application to electron impact ionisation of helium was first presented by \citet{joachain1992laser} of which the same group published a very comprehensive work on the subject a few years
	later \cite{Khalil1997}.
	In which, the authors themselves state that the primary driver for their extension of the theory from hydrogen to helium is to provide additional incentive to perform such laser assisted collisions
	experiments.

	This demand was filled by the work of \citet{hohr2005electron} and their follow up paper \cite{hohr2007laser} in which they 
	provided further experimental data and an additional soft photon approximation based theoretical comparison.
	However, from this comparison they found that not only was the soft photon approximation inadequate to describe the results of their experiment, but that it predicted diminution of the cross
	section when the experiment observed enhancement and conversely enhancement when the experiment observed diminution.
	For this reason, they have concluded that in such an ionisation event that there is a fundamental aspect of the physics that is missing from the existing theoretical descriptions
	and calls for comparison with more advanced theoretical models such as offered by $R$-matrix Floquet \cite{burke1991r}.
	For this experiment to challenge what was thought to be a well established theoretical grounding on the subject has come as a considerable surprise to many, and leaves both theorists and
	experimentalists with many interesting problems to consider.

	In order to further investigate this discrepancy we look to take the scattering framework that is the convergent close-coupling (CCC) method
	and extend its applicability to a soft photon approximation based method for 
	calculating field-assisted cross sections.
	The field-free scattering cross sections are calculated using the first Born approximation in the theoretical work presented 
	within \citet{hohr2007laser}. In this aspect through the application of CCC we 
	are well positioned to improve upon the strength of the theoretical description.
	Additionally, in the most comprehensive description available of the electron impact of helium \cite{Khalil1997} they also state that the most obvious limitation of their description is  
	the first Born treatment of the collisional stage of the calculation.
	More recent attempts of examining this discrepancy include incorporating second order Born terms for the field-free scattering event \cite{Makhoute2015}
	and target dressing effects \cite{GhoshDeb2010}, which although each find significant differences with the addition of these aspects, remain unsuccessful in rectifying the situation.
	Hence, the application of a more comprehensive collisional theory to this problem which treats the projectile-target interaction to all orders 
	will provide useful insight into the cause of this discrepancy.
	\subsection{Wigner Time Delay of \Hm and He near Threshold}
	The emission of an electron from an atom upon the absorption of an energetic photon (photoemission, or the photoelectric effect) is one of the most elementary quantum-mechanical phenomena.
	Up until recently, studies of photoemission mainly focused on energetics of the process and the temporal or dynamic aspects were ignored. 
	The fundamental reason for this is that the time scale involved for these processes (attoseconds 10$^{-18}$ s) is inaccessible for measurement.
	However, measurements on this time scale have now become possible with the invention of the so-called ``attosecond streak camera'' \cite{Baltuska2003,Kienberger2004}.  
	The camera makes use of a high harmonic generation (HHG) process which
	converts a driving near-infrared (NIR) femtosecond pulse into
	coherent extreme ultraviolet (XUV) bursts, at least one order of
	magnitude shorter than can be produced by conventional pulsed laser systems.
	The camera makes measurements through application of an attosecond XUV burst onto an atomic electron setting it in motion, while the
	same driving NIR pulse used to generate the attosecond pulse, after a carefully monitored time delay, is
	used to accelerate or decelerate the ionised electron.
	The effect of this interaction on the phase between the two pulsed sources then constitutes the measurement made by the camera.
	A key aspect of this process is the phase stabilisation of the driving NIR pulse with a shot-to-shot stability of a few attoseconds. 
	This stability allows the technique to be used as a temporal ruler on this time scale, which may then be applied to resolve various atomic processes in time.

	One such process is the time delay involved in atomic photoemission. 
	In this process it appears that the photoelectron leaves an atom with a short delay relative to the arrival of the ionising pulse. 
	Hence, the study of this process provides a mechanism for observing ultrafast electron dynamics \cite{RevModPhys.87.765}. 
	The first experimental observations of time delay in photoemission \cite{schultze2010delay,PhysRevLett.106.143002}
	gave rise to the rapidly developing field of attosecond chronoscopy. 
	The time delay in photoemission is
	interpreted in terms of the Wigner time delay introduced for a particle scattering in external potential \cite{Eisenbud1948,PhysRev.98.145,PhysRev.118.349}.  
	It is a delay, or advance, of a particle travelling through a potential landscape in
	comparison with the same particle travelling in a free space. 
	The Wigner time delay is calculated as an energy derivative of the
	scattering phase in a given partial wave (see \Cref{Wigner Time Delay}).  A similar definition is
	adapted in photoemission, where the time delay is related to the
	photoelectron group delay, and evaluated as an energy derivative of
	the phase of the ionisation amplitude \cite{schultze2010delay,PhysRevLett.105.233002}.

	If a single electron is set free when a multi-electron atom absorbs a photon, it is strictly speaking not a single-electron process. 
	Rather, it is the result of the correlated motion of all the electrons, and hence these correlations can have a significant influence on the properties of the emitted photoelectron. 
	To investigate the effect of inter-electron interactions on the Wigner time delay we consider the process of photodetachment of an electron from the negative hydrogen ion and compare it with
	that of elastic electron scattering on the hydrogen atom near the first excitation threshold. 
	The elastic scattering of an electron on hydrogen is the process that underpins the correlation in photodetachment of \Hm through the channel coupling in the ionisation continuum (see
	\Cref{Photoemission}). 
	Photodetachment of \Hm and electron scattering on H are therefore closely related processes, both of which involve a Wigner time delay that is strongly affected by their 
	inter-electron interaction.
	However, despite their similarities, there is a considerable difference in the lowest order interaction present in these systems causing them to exhibit contrasting behaviours with the
	opening of the $n=2$ excitation threshold.
	Additionally, we consider the photoionisation of helium and the associated scattering process of elastic scattering on He$^+$ to provide further comparison with the analysis of \Hm.
	Through this investigation \cite{PhysRevLett.117.143202} we gain considerable insight into the nature of the electronic interactions within these targets and provide 
	theoretical predictions for experimentalists looking to
	measure the time delay inherent in these processes.
	\chapter{Theory}
	In this chapter we provide an introduction to the theory required for an understanding of the problems we look to investigate and the approaches we utilise to do so.
	We begin providing background for the general scattering formalism present in all theory (\Cref{atomicscatteringbackground}), 
	electron impact ionisation (e,2e) (\Cref{e2e Geometry and Terminology}), and
	the considerations required for comparison with \citet{hohr2007laser} (\Cref{hohrcomp}).
	Next we consider the convergent close-coupling (CCC) approach to solving for field-free scattering amplitudes (\Cref{Convergent Close Coupling}),
	how you go about achieving convergence within the method (\Cref{Convergence Considerations}), and 
	an application of the theory to the (e,2e) process of helium (\Cref{Treatment of Helium}).
	Finally, we provide information pertaining to the treatment of laser assisted collision processes (\Cref{Laser Assisted Collisions}), 
	the soft photon approximation (\Cref{Soft Photon Approximation}),
	the application of the CCC method to photoemission (\Cref{Photoemission}),
	and the Wigner time delay of a scattering event (\Cref{Wigner Time Delay}).
	For additional derivations relevant to the following theory see the corresponding section in \Cref{Derivations}. 

	In this and subsequent chapters the system of atomic units (a.u.)\ will be used unless otherwise stated. 
	For those unfamiliar with the system of atomic units, see \Cref{atomic units}.
	However, note that the units of energy are an exceptional case, typically expressed in eV.
	Furthermore, time is typically given in terms of attoseconds ($\mathrm{as} =10^{-18}$ s) and intensity in W/cm$^2$.
	For the extent of this work the energetics of each species are sufficiently low such that no relativistic effects need be accounted for (see \Cref{hohrcomp}).
	Hence, what follows is a purely non-relativistic treatment of scattering theory. 
	Additionally, the mass to velocity ratios of the species are such that the centre of mass frame approximates the laboratory frame (see \Cref{hohrcomp}).
	Hence, no extra efforts are required in conversion between said frames for the purpose of comparison with experiment.
	\section{Atomic Scattering Background}
	\label{atomicscatteringbackground}
	Atomic scattering entails a projectile incident on a target atom, undergoing some interaction, and then leaving the system in some final state. 
	We now look to provide a description of this process derived from the foundations of quantum mechanics \cite{LLQuantum,mcdowell1970introduction}.
	Note that there is no explicit time dependence of the interaction potential and hence we may use a time independent formulation.
	Additionally, for the scope of this work we also assume that there is no explicit spin dependence on the potential. 
	This is a reasonable assumption as we are dealing with targets of low atomic charge ($Z$) such as helium and the spin-orbit interaction scales with $Z$.
	Consequently, spin only has an indirect effect through the Pauli exclusion principle.
	Furthermore, the following is only applicable for initially neutral targets.

	Let us denote the initial state of the projectile as $|\bm{k}_i\rangle$, which is described by its momentum $\bm{k}_i$.
	Asymptotically, this initial state is given as a plane wave
	\begin{alignat}{3}
		\lim_{\bm{r}\to\infty}\langle \bm{r}|\bm{k}_i\rangle=(2\pi)^{-3/2}e^{i\bm{k_i}\cdot \bm{r}},
	\end{alignat}
	where $\bm{r}$ is the spatial coordinate of the projectile.
	Let us describe the initial state of the target as $|\phi_i\rangle$ with corresponding energy $\epsilon_i$. 
	This can be equivalently described by the standard set of quantum numbers for each of the atomic electrons, but in the interest of generality we will remain with simply $|\phi_i\rangle$.
	For a specific treatment of electron scattering on helium see \Cref{Treatment of Helium}.
	Together we have the system in its initial state described by $|\phi_i\bm{k}_i\rangle$.

	\enlargethispage{0.5cm}
	Considering the final state of the system we similarly describe the projectile and target as $|\bm{k}_f\rangle$ and $|\phi_f\rangle$ respectively.
	However, the scattered projectile is now asymptotically described by the sum of a plane and spherical wave as
	\begin{alignat}{3}
		\lim_{\bm{r}\to\infty}\langle \bm{r}|\bm{k}_f\rangle&=(2\pi)^{-3/2}\left[e^{i\bm{k}_f\cdot\bm{r}}\delta_{fi}+f^S_{fi}(\bm{k}_f, \bm{k}_i)\frac{e^{ik_fr}}{r}\right],
	\end{alignat}
	where $f^S_{fi}(\bm{k}_f,\bm{k_i})$ is the scattering amplitude from state $i\to f$ of total spin $S$.
	This amplitude is related to the experimentally observable spin-resolved differential cross section via
	\begin{alignat}{3}
		\label{spinrescross}
		\dod{\sigma^S_{fi}(\theta,\phi)}{\Omega}&=\frac{k_f}{k_i}|f^S_{fi}(\bm{k}_f,\bm{k}_i)|^2,
	\end{alignat}
	where $\mathrm{d}\Omega$ is the element of solid angle in which the projectile is scattered. 
	The spin-resolved integrated cross section is then given by
	\begin{alignat}{3}
		\sigma^S_{fi}&=\int\mathrm{d}\Omega\dod{\sigma^S_{fi}(\theta,\phi)}{\Omega}\\
				&=\int\mathrm{d}\hat{k}_f\;\frac{k_f}{k_i}|f^S_{fi}(\bm{k}_f,\bm{k}_i)|^2.
	\end{alignat}
	The equivalent spin averaged quantities are calculated by averaging over initial spin states, and summing over final spin states such that
	\begin{alignat}{3}
		\dod{\sigma_{fi}}{\Omega}&=\frac{1}{2(2s_i+1)}\sum_S(2S+1)\dod{\sigma^S_{fi}}{\Omega},
	\end{alignat}
	where $s_i$ is the initial spin state of the projectile. 
	This is due to the nature of the final spin states being distinguishable whereas the initial spin states are indistinguishable.
	Similarly for the integrated cross section we have
	\begin{alignat}{3}
		\sigma_{fi}&=&&\int\mathrm{d}\Omega\dod{\sigma_{fi}(\theta,\phi)}{\Omega}.
	\end{alignat}
	The total cross section which is irrespective of the final state of the system is given by
	\begin{alignat}{3}
		\sigma^{\mathrm{tot}}_{i}&=\sum_f \sigma_{fi}.
	\end{alignat}

	The wavefunction describing the system $|\Psi^{S(+)}_i\rangle$ is a solution to the Schr\"odinger equation
	\begin{alignat}{3}
		(H-E)|\Psi^{S(+)}_i\rangle=0,
	\end{alignat}
	where $H$ is the total Hamiltonian of the system, $E=k_i^2/2+\epsilon_i=k_f^2/2+\epsilon_f$ is the total energy of the system and the superscript $^{(+)}$ is used to denote 
	outgoing spherical wave boundary conditions (see \Cref{Contourder}).
	Consequently, to satisfy the boundary condition of the scattered projectile the wavefunction must have the following asymptotic limit
	\begin{alignat}{3}
		\lim_{\bm{r}\to\infty}\langle \bm{r}|\Psi^{S(+)}_i\rangle&=(2\pi)^{-3/2}\left[e^{i\bm{k_i}\cdot \bm{r}}|\phi_i\rangle
		+\SumInt_{n}f^S_{ni}(\bm{k}_n,\bm{k}_i)\frac{e^{ik_nr}}{r}|\phi_n\rangle\right].
	\end{alignat}
	The goal of any scattering theory is therefore to calculate the $f^S_{fi}(\bm{k}_f,\bm{k_i})$ and subsequently the observable cross sections to compare with experiment.
	\subsubsection{$S$-Matrix}
	\enlargethispage{1cm}
	We may consider a scattering problem as taking an initial wavefunction $|\Psi_i\rangle$ to a final wavefunction $|\Psi_f\rangle$. 
	The scattering operator ($S$) (equivalently matrix when defined in terms of a basis) is then defined such that
	\begin{alignat}{3}
		|\Psi_f\rangle=S|\Psi_i\rangle.
	\end{alignat}
	Hence, the ${S}$ operator contains all the information as to the evolution of the system from its initial to final state.
	The $S$-matrix is then defined in terms of the bases determined by the initial and final states of the system such that
	\begin{alignat}{3}
		S_{fi}\equiv \langle \bm{k}_f \phi_f | {S}|\phi_i\bm{k}_i\rangle.
	\end{alignat}
	The above notation will be used to represent matrices defined on the basis defined by the initial and final states.
	It can be shown \cite{mcdowell1970introduction} that the $S$-matrix element from state $i\to f$ is given by
	\begin{alignat}{3}
		\label{Sfi}
		S_{fi}&=\delta_{fi}-\frac{2\pi i}{\sqrt{k_i k_f}}\langle \bm{k}_f\phi_f|V|\Psi_i\rangle
	\end{alignat}
	where $V$ contains the interaction potentials of the system.
	\subsubsection{$T$-Matrix}
	The transition matrix ($T$) is defined as the second term in \eqref{Sfi} such that
	\begin{alignat}{3}
		T_{fi}&=\langle \bm{k}_f\phi_f|V|\Psi_i\rangle.
	\end{alignat}
	Equivalently this definition in operator form is
	\begin{alignat}{3}
		T|\phi_i\bm{k_i}\rangle=V|\Psi_i\rangle.
	\end{alignat}
	The scattering amplitude in the case of non-breakup collisions such as elastic scattering or excitation is given exactly by the $T$-matrix element (see \Cref{Contourder} 
	for derivation) such that 
	\begin{alignat}{3}
		f_{fi}&=T_{fi}.
	\end{alignat}
	In the case of breakup collisions, such as ionisation, the derivation of the scattering amplitude has been a long standing problem, only recently being given formal grounding 
	\cite{Kadyrov20091516}. 
	When formulated utilising a set of square integrable states (see \Cref{Convergent Close Coupling}) the amplitude takes the form
	\begin{alignat}{3}
		\label{breakupamplitude}
		f_{fi}&=\langle \bm{q}_f|\phi_f\rangle T_{fi}\; ,
	\end{alignat}
	where $|\bm{q}_f\rangle$ is the continuum eigenfunction of the target Hamiltonian with energy $\epsilon_f$.
	In the case of a hydrogen target the $|\bm{q}_f\rangle$ are pure Coulomb waves.
	\subsubsection{Born Series}
	The Lippmann-Schwinger equation for the $T$ operator is given by 
	\begin{alignat}{3}
		\label{lippT}
		T=V+VG^{(+)}T
	\end{alignat}
	where $G^{(+)}$ is the Green's function corresponding to outgoing spherical wave boundary conditions (see \Cref{CCCder}).
	See \Cref{CCCT} for the momentum space form of \eqref{lippT} for the $T$-matrix elements within CCC.
	\enlargethispage{0.5cm}
	Iterating the application of \eqref{lippT} to itself generates the infinite series
	\begin{alignat}{3}
		\label{lippTseries}
		T=V+VG^{(+)}V + \left(VG^{(+)}\right)^2V + \left(VG^{(+)}\right)^3V+\ldots
	\end{alignat}
	known as the Born series.
	Truncating \eqref{lippTseries} to include $k$ terms is known as the $k$-th Born approximation which provides an increasingly accurate description of the coupling between reaction channels.
	Do note however, that the series expression \eqref{lippTseries} is not necessarily convergent.
	Most commonly encountered is the first Born approximation, valid when the interaction is weak, and is simply taking only the first term in \eqref{lippTseries} such that
	\begin{alignat}{3}
		T_{fi}=V_{fi}.
	\end{alignat}
	In the case of CCC method, \eqref{lippT} is solved directly for $T$ (see \Cref{Convergent Close Coupling}) and hence no such approximation is involved.
	\subsubsection{Optical Theorem}
	The optical theorem, first derived by \citet{sellmeier1871erklarung} in the context of refraction, is a consequence of either the conservation of energy or, in the context of quantum mechanics,
	probability.
	It states that the total cross section ($\sigma^{\mathrm{tot}}_i$) is related to the imaginary component of the forward elastic scattering amplitude ($f_{ii}$) such that
	\begin{alignat}{3}
	\label{Optical Theorem}
		\sigma^{\mathrm{tot}}_i&=-\frac{\mathrm{Im}\left(f_{ii}\right)}{\pi k_i}.
	\end{alignat}
	Do note that the proportionality constant varies with choice of normalisation and ours is designed to be consistent with the derivation presented in \citet{IgorLecture}.
	The consequence of this theorem is that as the total cross section is constrained by the elastic scattering cross section, all 
	cross sectional quantities that contribute to the total are interlinked.
	As such, convergence of any of these cross sectional quantities (such as excitation or ionisation) is indicative of the same occurring for the others as their sum is constrained.
	\subsection{Electron Impact Ionisation Geometry and \mbox{Terminology}}
	\label{e2e Geometry and Terminology}
\begin{figure}[H]
	\tdplotsetmaincoords{60}{110}
	\pgfmathsetmacro{\rvec}{.8}
	\pgfmathsetmacro{\thetavec}{30}
	\pgfmathsetmacro{\phivec}{60}
	\centering
		\begin{tikzpicture}[scale=5,tdplot_main_coords]

			\tdplotsetrotatedcoords{0}{270}{0}
			\draw[thick,->] (0,0,0) -- (1,0,0) node[anchor=north east]{$x$};
			\draw[thick,->] (0,0,0) -- (0,1,0) node[anchor=north west]{$y$};
			\draw[thick,->] (0,0,0) -- (0,0,0.8) node[anchor=south]{$z$};

			\draw[thick,color=red] (0,-1,0) node[anchor=south]{$k_i$} -- (0,0,0) ;
			\draw[thick,->,color=red] (0,-1,0) -- (0,-0.5,0) ;


			\tdplotsetrotatedcoords{90}{0}{0}
			\tdplotdrawarc[tdplot_rotated_coords,color=red]{(0,0,0)}{0.45}{0}{-25}{}{}

			\draw[color=red] (0.12,0.5,0) node{$\phi_a$};

			\tdplotsetrotatedcoords{90+45}{0}{0}
			\draw[thick,->,tdplot_rotated_coords,color=blue] (0,0,0) -- (\rvec,0,0) node[anchor=west]{$k_b$};
			\tdplotdrawarc[tdplot_rotated_coords,color=blue]{(0,0,0)}{0.5}{0}{-45}{anchor=west}{$\phi_b$}
			\tdplotsetrotatedcoords{90+75}{0}{0}
			\draw[thick,->,tdplot_rotated_coords,color=purple] (0,0,0) -- (0.4,0,0) node[anchor=south]{$q$};
			\tdplotsetrotatedcoords{90+25}{0}{0}
			\draw[thick,->,tdplot_rotated_coords,color=orange] (0,0,0) -- (-0.4,0,0) node[anchor=east]{$k_R$};

			\tdplotsetrotatedcoords{90-25}{0}{0}
			\draw[thick,->,tdplot_rotated_coords,color=red] (0,0,0) -- (1,0,0) node[anchor=west]{$k_a$};

			\draw[dashed] (\rvec,0,0) arc (0:360:\rvec);
		\end{tikzpicture}
		\caption[Standard (e,2e) process in an asymmetric coplanar geometry.]{Standard (e,2e) process in an asymmetric coplanar geometry. 
		The projectile initial ($k_i$) and final ($k_a$) momentum are given in {\textcolor{red}{red}} with scattering angle $\phi_a$, 
		the transferred momentum ($q=|\bm{k}_i-\bm{k}_a|$) in {\textcolor{purple}{purple}}, the ejected electron momentum ($k_b$) in {\textcolor{blue}{blue}} at an angle $\phi_b$, and the 
		momentum of the
		recoiling ion ($k_R$) in {\textcolor{orange}{orange}} such that $\bm{q}=\bm{k}_i-\bm{k}_a=\bm{k}_b+\bm{k}_R$. For the purposes of this work we will consider angles in the direction as denoted
		by $\phi_a$ to be positive and conversely those in the direction of $\phi_b$ to be negative, such that $\phi\in(-180^\circ,180^\circ]$.}
		\label{e2e fig}
	\end{figure}
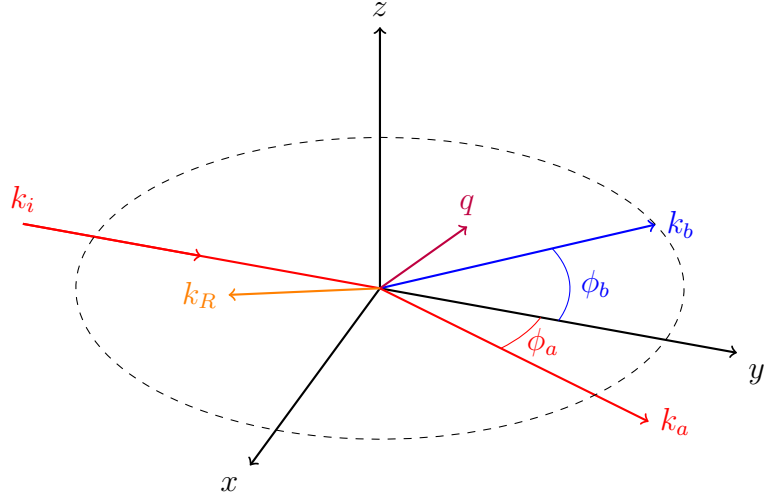

	Electron impact ionisation, dubbed the (e,2e) reaction, is an atomic scattering process in which an incoming electron ionises a target atom, resulting in two electrons leaving the system.
	Typically considered is a coplanar asymmetric geometry \cite{friedrich2012theoretical} where the projectile is scattered into the same plane as the ejected electron (see \Cref{e2e fig}) and the
	ejected electron is free to scatter at any angle within this plane.
	The projectile approaches with momentum $\bm{k}_i$, transfers momentum $\bm{q}$ to the target, and is scattered with a momentum of $\bm{k}_a$ such that $\bm{q}=\bm{k}_i-\bm{k}_a$.
	In terms of energetics we have for the final energy of the projectile $E(k_a)=k_a^2/2=(k_i^2-q^2)/2-\epsilon_i$, where $\epsilon_i$ is the ionisation energy of the target atom in its initial state.
	The transferred momentum $\bm{q}$ is distributed between the constituents of the target such that $\bm{q}=\bm{k}_R+\bm{k}_b$ where $\bm{k}_R$ is the recoil momentum of the resulting 
	ion and $\bm{k}_b$ is the momentum of the ejected electron.
	The azimuthal angle of scattering is denoted $\phi_a$ and $\phi_b$ for the projectile and ejected electron respectively.

	The relevant cross sectional quantity for such a process is known as the triply differential cross section (TDCS) \cite{TweedHe,BeatyHe}
	\begin{alignat}{3}
		\label{TDCS}
		\frac{\mathrm{d}^3\sigma}{\mathrm{d}\Omega_a\mathrm{d}\Omega_b\mathrm{d}E_b}&\propto\frac{k_a k_b}{k_i}\sum_{S}\frac{(2S+1)}{4}|f_{fi}^S(\bm{k}_a,\bm{k}_b,\bm{k}_i)|^2,
	\end{alignat}
	where $\mathrm{d}\Omega_a$ and $\mathrm{d}\Omega_b$ are the elements of solid angle in which the projectile and ejected electron are scattered and $f_{fi}^S(\bm{k}_a,\bm{k}_b,\bm{k}_i)$ is the
	scattering amplitude for a transition $i\to f$ such that $\epsilon_f>0$, i.e.\ ionisation.
	
	Note that in the case where $E(k_i)$ is very large the ejected electron is often referred to as the `slow' electron and correspondingly the projectile as the `fast' electron.
	This is due to large exchange of energy to the bound electron being a negligible reaction channel, causing them to be essentially distinguishable. 
	For example, in comparison with \citet{hohr2007laser} we consider $E(k_i)=1000$ eV with typical values of $E(k_b)$ in the order of $10$ eV.
	\subsection{Experimental Comparison Considerations}
	\label{hohrcomp}
\begin{figure}[H]
	\tdplotsetmaincoords{60}{110}
	\pgfmathsetmacro{\rvec}{.8}
	\pgfmathsetmacro{\thetavec}{30}
	\pgfmathsetmacro{\phivec}{60}
		\centering
		\begin{tikzpicture}[scale=5,tdplot_main_coords]

			\tdplotsetrotatedcoords{0}{270}{0}
			\draw[thick,->] (0,0,0) -- (1,0,0) node[anchor=north east]{$x$};
			\draw[thick,->] (0,0,0) -- (0,1,0) node[anchor=north west]{$y$};
			\draw[thick,->] (0,0,0) -- (0,0,0.8) node[anchor=south]{$z$};

			\draw[thick,<->,color=magenta] (0.4,-0.9,0) -- (-0.4,-0.9,0) node[anchor=south]{$F$};

			\draw[thick,color=red] (0,-1,0) node[anchor=south]{$k_i$} -- (0,0,0) ;
			\draw[thick,->,color=red] (0,-1,0) -- (0,-0.5,0) ;


			\tdplotsetrotatedcoords{90}{0}{0}
			\tdplotdrawarc[tdplot_rotated_coords,color=red]{(0,0,0)}{0.45}{0}{-25}{}{}

			\draw[color=red] (0.12,0.5,0) node{$\phi_a$};

			\tdplotsetrotatedcoords{90+45}{0}{0}
			\draw[thick,->,tdplot_rotated_coords,color=blue] (0,0,0) -- (\rvec,0,0) node[anchor=west]{$k_b$};
			\tdplotdrawarc[tdplot_rotated_coords,color=blue]{(0,0,0)}{0.5}{0}{-45}{anchor=west}{$\phi_b$}
			\tdplotsetrotatedcoords{90+75}{0}{0}
			\draw[thick,->,tdplot_rotated_coords,color=purple] (0,0,0) -- (0.4,0,0) node[anchor=south]{$q$};
			\tdplotsetrotatedcoords{90+25}{0}{0}
			\draw[thick,->,tdplot_rotated_coords,color=orange] (0,0,0) -- (-0.4,0,0) node[anchor=east]{$k_R$};

			\tdplotsetrotatedcoords{90-25}{0}{0}
			\draw[thick,->,tdplot_rotated_coords,color=red] (0,0,0) -- (1,0,0) node[anchor=west]{$k_a$};

			\draw[dashed] (\rvec,0,0) arc (0:360:\rvec);
		\end{tikzpicture}
		\caption[(e,2e) process in an asymmetric coplanar geometry with addition of the laser as in \citet{hohr2007laser}.] 
		{(e,2e) process in an asymmetric coplanar geometry with addition of the laser as in \citet{hohr2007laser}. 
		Symbols are denoted as in \Cref{e2e fig} with the addition of a linearly polarised electric field ($F$) given in \textcolor{magenta}{magenta}.}
		\label{e2e fig2}
	\end{figure}
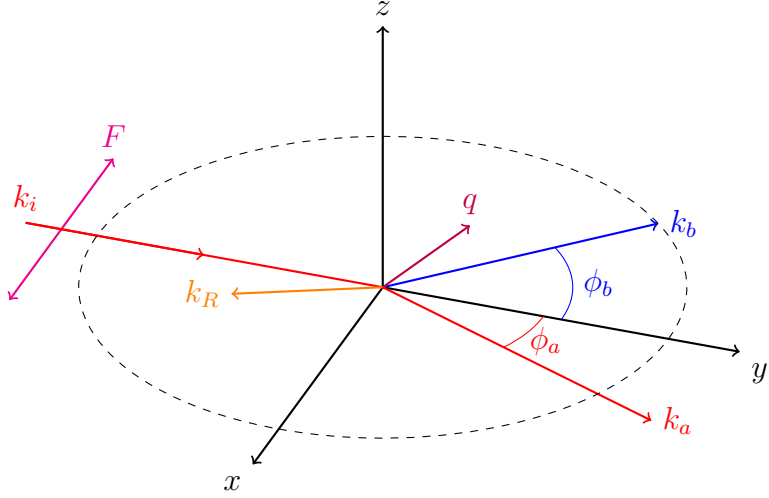

	\enlargethispage{0.5cm}
	The experimental data \cite{hohr2007laser} for the laser assisted electron impact ionisation of helium that
	we look to compare with are given for a projectile electron of energy $E(k_i)=1000$ eV in the presence of a laser field of photon energy
	$E_{\gamma}=1.15$ eV and intensity $I=4\times10^{14}$ W/cm$^2$ (see \Cref{exptparamtable}).
	This laser is oriented such that it produces a linearly polarised electric field $F$ parallel to the $x$-axis (see \Cref{e2e fig2}).

	The relation between the momentum of the electrons and their energy is given by the classical expression
	\vspace{-0.2cm}
	\begin{alignat}{3}
		\label{classicalk}
		k=\sqrt{2E}
	\end{alignat}
	\phantom{hi}\\[-1.0cm]
	and the relativistic expression
	\vspace{-0.2cm}
	\begin{alignat}{3}
		\label{relk}
		k=\frac{\sqrt{E(E+2c^2)}}{c}.
	\end{alignat}
	\phantom{hi}\\[-0.6cm]
	From \Cref{momentum fig} we can see the validity of the classical expression 
	still holds at energies around $1000$ eV and hence we are yet to need to include any relativistic considerations at these energies.
	It is unusual for the energy of the projectile to be sufficiently high to merit a relativistic treatment in atomic scattering, however it is often required when dealing with
	highly charged targets \cite{fursaRel}.

	The centre of mass velocity is given by
	\begin{alignat}{3}
		\bm{v}_{\mathrm{CM}}&=\frac{\bm{k}_a}{m_e+m_{\mathrm{He}}}\\
		&\approx\frac{\sqrt{2\times1000/27.21\; \mathrm{Hartree}}}{7.3\times 10^{3}\;m_e}\hat{\bm{y}}\\
		&\approx \bm{0}\;\mathrm{a.u.}\;.
	\end{alignat}
	Hence, we may consider there to be no difference between values calculated in laboratory or centre of mass frames.
	\noindent This is most often the case for atomic scattering problems.

	The cross sectional data is given in terms of momentum transfer, so we need to be able to freely convert between this and the scattering angle of the projectile.
	For a projectile with momentum $\bm{k}_i$ which is scattered by an angle $\phi_a$ and leaving with momentum $\bm{k}_a$ we have from the definition of $\bm{q}$ 
	that
	\begin{alignat}{3}
		&&\bm{q}&=\bm{k}_i-\bm{k}_a\\
		\implies && q&=|\bm{k}_i-\bm{k_a}|\\
		\label{qvsphia}
		&& &=\sqrt{k_i^2+k_a^2-2k_ik_a\cos\phi_a}.
	\end{alignat}
	\enlargethispage{4cm}
	\hspace{-0.25cm}
	The nature of this expression is demonstrated in \Cref{momtrans}.
	\vspace{-0.4cm}
	\begin{samepage}
	\begin{figure}[H]
		\centering
		\hspace{-0.6cm}
		\includegraphics[scale=0.86]{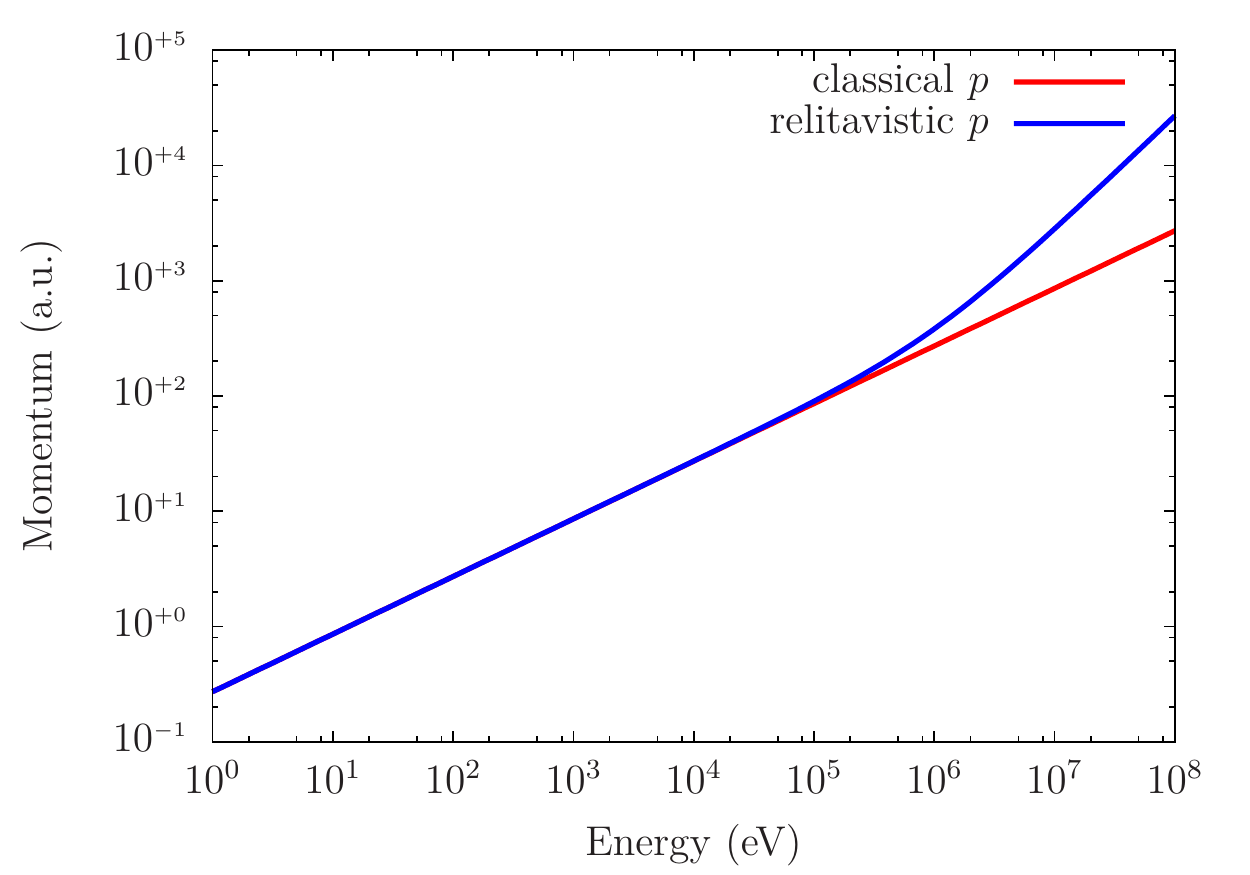}
		\caption[Comparison between classical \eqref{classicalk} and relativistic \eqref{relk} expressions for momentum as a function of energy.]
		{Comparison between classical \eqref{classicalk} and relativistic \eqref{relk} expressions for momentum as a function of energy.}
		\label{momentum fig}
	\end{figure}
	\vspace{-0.8cm}
	\begin{figure}[H]
		\centering
		\includegraphics[scale=0.82]{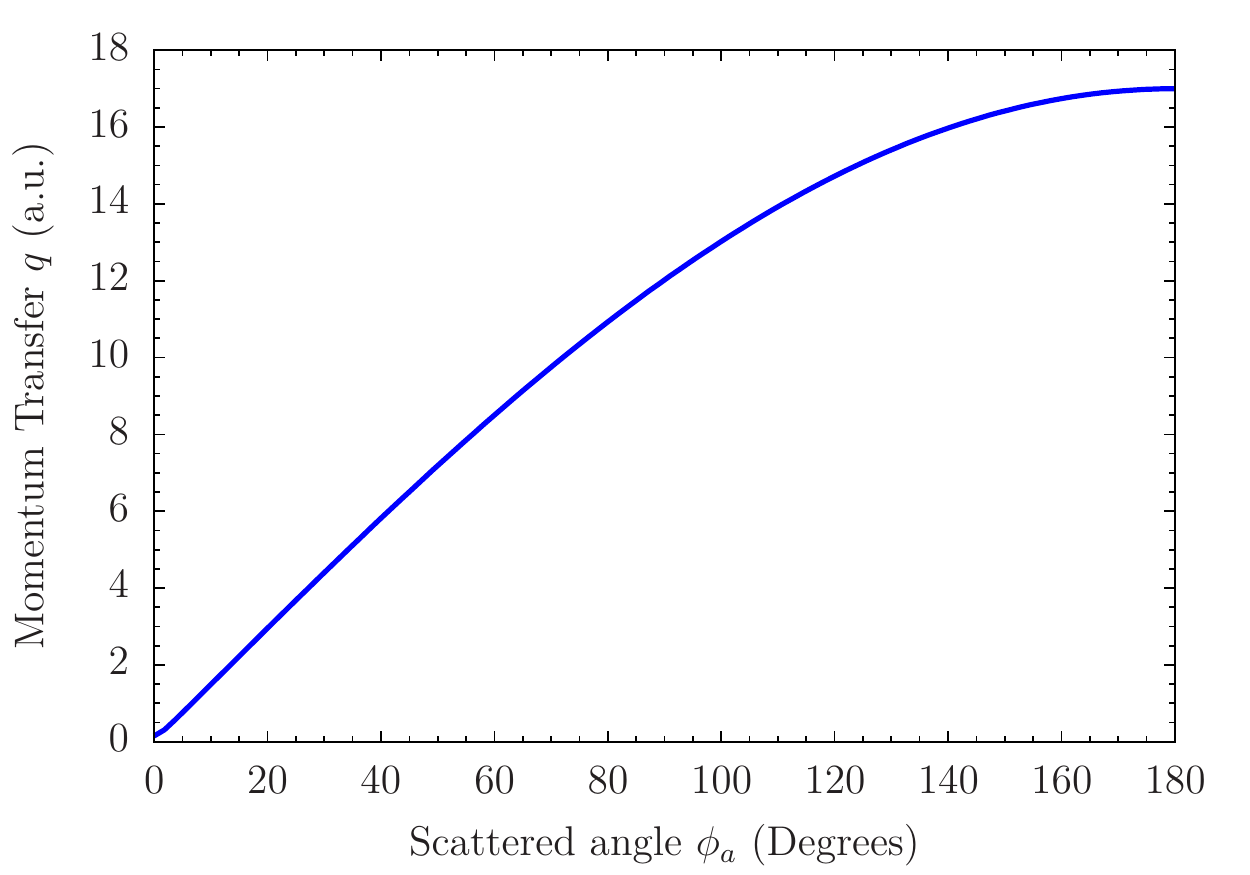}
		\caption[Magnitude of momentum transfer $q$ \eqref{qvsphia} as a function of scattering angle $\phi_a$.]
		{Magnitude of momentum transfer $q$ \eqref{qvsphia} as a function of scattering angle $\phi_a$.
		Here we have taken $E(k_i)=1000$ eV and $E(k_a)=10$ eV.}
		\label{momtrans}
	\end{figure}
	\end{samepage}
	\clearpage
	\section{Convergent Close-Coupling}
	\label{Convergent Close Coupling}
	\enlargethispage{2cm}
	The convergent close-coupling (CCC) is a general method of solving for scattering amplitudes and cross sections in atomic and molecular scattering.
	Initially developed for electron scattering on hydrogen \cite{BrayH} it has now been extended to the scattering of various projectiles on:
	hydrogen-like targets \cite{BrayHlike}, helium \cite{fursaHe} and helium-like targets \cite{fursaHelike}, positronium \cite{rawlinsPos}, simple molecules (H$_2$, H$_2^+$) \cite{zammitMol}, 
	noble gases \cite{fursaPosNob}, and a neon-like treatment of water (H$_2$O) \cite{abdurakhmanovH20}. 
	It is currently implemented for projectiles such as electrons, positrons \cite{kadyrovPos}, and heavy projectiles such as protons \cite{abdurakhmanovPro}, anti-protons
	\cite{abdurakhmanovHeAntiPro}, 
	and bare nuclei \cite{abdurakhmanovC6}.
	In the case of positive projectiles multi-centre calculations are available \cite{utamuratovMultiHe} due to the possibility of electron capture by the projectile, 
	and for highly charged targets a fully relativistic formulation is additionally implemented \cite{fursaRel}.
	Furthermore, CCC has been used to treat single \cite{brayelecandphoto} and double \cite{PhysRevA.65.012710,0953-4075-33-2-311,PhysRevLett.102.073006,kheifetsPhoto} 
	ionisation by photons (photoionisation).
	We present a short introduction of this application of CCC to photoionisation in \Cref{Photoemission}.
	When treating targets with low charge a $L-S$ coupling scheme is generally found to be more accurate \cite{fursaHeLS}, whereas for high charge targets $J-J$ coupling is utilised \cite{bostokRel}.

\begin{figure}[htbp]
		\centering
		\includegraphics[scale=0.9]{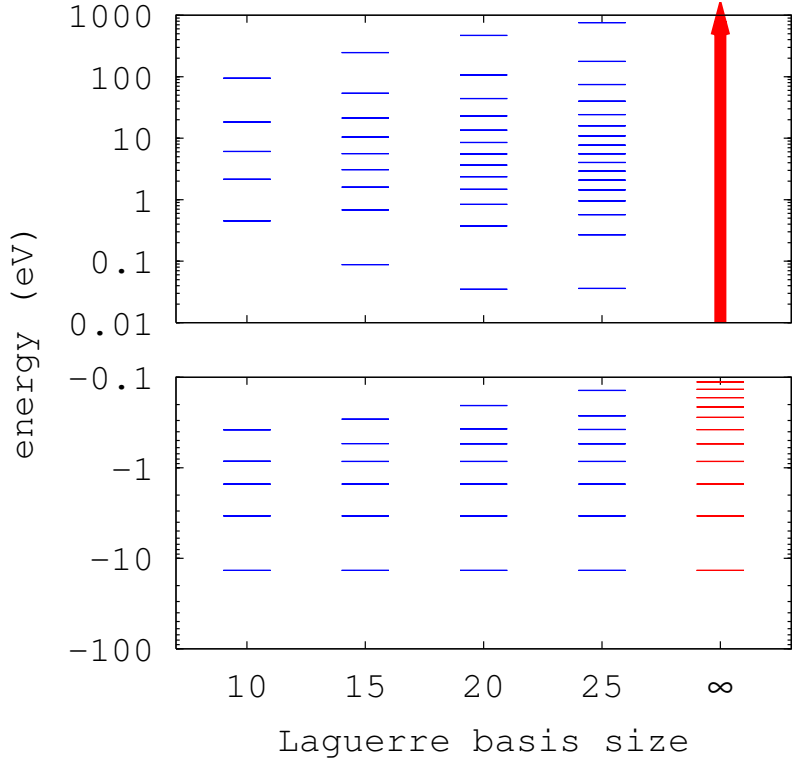}
		\caption[Energy eigenstates generated for increasing basis size $(N_0)$ demonstrating convergence to the true eigenstates of a $l=0$ hydrogen atom.]
		{Energy eigenstates generated for increasing basis size $(N_0)$ demonstrating convergence to the true eigenstates of a $l=0$ hydrogen atom.
		This figure is courtesy of \citet{IgorLecture}.}
		\label{Laguerre}
	\end{figure}

	One of the key mathematical complications in the formulation of atomic scattering is accounting for the true eigenstates of the target.
	This is a non-trivial task as there exists a countably infinite number of bound states (negative energy) and an uncountably infinite number of free states (positive energy).
	The defining characteristic of the CCC method is in this treatment, in which the states of the target are expanded through the use of a complete Laguerre basis
	\begin{alignat}{3}
		\label{laguerre}
		\xi_{jl}^{(\lambda_l)}(r)&=\left(\frac{\lambda_l(j-1)!}{(2l+1+j)!}\right)(\lambda_l r)^{l+1}\exp{(-\lambda_l r/2)}L_{j-1}^{2l+1}(\lambda_l r),
	\end{alignat}
	where $l$ is the angular momentum of the target state (s, p, d, $\ldots$), $\lambda_l$ is a corresponding free parameter that is chosen as best to fit the true target states,
	$N_l$ is the basis size for a particular value of $l$, $1\leq j\leq N_l$, and $L_{j-1}^{2l+1}(\lambda_l r)$ are the associated Laguerre polynomials
	\begin{alignat}{3}
		L_{j-1}^{2l+1}(\lambda_l r)&=\sum_{m=0}^{j-1}\frac{(-1)^m(j+2l)!(\lambda_l r)^m}{(j-1-m)!(2l+1+m)!m!}.
	\end{alignat}
	As for a complete orthonormal basis, the $|\xi_{jl}^{(\lambda)}\rangle$ satisfy
	\begin{alignat}{3}
		\big\langle\xi_{jl}^{(\lambda)}|\xi_{j'l}^{(\lambda)}\big\rangle&=\int_0^\infty \mathrm{d}r\;\xi_{jl}^{(\lambda)}(r)\xi_{j'l}^{(\lambda)}(r)\\
		&=\delta_{jj'}.
	\end{alignat}
	Utilizing this, we are able to express the identity operator, originally expressed in terms 
	of the true eigenstates ($|\phi_{nl}\rangle$), in terms of our Laguerre based states ($|\phi_{nl}^{({N_l})}\rangle$)
	by diagonalising the target Hamiltonian ($H_T$) such that
	\begin{alignat}{3}
		\label{phicreation}
		\phi_{nl}^{(N_l)}&=\sum_{j=1}^{N_l}|\xi_{jl}^{(\lambda_l)}\rangle\langle\xi_{jl}^{(\lambda_l)}|C_n\rangle
	\end{alignat}
	where the $|C_n\rangle$ are the eigenvectors of the matrix $A$ defined as
	\begin{alignat}{3}
		A_{jj'}=\langle\xi_{jl}^{(\lambda_l)}|H_T|\xi_{j'l}^{(\lambda_l)}\rangle.
	\end{alignat}
	Equivalently,
	\begin{alignat}{3}
		\sum_{j'=1}^{N_l}\langle\xi_{jl}^{(\lambda_l)}|H_T|\xi_{j'l}^{(\lambda_l)}\rangle\langle\xi_{j'l}^{(\lambda_l)}|C_n\rangle&=\epsilon_{nl}^{N_l}\langle\xi_{jl}^{(\lambda_l)}|C_n\rangle,
	\end{alignat}
	where the $\epsilon_{nl}^{({N_l})}$ are the energy eigenvalues generated by the diagonalisation. 
	This definition produces a set of orthonormal states $|\phi_{nl}^{(N_l)}\rangle$ which have the following property
	\begin{alignat}{3}
		\langle \phi_{fl}^{({N_l})}|H_T|\phi_{il}^{({N_l})}\rangle&=\epsilon_{fl}^{({N_l})}\delta_{fi}.
	\end{alignat}
	These states of corresponding energy $\epsilon_{fl}^{(N_l)}$ are those which the CCC calculation solves for in
	approximation to the true physical system.
	\Cref{Laguerre} demonstrates the energy levels generated in the diagonalisation of the hydrogen Hamiltonian with $l=0$, $\lambda_0=1$, of various basis sizes $N_0$ as compared to the true
	eigenstates of the target.
	Observe that for increasing basis size, the negative energy eigenstates better approximate the true eigenstates to higher $n$ and the positive energy eigenstates become increasingly dense.
	It addition, both the positive and negative states defined by this expansion are square integrable (their square, integrated across all space is finite).
	In fact we have for the identity operator given in terms of the true eigenstates that
	\begin{alignat}{3}
		I&=\SumInt_{n} |\phi_{n}\rangle\langle \phi_{n}|\\
		&=\lim_{{N}\to\infty}\sum_{n=1}^{N}|\phi_{n}^{({N})}\rangle\langle\phi_{n}^{({N})}|.
	\end{alignat}
	Here we have $N$ such that $N=\sum_{l}N_l$ and the subscript $n$ represents the full set of quantum numbers required to describe the state.
	This now allows us to expand the total wavefunction $|\Psi_i^{S(+)}\rangle$ in terms of the newly defined states via the following,
	\begin{alignat}{3}
		0&=(E-H)|\Psi_i^{S(+)}\rangle\\
		&\approx(E-H)\sum_{n=1}^{N} |\phi_{n}^{({N})}\rangle\langle\phi_{n}^{({N})}|\Psi_i^{S(+)}\rangle\\
		\label{CCCshrod}
		&\equiv(E-H)|\Psi_i^{S({N}+)}\rangle.
	\end{alignat}
	With this expression we can now formulate a set of coupled Lippmann-Schwinger equations for the transition amplitude (see \Cref{CCCder} for derivation) such that
	\begin{alignat}{3}
		\label{CCCTorig}
		\big\langle\bm{k}_f\phi_{n_f}^{({N})}\big|T^S\big|\phi_{n_i}^{({N})}\bm{k}_i\big\rangle&=\big\langle\bm{k}_f\phi_{n_f}^{({N})}\big|V^S\big|\phi_{n_i}^{({N})}\bm{k}_i\big\rangle\nonumber\\
		&+\sum_{n=1}^{N}\int\mathrm{d}^3k\;\frac{\big\langle\bm{k}_f\phi_{n_f}^{({N})}\big|V^S\big|\phi_{n}^{({N})}\bm{k}\big\rangle\big\langle\bm{k}\phi_{n}^{({N})}\big|T^S\big|\phi_{n_i}^{({N})}\bm{k}_i\big\rangle}{E+i0-\epsilon_{n}^{({N})}-k^2/2},
	\end{alignat}
	where $i0$ denotes an imaginary component added to the integral due to the singularity occurring when $k^2/2=E-\epsilon_{nl}^{({N_l})}$ (see \Cref{Contourder}),
	and the $V^S$ operator contains all the interaction potentials as well as the symmetrisation requirements of the wavefunction. 
	In the case of electron-hydrogen scattering, a two electron system, we have that
	\begin{alignat}{3}
		\label{VS}
		V^S&= V-(-1)^S(E-H)P_r,
	\end{alignat}
	where $P_r$ is the space exchange operator ($P_r|r_1r_2\rangle=|r_2r_1\rangle$), and $V$ contains the interaction potentials. 
	See \eqref{HeVs} for the equivalent electron-helium scattering expression.
	It is from \eqref{CCCTorig} that the convergent close-coupling approach gets its name, as it involves the solution of this set of coupled Lippmann-Schwinger equations.
	The original close coupling formalism was introduced by \citet{massey1931collision}, however the use of the Laguerre basis to discretise both the countably infinite bound states of the target 
	in addition to the uncountably infinite continuum of free states is unique to CCC.
	The motivation for the additional `convergence' in the name is the convergence with basis size that results from this discretisation process.

	Employing a partial wave expansion of \eqref{CCCTorig} we may reduce the problem into one dimension such that
	\begin{alignat}{3}
		\label{CCCT}
            \big\langle L_f &k_f l_f\phi_{n_fl_f}^{({N_l})}\big\| T^{SJ} \big\| \phi_{n_il_i}^{({N_l})} l_ik_i L_i\big\rangle = 
		\big\langle L_f k_f l_f\phi_{n_fl_f}^{({N_l})}\big\| V^{SJ} \big\|\phi_{n_il_i}^{({N_l})} l_ik_i L_i \big\rangle \nonumber\\
		&+ \sum_{l,L}\sum_{n=1}^{{N_l}}
            \int_0^\infty \! \mathrm{d} k \; k^2\frac{\big\langle L_f k_f l_f\phi_{n_fl_f}^{({N_l})} \big\| V^{SJ} \big\| \phi_{nl}^{({N_l})}l\;k\; L\big\rangle 
		\big\langle L\; k\; l\;\phi_{nl}^{({N_l})} \big\| T^{SJ} \big\|\phi_{n_il_i}^{({N_l})} l_ik_i L_i\big\rangle}
		{E+i0-\epsilon^{(N_l)}_{nl}-k^2/2} ,
	\end{alignat}
	where now the projectile is represented by its final linear momentum $k_f$ and orbital angular
	momentum $L_f$ relative to the target nucleus, and equivalently for its initial state. 
	Here $J$ is the total orbital angular momentum of the system and the subscripts of the target states $n$ and $l$ now refer to those as generated in \eqref{phicreation}.
	The $l_i$ and $l_f$ are the orbital angular momentum of the corresponding target state.
	Introduction of double bar $(\|)$ bra-ket notation is to distinguish from the angular dependent matrix elements.
	The original $T^{S}$ matrix elements may be restored from those of the partial wave expanded version $T^{SJ}$ via
	\begin{alignat}{3}
		\big\langle\bm{k}_f\phi_{n_f}^{({N})}\big|T^S\big|\phi_{n_i}^{({N})}\bm{k}_i\big\rangle\;
		&&=\sum_{L_f,M_f,J,K,L_i,M_i} &\Big[\big\la \hat{\bm{k}}_f \big| L_f M_f\big\rangle \big\la M_f L_f m_f l_f \big| J K\big\rangle\nonumber\\
		&& \times&\big\langle L_fk_fl_f\phi_{n_fl_f}^{({N})}\big\|T^{SJ}\big\|\phi_{n_il_i}^{({N})}l_ik_iL_i\big\rangle\nonumber \\
		\label{TPWT}
		&& \times &\big\la K J \big| l_i m_i L_i M_i \big\ra\big\la M_i L_i\big|\hat{\bm{k}}_i\big\rangle\Big].
	\end{alignat}
	Here $\la \hat{\bm{k}}_f | L_f M_f\rangle\equiv Y_{LM}(\hat{\bm{k}})$ are spherical harmonics, $C^{JK}_{LMlm}=\la M \;L\; m\; l | J K\rangle$ are Clebsch-Gordan coefficients,
	and $K=M+m$ is the total angular momentum projection in the $z$-quantisation direction.
	In \eqref{CCCT} a reaction channel $n$ is considered `open' (physically accessible) if $E-\epsilon_{nl}^{(N_l)}>0$, 
	i.e.\ the projectile has sufficient energy to leave the target in a state of energy $\epsilon_{nl}^{(N_l)}$
	and itself have momentum
		$k_{nl}=\sqrt{2\big(E-\epsilon^{(N_l)}_{nl}\big)}>0$.
	\enlargethispage{0.9cm}
	In the case of $k_{nl}$ being complex the channel is referred to as `closed' (physically inaccessible).
	For open channels the momentum integration in \eqref{CCCT} encounters a singularity when $k^2/2=E-\epsilon_{nl}^{(N_l)}$.
	With this knowledge we can rewrite \eqref{CCCT} as
	\enlargethispage{0.5cm}
	\begin{alignat}{3}
            \big\langle &L_f k_f l_f\phi_{n_fl_f}^{({N_l})}\big\| T^{SJ} \big\| \phi_{n_il_i}^{({N_l})} l_ik_i L_i\big\rangle = 
		\big\langle L_f k_f  l_f\phi_{n_fl_f}^{({N_l})}\big\| V^{SJ} \big\| \phi_{n_il_i}^{({N_l})} l_ik_i L_i \big\rangle \nonumber\\
		&+ \sum_{l,L}\Bigg[\sum_{n=1}^{{N_l}}
            \mathcal{P}\int_0^\infty \! \mathrm{d} k \; k^2\frac{\big\langle L_f k_f l_f\phi_{n_fl_f}^{({N_l})} \big\| V^{SJ} \big\| \phi_{nl}^{({N_l})}l\;k\; L\big\rangle 
		\big\langle L\; k\; l\;\phi_{nl}^{({N_l})} \big\| T^{SJ} \big\|\phi_{n_il_i}^{({N_l})} l_ik_i L_i\big\rangle}
		{E+i0-\epsilon^{(N_l)}_{nl}-k^2/2} \nonumber\\
		\label{CCCTSingular}
            &\qquad\quad-i\pi\sum_{n=1}^{{N_l^o}}k_{nl}\big\langle L_f k_f l_f\phi_{n_fl_f}^{({N_l})} \big\| V^{SJ} \big\| \phi_{nl}^{({N_l})}l\;k_{nl} L\big\rangle 
		\big\langle L\; k_{nl} l\;\phi_{nl}^{({N_l})} \big\| T^{SJ} \big\|\phi_{n_il_i}^{({N_l})} l_ik_i L_i\big\rangle\Bigg],
	\end{alignat}
	where $N_l^o$ is the number of open channels for a particular $l$, and for these open channels we have split the integration into a principle value component ($\mathcal{P}$) and residual
	contribution from the point of singularity.
	For closed channels the principle value term is identically equal to the original integral expression.
	We now define the $K$-matrix (unrelated to the $K$ in \eqref{TPWT}) as
	\begin{alignat}{3}
		\label{Kdef}
		\big\langle L_f k_fl_f\phi_{n_fl_f}^{({N_l})} \big\| K^{SJ} &\big\|
		\phi_{n_il_i}^{({N_l})}l_ik_iL_i\big\rangle=\sum_{l,L}\sum_{n=1}^{N_l^o}\Bigg[\big\langle L_f k_f l_f
		\phi_{n_fl_f}^{({N_l})} \big\| T^{SJ} \big\| \phi_{nl}^{({N_l})}l\;k_{nl}
		L\big\rangle\nonumber\\
		&\times\left(\delta_{l_il}\delta_{L_iL}\delta_{n_i n}+i\pi k_{nl}\big\langle L\;
		k_{nl}l\;\phi_{nl}^{({N_l})} \big\| K^{SJ} \big\|
		\phi_{n_il_i}^{({N_l})}l_ik_iL_i\big\rangle\right)\Bigg].
	\end{alignat}
	Using this definition we can express \eqref{CCCTSingular} as
	\begin{alignat}{3}
		\label{CCCK}
            \big\langle &L_f k_f l_f \phi_{n_fl_f}^{({N_l})}\big\| K^{SJ} \big\| \phi_{n_il_i}^{({N_l})} l_i k_i L_i\big\rangle = 
		\big\langle L_f k_f  l_f \phi_{n_fl_f}^{({N_l})}\big\| V^{SJ} \big\| \phi_{n_il_i}^{({N_l})} l_i k_i L_i \big\rangle \nonumber\\
		&+ \sum_{l,L}\sum_{n=1}^{{N_l}}
            \mathcal{P}\;\int_0^\infty \! \mathrm{d} k \; k^2\frac{\big\langle L_f k_f l_f \phi_{n_fl_f}^{({N_l})} \big\| V^{SJ} \big\| \phi_{nl}^{({N_l})} l \;k\; L\big\rangle 
		\big\langle L\; k\; l\;\phi_{nl}^{({N_l})} \big\| K^{SJ} \big\|\phi_{n_il_i}^{({N_l})} l_i k_i L_i\big\rangle}
		{E-\epsilon^{(N_l)}_{nl}-k^2/2},
	\end{alignat}
	where $\mathcal{P}$ again denotes the principle value of the integral.
	This expression for the $K$-matrix contains entirely real values and hence may be solved using purely real arithmetic.
	The $T$-matrix is then reconstructed by solving the considerably smaller set of equations \eqref{Kdef}.
	Interestingly the $T$-matrix is symmetric though not itself unitary. 
	However, it is directly related to the $S$-matrix via
	\begin{alignat}{3}
		\big\langle L_f k_f l_f \phi_{n_fl_f}^{({N_l})}\big\| S^{SJ} \big\| \phi_{n_il_i}^{({N_l})} l_i k_i
		L_i\big\rangle&=\delta_{fi}-\frac{2\pi i}{\sqrt{k_fk_i}}\big\langle L_f k_f l_f \phi_{n_fl_f}^{({N_l})}\big\| T^{SJ} \big\| \phi_{n_il_i}^{({N_l})} l_ik_iL_i\big\rangle
	\end{alignat}
	which is both symmetric and unitary. Finally, we have that the relationship between the $T$-matrix and the scattering amplitude is given by \cite{BrayRep}
	\begin{alignat}{3}
		\big\langle L_fk_fl_f\phi_{n_fl_f}^{({N_l})}\big\|f^{SJ}\big\|\phi_{n_il_i}^{({N_l})}l_ik_i L_i\big\rangle&=
		\big\langle L_f k_f l_f\phi_{n_fl_f}^{({N_l})}\big\| T^{SJ} \big\| \phi_{n_il_i}^{({N_l})} l_ik_i
		L_i\big\rangle,
	\end{alignat}
	in the case of elastic scattering or excitation. In the case of ionisation we have
	\begin{alignat}{3}
		\label{fionisation}
		\big\langle L_fk_fl_f\phi_{n_fl_f}^{({N_l})}\big\|f^{SJ}\big\|\phi_{n_il_i}^{({N_l})}l_ik_i L_i\big\rangle&=\big\langle q_f\big|\phi_{n_fl_f}^{({N_l})}\big\rangle
		\big\langle L_f k_f l_f\phi_{n_fl_f}^{({N_l})}\big\| T^{SJ} \big\| \phi_{n_il_i}^{({N_l})} l_i k_i
		L_i\big\rangle,
	\end{alignat}
	where $|q_f\rangle$ is the continuum eigenfunction of the target Hamiltonian with energy $\epsilon_f$.
	In the case of a hydrogen target the $|q_f\rangle$ are pure Coulomb waves.
	Do note that with respect to the notation used for (e,2e) (as in \Cref{e2e fig}) we have $\bm{k}_f=\bm{k}_a$ and $\bm{q}_f=\bm{k}_b$.
	\subsection{Convergence Considerations}
	\label{Convergence Considerations}
	Obtaining convergent results for $T^{SJ}$, and consequently the calculated cross sections,
	is achieved through including increasingly large angular momenta of target states $l$, number of partial waves $L$, and of the Laguerre basis size $N_l$ for each $l$.
	In the CCC method all states included in the calculation are coupled to one another (see \Cref{CCCTorig}), and as such each reaction channel is allocated `flux' in a manner that is
	affected by other channels.
	Convergence is achieved when adding further states to the basis set (either by allowing greater $l$ or increasing $N_l$) does not cause any significant redistribution of this flux.
	Convergence with partial waves is particularly straightforward as 
	in such a formulation each partial wave is essentially independent of one another and their combination involves a simple summation.
	As such, if the cross section corresponding to the final partial waves included in the calculation is negligible, then convergence with respect to this aspect has been achieved.
	In practice the tests of convergence are generally conducted by visual comparison of cross sections generated by various calculations involving different discretisations or number of partial waves.
	This form of convergence analysis for the electron impact ionisation calculations conducted as part of this work are presented in \Cref{Field Free}.

	Additionally, there are internal parameters that affect the numerics of the CCC calculation.
	Such parameters include the maximum radial coordinate in the system centred on the target nucleus, the various exponential fall off factors $\lambda_l$ (see \Cref{laguerre}), and those which
	define $k$-grid integration points.
	As these parameters have no physical significance and are purely features of the CCC numeric implementation results examining their effect will not be presented.
	However, it is interesting to note that modification of these internal numeric features has led to a recently developed alternative formulation \cite{bray2015solving,bray2016solving} 
	that has proven beneficial when performing calculations at an energy close to the threshold opening of reaction channels \cite{Fabrikant2016}.
	This formulation has been utilised for the near threshold calculations required in our photoemission study.
	\enlargethispage{2cm}
	\subsubsection{Born Subtraction}
	\label{Born Subtraction}
	The technique of Born subtraction, first proposed by \citet{McCarthyStelbovics}, allows for a numerically efficient method of extrapolation to high partial waves for which the first Born
	approximation becomes increasingly accurate.
	It involves writing the $T$ operator as
	\vspace{-0.3cm}
	\begin{alignat}{3}
		\label{BornSub}
		T=V+\sum_{J=0}^{J_{\mathrm{max}}} T_J-V_J
	\end{alignat}
	\vspace{-0.3cm}
	where
	\begin{alignat}{3}
		V=\sum_{J=0}^\infty V_J
	\end{alignat}
	and $J_{\mathrm{max}}$ is a freely chosen parameter.
	Observe that this is essentially adding and subtracting $V$ from the usual definition for $T$ where
	\begin{alignat}{3}
		T=\sum_{J=0}^\infty T_J.
	\end{alignat}
	However, in the implementation of \eqref{BornSub} $J_{\mathrm{max}}$ is chosen such that beyond this point the first Born approximation is sufficiently accurate so that 
	\begin{alignat}{3}
		\sum_{J>J_{\mathrm{max}}}T_J-V_J\approx 0
	\end{alignat}
	and hence for $J>J_{\mathrm{max}}$ we have
	\begin{alignat}{3}
		T_J=V_J.
	\end{alignat}
	The utility of this approach relies on the 
	closed form solution for the potential operator $V$ \cite{McCarthyStelbovics}, and this is used as the first term in \eqref{BornSub}.
	In doing this we can choose the number of partial waves (effectively choosing $J_{\mathrm{max}}$) for which the full CCC formulation applies
	and from this point onwards include an analytic tail accounting for an arbitrary number of partial waves beyond this point under the first Born approximation.
	This is particularly useful for problems involving a high incident energy where convergence with $L$ is slow.
	As such, we use this approach in our field-free calculations for 1 keV electrons on helium (see \Cref{BornSubtractionComp} and surrounding text).
	\subsection{Treatment of Helium}
	\label{Treatment of Helium}
	The Hamiltonian describing a target helium atom may be expressed as
	\begin{alignat}{3}
		H_{\mathrm{T}}&=H_1+H_2+V_{12}
	\end{alignat}
	where the target electrons are denoted as $1$ and $2$, with
	\begin{alignat}{3}
		\label{Hi}
		H_i&=K_i+V_i\\
		&=-\frac{1}{2}\nabla^2_i-\frac{2}{r_i}
	\end{alignat}
	for $i=1,2$ and
	\begin{alignat}{3}
		\label{V12}
		V_{12}&=\frac{1}{|\bm{r_1}-\bm{r_2}|}
	\end{alignat}
	is the electron-electron potential. If we reserve $0$ to refer to the projectile space, we have that the total Hamiltonian is given by
	\begin{alignat}{3}
		H=H_{\mathrm{T}}+H_0+V_{01}+V_{02},
	\end{alignat}
	where $H_0$, $V_{01}$, and $V_{02}$ are defined as in \eqref{Hi} and \eqref{V12} correspondingly.
	If we separate this Hamiltonian into asymptotic ($H_a$) and short ranged terms ($V$) we may express it as the their sum
	\vspace{-0.2cm}
	\begin{alignat}{3}
		H=H_a+V
	\end{alignat}
	\phantom{hi}\\[-1.0cm]
	where $H_a$ is given by
	\vspace{-0.2cm}
	\begin{alignat}{3}
		H_a=K_0+H_{\mathrm{T}}
	\end{alignat}
	\phantom{hi}\\[-1.0cm]
	and $V$ by
	\vspace{-0.3cm}
	\begin{alignat}{3}
		V=V_0+V_{01}+V_{02}.
	\end{alignat}
	The definition of $V^S$ containing the appropriate symmetrisation for the two electrons now becomes
	\begin{alignat}{3}
		\label{HeVs}
		V^S&=V-(-1)^S(E-H)(P_{01}+P_{02})
	\end{alignat}
	where $P_{ij}$ is the space exchange operator such that $P_{ij}|r_ir_j\rangle=|r_jr_i\rangle$.
	The target helium states may be expressed as \cite{fursaHeLS}
	\begin{alignat}{3}
		|\Phi_{n}\rangle=\sum_{\alpha,\beta}C_{\alpha\beta}^{(n)}|\varphi_\alpha\varphi_\beta\;:\;\pi_n l_ns_n\rangle
	\end{alignat}
	where $C_{\alpha\beta}^{(n)}$ are configuration interaction (CI) coefficients, the $|\varphi\rangle$ are single electron wavefunctions, and $\pi_n$, $l_n$, and $s_n$ are correspondingly 
	the resulting total parity,
	orbital angular momentum, and spin of the state. 
	The CI coefficients satisfy, 
	\begin{alignat}{3}
		C_{\alpha\beta}^{(n)}&=(-1)^{l_\alpha+\l_\beta-l_n-s_n}C_{\beta\alpha}^{(n)}
	\end{alignat}
	which ensures antisymmetry of the target states.
	The single electron wavefunctions are given by 
	\vspace{-0.3cm}
	\begin{alignat}{3}
		\label{singlewave}
		\langle \bm{r} | \varphi_\alpha\rangle&=\frac{1}{r}\phi_{n_\alpha l_\alpha}(r)Y_{l_\alpha m_\alpha}(\hat{\bm{r}})\xi(\sigma)
	\end{alignat}
	where $m_\alpha$ is the $z$-component projection of $l_\alpha$, $Y_{l_\alpha m_\alpha}$ is a spherical harmonic, $\sigma$ is the value of spin ($\pm 1/2$ in this case), and $\xi(\sigma)$ is the
	corresponding spin eigenfunction.
	Calculation of the $V^S$ matrix elements (via a Hartree-Fock approach) is considerably more complicated than for that of hydrogen. A detailed treatment is given in \citet{fursaHeLS}.
	From this point however, the solution is independent of the target, and we numerically solve \eqref{CCCK} 
	for the $T$-matrix elements and calculate the desired ionisation cross sections via \eqref{TDCS}.

	\enlargethispage{1.5cm}
	In cases where single electron processes are dominant, a considerably simpler treatment of the target structure known as the frozen core model has been shown to be sufficient \cite{cohenFrozen}.
	It is called such as one of the target electrons is always described by the He$^+$ 1$s$ orbital.
	For the single ionisation of helium, the problem we consider within this work, such a treatment is adequate, and as such is utilised to minimise computational resources 
	and for greater speed of calculation. 
	\clearpage
	\section{Laser Assisted Collisions}
	\label{Laser Assisted Collisions}
	Up until now we have considered collision processes that comprise of a projectile, a target, and in the case of ionisation, the ejected species.
	In the case of a laser assisted collision we consider an additional component to the system, the photon field.
	This introduces the following interactions to be considered within the treatment of the problem; the projectile-field, the target-field, and if applicable, the ejected-field interactions.
	Any theoretical description must adequately deal with these additional complexities, by either explicitly accounting for them, or working under suitable assumptions that allow their neglect.
	In this section we provide an introduction into each of these interactions and then elaborate on a theoretical treatment known as the soft photon approximation.
	
	The photon field is characterised by the parameters of frequency $\omega$, intensity $I$, and polarisation vector $\epsilon$.
	This polarisation of said field introduces a new physical axis to the system (see \Cref{e2e fig2} for example).
	However, in many circumstances considered, the primary influence of this field is through acting as an energy source (sink) via providing a mechanism of absorbing (emitting) photons 
	through the scattering process.
	Hence, the equivalent laser assisted collision processes are often denoted by the addition of the term $n\gamma$ such as ($n\gamma e,2e$) representing
	\begin{alignat}{3}
		e^-+\mathrm{A}\pm n\gamma\to 2e^-+\mathrm{A}^+,
	\end{alignat}
	where $\mathrm{A}$ represents some arbitrary neutral atom.
	Quantum mechanically, the explanation for the possibility of both absorption and emission through the introduction of the field is due to the Hermitian nature of the Hamiltonian, with the physical
	mechanism for emission being bremsstrahlung radiation.
	The reason for this influence on the energetics being considered the primary influence is that depending on the laser parameters it is often possible to neglect many of the 
	other effects of the field,
	whereas the absorption or emission characteristics are always present and have a considerable impact on the behaviour of the system.

	Let us now consider electron scattering on a helium atom in the presence of a laser field. 
	The total Hamiltonian may be expressed as \cite{laserass}
	\begin{alignat}{3}
		H=H_\mathrm{T}+H_0+H_\mathrm{F}+H_{0-\mathrm{T}}+H_{0-\mathrm{F}}+H_{\mathrm{T-F}},
	\end{alignat}
	where each $H$ is the partial Hamiltonian corresponding to the target (T), projectile (0), field (F), or an interaction involving a combination of these. 
	Firstly, we state that we will work within the Coulomb gauge, which is defined such that the vector potential $\bm{A}$ of the field satisfies $\nabla\cdot \bm{A}=0$.
	Under this condition, and with the additional fact that the field has no associated charge distribution, we have the scalar potential of the field $\varphi=0$.
	Hence, we need not include any additional potential in our Hamiltonian due to the field.
	The energy accrued by a free electron in a linearly polarised electromagnetic field (ponderomotive energy) produces the following Hamiltonian
	\begin{alignat}{3}
		H_{\mathrm{F}}=\frac{F_0^2}{4\omega^2},
	\end{alignat}
	where $F_0$ is the maximum amplitude and $\omega$ is the frequency of the field.
	This energy is typically very small compared to the other energetics involved in laser assisted scattering and is often omitted in the literature. 
	In this work, we will also omit this term from this point onwards.
	The Hamiltonian of the projectile electron in the presence of such an electromagnetic field (zero scalar potential) is given by
	\begin{alignat}{3}
		\label{Hprojlaser}
		H_0+H_{\mathrm{0-F}}&=\frac{1}{2}\left(-i\nabla_0+\frac{1}{c}\bm{A}\right)^2,
	\end{alignat}
	where the kinetic energy operator is ${K}={\bm{p}}-q\bm{A}/c$ \cite{eisberg1985quantum},
	${\bm{p}}=-i\nabla$ is the momentum operator,
	and $q$ is the charge of the projectile (in this case $-1$).
	The Hamiltonian of the projectile interaction with the target is given by
	\begin{alignat}{3}
		H_{\mathrm{0-T}}&=\frac{1}{|\bm{r}_0-\bm{r}_1|}+\frac{1}{|\bm{r}_0-\bm{r}_2|}-\frac{2}{r_0}
	\end{alignat}
	where 1 and 2 denote the two bound electrons of the target helium.
	The Hamiltonian of the target is given in a similar fashion to \eqref{Hprojlaser} as
	\begin{alignat}{3}
		H_{\mathrm{T}}+H_{\mathrm{T-F}}&=\frac{1}{2}\sum_{j=1}^2\left(-i\nabla_j+\frac{1}{c}\bm{A}\right)^2-\frac{2}{r_1}-\frac{2}{r_2}+\frac{1}{|\bm{r}_1-\bm{r}_2|}.
	\end{alignat}
	Here it is assumed that the nuclear core does not gain any appreciable kinetic energy due to the electromagnetic field.
	If we define a target potential term $V_{12}$ as
	\begin{alignat}{3}
		V_{12}=-\frac{2}{r_1}-\frac{2}{r_2}+\frac{1}{|\bm{r}_1-\bm{r}_2|}
	\end{alignat}
	and a projectile dependent term $W_{012}$ as
	\begin{alignat}{3}
		W_{012}=\frac{1}{|\bm{r}_0-\bm{r}_1|}+\frac{1}{|\bm{r}_0-\bm{r}_2|}-\frac{2}{r_0}
	\end{alignat}
	we may express the total Hamiltonian as
	\begin{alignat}{3}
		H&=\frac{1}{2}\sum_{j=1}^3\left(-i\nabla_j+\frac{1}{c}\bm{A}\right)^2+V_{12}+W_{012}.
	\end{alignat}
	It is clear that from this expression that a time dependence is introduced through that of $\bm{A}$ and hence we look to solve the time dependent Schr\"odinger equation
	\enlargethispage{1cm}
	\begin{alignat}{3}
		\label{lasershrod}
		\left(\frac{1}{2}\sum_{j=1}^3\left(-i\nabla_j+\frac{1}{c}\bm{A}\right)^2+V_{12}+W_{012}\right)|\Psi_i^{(+)}\rangle=i\frac{\partial|\Psi_i^{(+)}\rangle}{\partial t}.
	\end{alignat}
	Via the transformation \cite{cavaliere1980particle}
	\begin{alignat}{3}
		\label{transform}
		|\Phi_i^{(+)}\rangle&=\exp\left[-\frac{i}{c^2}\int^tA^2(\tau)\;\mathrm{d}\tau\right]|\Psi_i^{(+)}\rangle
	\end{alignat}
	we now express \eqref{lasershrod} as
	\begin{alignat}{3}
		\left(H'_0+H'_\mathrm{T}+W_{012}\right)|\Phi_i^{(+)}\rangle&=i\frac{\partial|\Phi_i^{(+)}\rangle}{\partial
		t},
	\end{alignat}
	where
	\begin{alignat}{3}
		H'_{\mathrm{0}}&=-\frac{1}{2}\nabla^2_0-\frac{i}{c}\bm{A}\cdot\nabla_0
	\end{alignat}
	and
	\begin{alignat}{3}
		H'_{\mathrm{T}}&=\sum_{j=1}^2\left(-\frac{1}{2}\nabla^2_j-\frac{i}{c}\bm{A}\cdot\nabla_j\right)+V_{12}.
	\end{alignat}
	In performing such a transformation we have used the definition of the Coulomb gauge to eliminate terms containing $\nabla\cdot\bm{A}$ and additionally the second fundamental theorem of calculus
	\begin{alignat}{3}
		\label{fundthm}
		&&F(t)&=\int^t f(\tau)\;\mathrm{d}\tau\\
		\implies && \frac{\partial F}{\partial t}&=f(t).
	\end{alignat}
	Note that the lower bound of the integral in both \eqref{transform} and \eqref{fundthm} are left blank as they are arbitrary (presuming they are independent of $t$).
	For asymptotic $r_0$ we have that $W_{012}\to 0$, and as such our Schr\"odinger equation becomes separable such that
	\begin{alignat}{3}
		\lim_{r_0\to\infty}|\Phi_i^{(+)}\rangle=|\xi_0\xi_\mathrm{T}\rangle,
	\end{alignat}
	where
	\begin{alignat}{3}
		\label{Hproj}
		H'_0|\xi_0\rangle&=i\frac{\partial|\xi_0\rangle}{\partial t}
	\end{alignat}
	and
	\begin{alignat}{3}
		\label{Htarget}
		H'_\mathrm{T}|\xi_\mathrm{T}\rangle&=i\frac{\partial|\xi_\mathrm{T}\rangle}{\partial t}.
	\end{alignat}
	\Cref{Hproj} has an exact solution \cite{Kroll1973}
	\begin{alignat}{3}
		\label{projstateint}
		\langle
		t\;\bm{r}_0|\xi_0\rangle&=(2\pi)^{-3/2}\exp\left(i\bm{k}\cdot\bm{r}_0\right)\exp\left[-\frac{i}{2}\int^t\left(k^2+\frac{2}{c}\;\bm{k}\cdot\bm{A}\;\mathrm{d}\tau\right)\right],
	\end{alignat}
	which is known as a Volkov state, whereas \eqref{Htarget} has no known solution. 
	It is here, in the solution of \eqref{Htarget}, where we now introduce some approximations.
	\subsection{Soft Photon Approximation}
	\label{Soft Photon Approximation}
	The soft photon approximation, that was first outlined by \citet{Kroll1973}, 
	describes a set of assumptions under which the scattering of a charged particle in the presence of a strong electromagnetic
	wave can be calculated using only field-free cross sections.
	The same result was proved much more succinctly the following year by \citet{Rahman1974}.
	Through the introduction of additional assumptions \citet{cavaliere1980particle} showed that the same form of result holds for ionising collisions.
	In this section we follow a similar argument to that given by Cavaliere in order to derive a relation for the field-assisted ionisation cross section.

	Firstly, we assume that the vector potential of the field takes the form
	\begin{alignat}{3}
		\label{laserform}
		\bm{A}=\frac{c\bm{F}_0}{\omega}\cos(\omega t),
	\end{alignat}
	where $c$ is the speed of light, $\bm{F}_0$ is the amplitude, and $\omega$ the frequency of the electric field.
	Observe that this corresponds to a linearly polarised field.
	This allows us to evaluate the integral in the description of projectile states \eqref{projstateint} such that
	\begin{alignat}{3}
		\langle t|\xi_0\rangle&=|\bm{k}_0\rangle\exp\left[-\frac{i}{2}\left(k^2t+\frac{2}{\omega^2}\;\bm{k}\cdot\bm{F}_0\sin(\omega t)\right)\right],
	\end{alignat}
	where we have represented the unperturbed plane wave solution of the projectile $|\bm{k}_0\rangle$.
	We now need to describe the bound initial state $|\xi_{\mathrm{T}i}\rangle$ and the free final state $|\xi_{\mathrm{T}f}\rangle$ of the target which are solutions of \eqref{Htarget}.
	As an approximate solution, we assume the frequency is sufficiently low compared to the internal electric field of the atom
	such that the initial state is described purely by the unperturbed
	target states $|\phi_i\rangle$ with the standard time dependence introduced
	\begin{alignat}{3}
		\langle t|\xi_{\mathrm{T}i}\rangle&=|\phi_i\rangle\exp\left\{-i\epsilon_i t\right\}.
	\end{alignat}
	For the final free state, the assumption is used that it can be given by 
	the same Coulomb wave solution as for the field-free ionisation of helium, with time modulation identical to that of the projectile
	\begin{alignat}{3}
		\label{freestateassumption}
		\langle t|\xi_{\mathrm{T}f}\rangle&=|\varphi_1\bm{k}_{{b}}\rangle\exp\left[-\frac{i}{2}\left(k_{{b}}^2t+\frac{2}{\omega^2}\;\bm{k}_{b}\cdot\bm{F}_0\sin(\omega t)\right)\right],
	\end{alignat}
	where $|\varphi_1\rangle$ is the ground state single electron wavefunction of helium \eqref{singlewave} and $|\bm{k}_{b}\rangle$ is the Coulomb wave solution for the ejected electron.
	Using these results we have for the field-assisted first order $T$-matrix element
	\begin{alignat}{3}
		^{\mathrm{FA}}T^{1\mathrm{st}}_{fi}&&=\int_{-\infty}^{+\infty}&\langle \xi_{0f}\xi_{\mathrm{T}f}|W_{012}|\xi_{\mathrm{T}i}\xi_{0i}\rangle\;\mathrm{d}t\\
		\label{nojacobi}
		&&=\int_{-\infty}^{+\infty}&\exp\left[{-i\left(\epsilon_i+\frac{k_i^2}{2}-\frac{k_a^2}{2}-\frac{k_{{b}}^2}{2}\right)t}-i\left(\bm{k}_i-\bm{k}_a-\bm{k}_{{b}}\right)\cdot
		\bm{F}_0\frac{\sin(\omega t)}{\omega^2}\right]\;\mathrm{d}t\nonumber\\
		&&\times&\langle \bm{k}_{0f}\phi_{f}|W_{012}|\phi_{i}\bm{k}_{0i}\rangle\\
		\label{withjacobi}
		&&=\int_{-\infty}^{+\infty}&e^{i\left(E_f-E_i\right)t}\sum_{n=-\infty}^\infty
		J_n\left(\frac{\left(\bm{k}_b-\bm{q}\right)\cdot\bm{F}_0}{\omega^2}\right)e^{in\omega t}\;\mathrm{d}t\times T^{1\mathrm{st}}_{fi}\\
		&&=\sum_{n=-\infty}^\infty &J_n\left(\alpha_n\right)\delta\left(E_f-E_i+n\omega\right) T^{1\mathrm{st}}_{fi},
	\end{alignat}
	where $T^{1\mathrm{st}}_{fi}=\langle \bm{k}_{0f}\phi_{f}|W_{012}|\phi_{i}\bm{k}_{0i}\rangle$ is the unperturbed $T$-matrix element in the first Born approximation, 
	$\bm{q}=\bm{k}_i-\bm{k}_a$ is the momentum transfer, $E_i=\epsilon_i+k_i^2/2$ and $E_f=(k_b^2+k_a^2)/2$ are the initial and final energy of the scattering system, 
	$J_n$ is the Bessel function of the first kind and 
	\begin{alignat}{3}
		\label{alphadef}
		\alpha_n=\left(\bm{k}_b-\bm{q}\right)\cdot\bm{F}_0/\omega^2.
	\end{alignat}
	Note that as absorption or emission of photons adjusts either $\bm{k}_b$ or $\bm{q}$ there is an implicit dependence on $n$ in \eqref{alphadef} and hence the subscript $n$ is present.
	In going from \eqref{nojacobi} to \eqref{withjacobi} we have employed the Jacobi-Anger expansion such that
	\begin{alignat}{3}
		e^{iz\sin\theta}&=\sum_{n=-\infty}^{\infty}J_n(z)e^{in\theta}.
	\end{alignat}
	Note that the above argument applies equally for each term in the Born series, and as such holds for the complete $T$-matrix elements \cite{Rahman1974}.
	Continuing via \eqref{breakupamplitude} and \eqref{TDCS} we find 
	\begin{alignat}{3}
		\label{SoftPhotonResult}
		\frac{\mathrm{d}\sigma^{\mathrm{FA}}}{\mathrm{d}\Omega_a \mathrm{d}\Omega_b \mathrm{d}E_b}&=\sum_{n=-\infty}^\infty J^2_n\left(\alpha_n\right)\frac{\mathrm{d}\sigma}{\mathrm{d}\Omega_a
		\mathrm{d}\Omega_b\mathrm{d} E_b}\Big|_{E_f=E_i+n\omega}\;\;,
	\end{alignat}
	where $E_b=k_b^2/2$.
	Note the equivalent expression for the free-free case derived by \citet{Kroll1973} takes the similar form
	\begin{alignat}{3}
		\label{SoftPhotonFree}
		{\frac{\mathrm{d}\sigma^{(n)}}{\mathrm{d}\Omega}}=J_n^2\left(\bm{q}\cdot \bm{F}_0/\omega^2\right)\frac{\mathrm{d}\sigma}{\mathrm{d}\Omega}\Big|_{E_f=E_i+n\omega}
	\end{alignat}
	for a process involving $n$ photons.

	This is an incredibly powerful result, as the coupling between the scattering system is entirely taken into account through the argument of the Bessel function $\alpha_n$ and the adjustment of the
	kinematics in an otherwise laser free scattering problem by $n$ photons of energy $\omega$.
	Furthermore, this sum is likely convergent as we have the following property of squared Bessel functions \cite{NIST:DLMF}
	\begin{alignat}{3}
		\label{Bessel property}
		\sum_{n=-\infty}^\infty J_n^2(z)=1,\quad \forall n\in\mathbb{Z},z\in\mathbb{C},
	\end{alignat}
	and that the cross sections themselves should not exhibit any divergent behaviour.
	Do note that a requirement of \eqref{Bessel property} is that the argument of the Bessel function is the same for each term in the sum whereas $\alpha$ has an implicit dependence on $n$ through
	$k_b$ and $q$ which are adjusted as appropriate to satisfy the energy conservation inherent in the adjustment of $E_f$ in \eqref{SoftPhotonResult}.
	Hence, we consider instead
	\begin{alignat}{3}
		\label{Bessel property alpha}
		\sum_{n=-\infty}^\infty J_n^2(\alpha_n)\approx 1.
	\end{alignat}
	A similar sum rule has been extensively studied in the case of free-free transitions (elastic scattering but with energy exchange with the laser field)
	\cite{Wein1977,Wein1979,Jung1979} and has been found to be valid in such cases.
	Furthermore, in the original \citet{cavaliere1980particle} paper, as part of their application to the ionisation of hydrogen in the presence of a strong laser field, 
	they report that \eqref{Bessel property alpha} is well satisfied, but nonetheless comment that considerable further investigation is required.
	Our own findings with regards to this sum rule in the case of helium are given in \Cref{e2eFieldAssisted}.

	A consequence that is unique to the soft photon approximation for ionisation is that $E_f$ contains two mechanisms for distributing the photon energy, 
	the projectile final momentum $k_a$, and the ejected electron momentum $k_b$, and in the general case it is unclear which term
	should be offset to calculate the physical field-assisted cross section.
	If both electrons have appreciable energy, then adjusting either term is equivalent, as the particles are indistinguishable. 
	However, for the kinematics that we are considering the ejected electron has a much lesser energy that the projectile (1 keV compared to $\sim10$ eV).
	In this case it is the slow outgoing electron that has its energy offset, as because of its low energy it is heavily influenced (in comparison to the projectile) 
	by the laser and resultant atomic fields.
	Additionally, an offset of a few eV to the 1 keV projectile results in a negligible difference to the cross section as at such a high energy no resonance effects due to atomic structure are
	present.
	This would allow the cross section to instead be treated as a slowly varying function with $n$ and 
	hence it may be taken outside the sum in \eqref{SoftPhotonResult}. 
	Then by \eqref{Bessel property} we would have that the field-assisted cross section is exactly equal to the field-free. 
\enlargethispage{0.5cm}
	However, the findings of the \citet{hohr2007laser} experiment with which we look to compare would suggest this to be unphysical.
	\subsubsection{Range of Validity}
	\label{Range of Validity}
	The preceding derivation is subject to a number of approximations, some explicit and others implicit, which are key to understanding the applicability of such a treatment.
	Firstly, it is interesting to note that in general the presence of the electromagnetic field causes the centre of mass frame to no longer be truly inertial.
	Although, unless the laser field is exceptionally strong this effect will be an exceedingly minor, and as such oscillations due to this interaction can be ignored.
	
	The description of the initial states of the target being purely the laser free states completely neglects target dressing effects.
	This is only a reasonable assumption when the electric field due to the laser is considerably smaller than the internal 
	atomic field ($5\times 10^{9}$ V/cm) and the photon energy is far from resonance with the atomic energy levels.
	In the case of the laser parameters we look to consider, the photon energy is $\omega=1.17$ eV and with a peak intensity of $4\times 10^{14}$ W/cm$^2$. 
	For a linearly polarised electromagnetic wave
	of the form \eqref{laserform} with these properties, we have a peak electrical field strength of $F_0=5.5\times 10^7$ V/cm.
	Hence, this approximation is thought to be appropriate for this system.
	The use of a Volkov state for the free electron in the laser field is an exact solution of \eqref{Hproj} and as such is a sufficient description of the projectile, contingent only 
	on this Hamiltonian being valid (do note that we have omitted the small ponderomotive energy term).

	The next notable assumption is involved in the description of the free state of the target after the collision.
	We have used the expression as in \eqref{freestateassumption}, which is a combination of the Coulomb wave solution of the free electron and modulated in time by the laser in the same manner as the
	incident plane wave.
	Such a form incorporates the interaction of the ejected electron with both the resultant field of the ionised target and electric field of the laser, and 
	furthermore behaves appropriately in the limit of $F_0\to 0$,
	yet is not a direct solution of \eqref{Htarget}.
	In the original \citet{cavaliere1980particle} paper they provide an analysis of this ansatz and conclude with the following inequality for its validity
	\begin{alignat}{3}
		\delta=\frac{p_\mathrm{L}}{q}\ll 1
	\end{alignat}
	where $p_\mathrm{L}$ is the momentum of the ejected electron due to the presence of the laser
	\begin{alignat}{3}
		\bm{p}_\mathrm{L}&=\frac{1}{c}\bm{A}\\
		&=\frac{F_0}{\omega}\cos(\omega t).
	\end{alignat}
	For the kinematics of interest within this work we have the peak momentum due to the laser being $\approx 0.25$ a.u., and the momentum transfer of the collision ranging from 0.5 to 1.0 a.u.
	Hence, for the system we consider the value of $\delta\approx 0.25$ to $0.5$. 

	Do note that experimentally the soft photon approximation has been found to be inadequate to describe free-free scattering events at small scattering angles (low $q$) 
	\cite{wallbank1993,geltman1996laser}. 
	Furthermore, the recent experiments of \citet{hohr2005electron,hohr2007laser} cast doubt onto its applicability to ionisation collisions as well.
\section{Photoemission}
\label{Photoemission}
	Interactions involving photons are often considered half-collisions with respect to other types of collision problems. 
	This is due to the lack of interaction between the photon and target electrons until absorption occurs.
	Hence, although they are in some ways simpler than the collision processes considered earlier in this work, they require a specific treatment within CCC \cite{KB98d}.
	We wish to analyse the differences in time delay (see \Cref{Wigner Time Delay}) for the photoemission of the H$^{-}$ ion and He, and as such we present theory relevant for two electron targets.

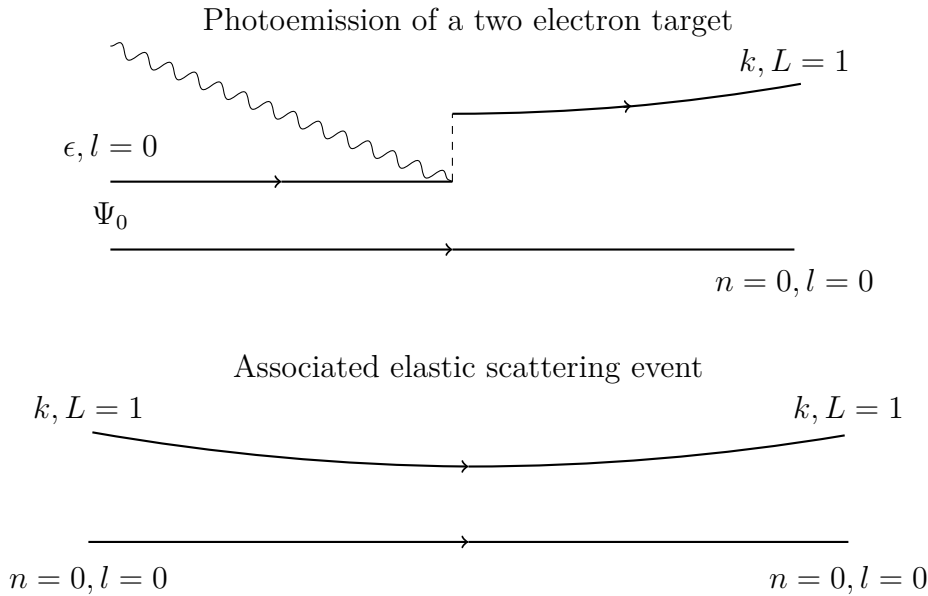
\begin{figure}[htbp]
	Photoemission of a two electron target
		\begin{tikzpicture}[scale=4.5]
			\tikzset{snake it/.style={decorate, decoration=snake}}
			\draw[thick,->] (0,0) -- (1,0) node[anchor=north east]{};
			\draw[thick]    (1,0) -- (2,0);
			\draw[thick,->] (0,0.2) -- (0.5,0.2) node[anchor=north west]{};
			\draw[thick]    (0.5,0.2) -- (1,0.2) node[anchor=north west]{};
			\draw[snake it] (0,0.6) -- (1,0.2);
			\draw[dashed]   (1,0.2) -- (1,0.4);
			\draw[thick,->] (1,0.4) arc (-90:-85:6);
			\draw[thick]    (1.5,0.42) arc (-85:-80:6);

			\node at (0,0.1) {$\Psi_0$};
			\node at (2,-0.1) {$n=0,l=0$};
			\node at (0, 0.3) {$\epsilon,l=0$};
			\node at (2,0.55) {$k,L=1$};

		\end{tikzpicture}\\[0.4cm]
		Associated elastic scattering event
		\begin{tikzpicture}[scale=5]
			\tikzset{snake it/.style={decorate, decoration=snake}}
			\draw[thick,->] (0,0) -- (1,0) node[anchor=north east]{};
			\draw[thick]    (1,0) -- (2,0);
			\draw[thick,<-] (1,0.2) arc (269.5:260:6);
			\draw[thick]    (1,0.2) arc (-90:-80.5:6);

			\node at (0,-0.1) {$n=0,l=0$};
			\node at (2,-0.1) {$n=0,l=0$};
			\node at (0,0.35) {$k,L=1$};
			\node at (2,0.35) {$k,L=1$};

		\end{tikzpicture}
		\caption[Schematic diagrams depicting the single photoemission of a two electron target (top) and the associated elastic scattering event (bottom).]
		{Schematic diagrams depicting the single photoemission of a two electron target (top) and the associated elastic scattering event (bottom).
		The solid lines represent electrons, and the undulatory line a photon.
		The upper diagram demonstrates the two electron ground state $\Psi_0$ with an electron of binding energy $\epsilon$ absorbing a photon and transitioning to an unbound state of momentum $k$ and
		orbital angular momentum $L=1$.
		The lower diagram demonstrates elastic scattering of an $L=1$ electron with momentum $k$ incident on a single electron target in the ground state.}
		\label{AlexFeynmann1}
	\end{figure}

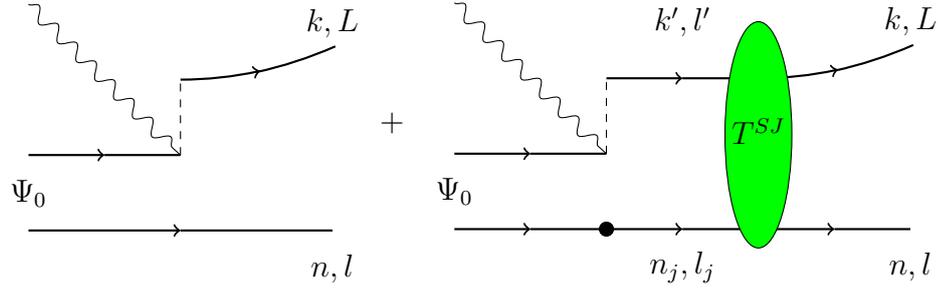
\begin{figure}[htbp]
		\begin{tikzpicture}[xscale=2,yscale=5]
			\tikzset{snake it/.style={decorate, decoration=snake}}
			\draw[thick,->] (0,0) -- (1,0) node[anchor=north east]{};
			\draw[thick]    (1,0) -- (2,0);
			\draw[thick,->] (0,0.2) -- (0.5,0.2) node[anchor=north west]{};
			\draw[thick]    (0.5,0.2) -- (1,0.2) node[anchor=north west]{};
			\draw[snake it] (0,0.6) -- (1,0.2);
			\draw[dashed]   (1,0.2) -- (1,0.4);
			\draw[thick,->] (1,0.4) arc (-90:-85:6);
			\draw[thick]    (1.5,0.42) arc (-85:-80:6);

			\node at (0,0.1) {$\Psi_0$};
			\node at (2,-0.1) {$n,l$};
			\node at (2,0.55) {$k,L$};
		\end{tikzpicture}
		\bp{0.6cm}+\vspace{4.3cm}\ep
		\begin{tikzpicture}[xscale=2,yscale=5]
			\tikzset{snake it/.style={decorate, decoration=snake}}
			\draw[thick,->] (0,0) -- (0.5,0) node[anchor=north east]{};
			\draw[thick,-] (0.5,0) -- (1.0,0) node[anchor=north east]{};
			\draw[thick,->]    (1,0) -- (1.5,0);
			\draw[thick,-]    (1.5,0) -- (2.0,0);
			\draw[thick,->]    (2,0) -- (2.5,0);
			\draw[thick,-]    (2.5,0) -- (3.0,0);
			\draw[thick,->] (0,0.2) -- (0.5,0.2) node[anchor=north west]{};
			\draw[thick]    (0.5,0.2) -- (1,0.2) node[anchor=north west]{};
			\draw[snake it] (0,0.6) -- (1,0.2);
			\draw[dashed]   (1,0.2) -- (1,0.4);
			\draw[thick,->] (1,0.4) -- (1.5,0.4);
			\draw[thick]    (1.5,0.4) -- (2.0,0.4);
			\draw[thick,->] (2,0.4) arc (-90:-85:6);
			\draw[thick]    (2.5,0.42) arc (-85:-80:6);

			\def\firstellipse{(2.0,0.25) ellipse (0.22 and 0.3)}
			\fill[green] \firstellipse;
			\draw \firstellipse;

			\fill[black] (1.0,0.0) ellipse (0.05 and 0.02);

			\node at (0,0.1) {$\Psi_0$};
			\node at (1.5,-0.1) {$n_j,l_j$};
			\node at (1.5,0.55) {$k',l'$};
			\node at (3,-0.1) {$n,l$};
			\node at (3,0.55) {$k,L$};
			\node at (2.0,0.26) {$T^{SJ}$};
		\end{tikzpicture}
		\vspace{-2.3cm}
		\caption[Graphical representation of the two-electron photoionisation
		amplitude in the CCC formalism.]
		{ Graphical representation of the two-electron photoionisation
		amplitude in the CCC formalism. 
		The first diagram represents the diving term and the second, the off-shell coupling term.
		Here a solid line represents an electron, an undulatory line a photon, and the shaded oval the half off-shell $T$-matrix.}
		\label{feynmann1}
	\end{figure}

	For a transition from the two electron ground state $|\Psi_0\rangle$ to that of an unbound photoelectron $|k L\rangle$ and target in the single electron state $|n l\rangle$, the 
	total photoemission amplitude (for light linearly polarised in the $z$ direction) is given by
	\begin{alignat}{3}
	\label{dipole}
	f^{\mathrm{ph}}(\bm{k})
	=(2\pi)^{3/2}k^{-1/2}
	\sum_{\genfrac{}{}{0pt}{}{L=l_i\pm1}{ M=m_i}}
	e^{i\delta_{L}(k)}i^{-L}Y_{LM}(\hat{\bm{k}})\, 
	\left(\begin{array}{rrr} L&1&l_i\\ M&0&m_i\\
	\end{array}\right) \la L kl n\| D \|\Psi_0\ra \ ,
	\end{alignat}
	where $\la L kln \| D \|\Psi_0\ra$ is the dipole matrix element
	stripped of all its angular momentum projections via a partial wave expansion (subsequently referred to as the reduced dipole matrix element), $Y_{LM}(\hat{\bm{k}})$ is a spherical harmonic,
	$\delta_{L}(k)$ is the phase shift associated with the $L$ partial wave, and $|\Psi_0\rangle$ (for either \Hm or He) in our case is calculated with a 20-term Hylleraas 
	expansion as in \citet{KB98d}.
	The target electron which absorbs the photon is described in its initial state by the quantum numbers $n_i$, and $l_i$ as per their usual
	definitions.
	The $z$-component of angular momentum $m$, is omitted from the state descriptions as it is eliminated through spherical symmetry. However, it does need to be explicitly considered when
	projecting to return angular dependencies (as in \eqref{TPWT} and \eqref{dipole}).
	Do note the change in notational convention away from $i\to f$ centric notation to that which 
	is more commonly used for elastic scattering and photoionisation, where final quantities are instead given no subscript.
	The cross section corresponding to this transition is given by
	\begin{alignat}{3}
		\sigma_{n_il_i\to kL}(\omega)=\frac{4\pi^2}{3}\alpha\omega\big|\la Lk l n\| D \|\Psi_0\ra\big|^2
	\end{alignat}
	where $\alpha$ is the fine structure constant and $\omega$ is the energy of the photon.
	Within the CCC formulation the reduced dipole matrix element is calculated via \cite{KIAB00}
	\begin{alignat}{3}
	\la Lk ln \| D \|\Psi_0\ra &=
	\langle Lk \, ln\|  d  \| \Psi_0 \rangle\nn\\
	&+\sum\limits_{l_jn_j}
	\sum_{L'}\isum\limits_{k'}
	\frac{\big\la L k\, ln\big\| T^{SJ}\big\| n_jl_j\,k'L'\big\ra}{E+i0-\epsilon_j-k'^2/2}
	\langle L'k'\,l_jn_j\|  d\|\Psi_0\rangle  \ ,
	\label{Tls}
	\end{alignat}
	where $\langle Lkln\|  d\|\Psi_0\rangle$ is the uncorrelated dipole matrix element.
	It is called such as it does not yet include the effect of electron-electron correlations.
	This element can be calculated within three different gauges known as the length, velocity, and acceleration forms.
	However, this choice is arbitrary as they are each equivalent. In the calculations that follow the velocity gauge is used.
	The $T^{SJ}$ term is the half off-shell $T$-matrix as calculated in the solution of \eqref{CCCT} for the associated photon free scattering process.
	For example, for the photodetachment of the H$^{-}$ ion the corresponding photon free scattering process is elastic electron scattering on hydrogen in the dipole singlet channel ($L=1$, $S=0$),
	with an incident energy corresponding to that of the emitted photoelectron.
	\enlargethispage{0.5cm}
	Both processes result in an outgoing electron of the same energy and angular momentum, 
	and the target as a ground state hydrogen atom, but the photodetachment process only contains `half' the collision.
	Note that the absorption of the photon imparts a unit of angular momentum to the electron and hence is the cause of the non-zero $L$ for the equivalent elastic scattering channel.
	Similarly for the photoionisation of helium, the associated photon free scattering process of which the half off-shell $T$-matrix is needed is elastic scattering on the $\ce{He^+}$ ion, 
	again in the dipole singlet channel.
	To illustrate the connection between these processes we provide the schematic diagrams of \Cref{AlexFeynmann1}.
	Furthermore, a graphical illustration of the physical meaning of the two terms in \eqref{Tls} is given in \Cref{feynmann1}. 
	\vspace{0.4cm}
	\clearpage
	\phantom{hellloooo}
\subsection{Wigner Time Delay}
\label{Wigner Time Delay}
	The Wigner time delay \cite{Eisenbud1948,wigner1955lower} (henceforth referred to as simply time delay) 
	is a measure of the difference between the time taken for a particle to travel through a potential landscape in comparison to free space.
	It is defined in terms of the energy derivative of the scattering phase in a given partial wave.
	To justify this definition we consider the derivation for the delay in the  
	formation of a photoelectron wavepacket emitted from an atomic target upon interaction with an extreme ultraviolet laser (XUV) pulse as given by \citet{kheifets2010delay}.
	See \Cref{Time Delay} for a simplified introduction to the concept of a scattering time delay as given in the review of \citet{de2002time}.
	For this system the time dependent Schr\"odinger equation can be written as
	\begin{alignat}{3}
		i\dpd{|\Psi(t)\rangle}{t}&=(H_{0}+H_{\mathrm{int}})|\Psi(t)\rangle
	\end{alignat}
	where $H_0$ is the Hamiltonian of the atomic target and $H_{\mathrm{int}}$ is that of the interaction with the laser field which is given in the velocity gauge as
	\begin{alignat}{3}
		H_{\mathrm{int}}&=\bm{A}(t)\cdot\sum_{j=1}^n \bm{k}_j.
	\end{alignat}
	Here $\bm{A}(t)$ is the vector potential of the laser field and $\bm{k}_j$ is the momentum of the $j$-th electron of total number $n$.
	The wavepacket of the emitted photoelectron is given as an expansion of the time dependent wavefunction over the set of scattering states such that
	\begin{alignat}{3}
		\langle \bm{r}|\Phi(t)\rangle&=\sum_{L}\int_0^\infty  \mathrm{d}k\;k^2\la \bm{r}|\bm{k}\rangle \la \bm{k}|\Psi(t)\rangle\\
		&=\sum_{L}\int_0^\infty  \mathrm{d}k\;k^2 a_{kL}(t)\la \bm{r}|\bm{k}\rangle e^{-iE_k t}
	\end{alignat}
	where the photoelectron in the continuum is written as $|\bm{k}\rangle$ with linear $k$ and orbital angular momentum $L$, and have defined the projection coefficients
	\begin{alignat}{3}
		a_{kL}(t)&=e^{iE_kt}\la \bm{k}|\Psi(t)\rangle
	\end{alignat}
	with $E_k=k^2/2$.
	This continuum state is given by 
	\begin{alignat}{3}
		\la \bm{r}|\bm{k}\rangle&=R_{kL}(r)Y_{Lm}(\hat{\bm{r}})
	\end{alignat}
	where for asymptotically large distances we have
	\begin{alignat}{3}
		\lim_{r\to\infty}R_{kL}(r)\propto \sin\left[kr +\delta_L(k)+\ln(2kr)/k-L\pi/2\right]
	\end{alignat}
	with $\delta_L(k)$ being the phase shift in the $L$-th partial wave.
	If we define a time $T$ such that for $|t|>T$ the XUV field is zero, we have that for $t>T$ the projection coefficients no longer depend on time. 
	Hence, for such times we can write
	\begin{alignat}{3}
		\label{projcoeff}
		a_{kL}(t>T)&=-i\int_{-T}^T \la \bm{k}|z|\Psi_0\rangle e^{i(E_k-E_0)t'}F(t')\;\mathrm{d}t'
	\end{alignat}
	where $\Psi_0$ is the ground state of the target with energy $E_0$, $F(t')$ is the electric field of the XUV laser pulse, and $\la \bm{k}|z|\Psi_0\rangle$ 
	is the angularly dependent dipole matrix element which is related
	to the reduced uncorrelated element via Clebsch-Gordan coefficients such that 
	\begin{alignat}{3}
		\la \bm{k}|z|\Psi_0\rangle\propto C^{LM}_{10 l_im_i}\la L k\|d\|\Psi_0\rangle.
	\end{alignat}
	Here we have assumed that the ejected electron was initially in the ground state ($l=1$ and $m=0$) and have denoted another bound electron in the initial state  
	as having orbital angular momentum $l_i$ and corresponding $z$-projection $m_i$.
	With this definition we can now write
	\begin{alignat}{3}
		a_{kL}(t>T)&\propto-i\la L k\|d\|\Psi_0\rangle \tilde{F}(E_k-E_0)
	\end{alignat}
	where
	\begin{alignat}{3}
		\tilde{F}(\omega)=\int_{-\infty}^{\infty}F(t')e^{i\omega t'}\mathrm{d}t'
	\end{alignat}
	is the Fourier transform of the XUV electric field.
	Note that we have been able to extend the integration limits in \eqref{projcoeff} to infinity as for $t>T$ the electric field $F(t)=0$ by definition.
	Therefore, we may now express the wavepacket for time $t>T$ and at an asymptotically large distance from the atomic target as
	\begin{alignat}{3}
		\lim_{r\to\infty}\langle \bm{r}|\Phi(t>T)\rangle\;&&\propto -i\sum_{L}\int_0^\infty  &\mathrm{d}k\;k^2\la L k\|d\|\Psi_0\rangle \tilde{F}(E_k-E_0) Y_{Lm}(\hat{\bm{r}})e^{-iE_k t}\nonumber\\
		&&&\times\sin\left[kr +\delta_L(k)+\ln(2kr)/k-L\pi/2\right].
	\end{alignat}
	If we now consider the sine term as a superposition of incoming and outgoing waves via 
	\vspace{-0.4cm}
	\begin{alignat}{3}
		\sin(z)&=\frac{e^{iz}-e^{-iz}}{2i}
	\end{alignat}
	and keep only the physically meaningful outgoing exponent we may instead write
	\begin{alignat}{3}
		\label{finalpacket}
		\lim_{r\to\infty}\langle \bm{r}|\Phi(t>T)\rangle\;&&\propto &-\frac{1}{2}\sum_{L}\int_0^\infty  \mathrm{d}k\;k^2\la L k\|d\|\Psi_0\rangle \tilde{F}(E_k-E_0) Y_{Lm}(\hat{\bm{r}})\nonumber\\
		&&&\times\exp\left[i\left(-k^2/2 t+kr +\delta_L(k)+\ln(2kr)/k-L\pi/2\right)\right].
	\end{alignat}
	Now we look to find how the peak of this packet moves with time in order to define a concept of time delay.
	Firstly, let us assume that for some $k=k_0$ the magnitude (terms not in the complex exponential) of the packet is at a maximum.
	Importantly each term in \eqref{finalpacket} is entirely real valued such that it can be unambiguously split into a magnitude and complex exponential phase.
	Considering \eqref{finalpacket} as a superposition of monochromatic waves (summed over $k$) for each $L$, we discern that large variation of phase will cause mainly destructive interference.
	Hence, the largest amplitude will occur when the maximum magnitude corresponds to a stationary point of the phase, such that for a given $L$
	\begin{alignat}{3}
		&&0&=\dod{}{k}\left[\mathrm{arg}\left(\lim_{r\to\infty}\langle \bm{r}|\Phi(t>T)\rangle\right)\right]\bigg|_{k=k_0}\\
		&&&=\dod{}{k}\left[-k^2/2 t+kr +\delta_L(k)+\ln(2kr)/k-L\pi/2\right]\bigg|_{k=k_0}\\
		\label{rclass}
		\implies && r&=k\left\{t-\dod{}{E}\left[\delta_L(k)+\ln(2kr)/k\right]\right\}\bigg|_{k=k_0}.
	\end{alignat}
	Here we have used that ${\mathrm{d}}/{\mathrm{d}k}=k\;{\mathrm{d}}/{\mathrm{d}E}$.
	We may interpret \eqref{rclass} as describing the quasi-classical motion of this wavepacket for an asymptotic distance and $t>T$.
	If we now consider the logarithm term to be sufficiently slowly varying with $k$ (equivalently $E$) such that it can absorbed into a constant, namely $r_0$, we may write \eqref{rclass} as
	\begin{alignat}{3}
		r=k_0(t-\tau_L)+r_0
	\end{alignat}
	where 
	\begin{alignat}{3}
		\tau_L=\frac{\mathrm{d}\delta_L(k)}{\mathrm{d}E}\Big|_{k=k_0}.
	\end{alignat}
	is the time delay of the packet in the $L$-th partial wave.

	\enlargethispage{0.8cm}
	To determine this time delay within the CCC formalism we extract the phase shift from the elastic $S$-matrix element via the definition
	\begin{alignat}{3}
		\label{Sphase}
		\la Lkln\|S^{SJ}\|nlkL\ra&=1-2\pi ik \la Lkln\|T^{SJ}\|nlkL\ra\\
		&=\exp\left\{2i\delta_{L}(k)\right\}.
	\end{alignat}
	Hence, we have
	\begin{alignat}{3}
		\delta_{L}(k)&=\frac{1}{2}\mathrm{arg}\left(\la Lkln\|S^{SJ}\|nlkL\ra\right).
	\end{alignat}
	However, there are two further considerations to make when employing such a definition of the time delay for this work.
	Firstly, in the case of elastic scattering on a charged targets for the ease of calculation a basis of Coulomb waves are used as opposed to the usual plane waves.
	These are continuous solutions to the Schr\"odinger equation at an asymptotic distance in the presence of long ranged Coulomb potential.
	In doing so, it removes the associated Coulomb phase ($\sigma_{L}$) from being part of the scattering event and is hence not included in the phase of the $S$-matrix.
	As such, for the total phase in a partial wave it must be added to the phase resulting from the scattering $S$-matrix.
	Hence, we define the elastic scattering time delay as
	\begin{alignat}{3}
		\label{tauel}
		\tau^\mathrm{el}_L &= \frac{\mathrm{d}}{\mathrm{d}E}\left[\sigma_{L}(k)+\frac{1}{2}\mathrm{arg}\left(\la Lkln\|S^{SJ}\|nlkL\ra\right)\right]\bigg|_{k=k_0}.
	\end{alignat}
	In the case of scattering on a neutral target the same definition is used with simply $\sigma_{L}(k)$ set to zero.
	Secondly, if instead we take into account the effect of electron-electron correlations and replace the uncorrelated dipole matrix element $\la L k\|d\|\Psi_0\rangle$ in \eqref{finalpacket} with the
	reduced dipole matrix element $\la L k\|D\|\Psi_0\rangle$ calculated via \eqref{Tls}, \eqref{finalpacket} now contains an additional complex component.
	To accommodate for this we split this matrix element into a real magnitude and complex phase via
	\begin{alignat}{3}
		\la L k\|D\|\Psi_0\rangle&=|\la L k\|D\|\Psi_0\rangle|\exp\left\{i\arg\left(\la L k\|D\|\Psi_0\rangle\right)\right\}.
	\end{alignat}
	The additional contribution to the phase results in the equivalent expression for the quasi-classical trajectory within partial wave $L$ as
	\begin{alignat}{3}
		r&=k\left\{t-\dod{}{E}\left[\delta_L(k)+\ln(2kr)/k+\arg\left(\la L k\|D\|\Psi_0\rangle\right)\right]\right\}\bigg|_{k=k_0},
	\end{alignat}
	and correspondingly the expression for the photoemission time delay becomes
	\begin{alignat}{3}
		\tau^{\mathrm{ph}}_L=\frac{\mathrm{d}}{\mathrm{d}E}\left(\delta_L(k)+\arg\left(\la L k\|D\|\Psi_0\rangle\right)\right)\Big|_{k=k_0}.
	\end{alignat}
	If instead we are interested in delay associated with the total photoemission process we use the amplitude \eqref{dipole} to define the total photoemission time delay as
	\begin{alignat}{3}
		\label{tauph}
		\tau_{\mathrm{tot}}^{\mathrm{ph}} &= \frac{\mathrm{d}}{\mathrm{d}E}\left[\mathrm{arg}\left(f^{\mathrm{ph}}(\bm{k})\right)\right]\bigg|_{k=k_0}.
	\end{alignat}
	It is this form of photoemission time delay which is to be compared with that of the associated elastic scattering event in the dipole singlet channel.

\chapter{Results}
	In this chapter we present the results of our analysis of laser assisted electron impact ionisation (e,2e) of helium utilising the CCC method under the soft photon approximation
	(\Cref{Electron Impact Ionisation of Helium}), and the Wigner time delay in the photoemission of \Hm and photoionisation of He (\Cref{Photoemission time delay and electron scattering time delay}).
\label{Results}
	\section{Laser Assisted Electron Impact Ionisation of Helium}
	\label{Electron Impact Ionisation of Helium}
	In this section we investigate the electron impact ionisation of helium in the presence of a strong laser field for parameters identical to that given in \citet{hohr2007laser}.
	These parameters are listed in \Cref{exptparamtable}.
	\begin{table}[b]
	\centering
	\begin{tabular}{l|rl}
		\hline
		Quantity& \multicolumn{2}{l}{Value}\\
		\hline
		Incident electron energy& 1& keV\\
		Ejected electron energy& 3 - 18& eV\\
		Momentum transfer&0.5 - 1.5 &a.u.\\
		Laser polarisation&$x$-direction&(see \Cref{e2e fig2})\\
		Photon energy &1.17&eV\\
		Peak intensity&$4\times 10^{14}$&W/cm$^2$\\
		Pulse duration&$7$&ns\\
		\hline
	\end{tabular}
	\caption{Experimental parameters as per \citet{hohr2007laser}.}
	\label{exptparamtable}
	\end{table}
	Note that for the extent of this work, we consider the target helium as a single electron system via the frozen-core model as described in \Cref{Treatment of Helium}.
	As such, each target state needs to be described only by the set of quantum numbers of the `active' electron, as the other is always considered to be in the $1$s orbital.
	As the data provided by \citet{hohr2007laser} is given over a range of ejected electron energies and momentum transfer we choose for comparison an ejected electron energy of $10$ eV 
	and momentum transfer of 0.5, 0.75 and 1.0 a.u.
	Using these values of momentum transfer, we can make a comparison with the data provided by H\"ohr \textit{et al}.\ 
	in their figure for momentum transfers of 0.7 - 1.0 a.u.\ and ejected electron energies of 7 - 12 eV.
	Additionally, experimental data provided by \citet{durr2008higher} is available for these parameters allowing us to make a comparison with our field-free calculations (see \Cref{ExpComp}).
	\subsection{Field-Free Calculations}
	\label{Field Free}
	We firstly look to establish convergent results for electron impact ionisation for the laser free parameters given in \Cref{exptparamtable}. 
	In the CCC method there are three factors pertaining to convergence (see \Cref{Convergence Considerations}): the number of partial waves ($L$), 
	the values of angular momentum included in the description of the target states ($l$), and the size of the Laguerre basis for each set of these states ($N_l$).
	Each of the above factors are independently convergent, such that they may each be checked individually.

\begin{figure}[htbp]
		\centering
		\includegraphics[scale=1.2]{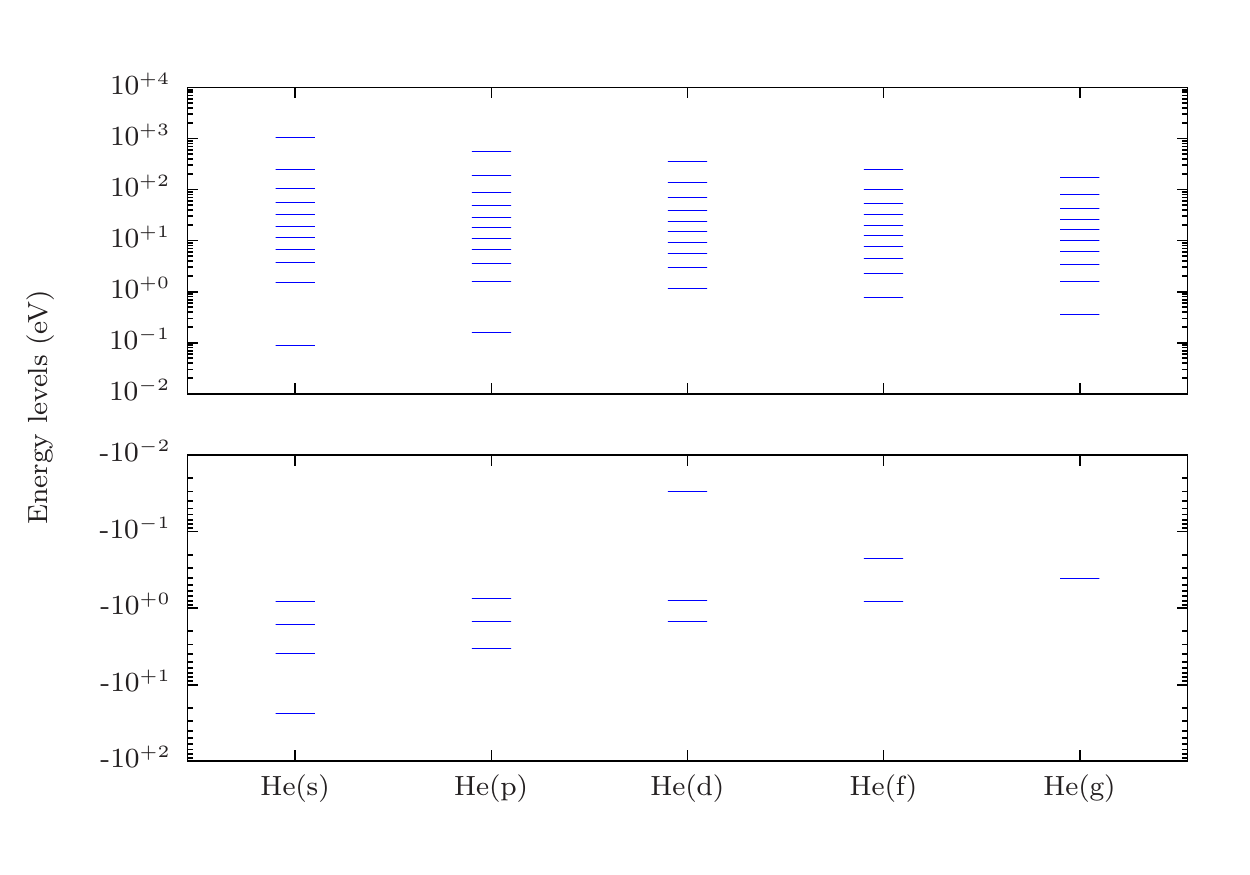}
		\caption[Energy levels of states generated from a Laguerre basis expansion of $N_l=15-l$ with $\lambda_l=2.0$.]
		{Energy levels of states generated from a Laguerre basis expansion of $N_l=15-l$ with $\lambda_l=2.0$. 
		This diagonalisation is that which is used in the initial convergence investigation with $L$.}
		\label{pLevelsHe}
	\end{figure}

	To obtain an indication of the number of partial waves required, a useful approach is to consider the total ionisation cross section (TICS) as a function of partial wave.
	As a result of the optical theorem (see \Cref{Optical Theorem}) each cross section is inherently linked with one another, and hence convergence of one quantity (i.e.\ TICS)
	is a good indicator of convergence elsewhere.
	For the initial calculations used in analysing convergence with partial waves, states with  $l\leq 4$ (s - g states) were included each with a corresponding Laguerre basis size of $N_l=15-l$.
	Additionally, only singlet states (total spin $S=0$) are included in the calculations as at an incident energy of 1 keV there is no mechanism for exchange. 
	This form of allocating basis size for each $l$ is commonly used as it removes the complexity of having to specify a basis size for each $l$. 
	Furthermore, such a form models the behaviour of bound states such that for each $l$ there is the same highest principle quantum number ($n$). 
	For example, if we want to consider the first 5 bound states ($n\leq 4$) for $l\leq3$ there are $n=1$ - $5$
	s-states, $n=2$ - $5$ p-states, $n=3$ - $5$ d-states, and $n=4$ - $5$ f-states.
	The energy eigenstates generated by this diagonalisation are given in \Cref{pLevelsHe}.
	Observe that the bound states are sparsely approximated in comparison to the positive energy free pseudostates. 
	For our purposes this is a desirable diagonalisation as we are interested in ionising cross sections as opposed to excitation, and this dense distribution 
	allows greater accuracy in interpolating from the positive energy discrete pseudostates within CCC onto the true continuum.

\begin{figure}[htbp]
		\centering
		\includegraphics[scale=1.0]{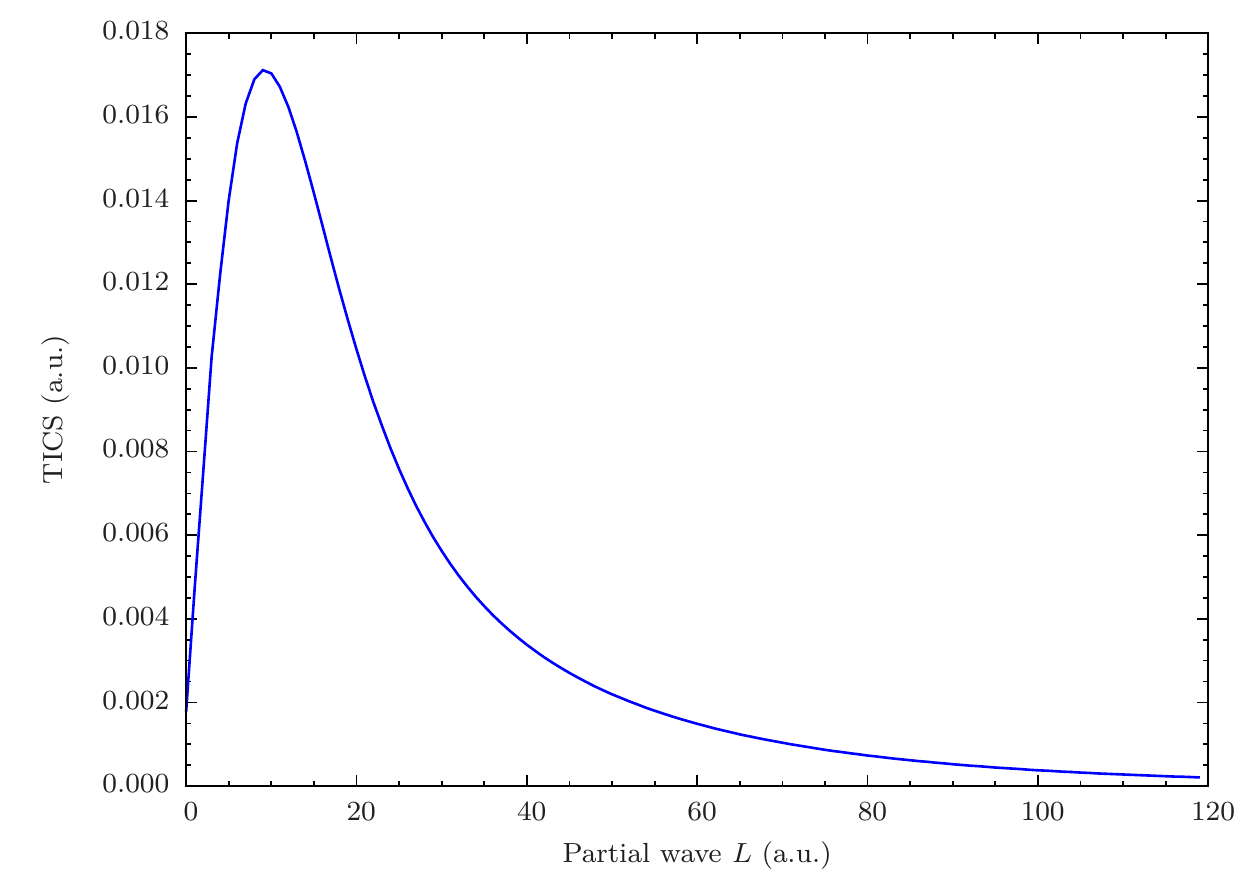}
		\caption[The total ionisation cross section (TICS) as a function of partial wave $L$ for 1 keV electrons incident on helium.]
		{The total ionisation cross section (TICS) as a function of partial wave $L$ for 1 keV electrons incident on helium as calculated by CCC.}
		\label{TICSvsPW}
	\end{figure}

\begin{figure}[htbp]
		\centering
		\includegraphics[scale=1.0]{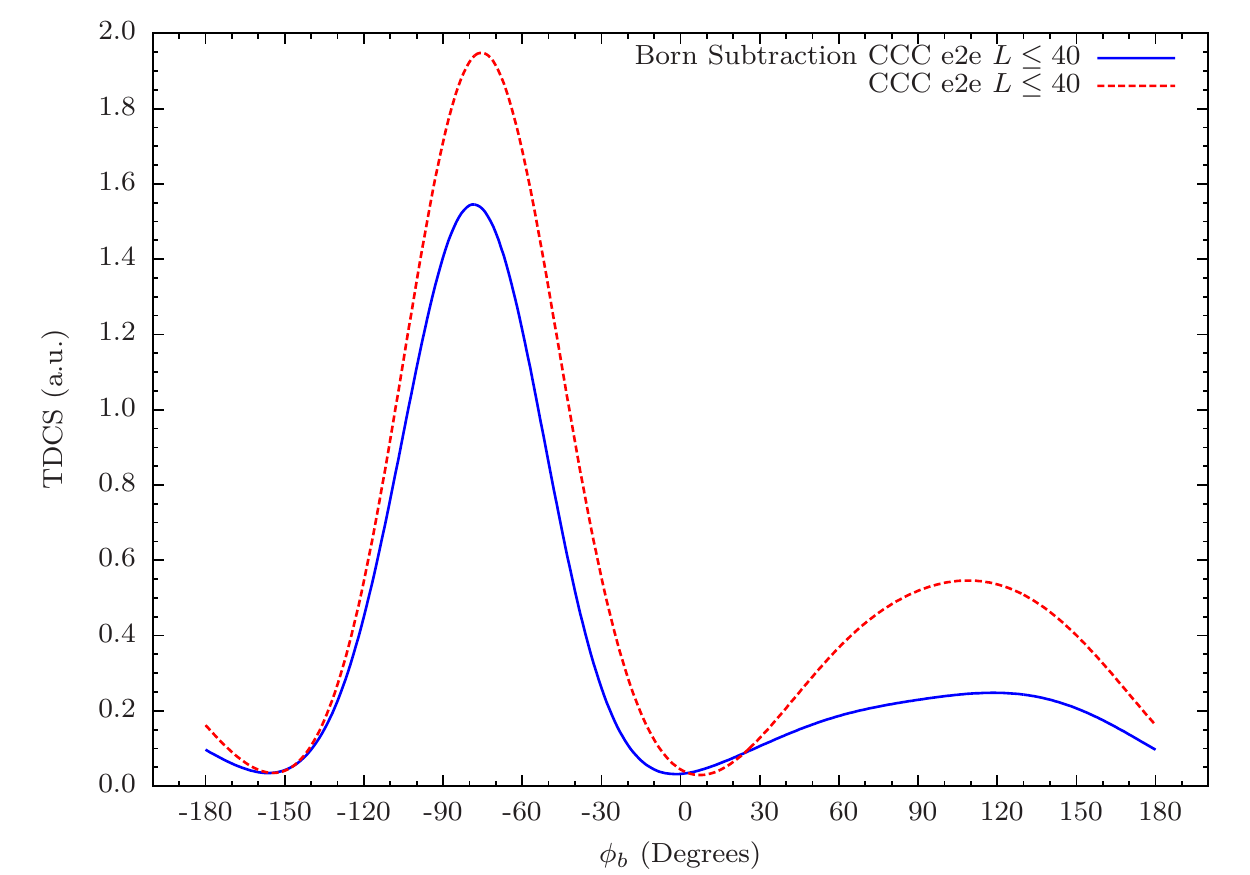}
		\caption[The triply differential cross section (TDCS) with ({blue}) and without ({red}) Born subtraction for 1 keV electrons incident on helium 
		within an asymmetric coplanar geometry of scattering angle $\phi_a=5^\circ$ ($q=0.756$ a.u.) and ejected electron energy $E_b=10$ eV.]
		{The triply differential cross section (TDCS) with ({blue}) and without ({red}) Born subtraction for 1 keV electrons incident on helium within an 
		asymmetric coplanar geometry of scattering angle $\phi_a=5^\circ$ ($q=0.756$ a.u.) and ejected electron energy $E_b=10$ eV. 
		Both calculations are comprised of 41 partial waves ($L\leq 40$) and contain states with $l\leq 4$ of corresponding Laguerre basis size $N_l=15-l$.}
		\label{BornSubtractionComp}
	\end{figure}

	The TICS as a function of partial wave for the calculational parameters as given above, is presented in \Cref{TICSvsPW}.
	From which, it is clear that convergence is very slow with $L$ as is typical for such a high incident energy.
	As such, we employ the method of Born subtraction described in \Cref{Born Subtraction}.
	This effectively assumes that beyond the final partial wave in the calculation the Born approximation is valid, and adds the corresponding analytic tail under this approximation.
	\Cref{BornSubtractionComp} contains the triply differential cross section (TDCS) generated from a calculation with and without Born subtraction that are otherwise identical.
	The large difference between the two results is indicative of the significant contribution from high partial waves (large $L$).
	The Born approximation is most appropriate when the strength of interaction is particularly weak, which is well satisfied at such a high incident energy.
	In this range, first order Born calculations typically produce results that are slightly larger than those of CCC but are otherwise the same. 
	Furthermore, this approximation becomes increasingly valid with larger $L$.
	The physical analogue of the partial wave $L$ is the impact parameter of the scattering event.
	Hence, a larger $L$ corresponds to a larger `distance' between the projectile and target and subsequently the interaction is weaker.
	Note that all results presented from this point onwards include Born subtraction.

\begin{figure}[htbp]
		\centering
		\includegraphics[scale=1.0]{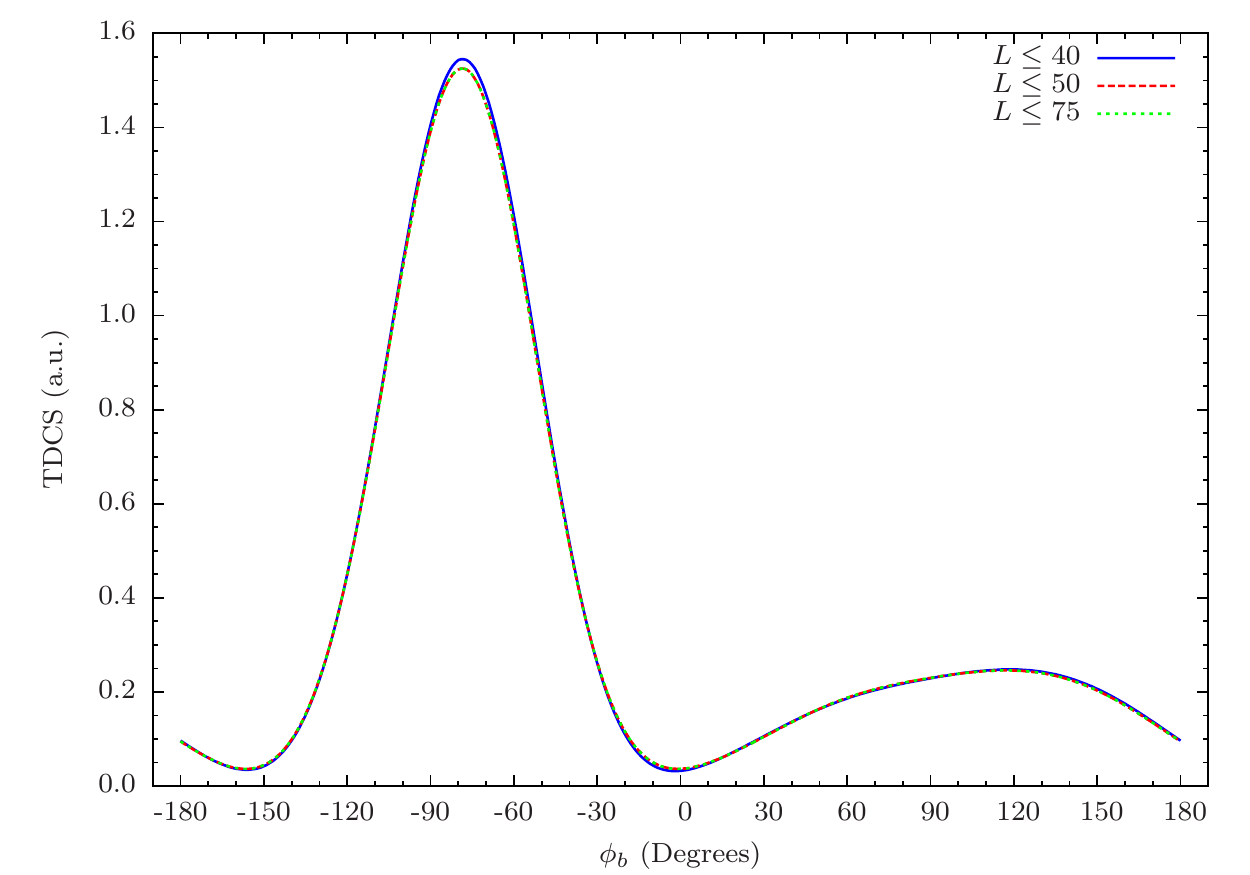}
		\caption[The triply differential cross section (TDCS) as calculated by CCC with $L\leq 40$, $50$, and $75$ for
		1 keV electrons incident on helium within an asymmetric coplanar geometry of scattering angle $\phi_a=5^\circ$ ($q=0.756$ a.u.) and ejected electron energy $E_b=10$ eV.]
		{The triply differential cross section (TDCS) as calculated by CCC with $L\leq 40$, $50$, and $75$ for
		1 keV electrons incident on helium within an asymmetric coplanar geometry of scattering angle $\phi_a=5^\circ$ ($q=0.756$ a.u.) and ejected electron energy $E_b=10$ eV.
		These calculations contain states of $l\leq 4$ each with a corresponding Laguerre basis of size $N_l=15-l$.}
		\label{LConvergence}
	\end{figure}

	\Cref{LConvergence} demonstrates convergence with $L$ by comparing calculations with $L\leq$ 40, 50, and 75.
	Observe that the $L\leq40$ calculation only exhibits subtle differences when compared to the 50 and 75, which themselves are indistinguishable at the precision inherent in the figure.
	As such, in subsequent calculations 51 partial waves ($L\leq 50$) are used.
	
\begin{figure}[htbp]
		\centering
		\includegraphics[scale=1.0]{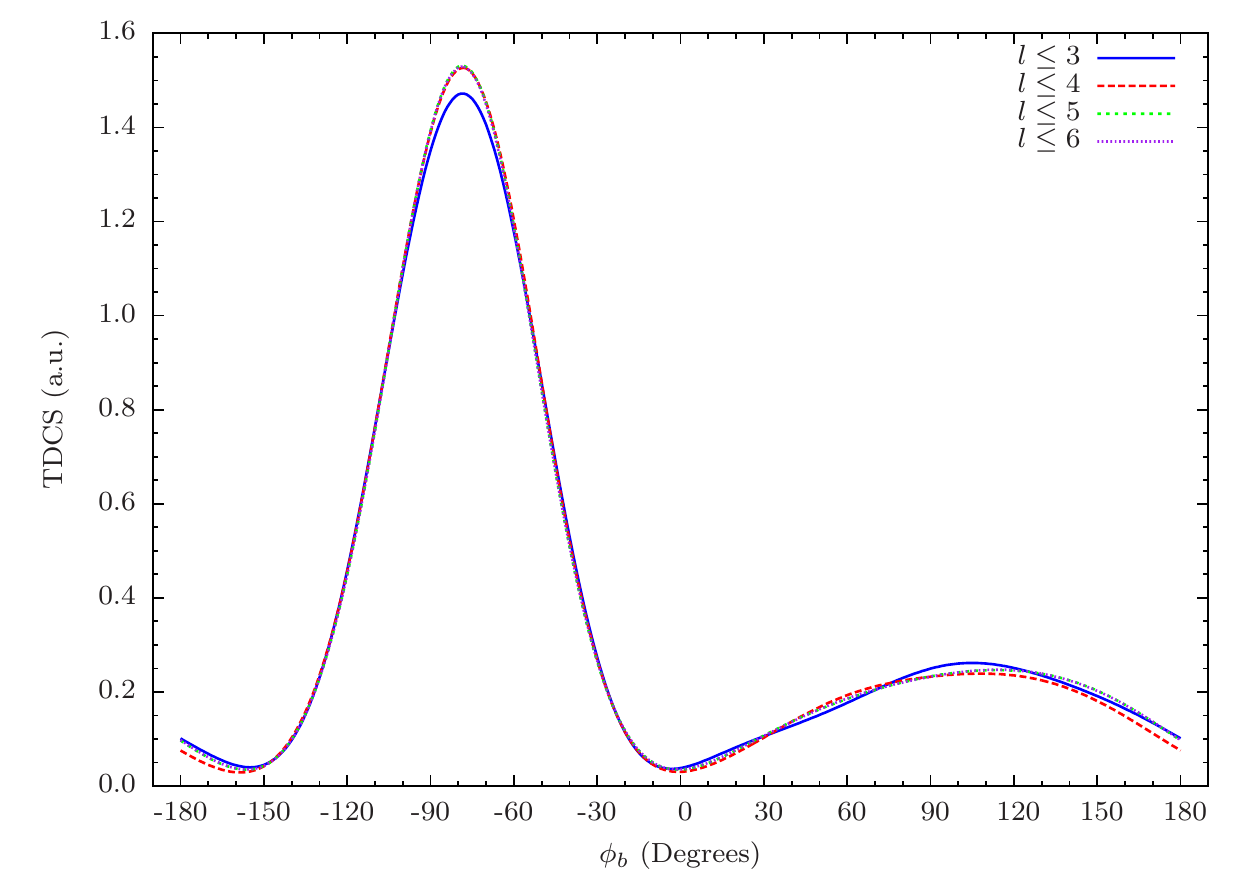}
		\caption[The triply differential cross section (TDCS) as calculated by CCC with $l\leq 3$, $4$, $5$, and $6$ for
		1 keV electrons incident on helium within an asymmetric coplanar geometry of scattering angle $\phi_a=5^\circ$ ($q=0.756$ a.u.) and ejected electron energy $E_b=10$ eV.]
		{The triply differential cross section (TDCS) as calculated by CCC with $l\leq 3$, $4$, $5$, and $6$ for
		1 keV electrons incident on helium within an asymmetric coplanar geometry of scattering angle $\phi_a=5^\circ$ ($q=0.756$ a.u.) and ejected electron energy $E_b=10$ eV.
		These calculations contain 51 partial waves ($L\leq 50$) with Laguerre bases of size $N_l=15-l$.}
		\label{LAConvergence}
	\end{figure}

	\enlargethispage{0.5cm}
	Now that we have determined an appropriate number of partial waves, we look to the $l$ of target states.
	\Cref{LAConvergence} demonstrates the behaviour of CCC calculations for including states of increasing $l$.
	There is a noticeable difference across both the binary and recoil peaks between the $l\leq 3$ and other results.
	The $l\leq 4$ result is in agreement across the binary peak but exhibits some discrepancy across the recoil peak compared to the $l\leq 5$ and $6$.
	Finally, the $l\leq 5$ is effectively identical to the $l\leq 6$ calculation and is thus considered to be converged for this value.
	Hence, for subsequent calculations states containing $l\leq 5$ are used.

\begin{figure}[htbp]
		\centering
		\includegraphics[scale=1.0]{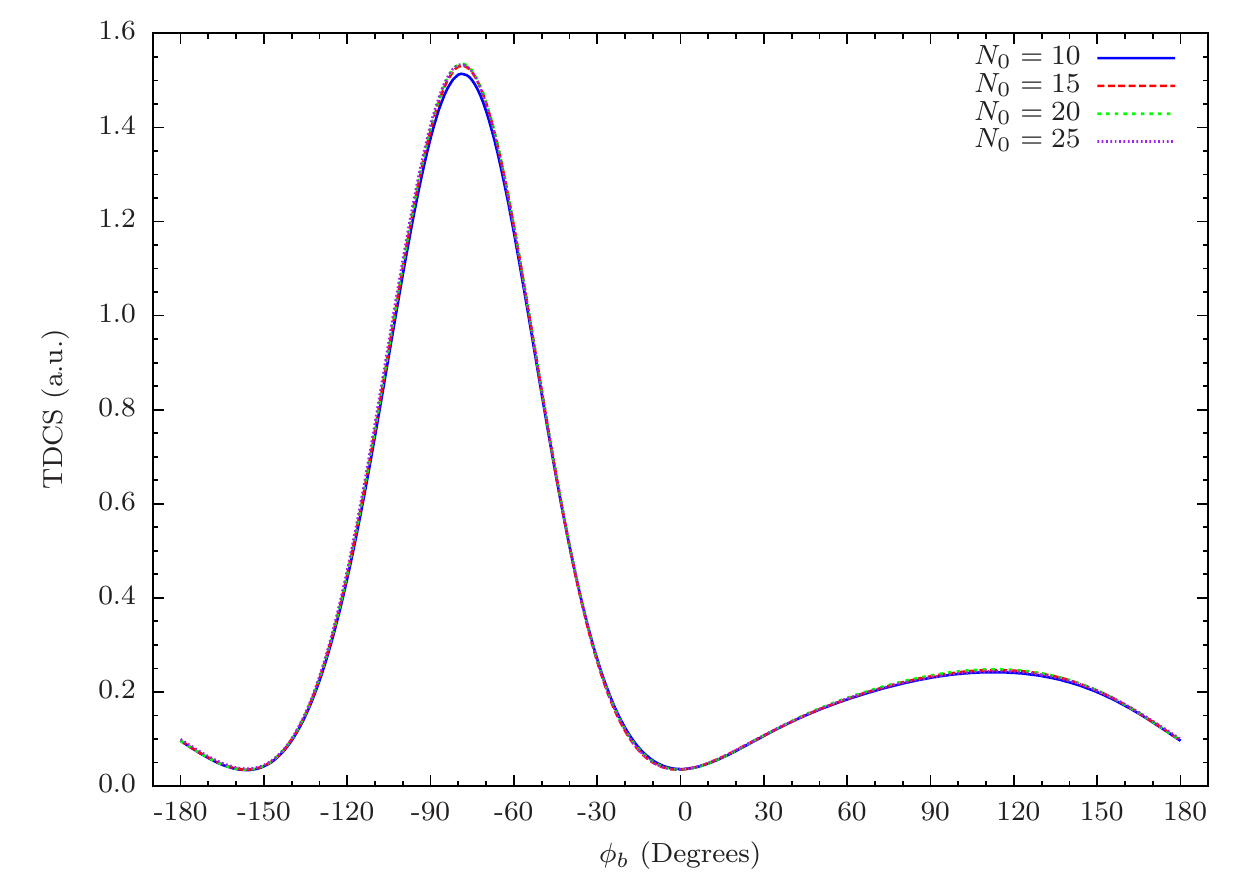}
		\caption[The triply differential cross section (TDCS) as calculated by CCC with Laguerre basis size of the form $N_l=N_0-l$ with $N_0=10$, $15$, $20$, and $25$ for
		1 keV electrons incident on helium within an asymmetric coplanar geometry of scattering angle $\phi_a=5^\circ$ ($q=0.756$ a.u.) and ejected electron energy $E_b=10$ eV.]
		{The triply differential cross section (TDCS) as calculated by CCC with Laguerre basis size of the form $N_l=N_0-l$ with $N_0=10$, $15$, $20$, and $25$ for
		1 keV electrons incident on helium within an asymmetric coplanar geometry of scattering angle $\phi_a=5^\circ$ ($q=0.756$ a.u.) and ejected electron energy $E_b=10$ eV.
		These calculations contain 51 partial waves ($L\leq 50$) and states with $l\leq 5$.}
		\label{NlConvergence}
	\end{figure}

	Now that we have determined both the number of partial waves, and the largest value of $l$ to include of the target states, we look to the size of Laguerre basis ($N_l$) for each value of $l$.
	We use the form of distributing states $N_l=N_0-l$ for the reasons discussed earlier in this section.
	With this definition, we only must check convergence with the choice of $N_0$, this being demonstrated in \Cref{NlConvergence}.
	Observe there is only minor variation with increasing $N_0$.
	The only noticeable difference occurs across the binary peak where the $N_0=10$ is lower than that predicted by the larger calculations.
	This is true to an even lesser extent for the $N_0=15$ calculation across this same peak.
	However, between the two largest calculations, the $N_0=20$ and $25$, there is no discernible difference at this precision.
	This observation leads us to conclude that convergence has been achieved for a value of $N_0=20$.
	Subsequently, for a basis of the form $N_l=N_0-l$, the value of $N_0$ to be used in further calculations is 20.
	\enlargethispage{0.5cm}
	\begin{figure}[htbp]
		\centering
		\includegraphics[scale=1.2]{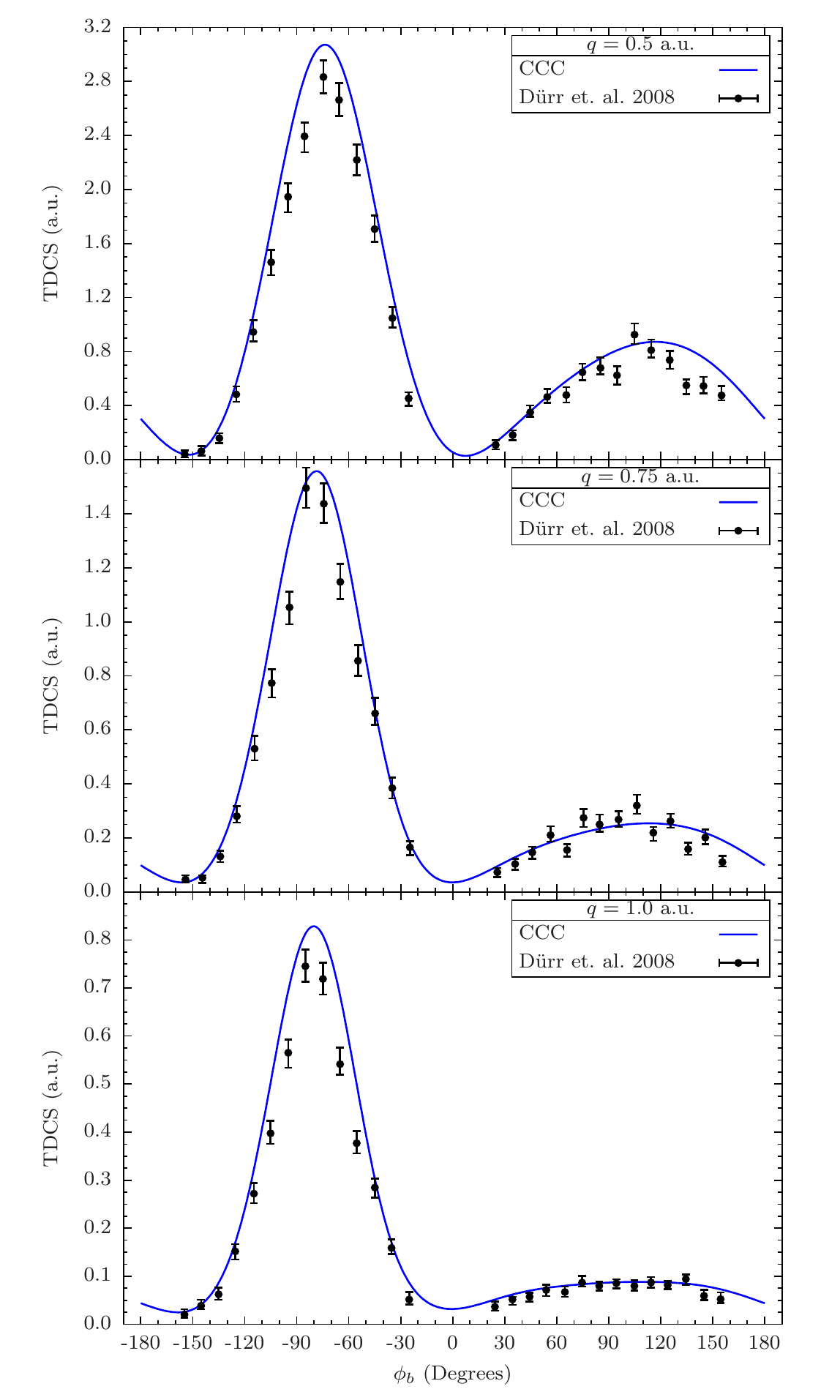}
		\caption[The triply differential cross section (TDCS) as calculated by CCC in comparison with experiment \cite{durr2008higher} for a momentum transfer of $q=0.5$, $0.75$, and $1.0$ a.u.\ of
		1 keV electrons incident on helium and an ejected electron energy of $E_b=10$ eV within an asymmetric coplanar geometry.]
		{The triply differential cross section (TDCS) as calculated by CCC in comparison with experiment \cite{durr2008higher} for a momentum transfer of $q=0.5$, $0.75$, and $1.0$ a.u.\ of
		1 keV electrons incident on helium and an ejected electron energy of $E_b=10$ eV within an asymmetric coplanar geometry.
		These calculations contain 51 partial waves ($L\leq 50$), states with $l\leq 5$, and a corresponding Laguerre basis size $N_l=20-l$.}
		\label{ExpComp}
	\end{figure}

	The final and ultimate check of convergence to physically relevant values is of course comparison with experiment.
	In \Cref{ExpComp} we present a comparison with the experimental values given by \citet{durr2008higher} with the CCC calculations of parameters determined above ($L\leq 50$, $l\leq 5$, and $N_0=20$)
	for three values of momentum transfer; $q=0.5$, $0.75$, and $1.0$ a.u.
	Cross sections of such forms are typical where there is a dominant binary peak in the direction of $\bm{q}$ and a smaller recoil peak in the direction of $-\bm{q}$, with their ratio increasingly
	favouring the binary peak for larger momentum transfer.
	The strong dependence of the overall magnitude of the cross sections with momentum transfer is indicative of the high incident energy projectile having a relatively low probability to interact
	strongly with the target, resulting in a significant deflection.
	The experiment is generally in good agreement with the calculations and is hence indicative of their convergence to that of the physical system.
	The most notable discrepancy is across the main binary peak where the calculation is slightly higher than the measured values.
	This was also found in the comparison with CCC made in the original paper \cite{durr2008higher}, and as such is not indicative that the results are not yet convergent.
	From these observations we conclude that our field-free calculations are now of a sufficient accuracy to move forward to the field-assisted case.
	\subsection{Field-Assisted Calculations}
	\label{e2eFieldAssisted}
	Ultimately we wish to evaluate
	\begin{alignat}{3}
		\frac{\mathrm{d}\sigma^{\mathrm{FA}}}{\mathrm{d}\Omega_a \mathrm{d}\Omega_b \mathrm{d}E_b}&=\sum_{n=-\infty}^\infty J^2_n\left(\alpha_n\right)\frac{\mathrm{d}\sigma}{\mathrm{d}\Omega_a
		\mathrm{d}\Omega_b\mathrm{d} E_b}\Big|_{E_f=E_i+n\omega}\;\;, \tag{\ref{SoftPhotonResult}}
	\end{alignat}
	and provide a comparison with the results of \citet{hohr2007laser}.
	The behaviour of the field-free cross section should not present any wild variation with $n$ (see \Cref{EbComp}), and as such we firstly examine the sum rule
	\begin{alignat}{3}
		\sum_{n=-\infty}^\infty J_n^2(\alpha_n)\approx 1 \tag{\ref{Bessel property alpha}}
	\end{alignat}
	which provides a useful first indication of the number of terms that are likely required to achieve convergence in \eqref{SoftPhotonResult}.
	For the geometries in the experiment by \citet{hohr2007laser} (see \Cref{e2e fig2}) $\alpha_n$, as defined in \eqref{alphadef}, is given by
	\begin{alignat}{3}
		\alpha_n=(k_b\sin\phi_b-k_a\sin\phi_a)F_0/\omega^2.
	\end{alignat}
	Recall that the $n$ dependence comes from the adjustment of $k_b$ or $k_a$ to satisfy energy conservation as discussed in \Cref{Soft Photon Approximation}.
	Note that $\alpha_n$ could equally be defined as $-\alpha_n$ depending on the sign of $\bm{F}_0$.
	\enlargethispage{0.5cm}
	However, this choice is arbitrary (as it should be) as we have the following property of Bessel functions \cite{NIST:DLMF}
	\begin{alignat}{3}
		J_n^2(z)&=J_n^2(-z),\quad \forall n\in\mathbb{Z},z\in\mathbb{C}.
	\end{alignat}

\begin{figure}[htbp]
		\centering
		\includegraphics[scale=1.2]{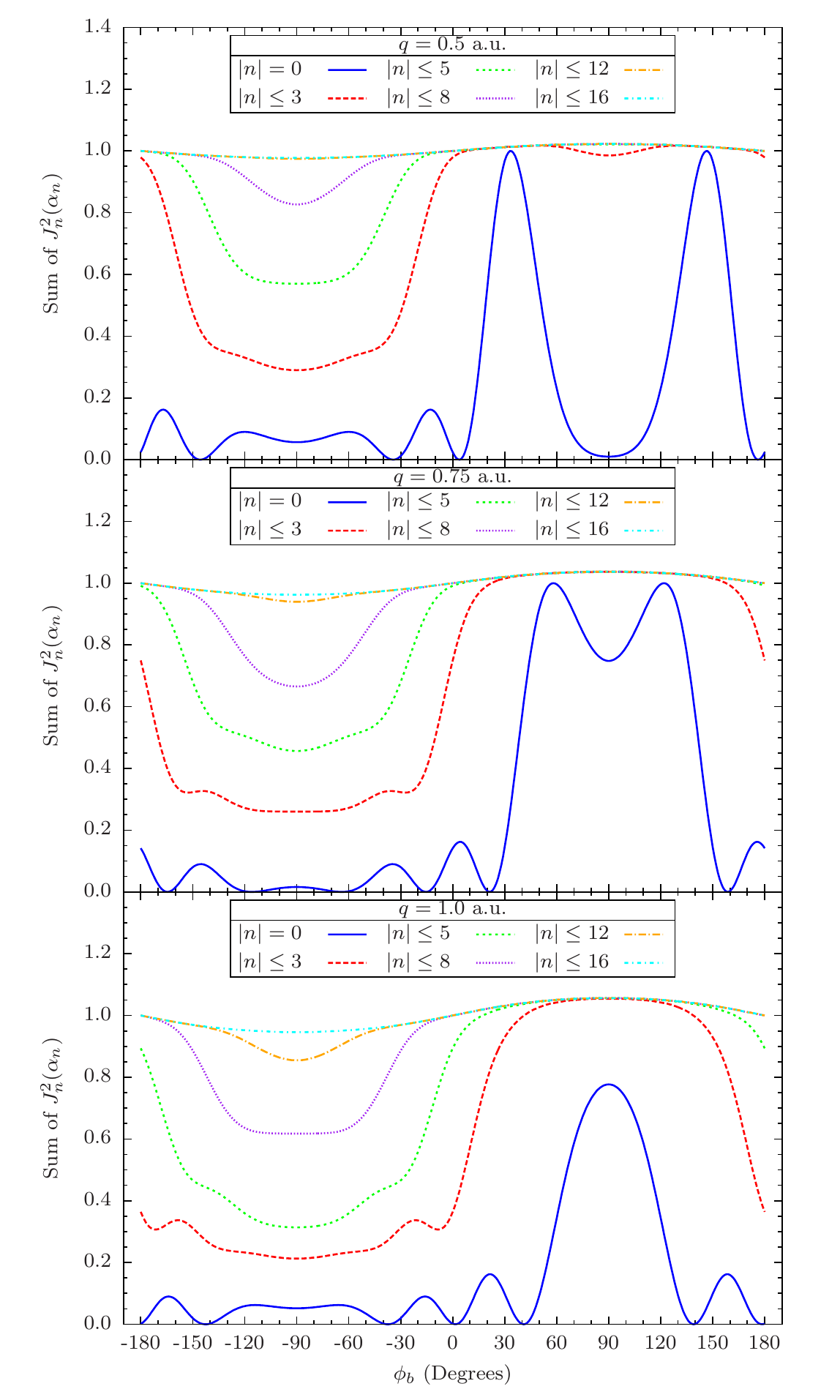}
		\caption[The sum of squared Bessel functions as in \eqref{Bessel property alpha} for $|n|\leq 0$, $3$, $8$, $12$, and $16$ of momentum transfers $q= 0.5$, $0.75$, and $1.0$
		a.u.]
		{The sum of squared Bessel functions as in \eqref{Bessel property alpha} for $|n|\leq 0$, $3$, $8$, $12$, and $16$ of momentum transfers $q= 0.5$, $0.75$, and $1.0$
		a.u. $\alpha_n$ \eqref{alphadef} is calculated by modifying the ejected electron energy by $n\omega$ with the exception for sufficiently large negative values of $n$ such that 
		further quanta of photon
		energies would cause it to become negative. For this case, further energy was removed from $k_a$ (see text).}
		\label{AlphaConvergence}
	\end{figure}

\begin{figure}[htbp]
		\centering
		\includegraphics[scale=1.2]{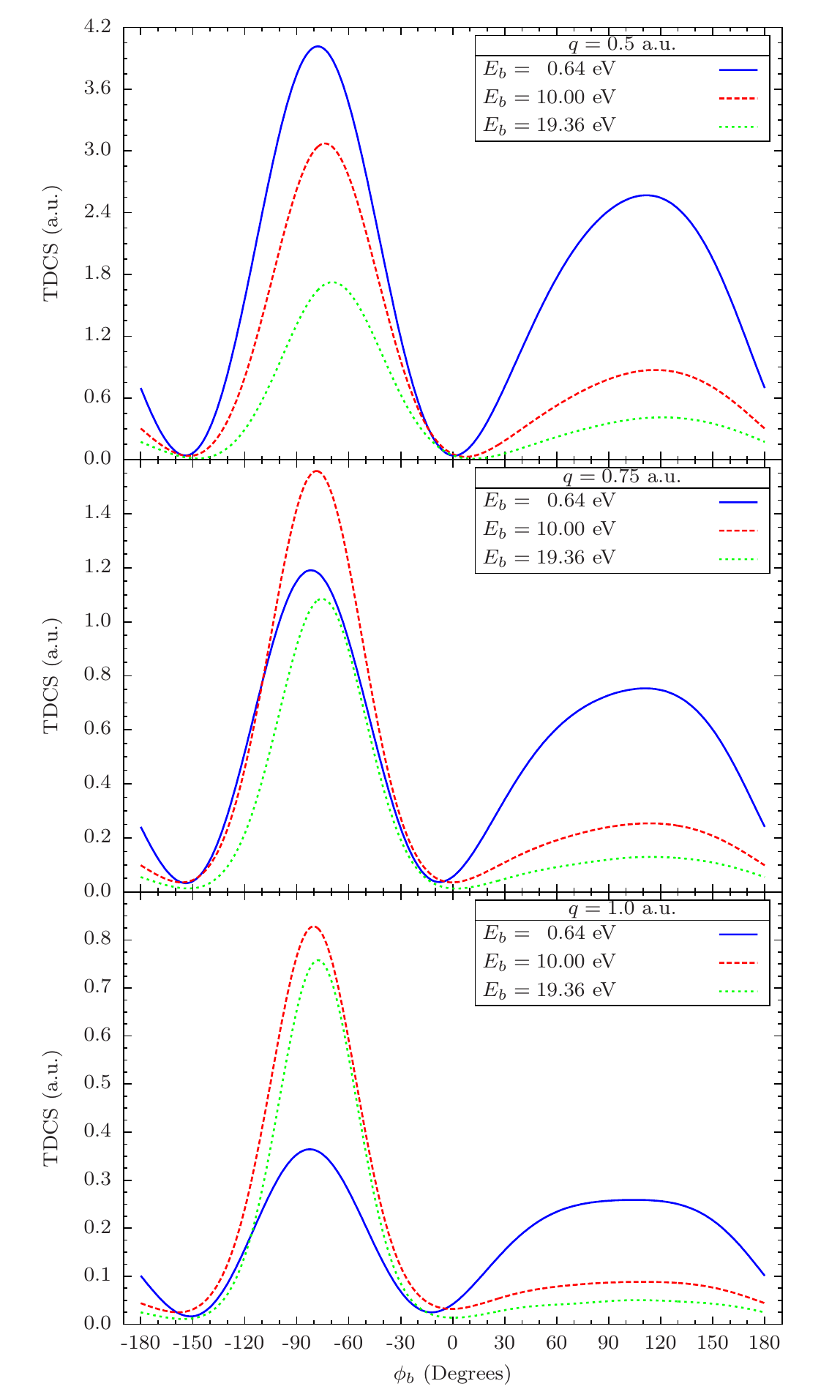}
		\caption[The triply differential cross section (TDCS) as calculated by CCC for a momentum transfer of $q$ = 0.5, 0.75, and 1.0 a.u.\ of 1 keV electrons
		incident on helium and an ejected electron energy of $E_b$ = 0.67, 10.0, and 19.36 eV within an asymmetric coplanar geometry, as used in the soft photon approximation sum 
		\eqref{SoftPhotonResult}
		for the terms $n=-8$, 0, and $8$ respectively.]
		{
		The triply differential cross section (TDCS) as calculated by CCC for a momentum transfer of $q=  0.5$, 0.75, and 1.0 a.u.\ of 1 keV electrons
		incident on helium and an ejected electron energy of $E_b$ = 0.67, 10.0, and 19.36 eV within an asymmetric coplanar geometry, as used in the soft photon approximation sum 
		\eqref{SoftPhotonResult}
		for the terms $n=-8$, 0, and $8$ respectively.
		}
		\label{EbComp}
	\end{figure}

	\Cref{AlphaConvergence} demonstrates the results of our convergence study of \eqref{Bessel property alpha} for the kinematics described in \Cref{exptparamtable}.
	For each value of $\alpha_n$ the ejected electron energy $E_b$ was adjusted by $n\omega$ with the exception for 
	sufficiently large negative values of $n$ such that further quanta of photon energies would cause it to become negative. 
	In this case, the further energy was removed from $k_a$.
	The ejected electron energy is adjusted preferentially for this system by the argument outlined in \Cref{Soft Photon Approximation} and $E_b$ is prevented from having negative values as the
	associated (e,2e) process is entirely unphysical.
	Observe that convergence is achieved in each case at a value of $|n|\leq 16$ to a sinusoidal function of $\phi_b$ about unity.
	A notable feature present in each case is that convergence is slowest across the angles corresponding to the binary peak (direction of $\bm{q}$) that is visible in the (e,2e) cross sections
	(see \Cref{ExpComp}).
	Though each Bessel function individually varies considerably, their sum compensates for these variations, producing a smooth and comparatively slowly varying function of $\phi_b$.
	It is interesting to note that the dependence of the argument with $n$ in this case preserves a sinusoidal behaviour with $\phi_b$ rather than being identically one as in \eqref{Bessel property}.
	Regardless, the sum is clearly convergent and hence suggests that for a similar number of terms \eqref{SoftPhotonResult} will likewise be convergent.

	Now we examine the behaviour of the (e,2e) ionisation cross section as a function of $n$ (equivalently $E_b$) to obtain further insight into the convergence of \eqref{SoftPhotonResult}.
	\Cref{EbComp} contains the results of calculations corresponding to $n=-8$, $0$, and $8$, with calculational parameters as determined in \Cref{Field Free}, for the three values of
	momentum transfer we are considering.
	Observe that over this large range of ejected projectile energies the magnitudes of the calculated cross sections do not vary considerably, providing yet further evidence suggesting that
	\eqref{SoftPhotonResult} should be convergent for the kinematics we consider.
	An interesting feature is that the recoil peak is considerably more pronounced for the lower ejected electron energy.
	This is indicative that when low amounts of energy are transferred to the ejected electron that being ejected in the $-\bm{q}$ direction and the residual ion being given $2\bm{q}$ 
	is a relatively more common process compared to a standard binary collision.
	Additionally, although the lower momentum transfer processes are the most likely, for a given momentum transfer the ejected electron energy which corresponds to the largest cross section varies.
	This is likely due the changing ratio of the energy of the ejected electron compared to the energy associated with the momentum transfer itself ($q^2/2$).
	In the case of large negative values of $n$ where further energy is removed from $k_a$, the cross section is changed negligibly as the incident energy of 1 keV is sufficiently high 
	such that removal of a few eV makes no significant difference to the scattering problem. 
	Hence, for the terms that require modification of $k_a$ the appropriate unadjusted cross section ($E_b$ is significantly lowered but importantly still positive) is used.
	
\begin{figure}[htbp]
		\centering
		\includegraphics[scale=1.2]{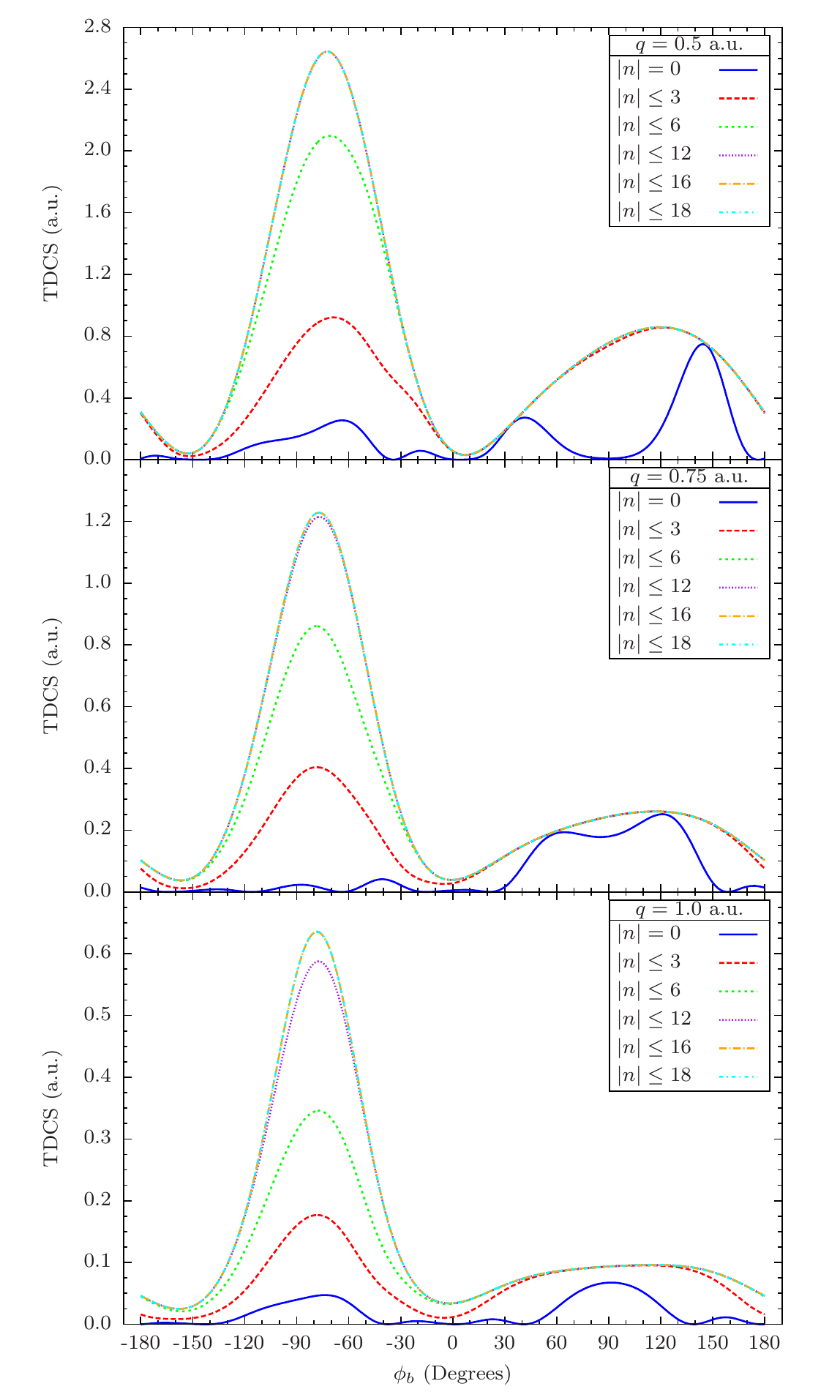}
		\caption[The sum of field-free triply differential cross sections weighted by squared Bessel functions as in the soft photon approximation for the field-assisted cross section
		\eqref{SoftPhotonResult} of momentum transfer $q=0.5$, $1.0$, and $1.0$ a.u., and including terms with $|n|\leq 0$, $3$, $6$, $12$, $16$, and $18$.]
		{The sum of field-free triply differential cross sections weighted by squared Bessel functions as in the soft photon approximation for the field-assisted cross section
		\eqref{SoftPhotonResult} of momentum transfer $q=0.5$, $1.0$, and $1.0$ a.u., and including terms with $|n|\leq 0$, $3$, $6$, $12$, $16$, and $18$.
		}
		\label{SoftPhotonConvergence}
	\end{figure}

	\clearpage
	We are now in a position to evaluate the field-assisted cross section via the soft photon approximation \eqref{SoftPhotonResult}.
	The results of our convergence study of \eqref{SoftPhotonResult} are given in \Cref{SoftPhotonConvergence}.
	Observe that in each case although the initial term is rather unrecognisable, including as few terms as in the $n\leq |3|$ results the expected double peak structure is again visible.
	With this many terms the recoil peak has already reached its converged value, whereas the main binary peak requires considerably more terms to do the same.
	The observation that large values of $|n|$ provide a significant contribution to the field-assisted cross section suggests that for this physical system multi-photon processes are prevalent,
	particularly for those scattering events with the electron ejected in the $\bm{q}$ direction.
	Note that these results indicate that convergence is slower with increasing momentum transfer, as was similarly found for the sum of the Bessel functions alone (see \Cref{AlphaConvergence}).
	In the case of $q=0.5$ a.u.\ convergence across the entire range of $\phi_b$ has occurred for $|n|\leq 12$ whereas for both the $q= 0.75$ and $1.0$ a.u.\ this has occurred for $|n|\leq 16$.
	Considering that for all three parameters we have convergence below $|n|\leq 18$ and that no additional calculation is necessary from this point onwards, we use the $|n|\leq 18$ results in
	subsequent comparisons.
	\begin{figure}[htbp]
		\centering
		\includegraphics[scale=1.2]{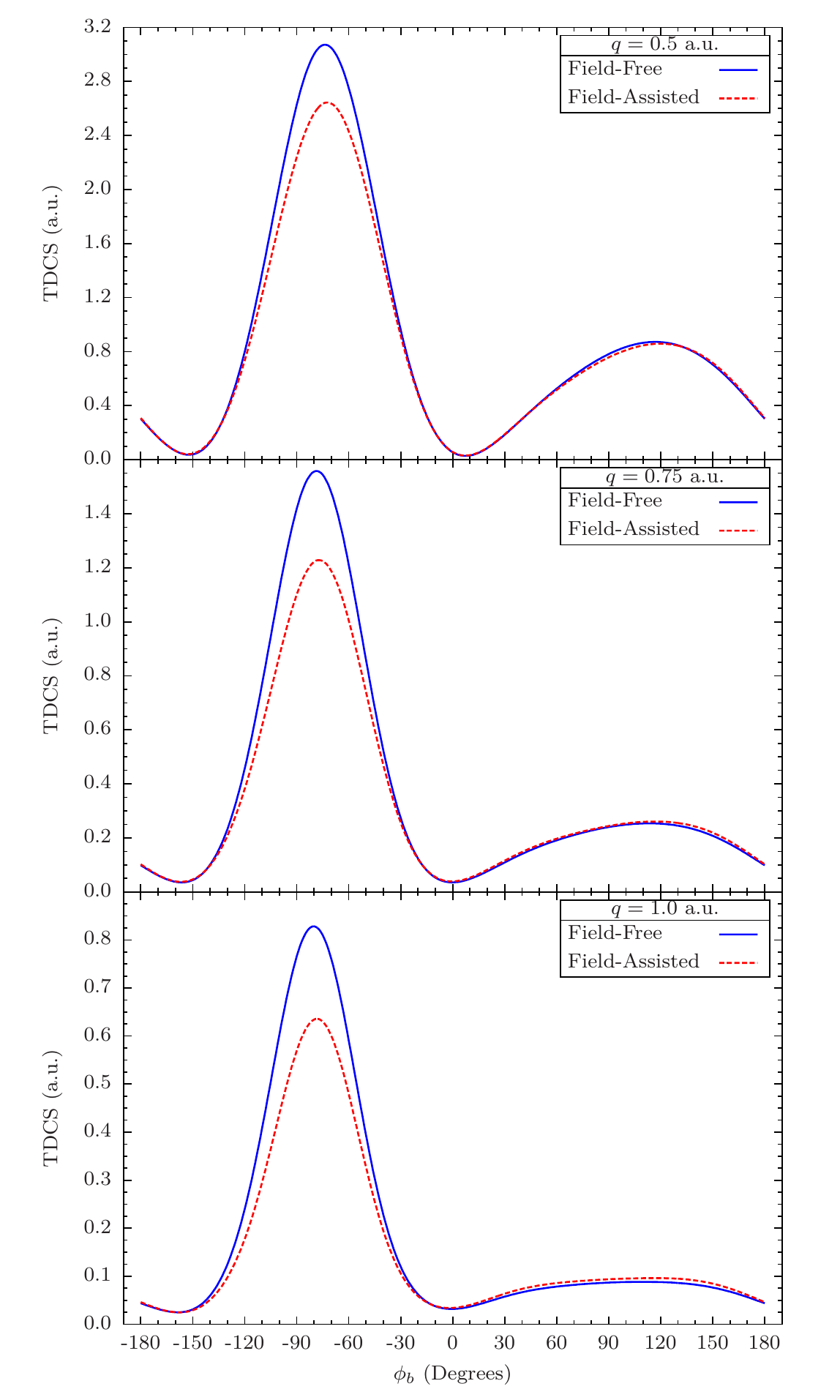}
		\caption[The field-free triply differential cross section (TDCS) as calculated by CCC and the field-assisted cross section as a result of our soft photon approximation calculations
		for a momentum transfer of $q$ = 0.5, 0.75, and 1.0 a.u.\ of 1 keV electrons
		incident on helium and an ejected electron energy of $E_b$ = 0.67, 10.0, and 19.36 eV within an asymmetric coplanar geometry.]
		{The field-free triply differential cross section (TDCS) as calculated by CCC and the field-assisted cross section as a result of our soft photon approximation calculations
		for a momentum transfer of $q$ = 0.5, 0.75, and 1.0 a.u.\ of 1 keV electrons
		incident on helium and an ejected electron energy of $E_b$ = 10.0 eV within an asymmetric coplanar geometry.}
		\label{SoftPhotons}
	\end{figure}

\begin{figure}[ht]
		\centering
		\includegraphics[scale=1.2]{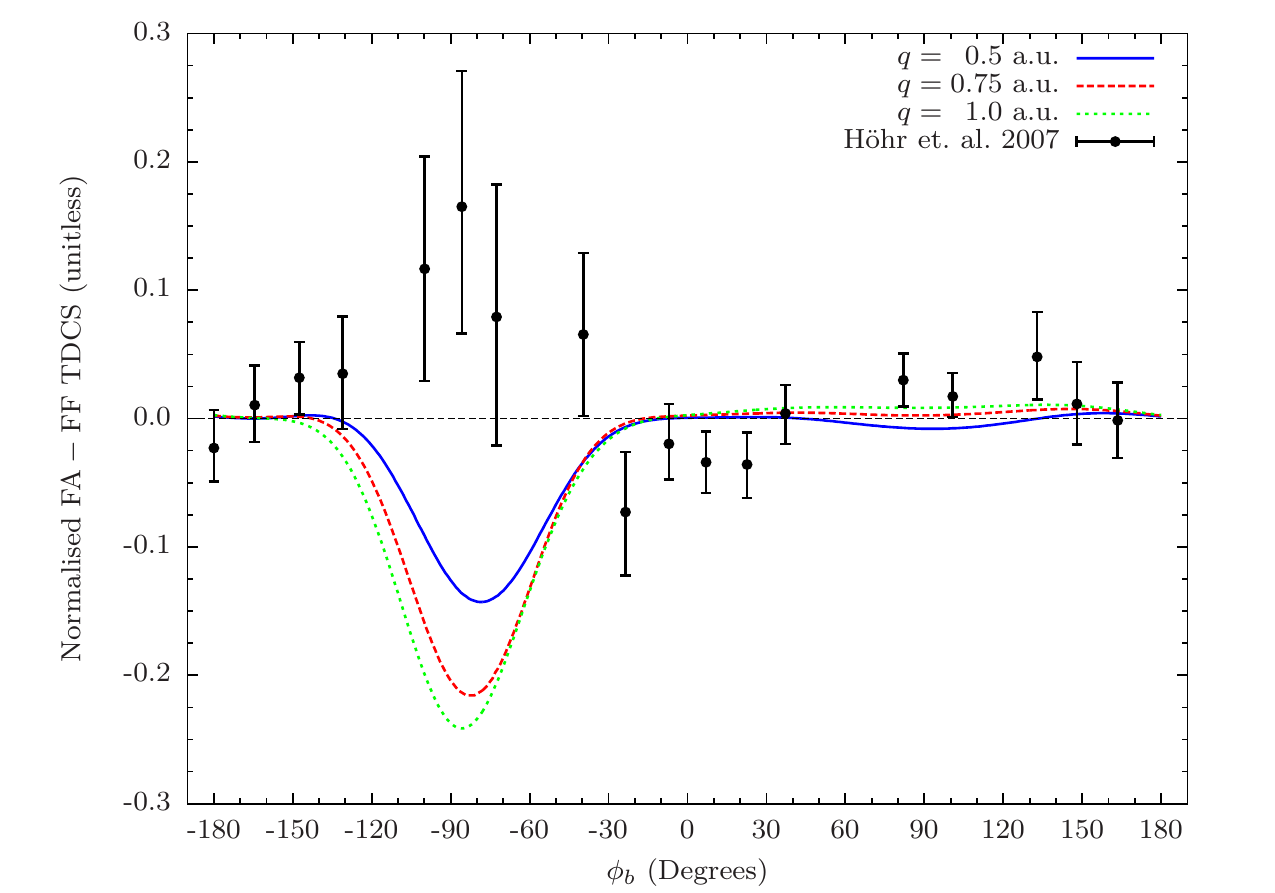}
		\caption
		[The normalised difference of the field-assisted (FA) and field-free (FF) TDCS of 1 keV electrons incident on helium 
		as calculated by CCC within the soft photon approximation for momentum transfers $q= 0.5$, $0.75$, and $1.0$ a.u.\ 
		with an ejected electron energy of $E_b=10$ eV
		in comparison with the experimental results of \citet{hohr2007laser} for momentum transfers of $q=0.7-1.0$ a.u.\ and ejected electron energies of $E_b=7-12$ eV
		within an asymmetric coplanar geometry.]
		{The normalised difference of the field-assisted (FA) and field-free (FF) TDCS of 1 keV electrons incident on helium 
		as calculated by CCC within the soft photon approximation for momentum transfers $q= 0.5$, $0.75$, and $1.0$ a.u.\ 
		with an ejected electron energy of $E_b=10$ eV
		in comparison with the experimental results of \citet{hohr2007laser} for momentum transfers of $q=0.7-1.0$ a.u.\ and ejected electron energies of $E_b=7-12$ eV
		within an asymmetric coplanar geometry.
		Each data set is normalised to their respective binary peak magnitude for comparison with one another.
		The experimental data is of arbitrary magnitude and scaled to fit the theory presented in the original paper \cite{hohr2007laser}.
		Here it is likewise normalised to its binary peak magnitude.
		The black dashed horizontal line across a difference of zero is to aid in discerning enhancement from diminution.
		}
		\label{SoftPhotonsExpComp}
	\end{figure}

	The converged results of the soft photon calculation in comparison to the field-free cross section as calculated by CCC are given in \Cref{SoftPhotons}.
	In all three cases there is a prominent diminution across the binary peak with little to no difference elsewhere.
	Otherwise, the forms of each laser assisted cross section retain the familiar form of the binary and recoil peaks.
	The influence of the laser is more clearly seen in \Cref{SoftPhotonsExpComp} which contains the difference between the field-assisted (FA) and field-free (FF) TDCS normalised to the 
	binary peak height in comparison with the experimental results of \citet{hohr2007laser}.
	These experimental values are originally on an arbitrary scale but have been scaled to fit their theory in their paper \cite{hohr2007laser}.
	Here they have likewise been normalised to the magnitude of the binary peak.
	Given that the experimental results are given over a range of momentum transfers, $q=0.7$ to $1.0$ a.u., and ejected electron energies, $E_b=7$ to $12$ eV, our calculations with $E_b=10$ eV and
	$q=0.5$, $0.75$, $1.0$ a.u.\ respectively should provide an appropriate comparison for this data set.
	Across the binary peak we see that the $q=0.5$ a.u.\ calculation has the smallest predicted relative decrease compared to those of the $q=0.75$ and 1.0 a.u.
	There is a further decrease between the $q=0.75$ and 1.0 a.u.\ but less so than before, suggesting that this effect decreases with increasing momentum transfer.
	The other region of interest occurs across the recoil peak, although the magnitude of the laser's influence is considerably lesser.
	Here we find that the $q=0.5$ a.u.\ suggests a small decrease but for the $q=0.75$ a.u., an increase, and for the $q=1.0$ a.u., a yet smaller increase.
	This appears to be the inverse behaviour as exhibited across the binary peak but on a considerably smaller scale.

	The most striking result however, is that the results of each calculation predict the opposite effect across the binary peak as that of the experiment.
	In this region despite large uncertainties in the presented data there is a strong indication that the presence of the laser is causing an enhancement of the cross section whereas the calculation 
	instead suggests diminution.
	This is identically the result of the theory presented within their paper \cite{hohr2007laser} for each set of parameters considered and which led them to conclude that the 
	existing theory is lacking a fundamental aspect of the physics.
	From our results we deduce that the area lacking in the description does not appear to be in the description of the field-free scattering as the cross sections calculated via CCC solve
	for this interaction to all orders.
	In many ways this is not a particular surprise as at such a high incident incident energy the Born approximation is typically found to be quite successful, and this was used as the basis for the
	field-free calculations within the original paper \cite{hohr2007laser}.
	Additionally, in the earlier theoretical work of \citet{Khalil1997} they state that including second order Born corrections is likely to produce a lesser effect than the approximations made in the
	description of the laser interaction.
	Essentially this is what we have confirmed however for instead including the effect of the entire Born series.
	Hence, we conclude that to reconcile this discrepancy a more nuanced description of the laser's influence upon the system is required and not in that of the field-free scattering.
	This is likely most necessary for the slow ejected electron of which is most heavily influenced.
	Finally, it is worth stating that the precision of the experimental measurements is unfortunately too coarse for a sensible comparison to be made with the behaviours across the recoil peak.
	If anything, the experiment suggests a small magnitude increase which is in better agreement with the larger $q$ calculations.

\section{Wigner Time Delay of \Hm and He}
\label{Photoemission time delay and electron scattering time delay}
	In this section we investigate the photoemission time delay \eqref{tauph} for the photodetachment of H$^-$ in comparison to the photoionisation of He.
	Additionally, we compare this delay to the elastic scattering time delay \eqref{tauel} of the associated photon free processes, this being elastic scattering on H and He$^+$ respectively in the
	dipole singlet channel.
	Particular attention is given to the behaviours across the $n=2$ threshold for the H$^-$ ion at $10.2$ eV (3/4 Ry) where the difference between these two time delays becomes significant due to the
	different first order inter-electron couplings.
	
	The CCC calculation for both targets included a Laguerre basis size allocated for each $l$ of the form $N_l=N_0-l$.
	The \Hm calculation (equivalently elastic scattering on H) included states with $l\leq 2$ and $N_0=20$.
	The He calculation (equivalently elastic scattering on He$^+$) included states with $l\leq 3$ and $N_0=15$.
	These values were determined for convergence in a similar manner to that demonstrated in \Cref{Field Free}.
	Do note that the incident energies considered here (0.01 - 100 eV) are considerably lower than previously (1 keV) and as such techniques including Born subtraction are no longer of any
	significant benefit.
	However, as we are conducting calculations near threshold the new numerical formulations of the treatment of the singularity within the CCC method \cite{bray2015solving,bray2016solving} 
	are utilised, including those developed for charged targets as part of this project.

\begin{figure}[htbp]

\vspace{-1.0cm}

\hspace{0.06cm}
\includegraphics[scale=2.45]{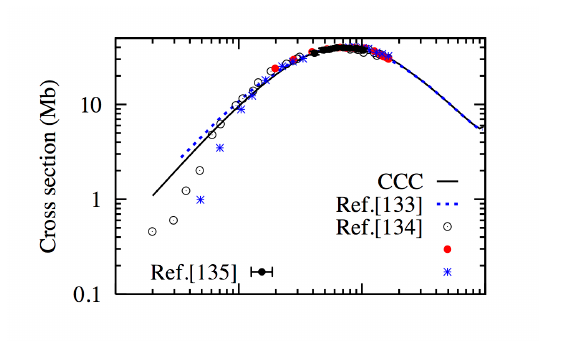}\\[-1.4cm]
\includegraphics[scale=2.0]{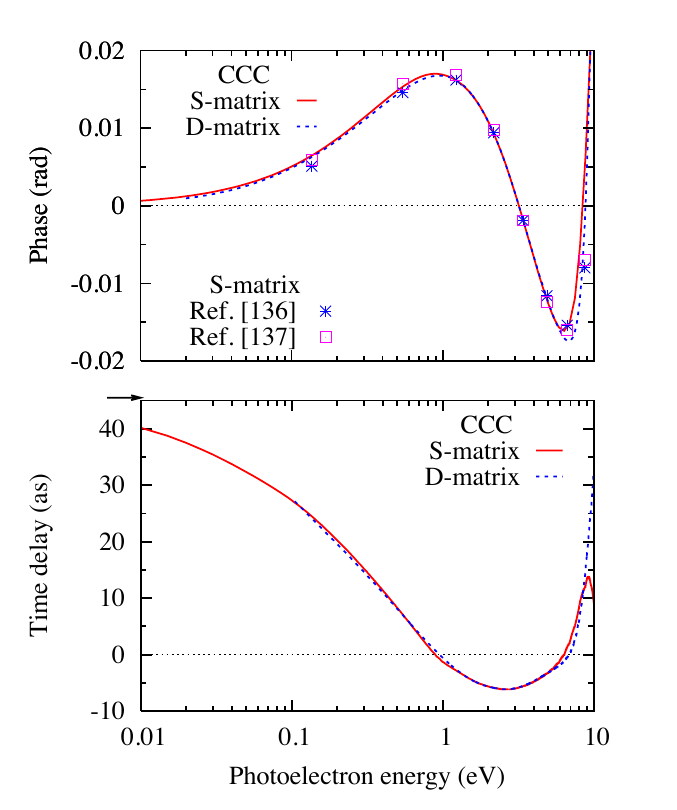}

\vspace{-0.4cm}

\caption[The photodetachment cross section (top), phase (middle) and time delay (bottom) of H$^{-}$ at photoelectron energies approaching the threshold at $10.2$ eV.]
	{The photodetachment cross section (top), phase (middle) and time delay (bottom) of H$^{-}$ at photoelectron energies approaching the threshold at $10.2$ eV.
	The cross sections calculated by CCC
  (black solid line) are compared with a $B$-spline
  calculation \cite{0953-4075-30-21-020} ({blue} dashed line), and the
  experimental data collated in \cite{POPP1976683} (original plotting symbols) and \cite{PhysRevA.91.033403} (black points).
  The $S$-matrix phase and time delay are for e-H scattering within the dipole singlet channel and are compared with literature values \cite{REGISTER1975431,PhysRevA.56.2435} for the phase.
  The $D$-matrix values are the calculated photoemission phase and time delay.
  The arrow indicates the limit of zero photoelectron energy.
}
\label{Fig1}
\end{figure}

	\Cref{Fig1} contains results for the photoemission cross section, phase, and time delay of \Hm at energies approaching the $n=2$ threshold at $10.2$ eV.
	In the top panel containing the cross sections we find that the CCC calculation agrees well with the $B$-spline calculation of \citet{0953-4075-30-21-020}.
	Generally good agreement is also seen with the experimental data collated in \citet{POPP1976683} with the exception of below $0.1$ eV.
	The newer set of experimental results \cite{PhysRevA.91.033403} additionally exhibits good agreement with the theoretical predictions. 
	The middle panel contains the phases as calculated from the $D$-matrix (photoemission phase) and $S$-matrix (electron scattering phase) in comparison with those calculated by the
	variational calculation of \citet{REGISTER1975431} and the hyperspherical close-coupling calculation of \citet{PhysRevA.56.2435}.
	All phase calculations are in good agreement.
	The indistinguishability of the photoemission and electron scattering phases across this energy range indicates that these are largely equivalent processes when there is yet to be an
	excitation channel with any significant contribution.
	Do note the variation that begins in the vicinity of the $n=2$ excitation threshold.
	The bottom panel contains the time delays of the corresponding phases which are simply their derivatives with respect to energy.
	The $D$-matrix results are for the photoemission time delay \eqref{tauph} and the $S$-matrix is the electron scattering time delay \eqref{tauel}, 
	both of which are again indistinguishable with the exception of the behaviour that begins to exhibit itself approaching a photoelectron energy of $10.2$ eV.

	The scattering phase shift is expected to follow the Wigner threshold law for a short range potential $\delta_l\propto E^{l+1/2}$ \cite{LLQuantum}.
	For a spherically symmetric target such as \Hm the $p$ wave ($\delta_1\propto E^{3/2}$) will define the photoemission time delay. As such 
	it is expected to vanish at the threshold of zero photoelectron energy as $\tau_{\mathrm{ph}}\propto E^{1/2}$.
	This was found to be the case for the frozen core Hartree-Fock calculation using the code of \citet{CCR79}.
	However, the inclusion of the polarisation potential $V(r)=\beta/r^4$ changes this result significantly.
	Doing so, yields a scattering phase of $\delta_1=(2\pi\beta/15)E$ and hence a constant time delay $\tau_{\mathrm{ph}}=2\pi\beta/15$ \cite{0953-4075-33-5-201}.
	Taking the CCC calculated value at a photoelectron energy of $10^{-5}$ eV (corresponding to the arrow in the bottom panel of \Cref{Fig1}) gives a time delay of 45.5 as. 
	This suggests $\beta\simeq 4.48$, which is in good agreement with expected analytical result of 9/2.

\begin{figure}[htbp]
\vspace{-1.0cm}
\hspace{-0.4cm}
\includegraphics[scale=2.2]{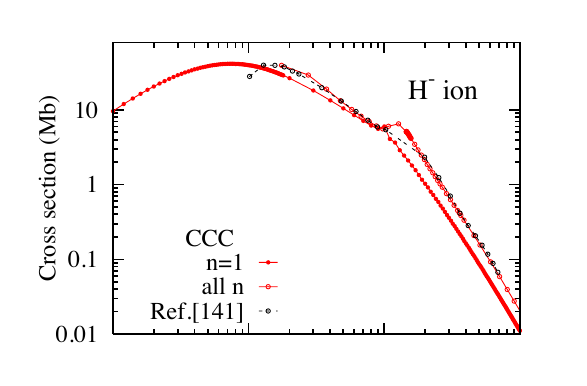}\\[-1.8cm]
\includegraphics[scale=2.49]{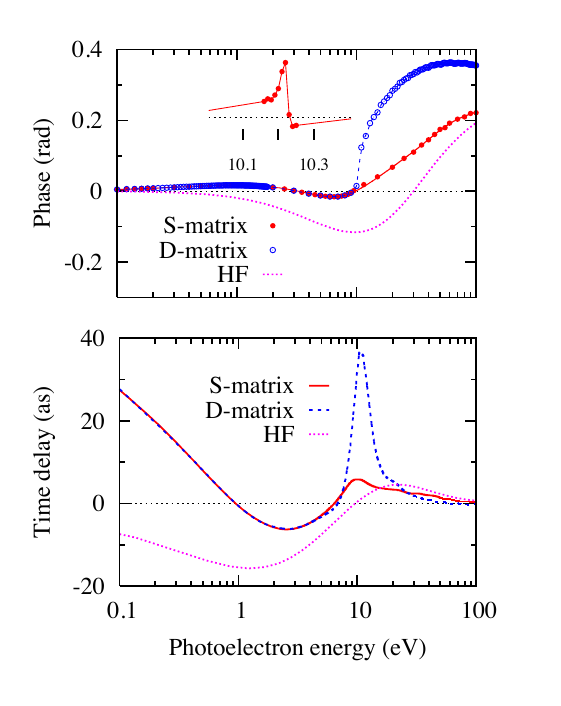}

\vspace{-0.2cm}
\caption[The photodetachment cross section (top), phase (middle), and time delay (bottom) of H$^{-}$ over an extended range of photoelectron energies.]
	{The photodetachment cross section (top), phase (middle), and time delay (bottom) of H$^{-}$ over an extended range of photoelectron energies.
	 Presented are the partial cross section leaving the target in the ground state $n_i=1$ ({red} filled circles) and the total cross section summed over $n_i<5$ ({red} 
	 open circles) 
	 in comparison with experiment \cite{BR76} (black circles). 
  	The $S$-matrix phase and time delay are for e-H scattering within the dipole singlet channel, the $D$-matrix values are the photoemission phase and time delay, and
	are compared with a Hartree-Fock frozen core calculation \cite{CCR79}.
	The inset highlights the $S$-matrix phase in the vicinity of the threshold for $n=2$ excitation at 10.2 eV.
}
\vspace{-1.0cm}
\label{Fig2}
\end{figure}

\begin{figure}[htbp]
\vspace{-0.43cm}
\hspace{-0.05cm}
\includegraphics[scale=2.16]{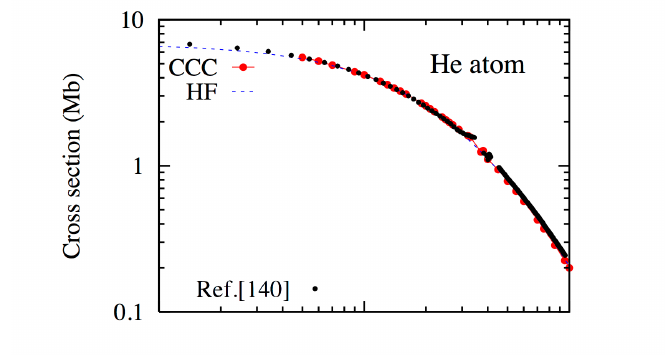}\\[-1.5cm]
\includegraphics[scale=2.47]{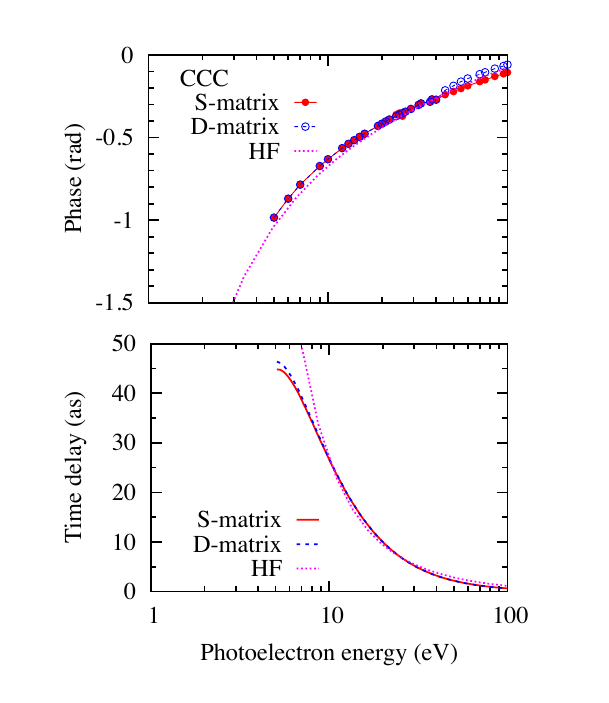}

\vspace{-1.0cm}
\caption[The photoionisation cross section (top), phase (middle) and time delay (bottom) of He over an extended range of photoelectron energies for a comparison with those of H$^-$.]
	{The photoionisation cross section (top), phase (middle) and time delay (bottom) of He over an extended range of photoelectron energies for a comparison with those of H$^-$.
	The total cross section that calculated for the ground state single ionisation of He by CCC ({red} circles), compared with experimental values \cite{SHYH94} (black points), 
	and a frozen core
	Hartree-Fock calculation \cite{CCR79} ({blue} dashed line).
  	The $S$-matrix phase and time delay are for e-He$^+$ scattering within the dipole singlet channel and the $D$-matrix values are the calculated photoemission phase and time delay.
}
\label{Fig3}
\end{figure}

	Now we look to examine the differences between the photoemission and electron scattering time delays when an additional excitation channel is present.
	We do so by considering both the \Hm and He targets over an extended energy range.
	This is because although they are both two electron systems, their ground state configurations are substantially different, yielding contrasting behaviours in terms of exhibited time delays.
	The photoemission cross section, phase, and time delay of H$^{-}$ and He are given in \Cref{Fig2} and \Cref{Fig3} respectively.
	We begin discussion with He as its results are altogether simpler. 
	Firstly note that the photoionisation cross section for He leaving the target in the $n=1$ ground state is indistinguishable from the total cross section (summed over all final states) 
	\cite{SHYH94}.
	This is indicative of the overwhelming dominance of this channel.
	In the top panel we compare the calculated cross sections of CCC, the frozen core Hartree-Fock calculation using the code of \citet{CCR79}, and the experimental values of \citet{SHYH94} all of
	which are in good agreement.
	Similarly, both definitions of phase and time delay along with the Hartree-Fock calculation produce highly similar results.
	This is indicative of the independent electron treatment within the Hatree-Fock basis being sufficient for both the photoionisation and elastic scattering processes due to the effect of 
	electron-electron correlations being negligible, as is often found to be the case when dealing with helium.
	This is a testament to the general applicability of the frozen core model to the helium atom.

	The results for \Hm are markedly different given in \Cref{Fig2}.
	The photodetachment channel leaving the hydrogen core in the $n=1$ is the largest contributor, but beyond the first excitation threshold it only accounts for
	approximately half of the total cross section.
	CCC results for the total cross sections are in good agreement with the experimental values of \citet{BR76}.
	However, the phase as calculated by the Hartree-Fock approach \cite{CCR79} exhibits considerable differences to those calculated from both the $D$- and $S$- matrices.
	This is indicative of a strong core polarisation effect that is unable to be accounted for within a frozen core model.
	Additionally, the photoemission phase and scattering phase begin to deviate significantly beyond the opening of the first excitation channel.
	Right on threshold the $S$-matrix phase experiences a narrow resonance in the dipole singlet channel of e-H elastic scattering 
	due to the quasi-bound state of the \Hm ion \cite{friedrich2012theoretical}.
	This resonance is magnified in the inset of the middle panel.
	Immediately above threshold both phases grow linearly with energy, however at a much greater rate for the $D$-matrix.
	This results in a sharp peak in the photoemission time delay as high as $\simeq 40$ as, a delay that is large enough to be readily measurable using existing attosecond chronoscopy techniques
	\cite{schultze2010delay,PhysRevLett.106.143002}.
	The equivalent peak in the scattering time delay is around a factor of 10 smaller.
	Each peak can be associated with the opening time of the newly accessible photoemission or inelastic scattering channel.
	With further increasing energy the photemission delay falls faster than the scattering delay, though at no point do they return to being indistinguishable.

\begin{figure}[htbp]
		\includegraphics[scale=0.8]{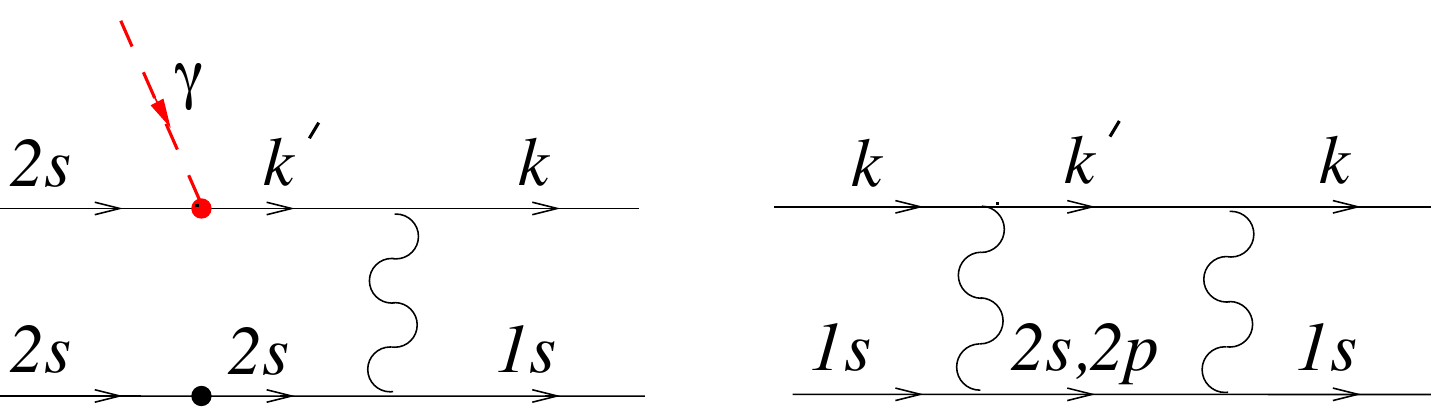}
		\caption
		[Graphical representation of the lowest order interactions involving virtual excitation to $n=2$ in the photodetachment of \Hm (left) and elastic scattering on H (right).]
		{Graphical representation of the lowest order interactions involving virtual excitation to $n=2$ in the photodetachment of \Hm (left) and elastic scattering on H (right).}
		\label{feynmann2}
	\end{figure}

	The large disparity between the photoemission and scattering time delays can be explained by the different lowest order electron-electron interaction involved in the 
	mixing of the ground and excited states.
	\Cref{feynmann2} contains two Feynmann diagrams which illustrate these interactions involving virtual excitation of the core to $n=2$ for photoemission (left) and elastic scattering (right).
	For elastic scattering on hydrogen, virtual excitation requires two successive interactions between the projection electron and the target; the first to excite and the second to de-excite.
	Hence, the only contribution to the time delay comes from a second order process.
	For photodetachment of the hydrogen ion however, the initial state contains a significant fraction ($\sim 20\%)$ of the 2$s^2$ configuration in its ground state of which only requires a single
	interaction to de-excite to the ground state.
	As such, there is a considerable contribution to the time delay in this case from a first order process, which produces the larger effect seen in the middle and bottom panels of \Cref{Fig2}
	across the opening of the $n=2$ excitation channel.
	For helium there is no significant fraction of 2s$^2$ in the ground state and as such we see no comparable effect in \Cref{Fig3}.

\chapter{Conclusions}
	Within this work we have investigated the discrepancy between theory and experiment presented by \citet{hohr2007laser} for the laser assisted electron impact ionisation of helium through
	implementing the soft photon approximation for the laser assisted cross section within the convergent close-coupling (CCC) method.
	In doing so we have shown that using the field-free cross sections as calculated by CCC is equally unable to reproduce the result of their experiment and is in fact consistent with 
	their theory which applies the first Born approximation to the field-free scattering.
	Hence, we have concluded that it is rather the description of the laser interaction that is the cause of the presented discrepancy, likely in that of the slow emitted electron which is heavily
	influenced by the laser field.
	This is the current consensus held within recent works \cite{GhoshDeb2010,Makhoute2015} which have employed more nuanced descriptions of the target dressing effects and field-free scattering than
	the original H\"ohr \textit{et al}.\ paper \cite{hohr2007laser}, although in the field-free case not to the extent offered by the CCC.
	Nonetheless, neither of these works are able to rectify the presented discrepancy.

	Additionally, we have investigated the photoemission time delay of the \Hm ion in comparison with the elastic scattering time delay of the associated photon free process of the elastic 
	elastic scattering of an electron on H within the dipole singlet channel. 
	Particular focus is given to the behaviours resulting from the opening of the $n=2$ excitation channel of \Hm at $10.2$ eV.
	For a comparison with these results for \Hm we also examined the photoionisation of He and the associated processes of elastic scattering on He$^+$.
	We found that across this excitation threshold there was considerable growth in the exhibited time delay which was approximately an order of magnitude larger in the photoemission case
	compared with that of the associated elastic scattering and attributed this effect to the ground state correlation present in \Hm.
	The peak delay was found to reach $\simeq 40$ as and is of a sufficiently high magnitude to be potentially measured using existing attosecond streaking \cite{schultze2010delay} and 
	interferometric \cite{PhysRevLett.106.143002} methods.
\section{Further Work}
\label{Further Work}
	Initially it was hoped to be able to apply the newly developed soft photon code with CCC to other problems such as to the elastic scattering problems 
	where this approximation has been already shown to work
	\cite{Wein1977,Wein1979,wallbank1987experimental} with the notable exceptions of small scattering angles \cite{wallbank1993,geltman1996laser}.
	However, typically the field-free to field-assisted ratio involving a specific number of photons is considered, which within this approximation is independent of the scattering target.
	Hence, there is no room for contribution from the use of the more comprehensive field-free scattering treatment.
	Additionally, we can potentially improve upon the approximations which omit target dressing effects of the laser atom interaction using existing perturbation techniques that have been applied to
	the problem \cite{Khalil1997}.
	Despite making these improvements however, it is unlikely to effectively tackle this discrepancy as this approach has already been applied in attempt to explain the same laser assisted electron 
	impact ionisation experiment \cite{GhoshDeb2010}.
	The consensus in the literature of which our results also support is that a more nuanced treatment is required in the description of the laser interaction 
	than that which is offered by the soft photon approximation.
	In order to provide such a treatment we hope either to utilise and extend the TDSE code developed by \citet{Patchkovskii2016153} for single electron systems 
	or develop our own code to incorporate multi-electron interactions. 
	We have seen as a result of our time delay calculations, the electron-electron correlations play a significant role in the time delay calculations. 
	As such, the development of
	a method that is able to take into this effect while solving for an atomic system interacting with an ultrashort laser pulse would constitute a significant breakthrough.

	With our discovery of the large enhancement of the time delay with the opening of a new reaction channel, particularly in the case of photodetachment (equivalently photoionisation), we expect to
	see such behaviours in a number of different targets other than those originally considered.
	For example, in the photodetachment of the Li$^-$ ion near the $2$ $^2P$ threshold \cite{PhysRevA.45.1544} and the photoionisation of metastable helium near the $n=3$ threshold \cite{KIAB00}.
	Similar effects attributed to electron correlation have already been observed in noble gases \cite{DAMOP2016}.
	Furthermore, in molecules near the threshold of new dissociation channels we again expect to see this kind of enhancement.
	Particularly ideal targets to observe this kind of enhancement however, are singly negative ions 
	as the measured time delay will not be affected by the laser coupling with the Coulomb field of the resulting ion \cite{Dahlstrom2012,C3FD00004D}.
	This field of attosecond spectroscopy is currently in its infancy and accordingly there are a vast number of interesting problems to consider both theoretically and experimentally.
	In addition to applying our established methodology to other atomic and molecular targets, the development of a time dependent code that is able to incorporate electron-electron correlations
	would potentially enable the solution of these time delay problems from another perspective.

	The most promising way for us to proceed is to devise a method based upon the numerical solution of the time dependent Schr\"odinger equation, as this offers 
	the most detailed and accurate theoretical description of laser driven electron dynamics.
      However, brute force solutions of this equation are only feasible for simple systems with one or two active electrons.
      Even for two electron systems driven by long wavelength laser radiation, finding such a
      solution presents a formidable challenge as the required computational resources increase very rapidly with the wavelength \cite{PhysRevA.93.023406}.
      More complex systems with many active electrons are outside the limits of such an approach.
      In order to overcome these complications, in our formulation we look to
      include the correlation effects into the ionisation and/or recombination stage of the laser driven electron dynamics, whereas the propagation of the ionised
	electrons in the laser field would remain restricted to one or two active particles.
	This propagation can be solved non-perturbatively by close-coupling methods already available for field-free propagation \cite{PhysRevA.57.318}. 
	In doing so we hope to take into account the many body effects of the electron correlations without causing the problem to become beyond the capabilities of available computational resources.
	If successful, this approach will allow us 
	to tackle such challenging problems as the non-sequential double ionisation of two electron systems (such as the He atom and H$_2$ molecule) by long wavelength radiation \cite{PhysRevA.93.023406},
	and that of resonantly enhanced harmonic generation in noble gas atoms \cite{0953-4075-39-13-S03} in addition to the problems investigated within this work. 
\clearpage
\thispagestyle{bibstyle}
\appendix
\chapter{Atomic Units}
\label{atomic units}
	The system of atomic units comes from setting the value of the electronic charge $e$, electronic mass $m_e$, reduced Planck constant $\hbar$, and Coulomb constant $k_e=1/(4\pi\epsilon_0)$ to unity.
	Accordingly we write 
	\begin{alignat}{3}
		e=m_e=\hbar=\frac{1}{4\pi\epsilon_0}=1.
	\end{alignat}
	Doing so allows the Hamiltonian of atomic systems to be written in a natural form.
	For example, the Hamiltonian of a hydrogen atom written in atomic units is simply
	\begin{alignat}{3}
		H=-\frac{1}{2}\nabla^2-\frac{1}{r}
	\end{alignat}
	and the resulting energy eigenstates are
	\begin{alignat}{3}
		E_n=\frac{1}{2n^2}
	\end{alignat}
	in units of Hartree (the atomic unit of energy) $E_h=27.2$ eV. 
	The ground state ($n=1$) energy defines the Rydberg as $1/2$ Hartree.
	The radii of the corresponding orbits within the Bohr model are given by 
	\begin{alignat}{3}
		r_n=n^2
	\end{alignat}
	in units of Bohr radii (the atomic unit of length) $a_0=5.29\times10^{-11}$ m.
	From the definition of these two units, through their combination and the addition of the constants set to unity, all other quantities in atomic units can be constructed.
	Cross sections are then given in units of $a_0^2$ which is equivalent to 28 Mb.
	Electric fields are given in units of $E_h/(ea_0)$ of which 1 a.u.\ is the equivalent of $5.14\times 10^{11}$ V/m.
	This is the strength of electric field in a hydrogen atom at the radii of orbit in the ground state.
	Similarly, time is given in units of $\hbar/E_h$ of which 1 a.u.\ is the equivalent of 24.2 as.

	A consequence of setting $\hbar$ to unity is that the de Broglie wave vector becomes analogous to momentum and similarly frequency with energy.
	For example, an electron with kinetic energy given in Hartrees has momentum in atomic units given by $p=k=\sqrt{2E}$.
	Furthermore, angular momentum becomes effectively unitless.

	The dimensionless fine structure constant $\alpha$ retains its magnitude in any system of units 
	\begin{alignat}{3}
		\alpha=\frac{e^2}{4\pi\epsilon_0 \hbar c}\approx\frac{1}{137}.
	\end{alignat}
	Hence, in atomic units $c\approx137$.
	
	\thispagestyle{bibstyle}
	The relation between laser intensity $I$ and peak electric field strength $F_0$ in a vacuum is given in atomic units by
	\begin{alignat}{3}
		I=\frac{F_0^2}{8\pi\alpha}.
	\end{alignat}
	The intensity that corresponds to an electric field strength of 1 a.u.\ in more common units is $3.5\times 10^{16}$ W/cm$^2$.
	At this intensity the force due to the laser becomes equal to that of the atomic field for a hydrogen atom.
	\begin{table}[htbp]
	\renewcommand{\arraystretch}{1.2}
	\centering
	\begin{tabular}{l|l|rl}
		\hline
		Quantity& Expression& \multicolumn{2}{l}{Value in other units}\\
		\hline
		Energy& $E_h=m_e e^4/(4\pi\epsilon_0\hbar)^2=m_e\alpha^2c^2$& 27.21 &eV\\
		Length& $a_0=4\pi\epsilon_0\hbar^2/(m_e e^2)=\hbar/(m_ec\alpha)$& 0.529 &\r{A}\\
		Cross Section& $a_0^2$& 28&Mb\\
		Time  & $\hbar/E_h$ &24.2 &as \\
		Velocity& $a_0E_h/\hbar=\alpha c$& 1/137 &c\\
		Momentum &$m_ea_0E_h/\hbar=\hbar/a_0$&$3.73$ &keV/c\\
		Angular Momentum&$ m_ea_0^2E_h/\hbar=\hbar$&$1.97$ &keV \r{A}/c\\
		Electric Field&$E_h/(ea_0)$&\hspace{0.8cm}$51.4$ &V/\r{A}\\
		Intensity&$E_h^2/(\hbar a_0^2)$&\hspace{0.8cm}$0.64$ &W/\r{A}$^2$\\
		\hline
	\end{tabular}
	\caption{Common quantities in the system of atomic units.}
	\label{atomictable}
	\end{table}

	\noindent In \Cref{atomictable} `\r{A}' are Angstroms (10$^{-10}$ m), `b' are barns $10^{-28}$ m$^2$, and `as' are attoseconds $10^{-18}$ s.
	\clearpage
	\thispagestyle{plain}
\chapter{Derivations}
\label{Derivations}
	Here we include some derivations that, although useful to the understanding of key equations, were deemed too lengthy and/or inappropriate for the main body of text.
	\section{Lippmann-Schwinger Equations}
	\label{CCCder}
	We begin with the Schr\"odinger equation
	\begin{alignat}{3}
		\tag{\ref{CCCshrod}}
		(E-H)|\Psi_i^{S(N+)}\rangle=0
	\end{alignat}
	and look to derive an expression for the transition amplitude given by
	\begin{alignat}{3}
		\big\langle\bm{k}_f\phi_{n_f}^{({N})}\big|T^S\big|\phi_{n_i}^{({N})}\bm{k}_i\big\rangle\equiv\big\langle\bm{k}_f\phi_{n_f}^{({N})}\big|V^S\big|\Psi_i^{S(N+)}\big\rangle.
	\end{alignat}
	Firstly we group terms in the Hamiltonian into the asymptotic part $H_a$ and the interaction potentials $V^S$.
	Note that the $S$ dependence of $V$ comes from symmetrisation terms included within it as in \eqref{VS} rather than the potentials themselves being spin dependent.
	In doing so we may now write \eqref{CCCshrod} as
	\begin{alignat}{3}
		V^S\big|\Psi_i^{S(N+)}\big\rangle&=(E-H_a)\big|\Psi_i^{S(N+)}\big\rangle.
	\end{alignat}
	Now projecting by $\big|\phi_n^{(N)}\big\rangle$ we have 
	\begin{alignat}{3}
		\big\langle \phi_n^{(N)}\big|V^S\big|\Psi_i^{S(N+)}\big\rangle&=\big\langle \phi_n\big|(E-H_a)\big|\Psi_i^{S(N+)}\big\rangle\\
		&=\big\langle \phi_n^{(N)}\big|(E-H_a)\sum_{m=1}^N \big|\phi^{(N)}_m\big\rangle\big\langle\phi^{(N)}_m\big|\Psi_i^{S(N+)}\big\rangle\\
		&=(E-\epsilon_n^{(N)}-k^2/2)\sum_{m=1}^N \delta_{mn}\big\langle\phi^{(N)}_m\big|\Psi_i^{S(N+)}\big\rangle\\
		&=(E-\epsilon_n^{(N)}-k^2/2)\big\langle\phi^{(N)}_n\big|\Psi_i^{S(N+)}\big\rangle.
	\end{alignat}
	Here we have used the fact that 
	\begin{alignat}{3}
		H_a\big|\phi^{(N)}_n\big\rangle=(\epsilon_n^{(N)}+k^2/2)\big|\phi_n^{(N)}\big\rangle
	\end{alignat}
	with $\epsilon_n^{(N)}$ being the asymptotic energy of the target in the $n$-th state and similarly $k^2/2$ for the projectile.
	Invoking the Green's function approach we can write
	\begin{alignat}{3}
		\big\langle\phi^{(N)}_n\big|\Psi_i^{S(N+)}\big\rangle&=\delta_{ni}\big|\bm{k}_i\big\rangle+\frac{\big\langle \phi_n^{(N)}\big|V^S\big|\Psi_i^{S(N+)}\big\rangle}{E+i0-\epsilon_n^{(N)}-k^2/2}\\
		\label{almostCCC}
		&=\delta_{ni}\big|\bm{k}_i\big\rangle+\int \mathrm{d}^3k \;\big|\bm{k}\big\rangle\frac{\big\langle \bm{k}\phi_n^{(N)}\big|V^S\big|\Psi_i^{S(N+)}\big\rangle}{E+i0-\epsilon_n^{(N)}-k^2/2}
	\end{alignat}
	where $i0\equiv\lim_{\epsilon\to 0^+}i\epsilon$ is added to ensure outgoing spherical wave boundary conditions (see \Cref{Contourder}).
	The term $\delta_{ni}\big|\bm{k}_i\big\rangle$ comes from the value of $\big\langle\phi^{(N)}_n\big|\Psi_i^{S(N+)}\big\rangle$ on the asymptotic boundary.
	Finally by multiplying both sides of \eqref{almostCCC} by 
	\begin{alignat}{3}
		\sum_{n=1}^N\big\langle \bm{k}_f\phi_f^{(N)}\big|V^S\big|\phi_n^{(N)}\big\rangle 
	\end{alignat}
	we have
	\begin{alignat}{3}
		\big\langle \bm{k}_f\phi_f^{(N)}\big|V^S\big|\Psi_i^{S(N+)}\big\rangle&=\sum_{n=1}^N\big\langle \bm{k}_f\phi_f^{(N)}\big|V^S\big|\phi_n^{(N)}\big\rangle\big
		\langle\phi^{(N)}_n\big|\Psi_i^{S(N+)}\big\rangle\\
		&=\big\langle \bm{k}_f\phi_f^{(N)}\big|V^S\big|\phi_i^{(N)}\bm{k}_i\big\rangle\nonumber\\
		&+\sum_{n=1}^N\int \mathrm{d}^3k \frac{\big\langle \bm{k}_f\phi_f\big|V^S\big|\phi_n^{(N)}\bm{k}\big\rangle
		\big\langle \bm{k}\phi_n^{(N)}\big|V^S\big|\Psi_i^{S(N+)}\big\rangle}{E+i0-\epsilon_n^{(N)}-k^2/2}
	\end{alignat}
	or equivalently
	\begin{alignat}{3}
		\big\langle \bm{k}_f\phi_f^{(N)}\big|T^S\big|\phi_i^{(N)}\bm{k}_i\big\rangle&=\big\langle \bm{k}_f\phi_f^{(N)}\big|V^S\big|\phi_i^{(N)}\bm{k}_i\big\rangle\nonumber\\
		&+\sum_{n=1}^N\int \mathrm{d}^3k \frac{\big\langle \bm{k}_f\phi_f\big|V^S\big|\phi_n^{(N)}\bm{k}\big\rangle\big\langle
		\bm{k}\phi_n^{(N)}\big|T^S\big|\phi_i^{(N)}\bm{k}_i\big\rangle}{E+i0-\epsilon_n^{(N)}-k^2/2}.
	\end{alignat}
	\thispagestyle{fancystyle}
	\clearpage
\section{Scattering Amplitude}
\label{Contourder}
	Here we look to provide a derivation for the relationship between the $T$-matrix and the scattering amplitude for elastic scattering.
	Let us consider a particle scattering on a short ranged potential $V$.
	The corresponding Schr\"odinger equation can be written as
	\begin{alignat}{3}
		(E-K)|\psi\rangle=V|\psi\rangle
	\end{alignat}
	where $E=k^2/2>0$ is the total energy of the system and $K$ is the kinetic energy of the projectile.
	Utilising the Green's function approach we may write the wavefunction as
	\begin{alignat}{3}
		|\psi\rangle&=|\bm{k}\rangle+\frac{V}{E-K}|\psi\rangle\\
		&=|\bm{k}\rangle+\int \mathrm{d}k'^3\frac{|\bm{k}'\rangle\langle\bm{k}'|V|\psi\rangle}{E-k'^2/2}
	\end{alignat}
	where $\bm{k}$ is defined such that $(E-K)|\bm{k}\rangle=0$.
	If we now project onto coordinate space and recognise that $\mathrm{d}k^3\equiv k^2 \mathrm{d}k\;\mathrm{d}\hat{\bm{k}}$ we can write for large $r$
	\begin{alignat}{3}
		\la\bm{r}|\psi\rangle&=\la\bm{r}|\bm{k}\rangle+\int_0^\infty \mathrm{d}k' \;k'^2\frac{\la\bm{r}|\bm{k}'\rangle\langle\bm{k}'|V|\psi\rangle}{E-k'^2/2}\\
		&=(2\pi)^{-3/2}\exp(i\bm{k}\cdot\bm{r})\nonumber\\
		&+(2\pi)^{-3/2}\int \mathrm{d}\hat{\bm{k}}\int_0^\infty \mathrm{d}k'\;k'^2\frac{\exp(i\bm{k}'\cdot\bm{r})}{k^2/2-k'^2/2}\langle\bm{k}'|V|\psi\rangle\\
		&=(2\pi)^{-3/2}\exp(i\bm{k}\cdot\bm{r})\nonumber\\
		\label{thing5}
		&+\frac{1}{ir\sqrt{2\pi}}
		\int_0^\infty \mathrm{d}k'\;2k'\frac{\exp(ik'r)\langle \hat{\bm{r}}k'|V|\psi\rangle-\langle -\hat{\bm{r}}k'|V|\psi\rangle\exp(-ik'r)}{(k+k')(k-k')}\\
		&=(2\pi)^{-3/2}\exp(i\bm{k}\cdot\bm{r})\nonumber\\
		&+\frac{1}{ir\sqrt{2\pi}}
		\int_{-\infty}^\infty \mathrm{d}k'\;2k'\frac{\exp(ik'r)\langle \hat{\bm{r}}k'|V|\psi\rangle}{(k+k')(k-k')}
	\end{alignat}
	where we have used
	\begin{alignat}{3}
		\la \bm{r}|\bm{k}\rangle =(2\pi)^{-3/2}\exp(i\bm{k}\cdot\bm{r}),
	\end{alignat}
	the expansion 
	\begin{alignat}{3}
		\lim_{r\to\infty}\exp(i\bm{k}\cdot\bm{r})&=\frac{2\pi}{ikr}\left(\delta(\hat{\bm{k}}-\hat{\bm{r}})\exp(ikr)-\delta(\hat{\bm{k}}+\hat{\bm{r}})\exp(-ikr)\right),
	\end{alignat}
	and the symmetry about $k'=0$ in \eqref{thing5} to extend the lower integration limit to $-\infty$.

\begin{figure}[htbp]
		\begin{tikzpicture}[scale=1.2]
			\draw[<->,thick] (-4,0) -- (4,0);
			\draw[->,thick] ( 0,0) -- (0,4);

			\draw[line width=1.5pt,color=blue]    (2.2,0) -- (3.0,0);
			\draw[line width=1.5pt,color=blue,->] (-1.8,0) -- (0.1,0);
			\draw[line width=1.5pt,color=blue]    (0,0) -- (1.8,0);
			\draw[line width=1.5pt,color=blue]    (-2.2,0) -- (-3.0,0);

			\draw[line width=1.5pt,color=blue,->]    (3,0) arc (0:90:3);
			\draw[line width=1.5pt,color=blue]    (-3,0) arc (180:90:3);
			\draw[line width=1.5pt,color=blue]    (1.8,0) arc (180:360:0.2);
			\draw[line width=1.5pt,color=blue]    (-1.8,0) arc (0:180:0.2);

			\draw[thick,->] (0.0,0.0) -- (-2.11,2.11);

			\node at (-1.05,1.45) {R};

			\fill[black] (2.0,0.0) circle (0.1);
			\fill[black] (-2.0,0.0) circle (0.1);

			\node at (4,-0.5) {Re($k$)};
			\node at (0.8,4) {Im($k$)};
			\node at (2.0,0.5) {$k'$};
			\node at (-2.0,-0.5) {$-k'$};
			\node at (0.8,4) {Im($k$)};
			\node at (2.2,2.7) {\textcolor{blue}{$C_1$}};
			\node at (1.0,-0.5) {\textcolor{blue}{$C_2$}};
		\end{tikzpicture}
		\caption
		[
		The closed contour $C=C_1+C_2$ that corresponds to a solution consistent with outgoing spherical wave boundary conditions.
		]
		{
		The closed contour $C=C_1+C_2$ that corresponds to a solution consistent with outgoing spherical wave boundary conditions.
		It contains the singularity occurring at $k'$ but avoids that of $-k'$. For incoming spherical wave boundary conditions this is reversed.
		}
		\label{contour}
	\end{figure}
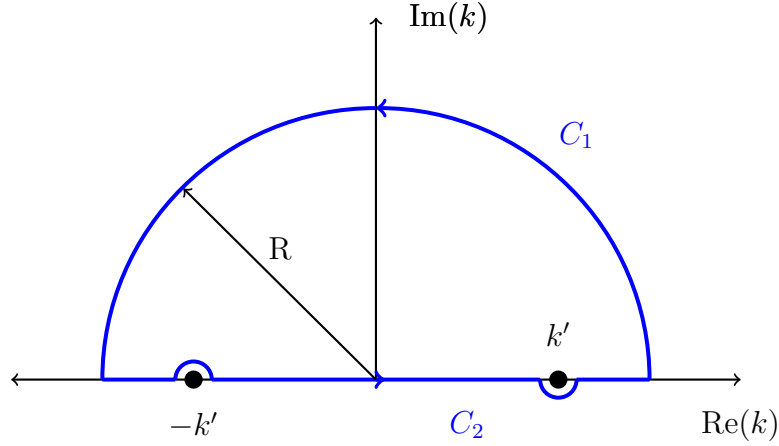

	Now let us consider the closed semi-circular contour $C=C_1+C_2$ of radius $R$ as shown in \Cref{contour}.
	The choice to use this contour which includes the pole at $k=k'$ and not that of $k=-k'$ is purely because it leads to an expression appropriate for an outgoing spherical wave.
	The addition of infinitesimally small imaginary components about these poles is commonly denoted by the addition of $+i0$ or superscript $^{(+)}$ .
	For the equivalent expression resulting in an incoming spherical wave the inclusion of these poles is reversed and accordingly the notation $-i0$ or $^{(-)}$ is used.
	Note that on this contour we have 
	\begin{alignat}{3}
		\lim_{R\to\infty}\int_{C_1} \mathrm{d}k'\;2k'\frac{\exp(ik'r)\langle \hat{\bm{r}}k'|V|\psi\rangle}{(k+k')(k-k')}&=0\\
		\lim_{R\to\infty}\int_{C_2} \mathrm{d}k'\;2k'\frac{\exp(ik'r)\langle \hat{\bm{r}}k'|V|\psi\rangle}{(k+k')(k-k')}&=
		\int_{-\infty}^{\infty} \mathrm{d}k'\;2k'\frac{\exp(ik'r)\langle \hat{\bm{r}}k'|V|\psi\rangle}{(k+k')(k-k')}.
	\end{alignat}
	Hence, we may write
	\begin{alignat}{3}
		\int_{-\infty}^{\infty} \mathrm{d}k'\;2k'\frac{\exp(ik'r)\langle \hat{\bm{r}}k'|V|\psi\rangle}{(k+k')(k-k')}&=
		\lim_{R\to\infty}\oint_{C} \mathrm{d}k'\;2k'\frac{\exp(ik'r)\langle \hat{\bm{r}}k'|V|\psi\rangle}{(k+k')(k-k')}.
	\end{alignat}
	Now using contour integration we may evaluate this integral using
	\begin{alignat}{3}
		\oint_{C}\mathrm{d}z\; f(z)&=2\pi i\sum_{i=1}^n \mathrm{Res}[f(z_i)]
	\end{alignat}
	where each $z_i$ corresponds to a pole of $f(z)$ and the residue is given by
	\begin{alignat}{3}
		\mathrm{Res}[f(z_i)]&=\frac{1}{n!}\lim_{z\to z_i}\frac{\mathrm{d^n}}{\mathrm{d}z^n}\left[(z-z_i)^{n+1}f(z)\right]
	\end{alignat}
	where $n$ is the order of the corresponding pole. 
	Doing so yields
	\begin{alignat}{3}
		\la\bm{r}|\psi\rangle&=(2\pi)^{-3/2}\exp(i\bm{k}\cdot\bm{r})\nonumber-\sqrt{2\pi}\langle \hat{\bm{r}}k|V|\psi\rangle\frac{\exp(ik'r)}{r}
	\end{alignat}
	and hence we have the scattering amplitude $f(\hat{\bm{r}}k)=-\sqrt{2\pi}\langle \hat{\bm{r}}k|V|\psi\rangle$.
	\clearpage
\section{Time Delay}
\label{Time Delay}
	Here we provide an introduction to the concept of a time delay in an elastic scattering collision through reproducing the comparatively 
	simple derivation summarised in the review of \citet{de2002time}.
	Consider a one dimensional wavepacket given by
	\begin{alignat}{3}
		\label{packet}
		\la t \;x |\Psi\ra&=\int |A(k)|\exp\Big\{i\big[kx-\omega(k) t+\eta(k)\big]\Big\}\;\mathrm{d}k
	\end{alignat}
	consisting of waves of magnitude $|A(k)|$, phase shift $\eta(k)$, and frequency $\omega(k)$.
	Assuming that $|A(k)|$ is at a maximum for $k_0$ it is a reasonable question to ask at what $x$ will the packet be peaked at a given time $t$.
	Considering \eqref{packet} as a superposition of monochromatic waves (summed over $k$), large variation of phase will cause mainly destructive interference.
	Hence, the largest amplitude will occur when the maximum of $|A(k)|$ ($k_0$ by definition) corresponds to a stationary point of the phase, i.e.\
	\begin{alignat}{3}
		&&\left[kx+\eta(k)-\omega(k)t\right]\big|_{k=k_0}&=\mathrm{const.}\\
		\implies && x+\left[\dod{\eta(k)}{k}-\dod{\omega(k)}{k}t\right]\bigg|_{k=k_0}&=0.
	\end{alignat}
	From which, it is clear that the point of constant phase (the maximum of the packet) travels with a velocity
	\begin{alignat}{3}
		v_g&=\dod{\omega(k)}{k}\bigg|_{k=k_0}.
	\end{alignat}
	For a dispersive medium ($\omega$ not linear with $k$) this will differ from the phase velocity
	\begin{alignat}{3}
		v_\phi&=\frac{\omega(k_0)}{k_0}.
	\end{alignat}

	If we now instead consider the wavepacket of s-wave ($L=0$) scattering on a spherically symmetric short ranged potential we can express the incoming packet via
	\begin{alignat}{3}
		\lim_{r\to\infty} \la r | \Psi_i\ra &= \int_0^\infty |A(E)|\exp\Big\{i\big[-kr-E t+\eta(E)\big]\Big\}\;\mathrm{d}E.
	\end{alignat}
	Note that here $k$ is simply shorthand via $k=\sqrt{2E}$.
	For this packet the point of constant phase occurs at a time
	\begin{alignat}{3}
		\label{tmax}
		t_{i\;\mathrm{max}}=\left[-\frac{r}{k}+\dod{\eta(E)}{E}\right]\bigg|_{E=E_0}.
	\end{alignat}
	This is correspondingly the point which the packet is centred about.
	Note that in coming to the expression \eqref{tmax} we require that $r$ is not a function of the projectile energy $E$ at an asymptotically large distance from the scattering centre.
	Similarly, considering the outgoing wavepacket we have
	\begin{alignat}{3}
		\lim_{r\to\infty} \la r | \Psi_f\ra &= \int_0^\infty |A(E)|\exp\Big\{i\big[kr-E t+\eta(E)+2\delta(E)\big]\Big\}\;\mathrm{d}E
	\end{alignat}
	where the $2\delta(E)$ term is the phase shift due to the scattering event associated with the elastic $S$-matrix element $S(E)=\exp\{2i\delta(E)\}$.
	The outgoing wavepacket is therefore centred about a time
	\begin{alignat}{3}
		t_{f\;\mathrm{max}}=\left[\frac{r}{k}+\dod{\eta(E)}{E}+2\dod{\delta(E)}{E}\right]\bigg|_{E=E_0}.
	\end{alignat}
	Interpreting this in comparison with the expression for the initial wavepacket \eqref{tmax} we define the time delay of this scattering event as
	\begin{alignat}{3}
		\tau&=\dod{\delta(E)}{E}\bigg|_{E=E_0}.
	\end{alignat}
	The same definition is appropriate for higher partial waves as for an asymptotic distance they simply differ by trivial factors.
	Hence, for a partial wave of given $L$ we have
	\begin{alignat}{3}
		\tau_{L}&=\dod{\delta_{L}(E)}{E}\bigg|_{E=E_0}.
	\end{alignat}
\clearpage
\pagestyle{bibstyle}
\nocite{apsrev41Control}

\begin{thebibliography}{149}%
\makeatletter
\providecommand \@ifxundefined [1]{%
 \@ifx{#1\undefined}
}%
\providecommand \@ifnum [1]{%
 \ifnum #1\expandafter \@firstoftwo
 \else \expandafter \@secondoftwo
 \fi
}%
\providecommand \@ifx [1]{%
 \ifx #1\expandafter \@firstoftwo
 \else \expandafter \@secondoftwo
 \fi
}%
\providecommand \natexlab [1]{#1}%
\providecommand \enquote  [1]{``#1''}%
\providecommand \bibnamefont  [1]{#1}%
\providecommand \bibfnamefont [1]{#1}%
\providecommand \citenamefont [1]{#1}%
\providecommand \href@noop [0]{\@secondoftwo}%
\providecommand \href [0]{\begingroup \@sanitize@url \@href}%
\providecommand \@href[1]{\@@startlink{#1}\@@href}%
\providecommand \@@href[1]{\endgroup#1\@@endlink}%
\providecommand \@sanitize@url [0]{\catcode `\\12\catcode `\$12\catcode
  `\&12\catcode `\#12\catcode `\^12\catcode `\_12\catcode `\%12\relax}%
\providecommand \@@startlink[1]{}%
\providecommand \@@endlink[0]{}%
\providecommand \url  [0]{\begingroup\@sanitize@url \@url }%
\providecommand \@url [1]{\endgroup\@href {#1}{\urlprefix }}%
\providecommand \urlprefix  [0]{URL }%
\providecommand \Eprint [0]{\href }%
\providecommand \doibase [0]{http://dx.doi.org/}%
\providecommand \selectlanguage [0]{\@gobble}%
\providecommand \bibinfo  [0]{\@secondoftwo}%
\providecommand \bibfield  [0]{\@secondoftwo}%
\providecommand \translation [1]{[#1]}%
\providecommand \BibitemOpen [0]{}%
\providecommand \bibitemStop [0]{}%
\providecommand \bibitemNoStop [0]{.\EOS\space}%
\providecommand \EOS [0]{\spacefactor3000\relax}%
\providecommand \BibitemShut  [1]{\csname bibitem#1\endcsname}%
\let\auto@bib@innerbib\@empty
\bibitem [{\citenamefont {Einstein}(1905)}]{einstein1905photoelectric}%
  \BibitemOpen
  \bibfield  {author} {\bibinfo {author} {\bibfnamefont {A.}~\bibnamefont
  {Einstein}},\ }\bibfield  {title} {\enquote {\bibinfo {title} {The
  photoelectric effect},}\ }\href
  {http://users.isy.liu.se/en/icg/jalar/kurser/QF/references/Einstein1905b.pdf}
  {\bibfield  {journal} {\bibinfo  {journal} {Ann. Phys.}\ }\textbf {\bibinfo
  {volume} {17}},\ \bibinfo {pages} {4} (\bibinfo {year} {1905})}\BibitemShut
  {NoStop}%
\bibitem [{\citenamefont {Born}(1924)}]{born1924quantenmechanik}%
  \BibitemOpen
  \bibfield  {author} {\bibinfo {author} {\bibfnamefont {M.}~\bibnamefont
  {Born}},\ }\bibfield  {title} {\enquote {\bibinfo {title} {{\"U}ber
  quantenmechanik},}\ }\href {\doibase 10.1007/BF01327341} {\bibfield
  {journal} {\bibinfo  {journal} {Z. Phys.}\ }\textbf {\bibinfo {volume}
  {26}},\ \bibinfo {pages} {379--395} (\bibinfo {year} {1924})}\BibitemShut
  {NoStop}%
\bibitem [{\citenamefont {Franck}\ and\ \citenamefont
  {Hertz}(1911)}]{franck1911zusammenhang}%
  \BibitemOpen
  \bibfield  {author} {\bibinfo {author} {\bibfnamefont {J.}~\bibnamefont
  {Franck}}\ and\ \bibinfo {author} {\bibfnamefont {G.}~\bibnamefont {Hertz}},\
  }\bibfield  {title} {\enquote {\bibinfo {title} {{\"U}ber einen zusammenhang
  zwischen quantenhypothese und ionisierungsspannung},}\ }\href@noop {}
  {\bibfield  {journal} {\bibinfo  {journal} {Verh. D. Phys. Ges.}\ }\textbf
  {\bibinfo {volume} {13}},\ \bibinfo {pages} {967--971} (\bibinfo {year}
  {1911})}\BibitemShut {NoStop}%
\bibitem [{\citenamefont {Ramsauer}(1921)}]{ramsauer1921wirkungsquerschnitt}%
  \BibitemOpen
  \bibfield  {author} {\bibinfo {author} {\bibfnamefont {C.}~\bibnamefont
  {Ramsauer}},\ }\bibfield  {title} {\enquote {\bibinfo {title} {{\"U}ber den
  wirkungsquerschnitt der gasmolek\"ule gegen\"uber langsamen elektronen},}\
  }\href {\doibase 10.1002/andp.19213690603} {\bibfield  {journal} {\bibinfo
  {journal} {Ann. d. Phys.}\ }\textbf {\bibinfo {volume} {369}},\ \bibinfo
  {pages} {513--540} (\bibinfo {year} {1921})}\BibitemShut {NoStop}%
\bibitem [{\citenamefont {Massey}\ and\ \citenamefont
  {Mohr}(1931)}]{massey1931collision}%
  \BibitemOpen
  \bibfield  {author} {\bibinfo {author} {\bibfnamefont {H.~S.~W.}\
  \bibnamefont {Massey}}\ and\ \bibinfo {author} {\bibfnamefont {C.~B.~O.}\
  \bibnamefont {Mohr}},\ }\bibfield  {title} {\enquote {\bibinfo {title} {The
  collision of electrons with simple atomic systems and electron exchange},}\
  }\href {http://www.jstor.org/stable/95680} {\bibfield  {journal} {\bibinfo
  {journal} {P. Roy. Soc. Lond. A Mat.}\ }\textbf {\bibinfo {volume} {132}},\
  \bibinfo {pages} {605--630} (\bibinfo {year} {1931})}\BibitemShut {NoStop}%
\bibitem [{\citenamefont {Born}(1926)}]{Born1926}%
  \BibitemOpen
  \bibfield  {author} {\bibinfo {author} {\bibfnamefont {M.}~\bibnamefont
  {Born}},\ }\bibfield  {title} {\enquote {\bibinfo {title} {Quantenmechanik
  der sto{\ss}vorg{\"a}nge},}\ }\href {\doibase 10.1007/BF01397184} {\bibfield
  {journal} {\bibinfo  {journal} {Z. Phys.}\ }\textbf {\bibinfo {volume}
  {38}},\ \bibinfo {pages} {803--827} (\bibinfo {year} {1926})}\BibitemShut
  {NoStop}%
\bibitem [{\citenamefont {Massey}(1956)}]{massey1956}%
  \BibitemOpen
  \bibfield  {author} {\bibinfo {author} {\bibfnamefont {H.~S.~W.}\
  \bibnamefont {Massey}},\ }\bibfield  {title} {\enquote {\bibinfo {title}
  {Theory of the scattering of slow electrons},}\ }\href {\doibase
  10.1103/RevModPhys.28.199} {\bibfield  {journal} {\bibinfo  {journal} {Rev.
  Mod. Phys.}\ }\textbf {\bibinfo {volume} {28}},\ \bibinfo {pages} {199--213}
  (\bibinfo {year} {1956})}\BibitemShut {NoStop}%
\bibitem [{\citenamefont {Weigold}\ \emph {et~al.}(1980)\citenamefont
  {Weigold}, \citenamefont {Frost},\ and\ \citenamefont
  {Nygaard}}]{weigold1980large}%
  \BibitemOpen
  \bibfield  {author} {\bibinfo {author} {\bibfnamefont {E.}~\bibnamefont
  {Weigold}}, \bibinfo {author} {\bibfnamefont {L.}~\bibnamefont {Frost}}, \
  and\ \bibinfo {author} {\bibfnamefont {K.~J.}\ \bibnamefont {Nygaard}},\
  }\bibfield  {title} {\enquote {\bibinfo {title} {Large-angle electron-photon
  coincidence experiment in atomic hydrogen},}\ }\href {\doibase
  10.1103/PhysRevA.21.1950} {\bibfield  {journal} {\bibinfo  {journal} {Phys.
  Rev. A}\ }\textbf {\bibinfo {volume} {21}},\ \bibinfo {pages} {1950}
  (\bibinfo {year} {1980})}\BibitemShut {NoStop}%
\bibitem [{\citenamefont {Williams}(1981)}]{williams1981electron}%
  \BibitemOpen
  \bibfield  {author} {\bibinfo {author} {\bibfnamefont {J.~F.}\ \bibnamefont
  {Williams}},\ }\bibfield  {title} {\enquote {\bibinfo {title}
  {Electron-photon angular correlations from the electron impact excitation of
  the 2s and 2p electronic configurations of atomic hydrogen},}\ }\href
  {http://stacks.iop.org/0022-3700/14/i=7/a=014} {\bibfield  {journal}
  {\bibinfo  {journal} {J. Phys. B}\ }\textbf {\bibinfo {volume} {14}},\
  \bibinfo {pages} {1197} (\bibinfo {year} {1981})}\BibitemShut {NoStop}%
\bibitem [{\citenamefont {Scholz}\ \emph {et~al.}(1991)\citenamefont {Scholz},
  \citenamefont {Walters}, \citenamefont {Burke},\ and\ \citenamefont
  {Scott}}]{scholz1991electron}%
  \BibitemOpen
  \bibfield  {author} {\bibinfo {author} {\bibfnamefont {T.~T.}\ \bibnamefont
  {Scholz}}, \bibinfo {author} {\bibfnamefont {H.~R.~J.}\ \bibnamefont
  {Walters}}, \bibinfo {author} {\bibfnamefont {P.~G.}\ \bibnamefont {Burke}},
  \ and\ \bibinfo {author} {\bibfnamefont {M.~P.}\ \bibnamefont {Scott}},\
  }\bibfield  {title} {\enquote {\bibinfo {title} {Electron scattering by
  atomic hydrogen at intermediate energies. ii. differential elastic, 1s-2s and
  1s-2p cross sections and 1s-2p electron-photon coincidence parameters},}\
  }\href {http://stacks.iop.org/0953-4075/24/i=8/a=023} {\bibfield  {journal}
  {\bibinfo  {journal} {J. Phys. B}\ }\textbf {\bibinfo {volume} {24}},\
  \bibinfo {pages} {2097} (\bibinfo {year} {1991})}\BibitemShut {NoStop}%
\bibitem [{\citenamefont {Van~Wyngaarden}\ and\ \citenamefont
  {Walters}(1986)}]{van1986elastic}%
  \BibitemOpen
  \bibfield  {author} {\bibinfo {author} {\bibfnamefont {W.~L.}\ \bibnamefont
  {Van~Wyngaarden}}\ and\ \bibinfo {author} {\bibfnamefont {H.~R.~J.}\
  \bibnamefont {Walters}},\ }\bibfield  {title} {\enquote {\bibinfo {title}
  {Elastic scattering and excitation of the 1s to 2s and 1s to 2p transitions
  in atomic hydrogen by electrons to medium to high energies},}\ }\href
  {http://stacks.iop.org/0022-3700/19/i=6/a=014} {\bibfield  {journal}
  {\bibinfo  {journal} {J. Phys. B}\ }\textbf {\bibinfo {volume} {19}},\
  \bibinfo {pages} {929} (\bibinfo {year} {1986})}\BibitemShut {NoStop}%
\bibitem [{\citenamefont {Bartschat}\ \emph {et~al.}(1996)\citenamefont
  {Bartschat}, \citenamefont {Hudson}, \citenamefont {Scott}, \citenamefont
  {Burke},\ and\ \citenamefont {Burke}}]{bartschat1996electron}%
  \BibitemOpen
  \bibfield  {author} {\bibinfo {author} {\bibfnamefont {K.}~\bibnamefont
  {Bartschat}}, \bibinfo {author} {\bibfnamefont {E.~T.}\ \bibnamefont
  {Hudson}}, \bibinfo {author} {\bibfnamefont {M.~P.}\ \bibnamefont {Scott}},
  \bibinfo {author} {\bibfnamefont {P.~G.}\ \bibnamefont {Burke}}, \ and\
  \bibinfo {author} {\bibfnamefont {V.~M.}\ \bibnamefont {Burke}},\ }\bibfield
  {title} {\enquote {\bibinfo {title} {Electron-atom scattering at low and
  intermediate energies using a pseudo-state/{$R$}-matrix basis},}\ }\href
  {http://stacks.iop.org/0953-4075/29/i=1/a=015} {\bibfield  {journal}
  {\bibinfo  {journal} {J. Phys. B}\ }\textbf {\bibinfo {volume} {29}},\
  \bibinfo {pages} {115} (\bibinfo {year} {1996})}\BibitemShut {NoStop}%
\bibitem [{\citenamefont {Rescigno}\ \emph {et~al.}(1999)\citenamefont
  {Rescigno}, \citenamefont {Baertschy}, \citenamefont {Isaacs},\ and\
  \citenamefont {McCurdy}}]{rescigno1999collisional}%
  \BibitemOpen
  \bibfield  {author} {\bibinfo {author} {\bibfnamefont {T.~N.}\ \bibnamefont
  {Rescigno}}, \bibinfo {author} {\bibfnamefont {M.}~\bibnamefont {Baertschy}},
  \bibinfo {author} {\bibfnamefont {W.~A.}\ \bibnamefont {Isaacs}}, \ and\
  \bibinfo {author} {\bibfnamefont {C.~W.}\ \bibnamefont {McCurdy}},\
  }\bibfield  {title} {\enquote {\bibinfo {title} {Collisional breakup in a
  quantum system of three charged particles},}\ }\href {\doibase
  10.1126/science.286.5449.2474} {\bibfield  {journal} {\bibinfo  {journal}
  {Science}\ }\textbf {\bibinfo {volume} {286}},\ \bibinfo {pages} {2474--2479}
  (\bibinfo {year} {1999})}\BibitemShut {NoStop}%
\bibitem [{\citenamefont {Pindzola}\ and\ \citenamefont
  {Schultz}(1996)}]{pindzola1996time}%
  \BibitemOpen
  \bibfield  {author} {\bibinfo {author} {\bibfnamefont {M.~S.}\ \bibnamefont
  {Pindzola}}\ and\ \bibinfo {author} {\bibfnamefont {D.~R.}\ \bibnamefont
  {Schultz}},\ }\bibfield  {title} {\enquote {\bibinfo {title} {Time-dependent
  close-coupling method for electron-impact ionization of hydrogen},}\ }\href
  {\doibase 10.1103/PhysRevA.53.1525} {\bibfield  {journal} {\bibinfo
  {journal} {Phys. Rev. A}\ }\textbf {\bibinfo {volume} {53}},\ \bibinfo
  {pages} {1525} (\bibinfo {year} {1996})}\BibitemShut {NoStop}%
\bibitem [{\citenamefont {Bray}\ and\ \citenamefont
  {Stelbovics}(1992{\natexlab{a}})}]{CCC1992}%
  \BibitemOpen
  \bibfield  {author} {\bibinfo {author} {\bibfnamefont {I.}~\bibnamefont
  {Bray}}\ and\ \bibinfo {author} {\bibfnamefont {A.~T.}\ \bibnamefont
  {Stelbovics}},\ }\bibfield  {title} {\enquote {\bibinfo {title} {Convergent
  close-coupling calculations of electron-hydrogen scattering},}\ }\href
  {\doibase 10.1103/PhysRevA.46.6995} {\bibfield  {journal} {\bibinfo
  {journal} {Phys. Rev. A}\ }\textbf {\bibinfo {volume} {46}},\ \bibinfo
  {pages} {6995--7011} (\bibinfo {year} {1992}{\natexlab{a}})}\BibitemShut
  {NoStop}%
\bibitem [{\citenamefont {Yalim}\ \emph {et~al.}(1997)\citenamefont {Yalim},
  \citenamefont {Cvejanovi{\'c}},\ and\ \citenamefont {Crowe}}]{yalim19971s}%
  \BibitemOpen
  \bibfield  {author} {\bibinfo {author} {\bibfnamefont {H.~A.}\ \bibnamefont
  {Yalim}}, \bibinfo {author} {\bibfnamefont {D.}~\bibnamefont
  {Cvejanovi{\'c}}}, \ and\ \bibinfo {author} {\bibfnamefont {A.}~\bibnamefont
  {Crowe}},\ }\bibfield  {title} {\enquote {\bibinfo {title} {1s-2p excitation
  of atomic hydrogen by electron impact studied using the angular correlation
  technique},}\ }\href {\doibase 10.1103/PhysRevLett.79.2951} {\bibfield
  {journal} {\bibinfo  {journal} {Phys. Rev. Lett.}\ }\textbf {\bibinfo
  {volume} {79}},\ \bibinfo {pages} {2951} (\bibinfo {year}
  {1997})}\BibitemShut {NoStop}%
\bibitem [{\citenamefont {O'Neill}\ \emph {et~al.}(1998)\citenamefont
  {O'Neill}, \citenamefont {Van~der Burgt}, \citenamefont {Dziczek},
  \citenamefont {Bowe}, \citenamefont {Chwirot},\ and\ \citenamefont
  {Slevin}}]{o1998polarization}%
  \BibitemOpen
  \bibfield  {author} {\bibinfo {author} {\bibfnamefont {R.~W.}\ \bibnamefont
  {O'Neill}}, \bibinfo {author} {\bibfnamefont {P.~J.~M.}\ \bibnamefont
  {Van~der Burgt}}, \bibinfo {author} {\bibfnamefont {D.}~\bibnamefont
  {Dziczek}}, \bibinfo {author} {\bibfnamefont {P.}~\bibnamefont {Bowe}},
  \bibinfo {author} {\bibfnamefont {S.}~\bibnamefont {Chwirot}}, \ and\
  \bibinfo {author} {\bibfnamefont {J.~A.}\ \bibnamefont {Slevin}},\ }\bibfield
   {title} {\enquote {\bibinfo {title} {Polarization correlation measurements
  of electron impact excitation of {H} (2 p) at 54.4 {eV}},}\ }\href {\doibase
  10.1103/PhysRevLett.80.1630} {\bibfield  {journal} {\bibinfo  {journal}
  {Phys. Rev. Lett.}\ }\textbf {\bibinfo {volume} {80}},\ \bibinfo {pages}
  {1630} (\bibinfo {year} {1998})}\BibitemShut {NoStop}%
\bibitem [{\citenamefont {Peterkop}(1962)}]{peterkop1962wave}%
  \BibitemOpen
  \bibfield  {author} {\bibinfo {author} {\bibfnamefont {R.~K.}\ \bibnamefont
  {Peterkop}},\ }\bibfield  {title} {\enquote {\bibinfo {title} {The wave
  function of the {eH} collision},}\ }\href
  {http://adsabs.harvard.edu/abs/1962OptSp..13...87P} {\bibfield  {journal}
  {\bibinfo  {journal} {Opt. Spectrosc.}\ }\textbf {\bibinfo {volume} {13}},\
  \bibinfo {pages} {87} (\bibinfo {year} {1962})}\BibitemShut {NoStop}%
\bibitem [{\citenamefont {Rudge}\ and\ \citenamefont
  {Seaton}(1965)}]{rudge1965ionization}%
  \BibitemOpen
  \bibfield  {author} {\bibinfo {author} {\bibfnamefont {M.~R.~H.}\
  \bibnamefont {Rudge}}\ and\ \bibinfo {author} {\bibfnamefont {M.~J.}\
  \bibnamefont {Seaton}},\ }\bibfield  {title} {\enquote {\bibinfo {title}
  {Ionization of atomic hydrogen by electron impact},}\ }\href {\doibase
  10.1098/rspa.1965.0020} {\bibfield  {journal} {\bibinfo  {journal} {P. Roy.
  Soc. Lond. A Mat.}\ }\textbf {\bibinfo {volume} {283}},\ \bibinfo {pages}
  {262--290} (\bibinfo {year} {1965})}\BibitemShut {NoStop}%
\bibitem [{\citenamefont {Bartschat}\ and\ \citenamefont
  {Burke}(1987)}]{bartschat1987r}%
  \BibitemOpen
  \bibfield  {author} {\bibinfo {author} {\bibfnamefont {K.}~\bibnamefont
  {Bartschat}}\ and\ \bibinfo {author} {\bibfnamefont {P.~G.}\ \bibnamefont
  {Burke}},\ }\bibfield  {title} {\enquote {\bibinfo {title} {The $r$-matrix
  method for electron impact ionisation},}\ }\href
  {http://stacks.iop.org/0022-3700/20/i=13/a=025} {\bibfield  {journal}
  {\bibinfo  {journal} {J. Phys. B}\ }\textbf {\bibinfo {volume} {20}},\
  \bibinfo {pages} {3191} (\bibinfo {year} {1987})}\BibitemShut {NoStop}%
\bibitem [{\citenamefont {Bray}\ and\ \citenamefont
  {Stelbovics}(1993)}]{bray1993calculation}%
  \BibitemOpen
  \bibfield  {author} {\bibinfo {author} {\bibfnamefont {I.}~\bibnamefont
  {Bray}}\ and\ \bibinfo {author} {\bibfnamefont {A.~T.}\ \bibnamefont
  {Stelbovics}},\ }\bibfield  {title} {\enquote {\bibinfo {title} {Calculation
  of the total ionization cross section and spin asymmetry in electron-hydrogen
  scattering from threshold to 500 {eV}},}\ }\href {\doibase
  10.1103/PhysRevLett.70.746} {\bibfield  {journal} {\bibinfo  {journal} {Phys.
  Rev. Lett.}\ }\textbf {\bibinfo {volume} {70}},\ \bibinfo {pages} {746}
  (\bibinfo {year} {1993})}\BibitemShut {NoStop}%
\bibitem [{\citenamefont {Baertschy}\ \emph {et~al.}(2001)\citenamefont
  {Baertschy}, \citenamefont {Rescigno}, \citenamefont {Isaacs}, \citenamefont
  {Li},\ and\ \citenamefont {McCurdy}}]{PhysRevA.63.022712}%
  \BibitemOpen
  \bibfield  {author} {\bibinfo {author} {\bibfnamefont {M.}~\bibnamefont
  {Baertschy}}, \bibinfo {author} {\bibfnamefont {T.~N.}\ \bibnamefont
  {Rescigno}}, \bibinfo {author} {\bibfnamefont {W.~A.}\ \bibnamefont
  {Isaacs}}, \bibinfo {author} {\bibfnamefont {X.}~\bibnamefont {Li}}, \ and\
  \bibinfo {author} {\bibfnamefont {C.~W.}\ \bibnamefont {McCurdy}},\
  }\bibfield  {title} {\enquote {\bibinfo {title} {Electron-impact ionization
  of atomic hydrogen},}\ }\href {\doibase 10.1103/PhysRevA.63.022712}
  {\bibfield  {journal} {\bibinfo  {journal} {Phys. Rev. A}\ }\textbf {\bibinfo
  {volume} {63}},\ \bibinfo {pages} {022712} (\bibinfo {year}
  {2001})}\BibitemShut {NoStop}%
\bibitem [{\citenamefont {Bray}(2002)}]{bray2002close}%
  \BibitemOpen
  \bibfield  {author} {\bibinfo {author} {\bibfnamefont {I.}~\bibnamefont
  {Bray}},\ }\bibfield  {title} {\enquote {\bibinfo {title} {Close-coupling
  approach to coulomb three-body problems},}\ }\href {\doibase
  10.1103/PhysRevLett.89.273201} {\bibfield  {journal} {\bibinfo  {journal}
  {Phys. Rev. Lett.}\ }\textbf {\bibinfo {volume} {89}},\ \bibinfo {pages}
  {273201} (\bibinfo {year} {2002})}\BibitemShut {NoStop}%
\bibitem [{\citenamefont {Bartschat}(2002)}]{bartschat2002convergent}%
  \BibitemOpen
  \bibfield  {author} {\bibinfo {author} {\bibfnamefont {K.}~\bibnamefont
  {Bartschat}},\ }\bibfield  {title} {\enquote {\bibinfo {title} {Convergent
  {$R$}-matrix with pseudostates calculations for electron-impact ionization of
  the $n = 2$ states in helium},}\ }\href
  {http://stacks.iop.org/0953-4075/35/i=23/a=104} {\bibfield  {journal}
  {\bibinfo  {journal} {J. Phys. B}\ }\textbf {\bibinfo {volume} {35}},\
  \bibinfo {pages} {L527} (\bibinfo {year} {2002})}\BibitemShut {NoStop}%
\bibitem [{\citenamefont {Colgan}\ and\ \citenamefont
  {Pindzola}(2006)}]{colgan2006double}%
  \BibitemOpen
  \bibfield  {author} {\bibinfo {author} {\bibfnamefont {J.}~\bibnamefont
  {Colgan}}\ and\ \bibinfo {author} {\bibfnamefont {M.~S.}\ \bibnamefont
  {Pindzola}},\ }\bibfield  {title} {\enquote {\bibinfo {title} {Double- and
  triple-differential cross sections for the low-energy electron-impact
  ionization of hydrogen},}\ }\href {\doibase 10.1103/PhysRevA.74.012713}
  {\bibfield  {journal} {\bibinfo  {journal} {Phys. Rev. A}\ }\textbf {\bibinfo
  {volume} {74}},\ \bibinfo {pages} {012713} (\bibinfo {year}
  {2006})}\BibitemShut {NoStop}%
\bibitem [{\citenamefont {Kadyrov}\ \emph {et~al.}(2003)\citenamefont
  {Kadyrov}, \citenamefont {Mukhamedzhanov}, \citenamefont {Stelbovics},\ and\
  \citenamefont {Bray}}]{kadyrov2003integral}%
  \BibitemOpen
  \bibfield  {author} {\bibinfo {author} {\bibfnamefont {A.~S.}\ \bibnamefont
  {Kadyrov}}, \bibinfo {author} {\bibfnamefont {A.~M.}\ \bibnamefont
  {Mukhamedzhanov}}, \bibinfo {author} {\bibfnamefont {A.~T.}\ \bibnamefont
  {Stelbovics}}, \ and\ \bibinfo {author} {\bibfnamefont {I.}~\bibnamefont
  {Bray}},\ }\bibfield  {title} {\enquote {\bibinfo {title} {Integral
  representation for the electron-atom ionization amplitude which is free of
  ambiguity and divergence problems},}\ }\href {\doibase
  10.1103/PhysRevLett.91.253202} {\bibfield  {journal} {\bibinfo  {journal}
  {Phys. Rev. Lett.}\ }\textbf {\bibinfo {volume} {91}},\ \bibinfo {pages}
  {253202} (\bibinfo {year} {2003})}\BibitemShut {NoStop}%
\bibitem [{\citenamefont {Kadyrov}\ \emph {et~al.}(2004)\citenamefont
  {Kadyrov}, \citenamefont {Mukhamedzhanov}, \citenamefont {Stelbovics},\ and\
  \citenamefont {Bray}}]{kadyrov2004theory}%
  \BibitemOpen
  \bibfield  {author} {\bibinfo {author} {\bibfnamefont {A.~S.}\ \bibnamefont
  {Kadyrov}}, \bibinfo {author} {\bibfnamefont {A.~M.}\ \bibnamefont
  {Mukhamedzhanov}}, \bibinfo {author} {\bibfnamefont {A.~T.}\ \bibnamefont
  {Stelbovics}}, \ and\ \bibinfo {author} {\bibfnamefont {I.}~\bibnamefont
  {Bray}},\ }\bibfield  {title} {\enquote {\bibinfo {title} {Theory of
  electron-impact ionization of atoms},}\ }\href {\doibase
  10.1103/PhysRevA.70.062703} {\bibfield  {journal} {\bibinfo  {journal} {Phys.
  Rev. A}\ }\textbf {\bibinfo {volume} {70}},\ \bibinfo {pages} {062703}
  (\bibinfo {year} {2004})}\BibitemShut {NoStop}%
\bibitem [{\citenamefont {Kadyrov}\ \emph {et~al.}(2008)\citenamefont
  {Kadyrov}, \citenamefont {Bray}, \citenamefont {Mukhamedzhanov},\ and\
  \citenamefont {Stelbovics}}]{kadyrov2008coulomb}%
  \BibitemOpen
  \bibfield  {author} {\bibinfo {author} {\bibfnamefont {A.~S.}\ \bibnamefont
  {Kadyrov}}, \bibinfo {author} {\bibfnamefont {I.}~\bibnamefont {Bray}},
  \bibinfo {author} {\bibfnamefont {A.~M.}\ \bibnamefont {Mukhamedzhanov}}, \
  and\ \bibinfo {author} {\bibfnamefont {A.~T.}\ \bibnamefont {Stelbovics}},\
  }\bibfield  {title} {\enquote {\bibinfo {title} {Coulomb breakup problem},}\
  }\href {\doibase 10.1103/PhysRevLett.101.230405} {\bibfield  {journal}
  {\bibinfo  {journal} {Phys. Rev. Lett.}\ }\textbf {\bibinfo {volume} {101}},\
  \bibinfo {pages} {230405} (\bibinfo {year} {2008})}\BibitemShut {NoStop}%
\bibitem [{\citenamefont {Kadyrov}\ \emph {et~al.}(2009)\citenamefont
  {Kadyrov}, \citenamefont {Bray}, \citenamefont {Mukhamedzhanov},\ and\
  \citenamefont {Stelbovics}}]{Kadyrov20091516}%
  \BibitemOpen
  \bibfield  {author} {\bibinfo {author} {\bibfnamefont {A.~S.}\ \bibnamefont
  {Kadyrov}}, \bibinfo {author} {\bibfnamefont {I.}~\bibnamefont {Bray}},
  \bibinfo {author} {\bibfnamefont {A.~M.}\ \bibnamefont {Mukhamedzhanov}}, \
  and\ \bibinfo {author} {\bibfnamefont {A.~T.}\ \bibnamefont {Stelbovics}},\
  }\bibfield  {title} {\enquote {\bibinfo {title} {Surface-integral formulation
  of scattering theory},}\ }\href {\doibase
  http://dx.doi.org/10.1016/j.aop.2009.02.003} {\bibfield  {journal} {\bibinfo
  {journal} {Ann. Phys.}\ }\textbf {\bibinfo {volume} {324}},\ \bibinfo {pages}
  {1516 -- 1546} (\bibinfo {year} {2009})},\ \bibinfo {note} {july 2009 Special
  Issue}\BibitemShut {NoStop}%
\bibitem [{\citenamefont {Sawey}\ \emph {et~al.}(1990)\citenamefont {Sawey},
  \citenamefont {Berrington}, \citenamefont {Burke},\ and\ \citenamefont
  {Kingston}}]{sawey1990electron}%
  \BibitemOpen
  \bibfield  {author} {\bibinfo {author} {\bibfnamefont {P.~M.~J.}\
  \bibnamefont {Sawey}}, \bibinfo {author} {\bibfnamefont {K.~A.}\ \bibnamefont
  {Berrington}}, \bibinfo {author} {\bibfnamefont {P.~G.}\ \bibnamefont
  {Burke}}, \ and\ \bibinfo {author} {\bibfnamefont {A.~E.}\ \bibnamefont
  {Kingston}},\ }\bibfield  {title} {\enquote {\bibinfo {title} {Electron
  scattering in helium at low energies: a 29-state {$R$}-matrix calculation},}\
  }\href {http://stacks.iop.org/0953-4075/23/i=23/a=014} {\bibfield  {journal}
  {\bibinfo  {journal} {J. Phys. B}\ }\textbf {\bibinfo {volume} {23}},\
  \bibinfo {pages} {4321} (\bibinfo {year} {1990})}\BibitemShut {NoStop}%
\bibitem [{\citenamefont {Pindzola}\ and\ \citenamefont
  {Robicheaux}(2000)}]{pindzola2000time}%
  \BibitemOpen
  \bibfield  {author} {\bibinfo {author} {\bibfnamefont {M.~S.}\ \bibnamefont
  {Pindzola}}\ and\ \bibinfo {author} {\bibfnamefont {F.~J.}\ \bibnamefont
  {Robicheaux}},\ }\bibfield  {title} {\enquote {\bibinfo {title}
  {Time-dependent close-coupling calculations for the electron-impact
  ionization of helium},}\ }\href {\doibase 10.1103/PhysRevA.61.052707}
  {\bibfield  {journal} {\bibinfo  {journal} {Phys. Rev. A}\ }\textbf {\bibinfo
  {volume} {61}},\ \bibinfo {pages} {052707} (\bibinfo {year}
  {2000})}\BibitemShut {NoStop}%
\bibitem [{\citenamefont {Kwan}\ \emph {et~al.}(1991)\citenamefont {Kwan},
  \citenamefont {Kauppila}, \citenamefont {Lukaszew}, \citenamefont {Parikh},
  \citenamefont {Stein}, \citenamefont {Wan},\ and\ \citenamefont
  {Dababneh}}]{kwan1991total}%
  \BibitemOpen
  \bibfield  {author} {\bibinfo {author} {\bibfnamefont {C.~K.}\ \bibnamefont
  {Kwan}}, \bibinfo {author} {\bibfnamefont {W.~E.}\ \bibnamefont {Kauppila}},
  \bibinfo {author} {\bibfnamefont {R.~A.}\ \bibnamefont {Lukaszew}}, \bibinfo
  {author} {\bibfnamefont {S.~P.}\ \bibnamefont {Parikh}}, \bibinfo {author}
  {\bibfnamefont {T.~S.}\ \bibnamefont {Stein}}, \bibinfo {author}
  {\bibfnamefont {Y.~J.}\ \bibnamefont {Wan}}, \ and\ \bibinfo {author}
  {\bibfnamefont {M.~S.}\ \bibnamefont {Dababneh}},\ }\bibfield  {title}
  {\enquote {\bibinfo {title} {Total cross-section measurements for positrons
  and electrons scattered by sodium and potassium atoms},}\ }\href {\doibase
  10.1103/PhysRevA.44.1620} {\bibfield  {journal} {\bibinfo  {journal} {Phys.
  Rev. A}\ }\textbf {\bibinfo {volume} {44}},\ \bibinfo {pages} {1620}
  (\bibinfo {year} {1991})}\BibitemShut {NoStop}%
\bibitem [{\citenamefont {Fursa}\ and\ \citenamefont
  {Bray}(1997{\natexlab{a}})}]{fursaHelike}%
  \BibitemOpen
  \bibfield  {author} {\bibinfo {author} {\bibfnamefont {D.~V.}\ \bibnamefont
  {Fursa}}\ and\ \bibinfo {author} {\bibfnamefont {I.}~\bibnamefont {Bray}},\
  }\bibfield  {title} {\enquote {\bibinfo {title} {Convergent close-coupling
  calculations of electron scattering on helium-like atoms and ions:
  electron-beryllium scattering},}\ }\href
  {http://stacks.iop.org/0953-4075/30/i=24/a=023} {\bibfield  {journal}
  {\bibinfo  {journal} {J. Phys. B}\ }\textbf {\bibinfo {volume} {30}},\
  \bibinfo {pages} {5895} (\bibinfo {year} {1997}{\natexlab{a}})}\BibitemShut
  {NoStop}%
\bibitem [{\citenamefont {Dzuba}\ \emph {et~al.}(1996)\citenamefont {Dzuba},
  \citenamefont {Flambaum}, \citenamefont {Gribakin},\ and\ \citenamefont
  {King}}]{dzuba1996many}%
  \BibitemOpen
  \bibfield  {author} {\bibinfo {author} {\bibfnamefont {V.~A.}\ \bibnamefont
  {Dzuba}}, \bibinfo {author} {\bibfnamefont {V.~V.}\ \bibnamefont {Flambaum}},
  \bibinfo {author} {\bibfnamefont {G.~F.}\ \bibnamefont {Gribakin}}, \ and\
  \bibinfo {author} {\bibfnamefont {W.~A.}\ \bibnamefont {King}},\ }\bibfield
  {title} {\enquote {\bibinfo {title} {Many-body calculations of positron
  scattering and annihilation from noble-gas atoms},}\ }\href
  {http://stacks.iop.org/0953-4075/29/i=14/a=024} {\bibfield  {journal}
  {\bibinfo  {journal} {J. Phys. B}\ }\textbf {\bibinfo {volume} {29}},\
  \bibinfo {pages} {3151} (\bibinfo {year} {1996})}\BibitemShut {NoStop}%
\bibitem [{\citenamefont {Hylleraas}(1929)}]{hylleraas1929neue}%
  \BibitemOpen
  \bibfield  {author} {\bibinfo {author} {\bibfnamefont {E.~A.}\ \bibnamefont
  {Hylleraas}},\ }\bibfield  {title} {\enquote {\bibinfo {title} {Neue
  berechnung der energie des heliums im grundzustande, sowie des tiefsten terms
  von ortho-helium},}\ }\href {\doibase 10.1007/BF01375457} {\bibfield
  {journal} {\bibinfo  {journal} {Z. Phys.}\ }\textbf {\bibinfo {volume}
  {54}},\ \bibinfo {pages} {347--366} (\bibinfo {year} {1929})}\BibitemShut
  {NoStop}%
\bibitem [{\citenamefont {Pekeris}(1958)}]{pekeris1958ground}%
  \BibitemOpen
  \bibfield  {author} {\bibinfo {author} {\bibfnamefont {C.~L.}\ \bibnamefont
  {Pekeris}},\ }\bibfield  {title} {\enquote {\bibinfo {title} {Ground state of
  two-electron atoms},}\ }\href {\doibase 10.1103/PhysRev.112.1649} {\bibfield
  {journal} {\bibinfo  {journal} {Phys. Rev.}\ }\textbf {\bibinfo {volume}
  {112}},\ \bibinfo {pages} {1649} (\bibinfo {year} {1958})}\BibitemShut
  {NoStop}%
\bibitem [{\citenamefont {Hanssen}\ \emph {et~al.}(1996)\citenamefont
  {Hanssen}, \citenamefont {Joulakian}, \citenamefont {Rivarola},\ and\
  \citenamefont {Allan}}]{hanssen1996differential}%
  \BibitemOpen
  \bibfield  {author} {\bibinfo {author} {\bibfnamefont {J.}~\bibnamefont
  {Hanssen}}, \bibinfo {author} {\bibfnamefont {B.}~\bibnamefont {Joulakian}},
  \bibinfo {author} {\bibfnamefont {R.~D.}\ \bibnamefont {Rivarola}}, \ and\
  \bibinfo {author} {\bibfnamefont {R.~J.}\ \bibnamefont {Allan}},\ }\bibfield
  {title} {\enquote {\bibinfo {title} {Differential cross section of the
  dissociative ionization of {H}$^{2+}$ by fast electron impact},}\ }\href
  {http://stacks.iop.org/1402-4896/53/i=1/a=007} {\bibfield  {journal}
  {\bibinfo  {journal} {Phys. Scripta}\ }\textbf {\bibinfo {volume} {53}},\
  \bibinfo {pages} {41} (\bibinfo {year} {1996})}\BibitemShut {NoStop}%
\bibitem [{\citenamefont {Zammit}\ \emph {et~al.}(2013)\citenamefont {Zammit},
  \citenamefont {Fursa},\ and\ \citenamefont {Bray}}]{zammitMol}%
  \BibitemOpen
  \bibfield  {author} {\bibinfo {author} {\bibfnamefont {M.~C.}\ \bibnamefont
  {Zammit}}, \bibinfo {author} {\bibfnamefont {D.~V.}\ \bibnamefont {Fursa}}, \
  and\ \bibinfo {author} {\bibfnamefont {I.}~\bibnamefont {Bray}},\ }\bibfield
  {title} {\enquote {\bibinfo {title} {Convergent-close-coupling formalism for
  positron scattering from molecules},}\ }\href {\doibase
  10.1103/PhysRevA.87.020701} {\bibfield  {journal} {\bibinfo  {journal} {Phys.
  Rev. A}\ }\textbf {\bibinfo {volume} {87}},\ \bibinfo {pages} {020701}
  (\bibinfo {year} {2013})}\BibitemShut {NoStop}%
\bibitem [{\citenamefont {Abdurakhmanov}\ \emph {et~al.}(2015)\citenamefont
  {Abdurakhmanov}, \citenamefont {Kadyrov}, \citenamefont {Fursa},
  \citenamefont {Avazbaev}, \citenamefont {Bailey},\ and\ \citenamefont
  {Bray}}]{abdurakhmanovH20}%
  \BibitemOpen
  \bibfield  {author} {\bibinfo {author} {\bibfnamefont {I.~B.}\ \bibnamefont
  {Abdurakhmanov}}, \bibinfo {author} {\bibfnamefont {A.~S.}\ \bibnamefont
  {Kadyrov}}, \bibinfo {author} {\bibfnamefont {D.~V.}\ \bibnamefont {Fursa}},
  \bibinfo {author} {\bibfnamefont {S.~K.}\ \bibnamefont {Avazbaev}}, \bibinfo
  {author} {\bibfnamefont {J.~J.}\ \bibnamefont {Bailey}}, \ and\ \bibinfo
  {author} {\bibfnamefont {I.}~\bibnamefont {Bray}},\ }\bibfield  {title}
  {\enquote {\bibinfo {title} {Antiproton-impact ionization of {Ne}, {Ar},
  {Kr}, {Xe}, and {H$_2$O}},}\ }\href {\doibase 10.1103/PhysRevA.91.022712}
  {\bibfield  {journal} {\bibinfo  {journal} {Phys. Rev. A}\ }\textbf {\bibinfo
  {volume} {91}},\ \bibinfo {pages} {022712} (\bibinfo {year}
  {2015})}\BibitemShut {NoStop}%
\bibitem [{\citenamefont {Bartschat}(1998)}]{bartschat1998r}%
  \BibitemOpen
  \bibfield  {author} {\bibinfo {author} {\bibfnamefont {K.}~\bibnamefont
  {Bartschat}},\ }\bibfield  {title} {\enquote {\bibinfo {title} {The
  {$R$}-matrix with pseudo-states method: Theory and applications to electron
  scattering and photoionization},}\ }\href {\doibase
  http://dx.doi.org/10.1016/S0010-4655(98)00057-5} {\bibfield  {journal}
  {\bibinfo  {journal} {Comput. Phys. Commun.}\ }\textbf {\bibinfo {volume}
  {114}},\ \bibinfo {pages} {168--182} (\bibinfo {year} {1998})}\BibitemShut
  {NoStop}%
\bibitem [{\citenamefont {Kernoghan}\ \emph {et~al.}(1996)\citenamefont
  {Kernoghan}, \citenamefont {Robinson}, \citenamefont {McAlinden},\ and\
  \citenamefont {Walters}}]{kernoghan1996positron}%
  \BibitemOpen
  \bibfield  {author} {\bibinfo {author} {\bibfnamefont {A.~A.}\ \bibnamefont
  {Kernoghan}}, \bibinfo {author} {\bibfnamefont {D.~J.~R.}\ \bibnamefont
  {Robinson}}, \bibinfo {author} {\bibfnamefont {M.~T.}\ \bibnamefont
  {McAlinden}}, \ and\ \bibinfo {author} {\bibfnamefont {H.~R.~J.}\
  \bibnamefont {Walters}},\ }\bibfield  {title} {\enquote {\bibinfo {title}
  {Positron scattering by atomic hydrogen},}\ }\href
  {http://stacks.iop.org/0953-4075/29/i=10/a=017} {\bibfield  {journal}
  {\bibinfo  {journal} {J. Phys. B}\ }\textbf {\bibinfo {volume} {29}},\
  \bibinfo {pages} {2089} (\bibinfo {year} {1996})}\BibitemShut {NoStop}%
\bibitem [{\citenamefont {Abdurakhmanov}\ \emph {et~al.}(2016)\citenamefont
  {Abdurakhmanov}, \citenamefont {Kadyrov},\ and\ \citenamefont
  {Bray}}]{abdurakhmanovPro}%
  \BibitemOpen
  \bibfield  {author} {\bibinfo {author} {\bibfnamefont {I.~B.}\ \bibnamefont
  {Abdurakhmanov}}, \bibinfo {author} {\bibfnamefont {A.~S.}\ \bibnamefont
  {Kadyrov}}, \ and\ \bibinfo {author} {\bibfnamefont {I.}~\bibnamefont
  {Bray}},\ }\bibfield  {title} {\enquote {\bibinfo {title} {Accurate solution
  of the proton-hydrogen three-body scattering problem},}\ }\href
  {http://stacks.iop.org/0953-4075/49/i=3/a=03LT01} {\bibfield  {journal}
  {\bibinfo  {journal} {J. Phys. B}\ }\textbf {\bibinfo {volume} {49}},\
  \bibinfo {pages} {03LT01} (\bibinfo {year} {2016})}\BibitemShut {NoStop}%
\bibitem [{\citenamefont {Abdurakhmanov}\ \emph {et~al.}(2011)\citenamefont
  {Abdurakhmanov}, \citenamefont {Kadyrov}, \citenamefont {Fursa},
  \citenamefont {Bray},\ and\ \citenamefont
  {Stelbovics}}]{abdurakhmanovHeAntiPro}%
  \BibitemOpen
  \bibfield  {author} {\bibinfo {author} {\bibfnamefont {I.~B.}\ \bibnamefont
  {Abdurakhmanov}}, \bibinfo {author} {\bibfnamefont {A.~S.}\ \bibnamefont
  {Kadyrov}}, \bibinfo {author} {\bibfnamefont {D.~V.}\ \bibnamefont {Fursa}},
  \bibinfo {author} {\bibfnamefont {I.}~\bibnamefont {Bray}}, \ and\ \bibinfo
  {author} {\bibfnamefont {A.~T.}\ \bibnamefont {Stelbovics}},\ }\bibfield
  {title} {\enquote {\bibinfo {title} {Convergent close-coupling calculations
  of helium single ionization by antiproton impact},}\ }\href {\doibase
  10.1103/PhysRevA.84.062708} {\bibfield  {journal} {\bibinfo  {journal} {Phys.
  Rev. A}\ }\textbf {\bibinfo {volume} {84}},\ \bibinfo {pages} {062708}
  (\bibinfo {year} {2011})}\BibitemShut {NoStop}%
\bibitem [{\citenamefont {Abdurakhmanov}\ \emph {et~al.}(2012)\citenamefont
  {Abdurakhmanov}, \citenamefont {Bray}, \citenamefont {Fursa}, \citenamefont
  {Kadyrov},\ and\ \citenamefont {Stelbovics}}]{abdurakhmanovC6}%
  \BibitemOpen
  \bibfield  {author} {\bibinfo {author} {\bibfnamefont {I.~B.}\ \bibnamefont
  {Abdurakhmanov}}, \bibinfo {author} {\bibfnamefont {I.}~\bibnamefont {Bray}},
  \bibinfo {author} {\bibfnamefont {D.~V.}\ \bibnamefont {Fursa}}, \bibinfo
  {author} {\bibfnamefont {A.~S.}\ \bibnamefont {Kadyrov}}, \ and\ \bibinfo
  {author} {\bibfnamefont {A.~T.}\ \bibnamefont {Stelbovics}},\ }\bibfield
  {title} {\enquote {\bibinfo {title} {Fully differential cross section for
  single ionization in energetic {C$^{6+}$}-{He} collisions},}\ }\href
  {\doibase 10.1103/PhysRevA.86.034701} {\bibfield  {journal} {\bibinfo
  {journal} {Phys. Rev. A}\ }\textbf {\bibinfo {volume} {86}},\ \bibinfo
  {pages} {034701} (\bibinfo {year} {2012})}\BibitemShut {NoStop}%
\bibitem [{\citenamefont {Fainstein}\ \emph {et~al.}(1991)\citenamefont
  {Fainstein}, \citenamefont {Ponce},\ and\ \citenamefont
  {Rivarola}}]{fainstein1991two}%
  \BibitemOpen
  \bibfield  {author} {\bibinfo {author} {\bibfnamefont {P.~D.}\ \bibnamefont
  {Fainstein}}, \bibinfo {author} {\bibfnamefont {V.~H.}\ \bibnamefont
  {Ponce}}, \ and\ \bibinfo {author} {\bibfnamefont {R.~D.}\ \bibnamefont
  {Rivarola}},\ }\bibfield  {title} {\enquote {\bibinfo {title} {Two-centre
  effects in ionization by ion impact},}\ }\href
  {http://stacks.iop.org/0953-4075/24/i=14/a=005} {\bibfield  {journal}
  {\bibinfo  {journal} {J. Phys. B}\ }\textbf {\bibinfo {volume} {24}},\
  \bibinfo {pages} {3091} (\bibinfo {year} {1991})}\BibitemShut {NoStop}%
\bibitem [{\citenamefont {Fabrikant}\ \emph {et~al.}(2016)\citenamefont
  {Fabrikant}, \citenamefont {Bray}, \citenamefont {Kadyrov},\ and\
  \citenamefont {Bray}}]{Fabrikant2016}%
  \BibitemOpen
  \bibfield  {author} {\bibinfo {author} {\bibfnamefont {I.~I.}\ \bibnamefont
  {Fabrikant}}, \bibinfo {author} {\bibfnamefont {A.~W.}\ \bibnamefont {Bray}},
  \bibinfo {author} {\bibfnamefont {A.~S.}\ \bibnamefont {Kadyrov}}, \ and\
  \bibinfo {author} {\bibfnamefont {I.}~\bibnamefont {Bray}},\ }\bibfield
  {title} {\enquote {\bibinfo {title} {Near-threshold behavior of
  positronium-antiproton scattering},}\ }\href {\doibase
  10.1103/PhysRevA.94.012701} {\bibfield  {journal} {\bibinfo  {journal} {Phys.
  Rev. A}\ }\textbf {\bibinfo {volume} {94}},\ \bibinfo {pages} {012701}
  (\bibinfo {year} {2016})}\BibitemShut {NoStop}%
\bibitem [{\citenamefont {Bailey}\ \emph {et~al.}(2015)\citenamefont {Bailey},
  \citenamefont {Kadyrov}, \citenamefont {Abdurakhmanov}, \citenamefont
  {Fursa},\ and\ \citenamefont {Bray}}]{bailey2015antiproton}%
  \BibitemOpen
  \bibfield  {author} {\bibinfo {author} {\bibfnamefont {J.~J.}\ \bibnamefont
  {Bailey}}, \bibinfo {author} {\bibfnamefont {A.~S.}\ \bibnamefont {Kadyrov}},
  \bibinfo {author} {\bibfnamefont {I.~B.}\ \bibnamefont {Abdurakhmanov}},
  \bibinfo {author} {\bibfnamefont {D.~V.}\ \bibnamefont {Fursa}}, \ and\
  \bibinfo {author} {\bibfnamefont {I.}~\bibnamefont {Bray}},\ }\bibfield
  {title} {\enquote {\bibinfo {title} {Antiproton stopping in {H}$_2$ and
  {H}$_2${O}},}\ }\href {\doibase 10.1103/PhysRevA.92.052711} {\bibfield
  {journal} {\bibinfo  {journal} {Phys. Rev. A}\ }\textbf {\bibinfo {volume}
  {92}},\ \bibinfo {pages} {052711} (\bibinfo {year} {2015})}\BibitemShut
  {NoStop}%
\bibitem [{\citenamefont {Maiman}(1960)}]{maiman1960stimulated}%
  \BibitemOpen
  \bibfield  {author} {\bibinfo {author} {\bibfnamefont {T.~H.}\ \bibnamefont
  {Maiman}},\ }\bibfield  {title} {\enquote {\bibinfo {title} {Stimulated
  optical radiation in ruby},}\ }\href {http://dx.doi.org/10.1038/187493a0}
  {\bibfield  {journal} {\bibinfo  {journal} {Nature}\ }\textbf {\bibinfo
  {volume} {187}},\ \bibinfo {pages} {493--494} (\bibinfo {year}
  {1960})}\BibitemShut {NoStop}%
\bibitem [{\citenamefont {Gordon}\ \emph {et~al.}(1955)\citenamefont {Gordon},
  \citenamefont {Zeiger},\ and\ \citenamefont {Townes}}]{maser}%
  \BibitemOpen
  \bibfield  {author} {\bibinfo {author} {\bibfnamefont {J.~P.}\ \bibnamefont
  {Gordon}}, \bibinfo {author} {\bibfnamefont {H.~J.}\ \bibnamefont {Zeiger}},
  \ and\ \bibinfo {author} {\bibfnamefont {C.~H.}\ \bibnamefont {Townes}},\
  }\bibfield  {title} {\enquote {\bibinfo {title} {The maser-new type of
  microwave amplifier, frequency standard, and spectrometer},}\ }\href
  {\doibase 10.1103/PhysRev.99.1264} {\bibfield  {journal} {\bibinfo  {journal}
  {Phys. Rev.}\ }\textbf {\bibinfo {volume} {99}},\ \bibinfo {pages}
  {1264--1274} (\bibinfo {year} {1955})}\BibitemShut {NoStop}%
\bibitem [{\citenamefont {Voronov}\ and\ \citenamefont
  {Delone}(1965)}]{voronov1965ionization}%
  \BibitemOpen
  \bibfield  {author} {\bibinfo {author} {\bibfnamefont {G.~S.}\ \bibnamefont
  {Voronov}}\ and\ \bibinfo {author} {\bibfnamefont {N.~B.}\ \bibnamefont
  {Delone}},\ }\bibfield  {title} {\enquote {\bibinfo {title} {Ionization of
  the xenon atom by the electric field of ruby laser emission},}\ }\href
  {http://www.jetpletters.ac.ru/ps/1590/article_24398.shtml} {\bibfield
  {journal} {\bibinfo  {journal} {J. Exp. Theor. Phys.}\ }\textbf {\bibinfo
  {volume} {1}},\ \bibinfo {pages} {66} (\bibinfo {year} {1965})}\BibitemShut
  {NoStop}%
\bibitem [{\citenamefont {Tozer}(1965)}]{tozer1965theory}%
  \BibitemOpen
  \bibfield  {author} {\bibinfo {author} {\bibfnamefont {B.~A.}\ \bibnamefont
  {Tozer}},\ }\bibfield  {title} {\enquote {\bibinfo {title} {Theory of the
  ionization of gases by laser beams},}\ }\href {\doibase
  10.1103/PhysRev.137.A1665} {\bibfield  {journal} {\bibinfo  {journal} {Phys.
  Rev.}\ }\textbf {\bibinfo {volume} {137}},\ \bibinfo {pages} {A1665}
  (\bibinfo {year} {1965})}\BibitemShut {NoStop}%
\bibitem [{\citenamefont {Agostini}\ \emph {et~al.}(1968)\citenamefont
  {Agostini}, \citenamefont {Barjot}, \citenamefont {Bonnal}, \citenamefont
  {Mainfray}, \citenamefont {Manus},\ and\ \citenamefont
  {Morellec}}]{agostini1968multiphoton}%
  \BibitemOpen
  \bibfield  {author} {\bibinfo {author} {\bibfnamefont {P.}~\bibnamefont
  {Agostini}}, \bibinfo {author} {\bibfnamefont {G.}~\bibnamefont {Barjot}},
  \bibinfo {author} {\bibfnamefont {J.}~\bibnamefont {Bonnal}}, \bibinfo
  {author} {\bibfnamefont {G.}~\bibnamefont {Mainfray}}, \bibinfo {author}
  {\bibfnamefont {C.}~\bibnamefont {Manus}}, \ and\ \bibinfo {author}
  {\bibfnamefont {J.}~\bibnamefont {Morellec}},\ }\bibfield  {title} {\enquote
  {\bibinfo {title} {Multiphoton ionization of hydrogen and rare gases},}\
  }\href {\doibase 10.1109/JQE.1968.1074955} {\bibfield  {journal} {\bibinfo
  {journal} {IEEE J. Quantum Elect.}\ }\textbf {\bibinfo {volume} {4}},\
  \bibinfo {pages} {667--669} (\bibinfo {year} {1968})}\BibitemShut {NoStop}%
\bibitem [{\citenamefont {Faisal}(1973)}]{Faisal1973}%
  \BibitemOpen
  \bibfield  {author} {\bibinfo {author} {\bibfnamefont {F.~H.~M.}\
  \bibnamefont {Faisal}},\ }\bibfield  {title} {\enquote {\bibinfo {title}
  {Multiple absorption of laser photons by atoms},}\ }\href
  {http://stacks.iop.org/0022-3700/6/i=4/a=011} {\bibfield  {journal} {\bibinfo
   {journal} {J. Phys. B}\ }\textbf {\bibinfo {volume} {6}},\ \bibinfo {pages}
  {L89} (\bibinfo {year} {1973})}\BibitemShut {NoStop}%
\bibitem [{\citenamefont {Deutsch}(1965)}]{DeutschRuby}%
  \BibitemOpen
  \bibfield  {author} {\bibinfo {author} {\bibfnamefont {T.}~\bibnamefont
  {Deutsch}},\ }\bibfield  {title} {\enquote {\bibinfo {title} {Mode-locking
  effects in an internally modulated ruby laser},}\ }\href {\doibase
  http://dx.doi.org/10.1063/1.1754321} {\bibfield  {journal} {\bibinfo
  {journal} {Appl. Phys. Lett.}\ }\textbf {\bibinfo {volume} {7}},\ \bibinfo
  {pages} {80--82} (\bibinfo {year} {1965})}\BibitemShut {NoStop}%
\bibitem [{\citenamefont {DiDomenico}\ \emph {et~al.}(1966)\citenamefont
  {DiDomenico}, \citenamefont {Geusic}, \citenamefont {Marcos},\ and\
  \citenamefont {Smith}}]{DiDomenicoNd}%
  \BibitemOpen
  \bibfield  {author} {\bibinfo {author} {\bibfnamefont {M.}~\bibnamefont
  {DiDomenico}}, \bibinfo {author} {\bibfnamefont {J.~E.}\ \bibnamefont
  {Geusic}}, \bibinfo {author} {\bibfnamefont {H.~M.}\ \bibnamefont {Marcos}},
  \ and\ \bibinfo {author} {\bibfnamefont {R.~G.}\ \bibnamefont {Smith}},\
  }\bibfield  {title} {\enquote {\bibinfo {title} {Generation of ultrashort
  optical pulses by mode locking the {YAIG}: Nd laser},}\ }\href {\doibase
  http://dx.doi.org/10.1063/1.1754544} {\bibfield  {journal} {\bibinfo
  {journal} {Appl. Phys. Lett.}\ }\textbf {\bibinfo {volume} {8}},\ \bibinfo
  {pages} {180--183} (\bibinfo {year} {1966})}\BibitemShut {NoStop}%
\bibitem [{\citenamefont {Moulton}(1982)}]{moulton1982ti}%
  \BibitemOpen
  \bibfield  {author} {\bibinfo {author} {\bibfnamefont {P.}~\bibnamefont
  {Moulton}},\ }\bibfield  {title} {\enquote {\bibinfo {title} {Ti-doped
  sapphire: tunable solid-state laser},}\ }\href {\doibase
  10.1364/ON.8.6.000009} {\bibfield  {journal} {\bibinfo  {journal} {Opt.
  News}\ }\textbf {\bibinfo {volume} {8}},\ \bibinfo {pages} {9--9} (\bibinfo
  {year} {1982})}\BibitemShut {NoStop}%
\bibitem [{\citenamefont {Yu}\ \emph {et~al.}(2000)\citenamefont {Yu},
  \citenamefont {Babzien}, \citenamefont {Ben-Zvi}, \citenamefont {DiMauro},
  \citenamefont {Doyuran}, \citenamefont {Graves}, \citenamefont {Johnson},
  \citenamefont {Krinsky}, \citenamefont {Malone}, \citenamefont {Pogorelsky},
  \citenamefont {Skaritka}, \citenamefont {Rakowsky}, \citenamefont {Solomon},
  \citenamefont {Wang}, \citenamefont {Woodle}, \citenamefont {Yakimenko},
  \citenamefont {Biedron}, \citenamefont {Galayda}, \citenamefont {Gluskin},
  \citenamefont {Jagger}, \citenamefont {Sajaev},\ and\ \citenamefont
  {Vasserman}}]{YuFEL}%
  \BibitemOpen
  \bibfield  {author} {\bibinfo {author} {\bibfnamefont {L.~H.}\ \bibnamefont
  {Yu}}, \bibinfo {author} {\bibfnamefont {M.}~\bibnamefont {Babzien}},
  \bibinfo {author} {\bibfnamefont {I.}~\bibnamefont {Ben-Zvi}}, \bibinfo
  {author} {\bibfnamefont {L.~F.}\ \bibnamefont {DiMauro}}, \bibinfo {author}
  {\bibfnamefont {A.}~\bibnamefont {Doyuran}}, \bibinfo {author} {\bibfnamefont
  {W.}~\bibnamefont {Graves}}, \bibinfo {author} {\bibfnamefont
  {E.}~\bibnamefont {Johnson}}, \bibinfo {author} {\bibfnamefont
  {S.}~\bibnamefont {Krinsky}}, \bibinfo {author} {\bibfnamefont
  {R.}~\bibnamefont {Malone}}, \bibinfo {author} {\bibfnamefont
  {I.}~\bibnamefont {Pogorelsky}}, \bibinfo {author} {\bibfnamefont
  {J.}~\bibnamefont {Skaritka}}, \bibinfo {author} {\bibfnamefont
  {G.}~\bibnamefont {Rakowsky}}, \bibinfo {author} {\bibfnamefont
  {L.}~\bibnamefont {Solomon}}, \bibinfo {author} {\bibfnamefont {X.~J.}\
  \bibnamefont {Wang}}, \bibinfo {author} {\bibfnamefont {M.}~\bibnamefont
  {Woodle}}, \bibinfo {author} {\bibfnamefont {V.}~\bibnamefont {Yakimenko}},
  \bibinfo {author} {\bibfnamefont {S.~G.}\ \bibnamefont {Biedron}}, \bibinfo
  {author} {\bibfnamefont {J.~N.}\ \bibnamefont {Galayda}}, \bibinfo {author}
  {\bibfnamefont {E.}~\bibnamefont {Gluskin}}, \bibinfo {author} {\bibfnamefont
  {J.}~\bibnamefont {Jagger}}, \bibinfo {author} {\bibfnamefont
  {V.}~\bibnamefont {Sajaev}}, \ and\ \bibinfo {author} {\bibfnamefont
  {I.}~\bibnamefont {Vasserman}},\ }\bibfield  {title} {\enquote {\bibinfo
  {title} {High-gain harmonic-generation free-electron laser},}\ }\href
  {\doibase 10.1126/science.289.5481.932} {\bibfield  {journal} {\bibinfo
  {journal} {Science}\ }\textbf {\bibinfo {volume} {289}},\ \bibinfo {pages}
  {932--934} (\bibinfo {year} {2000})}\BibitemShut {NoStop}%
\bibitem [{\citenamefont {Zewail}(1988)}]{zewail1988laser}%
  \BibitemOpen
  \bibfield  {author} {\bibinfo {author} {\bibfnamefont {A.~H.}\ \bibnamefont
  {Zewail}},\ }\bibfield  {title} {\enquote {\bibinfo {title} {Laser
  femtochemistry},}\ }\href
  {http://search.proquest.com/docview/213537780?accountid=8330} {\bibfield
  {journal} {\bibinfo  {journal} {Science}\ }\textbf {\bibinfo {volume}
  {242}},\ \bibinfo {pages} {1645--1653} (\bibinfo {year} {1988})}\BibitemShut
  {NoStop}%
\bibitem [{\citenamefont {Zewail}\ and\ \citenamefont
  {Bernstein}(1988)}]{zewail1988real}%
  \BibitemOpen
  \bibfield  {author} {\bibinfo {author} {\bibfnamefont {A.~H.}\ \bibnamefont
  {Zewail}}\ and\ \bibinfo {author} {\bibfnamefont {R.~B.}\ \bibnamefont
  {Bernstein}},\ }\bibfield  {title} {\enquote {\bibinfo {title} {Real-time
  laser femtochemistry: Viewing the transition from reagents to products},}\
  }\href {\doibase 10.1021/cen-v066n045.p024} {\bibfield  {journal} {\bibinfo
  {journal} {Chem. Eng. News}\ }\textbf {\bibinfo {volume} {66}},\ \bibinfo
  {pages} {24--43} (\bibinfo {year} {1988})}\BibitemShut {NoStop}%
\bibitem [{\citenamefont {Lewenstein}\ \emph {et~al.}(1994)\citenamefont
  {Lewenstein}, \citenamefont {Balcou}, \citenamefont {Ivanov}, \citenamefont
  {L'Huillier},\ and\ \citenamefont {Corkum}}]{LewensteinHHG}%
  \BibitemOpen
  \bibfield  {author} {\bibinfo {author} {\bibfnamefont {M.}~\bibnamefont
  {Lewenstein}}, \bibinfo {author} {\bibfnamefont {P.}~\bibnamefont {Balcou}},
  \bibinfo {author} {\bibfnamefont {M.~Y.}\ \bibnamefont {Ivanov}}, \bibinfo
  {author} {\bibfnamefont {A.}~\bibnamefont {L'Huillier}}, \ and\ \bibinfo
  {author} {\bibfnamefont {P.~B.}\ \bibnamefont {Corkum}},\ }\bibfield  {title}
  {\enquote {\bibinfo {title} {Theory of high-harmonic generation by
  low-frequency laser fields},}\ }\href {\doibase 10.1103/PhysRevA.49.2117}
  {\bibfield  {journal} {\bibinfo  {journal} {Phys. Rev. A}\ }\textbf {\bibinfo
  {volume} {49}},\ \bibinfo {pages} {2117--2132} (\bibinfo {year}
  {1994})}\BibitemShut {NoStop}%
\bibitem [{\citenamefont {Antoine}\ \emph {et~al.}(1996)\citenamefont
  {Antoine}, \citenamefont {L'Huillier},\ and\ \citenamefont
  {Lewenstein}}]{AntoineAttosecond}%
  \BibitemOpen
  \bibfield  {author} {\bibinfo {author} {\bibfnamefont {P.}~\bibnamefont
  {Antoine}}, \bibinfo {author} {\bibfnamefont {A.}~\bibnamefont {L'Huillier}},
  \ and\ \bibinfo {author} {\bibfnamefont {M.}~\bibnamefont {Lewenstein}},\
  }\bibfield  {title} {\enquote {\bibinfo {title} {Attosecond pulse trains
  using high\char21{}order harmonics},}\ }\href {\doibase
  10.1103/PhysRevLett.77.1234} {\bibfield  {journal} {\bibinfo  {journal}
  {Phys. Rev. Lett.}\ }\textbf {\bibinfo {volume} {77}},\ \bibinfo {pages}
  {1234--1237} (\bibinfo {year} {1996})}\BibitemShut {NoStop}%
\bibitem [{\citenamefont {Itatani}\ \emph {et~al.}(2002)\citenamefont
  {Itatani}, \citenamefont {Qu{\'e}r{\'e}}, \citenamefont {Yudin},
  \citenamefont {Ivanov}, \citenamefont {Krausz},\ and\ \citenamefont
  {Corkum}}]{itatani2002attosecond}%
  \BibitemOpen
  \bibfield  {author} {\bibinfo {author} {\bibfnamefont {J.}~\bibnamefont
  {Itatani}}, \bibinfo {author} {\bibfnamefont {F.}~\bibnamefont
  {Qu{\'e}r{\'e}}}, \bibinfo {author} {\bibfnamefont {G.~L.}\ \bibnamefont
  {Yudin}}, \bibinfo {author} {\bibfnamefont {M.~Y.}\ \bibnamefont {Ivanov}},
  \bibinfo {author} {\bibfnamefont {F.}~\bibnamefont {Krausz}}, \ and\ \bibinfo
  {author} {\bibfnamefont {P.~B.}\ \bibnamefont {Corkum}},\ }\bibfield  {title}
  {\enquote {\bibinfo {title} {Attosecond streak camera},}\ }\href {\doibase
  https://doi.org/10.1103/PhysRevLett.88.173903} {\bibfield  {journal}
  {\bibinfo  {journal} {Phys. Rev. Lett.}\ }\textbf {\bibinfo {volume} {88}},\
  \bibinfo {pages} {173903} (\bibinfo {year} {2002})}\BibitemShut {NoStop}%
\bibitem [{\citenamefont {Corkum}\ and\ \citenamefont
  {Krausz}(2007)}]{corkum2007attosecond}%
  \BibitemOpen
  \bibfield  {author} {\bibinfo {author} {\bibfnamefont {P.~B.}\ \bibnamefont
  {Corkum}}\ and\ \bibinfo {author} {\bibfnamefont {F.}~\bibnamefont
  {Krausz}},\ }\bibfield  {title} {\enquote {\bibinfo {title} {Attosecond
  science},}\ }\href {\doibase 10.1038/nphys620} {\bibfield  {journal}
  {\bibinfo  {journal} {Nat. Phys.}\ }\textbf {\bibinfo {volume} {3}},\
  \bibinfo {pages} {381--387} (\bibinfo {year} {2007})}\BibitemShut {NoStop}%
\bibitem [{\citenamefont {Ivanov}\ and\ \citenamefont
  {Smirnova}(2011)}]{ivanov2011accurate}%
  \BibitemOpen
  \bibfield  {author} {\bibinfo {author} {\bibfnamefont {M.}~\bibnamefont
  {Ivanov}}\ and\ \bibinfo {author} {\bibfnamefont {O.}~\bibnamefont
  {Smirnova}},\ }\bibfield  {title} {\enquote {\bibinfo {title} {How accurate
  is the attosecond streak camera?}}\ }\href {\doibase
  10.1103/PhysRevLett.107.213605} {\bibfield  {journal} {\bibinfo  {journal}
  {Phys. Rev. Lett.}\ }\textbf {\bibinfo {volume} {107}},\ \bibinfo {pages}
  {213605} (\bibinfo {year} {2011})}\BibitemShut {NoStop}%
\bibitem [{\citenamefont {Kroll}\ and\ \citenamefont
  {Watson}(1973)}]{Kroll1973}%
  \BibitemOpen
  \bibfield  {author} {\bibinfo {author} {\bibfnamefont {N.~M.}\ \bibnamefont
  {Kroll}}\ and\ \bibinfo {author} {\bibfnamefont {K.~M.}\ \bibnamefont
  {Watson}},\ }\bibfield  {title} {\enquote {\bibinfo {title} {Charged-particle
  scattering in the presence of a strong electromagnetic wave},}\ }\href
  {\doibase 10.1103/PhysRevA.8.804} {\bibfield  {journal} {\bibinfo  {journal}
  {Phys. Rev. A}\ }\textbf {\bibinfo {volume} {8}},\ \bibinfo {pages}
  {804--809} (\bibinfo {year} {1973})}\BibitemShut {NoStop}%
\bibitem [{\citenamefont {Ladino}\ \emph {et~al.}(2011)\citenamefont {Ladino},
  \citenamefont {MacAdam}, \citenamefont {Martin} \emph
  {et~al.}}]{ladino2011high}%
  \BibitemOpen
  \bibfield  {author} {\bibinfo {author} {\bibfnamefont {L.}~\bibnamefont
  {Ladino}}, \bibinfo {author} {\bibfnamefont {K.~B.}\ \bibnamefont {MacAdam}},
  \bibinfo {author} {\bibfnamefont {N.~L.~S.}\ \bibnamefont {Martin}},  \emph
  {et~al.},\ }\bibfield  {title} {\enquote {\bibinfo {title} {High-energy
  electron-helium scattering in a {Nd}: {YAG} laser field},}\ }\href {\doibase
  10.1103/PhysRevA.83.022706} {\bibfield  {journal} {\bibinfo  {journal} {Phys.
  Rev. A}\ }\textbf {\bibinfo {volume} {83}},\ \bibinfo {pages} {022706}
  (\bibinfo {year} {2011})}\BibitemShut {NoStop}%
\bibitem [{\citenamefont {Rahman}(1974)}]{Rahman1974}%
  \BibitemOpen
  \bibfield  {author} {\bibinfo {author} {\bibfnamefont {N.~K.}\ \bibnamefont
  {Rahman}},\ }\bibfield  {title} {\enquote {\bibinfo {title} {Effect of
  intense electromagnetic wave in charged-particle scattering},}\ }\href
  {\doibase 10.1103/PhysRevA.10.440} {\bibfield  {journal} {\bibinfo  {journal}
  {Phys. Rev. A}\ }\textbf {\bibinfo {volume} {10}},\ \bibinfo {pages}
  {440--441} (\bibinfo {year} {1974})}\BibitemShut {NoStop}%
\bibitem [{\citenamefont {Weingartshofer}\ \emph {et~al.}(1977)\citenamefont
  {Weingartshofer}, \citenamefont {Holmes}, \citenamefont {Caudle},
  \citenamefont {Clarke},\ and\ \citenamefont {Kr\"uger}}]{Wein1977}%
  \BibitemOpen
  \bibfield  {author} {\bibinfo {author} {\bibfnamefont {A.}~\bibnamefont
  {Weingartshofer}}, \bibinfo {author} {\bibfnamefont {J.~K.}\ \bibnamefont
  {Holmes}}, \bibinfo {author} {\bibfnamefont {G.}~\bibnamefont {Caudle}},
  \bibinfo {author} {\bibfnamefont {E.~M.}\ \bibnamefont {Clarke}}, \ and\
  \bibinfo {author} {\bibfnamefont {H.}~\bibnamefont {Kr\"uger}},\ }\bibfield
  {title} {\enquote {\bibinfo {title} {Direct observation of multiphoton
  processes in laser-induced free-free transitions},}\ }\href {\doibase
  10.1103/PhysRevLett.39.269} {\bibfield  {journal} {\bibinfo  {journal} {Phys.
  Rev. Lett.}\ }\textbf {\bibinfo {volume} {39}},\ \bibinfo {pages} {269--270}
  (\bibinfo {year} {1977})}\BibitemShut {NoStop}%
\bibitem [{\citenamefont {Weingartshofer}\ \emph {et~al.}(1979)\citenamefont
  {Weingartshofer}, \citenamefont {Clarke}, \citenamefont {Holmes},\ and\
  \citenamefont {Jung}}]{Wein1979}%
  \BibitemOpen
  \bibfield  {author} {\bibinfo {author} {\bibfnamefont {A.}~\bibnamefont
  {Weingartshofer}}, \bibinfo {author} {\bibfnamefont {E.~M.}\ \bibnamefont
  {Clarke}}, \bibinfo {author} {\bibfnamefont {J.~K.}\ \bibnamefont {Holmes}},
  \ and\ \bibinfo {author} {\bibfnamefont {C.}~\bibnamefont {Jung}},\
  }\bibfield  {title} {\enquote {\bibinfo {title} {Experiments on multiphoton
  free-free transitions},}\ }\href {\doibase 10.1103/PhysRevA.19.2371}
  {\bibfield  {journal} {\bibinfo  {journal} {Phys. Rev. A}\ }\textbf {\bibinfo
  {volume} {19}},\ \bibinfo {pages} {2371--2376} (\bibinfo {year}
  {1979})}\BibitemShut {NoStop}%
\bibitem [{\citenamefont {Wallbank}\ \emph {et~al.}(1987)\citenamefont
  {Wallbank}, \citenamefont {Holmes},\ and\ \citenamefont
  {Weingartshofer}}]{wallbank1987experimental}%
  \BibitemOpen
  \bibfield  {author} {\bibinfo {author} {\bibfnamefont {B.}~\bibnamefont
  {Wallbank}}, \bibinfo {author} {\bibfnamefont {J.~K.}\ \bibnamefont
  {Holmes}}, \ and\ \bibinfo {author} {\bibfnamefont {A.}~\bibnamefont
  {Weingartshofer}},\ }\bibfield  {title} {\enquote {\bibinfo {title}
  {Experimental differential cross sections for multiphoton free-free
  transitions},}\ }\href {http://stacks.iop.org/0022-3700/20/i=22/a=021}
  {\bibfield  {journal} {\bibinfo  {journal} {J. Phys. B}\ }\textbf {\bibinfo
  {volume} {20}},\ \bibinfo {pages} {6121} (\bibinfo {year}
  {1987})}\BibitemShut {NoStop}%
\bibitem [{\citenamefont {Wallbank}\ and\ \citenamefont
  {Holmes}(1993)}]{wallbank1993}%
  \BibitemOpen
  \bibfield  {author} {\bibinfo {author} {\bibfnamefont {B.}~\bibnamefont
  {Wallbank}}\ and\ \bibinfo {author} {\bibfnamefont {J.~K.}\ \bibnamefont
  {Holmes}},\ }\bibfield  {title} {\enquote {\bibinfo {title} {Laser-assisted
  elastic electron-atom collisions},}\ }\href {\doibase
  10.1103/PhysRevA.48.R2515} {\bibfield  {journal} {\bibinfo  {journal} {Phys.
  Rev. A}\ }\textbf {\bibinfo {volume} {48}},\ \bibinfo {pages} {R2515--R2518}
  (\bibinfo {year} {1993})}\BibitemShut {NoStop}%
\bibitem [{\citenamefont {Geltman}(1996)}]{geltman1996laser}%
  \BibitemOpen
  \bibfield  {author} {\bibinfo {author} {\bibfnamefont {S.}~\bibnamefont
  {Geltman}},\ }\bibfield  {title} {\enquote {\bibinfo {title} {Laser-assisted
  collisions: The {Kroll}-{Watson} formula and bremsstrahlung theory},}\ }\href
  {\doibase 10.1103/PhysRevA.53.3473} {\bibfield  {journal} {\bibinfo
  {journal} {Phys. Rev. A}\ }\textbf {\bibinfo {volume} {53}},\ \bibinfo
  {pages} {3473} (\bibinfo {year} {1996})}\BibitemShut {NoStop}%
\bibitem [{\citenamefont {Byron~Jr}\ and\ \citenamefont
  {Joachain}(1984)}]{byron1984electron}%
  \BibitemOpen
  \bibfield  {author} {\bibinfo {author} {\bibfnamefont {F.~W.}\ \bibnamefont
  {Byron~Jr}}\ and\ \bibinfo {author} {\bibfnamefont {C.~J.}\ \bibnamefont
  {Joachain}},\ }\bibfield  {title} {\enquote {\bibinfo {title} {Electron-atom
  collisions in a strong laser field},}\ }\href
  {http://stacks.iop.org/0022-3700/17/i=9/a=006} {\bibfield  {journal}
  {\bibinfo  {journal} {J. Phys. B}\ }\textbf {\bibinfo {volume} {17}},\
  \bibinfo {pages} {L295} (\bibinfo {year} {1984})}\BibitemShut {NoStop}%
\bibitem [{\citenamefont {Cavaliere}\ \emph {et~al.}(1980)\citenamefont
  {Cavaliere}, \citenamefont {Ferrante},\ and\ \citenamefont
  {Leone}}]{cavaliere1980particle}%
  \BibitemOpen
  \bibfield  {author} {\bibinfo {author} {\bibfnamefont {P.}~\bibnamefont
  {Cavaliere}}, \bibinfo {author} {\bibfnamefont {G.}~\bibnamefont {Ferrante}},
  \ and\ \bibinfo {author} {\bibfnamefont {C.}~\bibnamefont {Leone}},\
  }\bibfield  {title} {\enquote {\bibinfo {title} {Particle-atom ionising
  collisions in the presence of a laser radiation field},}\ }\href
  {http://stacks.iop.org/0022-3700/13/i=22/a=021} {\bibfield  {journal}
  {\bibinfo  {journal} {J. Phys. B}\ }\textbf {\bibinfo {volume} {13}},\
  \bibinfo {pages} {4495} (\bibinfo {year} {1980})}\BibitemShut {NoStop}%
\bibitem [{\citenamefont {Cavaliere}\ \emph {et~al.}(1981)\citenamefont
  {Cavaliere}, \citenamefont {Leone}, \citenamefont {Zangara},\ and\
  \citenamefont {Ferrante}}]{cavaliere1981effects}%
  \BibitemOpen
  \bibfield  {author} {\bibinfo {author} {\bibfnamefont {P.}~\bibnamefont
  {Cavaliere}}, \bibinfo {author} {\bibfnamefont {C.}~\bibnamefont {Leone}},
  \bibinfo {author} {\bibfnamefont {R.}~\bibnamefont {Zangara}}, \ and\
  \bibinfo {author} {\bibfnamefont {G.}~\bibnamefont {Ferrante}},\ }\bibfield
  {title} {\enquote {\bibinfo {title} {Effects of a laser field on
  electron-atom ionizing collisions},}\ }\href {\doibase
  10.1103/PhysRevA.24.910} {\bibfield  {journal} {\bibinfo  {journal} {Phys.
  Rev. A}\ }\textbf {\bibinfo {volume} {24}},\ \bibinfo {pages} {910} (\bibinfo
  {year} {1981})}\BibitemShut {NoStop}%
\bibitem [{\citenamefont {Banerji}\ and\ \citenamefont
  {Mittleman}(1981)}]{banerji1981electron}%
  \BibitemOpen
  \bibfield  {author} {\bibinfo {author} {\bibfnamefont {J.}~\bibnamefont
  {Banerji}}\ and\ \bibinfo {author} {\bibfnamefont {M.~H.}\ \bibnamefont
  {Mittleman}},\ }\bibfield  {title} {\enquote {\bibinfo {title} {Electron-atom
  ionising collisions in the presence of a low-frequency laser field},}\ }\href
  {http://stacks.iop.org/0022-3700/14/i=19/a=020} {\bibfield  {journal}
  {\bibinfo  {journal} {J. Phys. B}\ }\textbf {\bibinfo {volume} {14}},\
  \bibinfo {pages} {3717} (\bibinfo {year} {1981})}\BibitemShut {NoStop}%
\bibitem [{\citenamefont {Joachain}\ \emph {et~al.}(1988)\citenamefont
  {Joachain}, \citenamefont {Francken}, \citenamefont {Maquet}, \citenamefont
  {Martin},\ and\ \citenamefont {Veniard}}]{Joachain1988}%
  \BibitemOpen
  \bibfield  {author} {\bibinfo {author} {\bibfnamefont {C.~J.}\ \bibnamefont
  {Joachain}}, \bibinfo {author} {\bibfnamefont {P.}~\bibnamefont {Francken}},
  \bibinfo {author} {\bibfnamefont {A.}~\bibnamefont {Maquet}}, \bibinfo
  {author} {\bibfnamefont {P.}~\bibnamefont {Martin}}, \ and\ \bibinfo {author}
  {\bibfnamefont {V.}~\bibnamefont {Veniard}},\ }\bibfield  {title} {\enquote
  {\bibinfo {title} {(e,2e) collisions in the presence of a laser field},}\
  }\href {\doibase 10.1103/PhysRevLett.61.165} {\bibfield  {journal} {\bibinfo
  {journal} {Phys. Rev. Lett.}\ }\textbf {\bibinfo {volume} {61}},\ \bibinfo
  {pages} {165--168} (\bibinfo {year} {1988})}\BibitemShut {NoStop}%
\bibitem [{\citenamefont {Martin}\ \emph {et~al.}(1989)\citenamefont {Martin},
  \citenamefont {Veniard}, \citenamefont {Maquet}, \citenamefont {Francken},\
  and\ \citenamefont {Joachain}}]{Martin1989}%
  \BibitemOpen
  \bibfield  {author} {\bibinfo {author} {\bibfnamefont {P.}~\bibnamefont
  {Martin}}, \bibinfo {author} {\bibfnamefont {V.}~\bibnamefont {Veniard}},
  \bibinfo {author} {\bibfnamefont {A.}~\bibnamefont {Maquet}}, \bibinfo
  {author} {\bibfnamefont {P.}~\bibnamefont {Francken}}, \ and\ \bibinfo
  {author} {\bibfnamefont {C.~J.}\ \bibnamefont {Joachain}},\ }\bibfield
  {title} {\enquote {\bibinfo {title} {Electron-impact ionization of atomic
  hydrogen in the presence of a laser field},}\ }\href {\doibase
  10.1103/PhysRevA.39.6178} {\bibfield  {journal} {\bibinfo  {journal} {Phys.
  Rev. A}\ }\textbf {\bibinfo {volume} {39}},\ \bibinfo {pages} {6178--6189}
  (\bibinfo {year} {1989})}\BibitemShut {NoStop}%
\bibitem [{\citenamefont {Joachain}\ \emph {et~al.}(1992)\citenamefont
  {Joachain}, \citenamefont {Makhoute}, \citenamefont {Maquet},\ and\
  \citenamefont {Taieb}}]{joachain1992laser}%
  \BibitemOpen
  \bibfield  {author} {\bibinfo {author} {\bibfnamefont {C.~J.}\ \bibnamefont
  {Joachain}}, \bibinfo {author} {\bibfnamefont {A.}~\bibnamefont {Makhoute}},
  \bibinfo {author} {\bibfnamefont {A.}~\bibnamefont {Maquet}}, \ and\ \bibinfo
  {author} {\bibfnamefont {R.}~\bibnamefont {Taieb}},\ }\bibfield  {title}
  {\enquote {\bibinfo {title} {Laser-assisted (e,2e) collisions in helium},}\
  }\href {\doibase 10.1007/BF01429264} {\bibfield  {journal} {\bibinfo
  {journal} {Z. Phys. D}\ }\textbf {\bibinfo {volume} {23}},\ \bibinfo {pages}
  {397--401} (\bibinfo {year} {1992})}\BibitemShut {NoStop}%
\bibitem [{\citenamefont {Khalil}\ \emph {et~al.}(1997)\citenamefont {Khalil},
  \citenamefont {Maquet}, \citenamefont {Ta\"{\i}eb}, \citenamefont
  {Joachain},\ and\ \citenamefont {Makhoute}}]{Khalil1997}%
  \BibitemOpen
  \bibfield  {author} {\bibinfo {author} {\bibfnamefont {D.}~\bibnamefont
  {Khalil}}, \bibinfo {author} {\bibfnamefont {A.}~\bibnamefont {Maquet}},
  \bibinfo {author} {\bibfnamefont {R.}~\bibnamefont {Ta\"{\i}eb}}, \bibinfo
  {author} {\bibfnamefont {C.~J.}\ \bibnamefont {Joachain}}, \ and\ \bibinfo
  {author} {\bibfnamefont {A.}~\bibnamefont {Makhoute}},\ }\bibfield  {title}
  {\enquote {\bibinfo {title} {Laser-assisted (e,2e) collisions in helium},}\
  }\href {\doibase 10.1103/PhysRevA.56.4918} {\bibfield  {journal} {\bibinfo
  {journal} {Phys. Rev. A}\ }\textbf {\bibinfo {volume} {56}},\ \bibinfo
  {pages} {4918--4928} (\bibinfo {year} {1997})}\BibitemShut {NoStop}%
\bibitem [{\citenamefont {H{\"o}hr}\ \emph {et~al.}(2005)\citenamefont
  {H{\"o}hr}, \citenamefont {Dorn}, \citenamefont {Najjari}, \citenamefont
  {Fischer}, \citenamefont {Schr{\"o}ter},\ and\ \citenamefont
  {Ullrich}}]{hohr2005electron}%
  \BibitemOpen
  \bibfield  {author} {\bibinfo {author} {\bibfnamefont {C.}~\bibnamefont
  {H{\"o}hr}}, \bibinfo {author} {\bibfnamefont {A.}~\bibnamefont {Dorn}},
  \bibinfo {author} {\bibfnamefont {B.}~\bibnamefont {Najjari}}, \bibinfo
  {author} {\bibfnamefont {D.}~\bibnamefont {Fischer}}, \bibinfo {author}
  {\bibfnamefont {C.~D.}\ \bibnamefont {Schr{\"o}ter}}, \ and\ \bibinfo
  {author} {\bibfnamefont {J.}~\bibnamefont {Ullrich}},\ }\bibfield  {title}
  {\enquote {\bibinfo {title} {Electron impact ionization in the presence of a
  laser field: A kinematically complete (n $\gamma$ e, 2 e) experiment},}\
  }\href {\doibase 10.1103/PhysRevLett.94.153201} {\bibfield  {journal}
  {\bibinfo  {journal} {Phys. Rev. Lett.}\ }\textbf {\bibinfo {volume} {94}},\
  \bibinfo {pages} {153201} (\bibinfo {year} {2005})}\BibitemShut {NoStop}%
\bibitem [{\citenamefont {H{\"o}hr}\ \emph {et~al.}(2007)\citenamefont
  {H{\"o}hr}, \citenamefont {Dorn}, \citenamefont {Najjari}, \citenamefont
  {Fischer}, \citenamefont {Schr{\"o}ter},\ and\ \citenamefont
  {Ullrich}}]{hohr2007laser}%
  \BibitemOpen
  \bibfield  {author} {\bibinfo {author} {\bibfnamefont {C.}~\bibnamefont
  {H{\"o}hr}}, \bibinfo {author} {\bibfnamefont {A.}~\bibnamefont {Dorn}},
  \bibinfo {author} {\bibfnamefont {B.}~\bibnamefont {Najjari}}, \bibinfo
  {author} {\bibfnamefont {D.}~\bibnamefont {Fischer}}, \bibinfo {author}
  {\bibfnamefont {C.~D.}\ \bibnamefont {Schr{\"o}ter}}, \ and\ \bibinfo
  {author} {\bibfnamefont {J.}~\bibnamefont {Ullrich}},\ }\bibfield  {title}
  {\enquote {\bibinfo {title} {Laser-assisted electron-impact ionization of
  atoms},}\ }\href {\doibase http://dx.doi.org/10.1016/j.elspec.2007.02.001}
  {\bibfield  {journal} {\bibinfo  {journal} {J. Electron Spectrosc. Relat.
  Phenom.}\ }\textbf {\bibinfo {volume} {161}},\ \bibinfo {pages} {172--177}
  (\bibinfo {year} {2007})}\BibitemShut {NoStop}%
\bibitem [{\citenamefont {Burke}\ \emph {et~al.}(1991)\citenamefont {Burke},
  \citenamefont {Francken},\ and\ \citenamefont {Joachain}}]{burke1991r}%
  \BibitemOpen
  \bibfield  {author} {\bibinfo {author} {\bibfnamefont {P.~G.}\ \bibnamefont
  {Burke}}, \bibinfo {author} {\bibfnamefont {P.}~\bibnamefont {Francken}}, \
  and\ \bibinfo {author} {\bibfnamefont {C.~J.}\ \bibnamefont {Joachain}},\
  }\bibfield  {title} {\enquote {\bibinfo {title} {{$R$}-matrix-floquet theory
  of multiphoton processes},}\ }\href
  {http://stacks.iop.org/0953-4075/24/i=4/a=005} {\bibfield  {journal}
  {\bibinfo  {journal} {J. Phys. B}\ }\textbf {\bibinfo {volume} {24}},\
  \bibinfo {pages} {761} (\bibinfo {year} {1991})}\BibitemShut {NoStop}%
\bibitem [{\citenamefont {Makhoute}\ \emph {et~al.}(2015)\citenamefont
  {Makhoute}, \citenamefont {Ajana}, \citenamefont {Khalil},\ and\
  \citenamefont {Chaddou}}]{Makhoute2015}%
  \BibitemOpen
  \bibfield  {author} {\bibinfo {author} {\bibfnamefont {A.}~\bibnamefont
  {Makhoute}}, \bibinfo {author} {\bibfnamefont {I.}~\bibnamefont {Ajana}},
  \bibinfo {author} {\bibfnamefont {D.}~\bibnamefont {Khalil}}, \ and\ \bibinfo
  {author} {\bibfnamefont {S.}~\bibnamefont {Chaddou}},\ }\bibfield  {title}
  {\enquote {\bibinfo {title} {Second-order born calculation of laser-assisted
  single ionization of helium by electrons},}\ }\href {\doibase
  10.1140/epjd/e2015-60141-5} {\bibfield  {journal} {\bibinfo  {journal} {Eur.
  Phys. J. D.}\ }\textbf {\bibinfo {volume} {69}},\ \bibinfo {pages} {160}
  (\bibinfo {year} {2015})}\BibitemShut {NoStop}%
\bibitem [{\citenamefont {Ghosh~Deb}\ and\ \citenamefont
  {Sinha}(2010)}]{GhoshDeb2010}%
  \BibitemOpen
  \bibfield  {author} {\bibinfo {author} {\bibfnamefont {S.}~\bibnamefont
  {Ghosh~Deb}}\ and\ \bibinfo {author} {\bibfnamefont {C.}~\bibnamefont
  {Sinha}},\ }\bibfield  {title} {\enquote {\bibinfo {title} {Multiphoton
  effects in laser-assisted ionization of a helium atom by electron impact},}\
  }\href {\doibase 10.1140/epjd/e2010-00209-2} {\bibfield  {journal} {\bibinfo
  {journal} {Eur. Phys. J. D.}\ }\textbf {\bibinfo {volume} {60}},\ \bibinfo
  {pages} {287--294} (\bibinfo {year} {2010})}\BibitemShut {NoStop}%
\bibitem [{\citenamefont {{Baltu\u{s}ka {\em et al}}}(2003)}]{Baltuska2003}%
  \BibitemOpen
  \bibfield  {author} {\bibinfo {author} {\bibfnamefont {A.}~\bibnamefont
  {{Baltu\u{s}ka {\em et al}}}},\ }\bibfield  {title} {\enquote {\bibinfo
  {title} {Attosecond control of electronic processes by intense light
  fields},}\ }\href {http://dx.doi.org/10.1038/nature01414} {\bibfield
  {journal} {\bibinfo  {journal} {Nature}\ }\textbf {\bibinfo {volume} {421}},\
  \bibinfo {pages} {611} (\bibinfo {year} {2003})}\BibitemShut {NoStop}%
\bibitem [{\citenamefont {{Kienberger {\em et~al}}}(2004)}]{Kienberger2004}%
  \BibitemOpen
  \bibfield  {author} {\bibinfo {author} {\bibfnamefont {R.}~\bibnamefont
  {{Kienberger {\em et~al}}}},\ }\bibfield  {title} {\enquote {\bibinfo {title}
  {Atomic transient recorder},}\ }\href {http://dx.doi.org/10.1038/nature02277}
  {\bibfield  {journal} {\bibinfo  {journal} {Nature}\ }\textbf {\bibinfo
  {volume} {427}},\ \bibinfo {pages} {817--821} (\bibinfo {year}
  {2004})}\BibitemShut {NoStop}%
\bibitem [{\citenamefont {Pazourek}\ \emph {et~al.}(2015)\citenamefont
  {Pazourek}, \citenamefont {Nagele},\ and\ \citenamefont
  {Burgd\"orfer}}]{RevModPhys.87.765}%
  \BibitemOpen
  \bibfield  {author} {\bibinfo {author} {\bibfnamefont {R.}~\bibnamefont
  {Pazourek}}, \bibinfo {author} {\bibfnamefont {S.}~\bibnamefont {Nagele}}, \
  and\ \bibinfo {author} {\bibfnamefont {J.}~\bibnamefont {Burgd\"orfer}},\
  }\bibfield  {title} {\enquote {\bibinfo {title} {Attosecond chronoscopy of
  photoemission},}\ }\href {\doibase 10.1103/RevModPhys.87.765} {\bibfield
  {journal} {\bibinfo  {journal} {Rev. Mod. Phys.}\ }\textbf {\bibinfo {volume}
  {87}},\ \bibinfo {pages} {765} (\bibinfo {year} {2015})}\BibitemShut
  {NoStop}%
\bibitem [{\citenamefont {Schultze}\ \emph {et~al.}(2010)\citenamefont
  {Schultze}, \citenamefont {Fie{\ss}}, \citenamefont {Karpowicz},
  \citenamefont {Gagnon}, \citenamefont {Korbman}, \citenamefont {Hofstetter},
  \citenamefont {Neppl}, \citenamefont {Cavalieri}, \citenamefont {Komninos},
  \citenamefont {Mercouris} \emph {et~al.}}]{schultze2010delay}%
  \BibitemOpen
  \bibfield  {author} {\bibinfo {author} {\bibfnamefont {M.}~\bibnamefont
  {Schultze}}, \bibinfo {author} {\bibfnamefont {M.}~\bibnamefont {Fie{\ss}}},
  \bibinfo {author} {\bibfnamefont {N.}~\bibnamefont {Karpowicz}}, \bibinfo
  {author} {\bibfnamefont {J.}~\bibnamefont {Gagnon}}, \bibinfo {author}
  {\bibfnamefont {M.}~\bibnamefont {Korbman}}, \bibinfo {author} {\bibfnamefont
  {M.}~\bibnamefont {Hofstetter}}, \bibinfo {author} {\bibfnamefont
  {S.}~\bibnamefont {Neppl}}, \bibinfo {author} {\bibfnamefont {A.~L.}\
  \bibnamefont {Cavalieri}}, \bibinfo {author} {\bibfnamefont {Y.}~\bibnamefont
  {Komninos}}, \bibinfo {author} {\bibfnamefont {T.}~\bibnamefont {Mercouris}},
   \emph {et~al.},\ }\bibfield  {title} {\enquote {\bibinfo {title} {Delay in
  photoemission},}\ }\href {\doibase 10.1126/science.1189401} {\bibfield
  {journal} {\bibinfo  {journal} {Science}\ }\textbf {\bibinfo {volume}
  {328}},\ \bibinfo {pages} {1658--1662} (\bibinfo {year} {2010})}\BibitemShut
  {NoStop}%
\bibitem [{\citenamefont {Kl\"under}\ \emph {et~al.}(2011)\citenamefont
  {Kl\"under}, \citenamefont {Dahlstr\"om}, \citenamefont {Gisselbrecht},
  \citenamefont {Fordell}, \citenamefont {Swoboda}, \citenamefont {Gu\'enot},
  \citenamefont {Johnsson}, \citenamefont {Caillat}, \citenamefont
  {Mauritsson}, \citenamefont {Maquet}, \citenamefont {Ta\"\i{}eb},\ and\
  \citenamefont {L'Huillier}}]{PhysRevLett.106.143002}%
  \BibitemOpen
  \bibfield  {author} {\bibinfo {author} {\bibfnamefont {K.}~\bibnamefont
  {Kl\"under}}, \bibinfo {author} {\bibfnamefont {J.~M.}\ \bibnamefont
  {Dahlstr\"om}}, \bibinfo {author} {\bibfnamefont {M.}~\bibnamefont
  {Gisselbrecht}}, \bibinfo {author} {\bibfnamefont {T.}~\bibnamefont
  {Fordell}}, \bibinfo {author} {\bibfnamefont {M.}~\bibnamefont {Swoboda}},
  \bibinfo {author} {\bibfnamefont {D.}~\bibnamefont {Gu\'enot}}, \bibinfo
  {author} {\bibfnamefont {P.}~\bibnamefont {Johnsson}}, \bibinfo {author}
  {\bibfnamefont {J.}~\bibnamefont {Caillat}}, \bibinfo {author} {\bibfnamefont
  {J.}~\bibnamefont {Mauritsson}}, \bibinfo {author} {\bibfnamefont
  {A.}~\bibnamefont {Maquet}}, \bibinfo {author} {\bibfnamefont
  {R.}~\bibnamefont {Ta\"\i{}eb}}, \ and\ \bibinfo {author} {\bibfnamefont
  {A.}~\bibnamefont {L'Huillier}},\ }\bibfield  {title} {\enquote {\bibinfo
  {title} {Probing single-photon ionization on the attosecond time scale},}\
  }\href {http://dx.doi.org/10.1103/PhysRevLett.106.143002} {\bibfield
  {journal} {\bibinfo  {journal} {Phys. Rev. Lett.}\ }\textbf {\bibinfo
  {volume} {106}},\ \bibinfo {pages} {143002} (\bibinfo {year}
  {2011})}\BibitemShut {NoStop}%
\bibitem [{\citenamefont {Eisenbud}(1948)}]{Eisenbud1948}%
  \BibitemOpen
  \bibfield  {author} {\bibinfo {author} {\bibfnamefont {L.}~\bibnamefont
  {Eisenbud}},\ }\emph {\bibinfo {title} {Formal properties of nuclear
  collisions}},\ \href@noop {} {Ph.D. thesis},\ \bibinfo  {school} {Princeton
  University} (\bibinfo {year} {1948})\BibitemShut {NoStop}%
\bibitem [{\citenamefont {Wigner}(1955{\natexlab{a}})}]{PhysRev.98.145}%
  \BibitemOpen
  \bibfield  {author} {\bibinfo {author} {\bibfnamefont {E.~P.}\ \bibnamefont
  {Wigner}},\ }\bibfield  {title} {\enquote {\bibinfo {title} {Lower limit for
  the energy derivative of the scattering phase shift},}\ }\href {\doibase
  http://dx.doi.org/10.1103/PhysRev.98.145} {\bibfield  {journal} {\bibinfo
  {journal} {Phys. Rev.}\ }\textbf {\bibinfo {volume} {98}},\ \bibinfo {pages}
  {145--147} (\bibinfo {year} {1955}{\natexlab{a}})}\BibitemShut {NoStop}%
\bibitem [{\citenamefont {Smith}(1960)}]{PhysRev.118.349}%
  \BibitemOpen
  \bibfield  {author} {\bibinfo {author} {\bibfnamefont {F.~T.}\ \bibnamefont
  {Smith}},\ }\bibfield  {title} {\enquote {\bibinfo {title} {Lifetime matrix
  in collision theory},}\ }\href
  {http://link.aps.org/doi/10.1103/PhysRev.118.349} {\bibfield  {journal}
  {\bibinfo  {journal} {Phys. Rev.}\ }\textbf {\bibinfo {volume} {118}},\
  \bibinfo {pages} {349--356} (\bibinfo {year} {1960})}\BibitemShut {NoStop}%
\bibitem [{\citenamefont {Kheifets}\ and\ \citenamefont
  {Ivanov}(2010{\natexlab{a}})}]{PhysRevLett.105.233002}%
  \BibitemOpen
  \bibfield  {author} {\bibinfo {author} {\bibfnamefont {A.~S.}\ \bibnamefont
  {Kheifets}}\ and\ \bibinfo {author} {\bibfnamefont {I.~A.}\ \bibnamefont
  {Ivanov}},\ }\bibfield  {title} {\enquote {\bibinfo {title} {Delay in atomic
  photoionization},}\ }\href {\doibase 10.1103/PhysRevLett.105.233002}
  {\bibfield  {journal} {\bibinfo  {journal} {Phys. Rev. Lett.}\ }\textbf
  {\bibinfo {volume} {105}},\ \bibinfo {pages} {233002} (\bibinfo {year}
  {2010}{\natexlab{a}})}\BibitemShut {NoStop}%
\bibitem [{\citenamefont {Kheifets}\ \emph {et~al.}(2016)\citenamefont
  {Kheifets}, \citenamefont {Bray},\ and\ \citenamefont
  {Bray}}]{PhysRevLett.117.143202}%
  \BibitemOpen
  \bibfield  {author} {\bibinfo {author} {\bibfnamefont {A.~S.}\ \bibnamefont
  {Kheifets}}, \bibinfo {author} {\bibfnamefont {A.~W.}\ \bibnamefont {Bray}},
  \ and\ \bibinfo {author} {\bibfnamefont {I.}~\bibnamefont {Bray}},\
  }\bibfield  {title} {\enquote {\bibinfo {title} {Attosecond time delay in
  photoemission and electron scattering near threshold},}\ }\href {\doibase
  10.1103/PhysRevLett.117.143202} {\bibfield  {journal} {\bibinfo  {journal}
  {Phys. Rev. Lett.}\ }\textbf {\bibinfo {volume} {117}},\ \bibinfo {pages}
  {143202} (\bibinfo {year} {2016})}\BibitemShut {NoStop}%
\bibitem [{\citenamefont {Landau}\ and\ \citenamefont
  {Lifshitz}(1985)}]{LLQuantum}%
  \BibitemOpen
  \bibfield  {author} {\bibinfo {author} {\bibfnamefont {L.}~\bibnamefont
  {Landau}}\ and\ \bibinfo {author} {\bibfnamefont {E.~M.}\ \bibnamefont
  {Lifshitz}},\ }\href@noop {} {\emph {\bibinfo {title} {Quantum Mechanics
  (Non-relativistic theory)}}},\ \bibinfo {edition} {3rd}\ ed.,\ \bibinfo
  {series} {Course of theoretical physics}, Vol.~\bibinfo {volume} {3}\
  (\bibinfo  {publisher} {Pergamon press, Oxford},\ \bibinfo {year}
  {1985})\BibitemShut {NoStop}%
\bibitem [{\citenamefont {McDowell}\ and\ \citenamefont
  {Coleman}(1970)}]{mcdowell1970introduction}%
  \BibitemOpen
  \bibfield  {author} {\bibinfo {author} {\bibfnamefont {M.~R.~C.}\
  \bibnamefont {McDowell}}\ and\ \bibinfo {author} {\bibfnamefont {J.~P.}\
  \bibnamefont {Coleman}},\ }\href
  {https://books.google.com.au/books?id=MaB9AAAAIAAJ} {\emph {\bibinfo {title}
  {Introduction to the theory of ion-atom collisions}}}\ (\bibinfo  {publisher}
  {North-Holland Pub. Co.},\ \bibinfo {year} {1970})\BibitemShut {NoStop}%
\bibitem [{\citenamefont {Sellmeier}(1871)}]{sellmeier1871erklarung}%
  \BibitemOpen
  \bibfield  {author} {\bibinfo {author} {\bibfnamefont {W.}~\bibnamefont
  {Sellmeier}},\ }\bibfield  {title} {\enquote {\bibinfo {title} {Zur
  erkl{\"a}rung der abnormen farbenfolge im spectrum einiger substanzen},}\
  }\href
  {http://onlinelibrary.wiley.com/store/10.1002/andp.18712190612/asset/18712190612_ftp.pdf?v=1&t=iui7sfqq&s=69bb6a20ca7bd978c957c99294246d71b8cc9456}
  {\bibfield  {journal} {\bibinfo  {journal} {Ann. d. Phys. u. Chem.}\ }\textbf
  {\bibinfo {volume} {219}},\ \bibinfo {pages} {272--282} (\bibinfo {year}
  {1871})}\BibitemShut {NoStop}%
\bibitem [{\citenamefont {Bray}(2016)}]{IgorLecture}%
  \BibitemOpen
  \bibfield  {author} {\bibinfo {author} {\bibfnamefont {I.}~\bibnamefont
  {Bray}},\ }\href@noop {} {\enquote {\bibinfo {title} {Lecture notes on
  computational quantum mechanics},}\ } (\bibinfo {year} {2016})\BibitemShut
  {NoStop}%
\bibitem [{\citenamefont {Friedrich}(2012)}]{friedrich2012theoretical}%
  \BibitemOpen
  \bibfield  {author} {\bibinfo {author} {\bibfnamefont {H.}~\bibnamefont
  {Friedrich}},\ }\href {https://books.google.com.au/books?id=DTnvCAAAQBAJ}
  {\emph {\bibinfo {title} {Theoretical Atomic Physics}}}\ (\bibinfo
  {publisher} {Springer Berlin Heidelberg},\ \bibinfo {year}
  {2012})\BibitemShut {NoStop}%
\bibitem [{\citenamefont {Tweed}(1980)}]{TweedHe}%
  \BibitemOpen
  \bibfield  {author} {\bibinfo {author} {\bibfnamefont {R.~J.}\ \bibnamefont
  {Tweed}},\ }\bibfield  {title} {\enquote {\bibinfo {title} {Triple
  differential cross section for the ionisation of helium by electronic
  impact},}\ }\href {http://stacks.iop.org/0022-3700/13/i=22/a=019} {\bibfield
  {journal} {\bibinfo  {journal} {J. Phys. B}\ }\textbf {\bibinfo {volume}
  {13}},\ \bibinfo {pages} {4467} (\bibinfo {year} {1980})}\BibitemShut
  {NoStop}%
\bibitem [{\citenamefont {Beaty}\ \emph {et~al.}(1977)\citenamefont {Beaty},
  \citenamefont {Hesselbacher}, \citenamefont {Hong},\ and\ \citenamefont
  {Moore}}]{BeatyHe}%
  \BibitemOpen
  \bibfield  {author} {\bibinfo {author} {\bibfnamefont {E.~C.}\ \bibnamefont
  {Beaty}}, \bibinfo {author} {\bibfnamefont {K.~H.}\ \bibnamefont
  {Hesselbacher}}, \bibinfo {author} {\bibfnamefont {S.~P.}\ \bibnamefont
  {Hong}}, \ and\ \bibinfo {author} {\bibfnamefont {J.~H.}\ \bibnamefont
  {Moore}},\ }\bibfield  {title} {\enquote {\bibinfo {title}
  {Triple-differential three-dimensional cross sections for low-energy electron
  impact ionization of helium},}\ }\href
  {http://stacks.iop.org/0022-3700/10/i=4/a=015} {\bibfield  {journal}
  {\bibinfo  {journal} {J. Phys. B}\ }\textbf {\bibinfo {volume} {10}},\
  \bibinfo {pages} {611} (\bibinfo {year} {1977})}\BibitemShut {NoStop}%
\bibitem [{\citenamefont {Fursa}\ and\ \citenamefont {Bray}(2008)}]{fursaRel}%
  \BibitemOpen
  \bibfield  {author} {\bibinfo {author} {\bibfnamefont {D.~V.}\ \bibnamefont
  {Fursa}}\ and\ \bibinfo {author} {\bibfnamefont {I.}~\bibnamefont {Bray}},\
  }\bibfield  {title} {\enquote {\bibinfo {title} {Fully relativistic
  convergent close-coupling method for excitation and ionization processes in
  electron collisions with atoms and ions},}\ }\href {\doibase
  10.1103/PhysRevLett.100.113201} {\bibfield  {journal} {\bibinfo  {journal}
  {Phys. Rev. Lett.}\ }\textbf {\bibinfo {volume} {100}},\ \bibinfo {pages}
  {113201} (\bibinfo {year} {2008})}\BibitemShut {NoStop}%
\bibitem [{\citenamefont {Bray}\ and\ \citenamefont
  {Stelbovics}(1992{\natexlab{b}})}]{BrayH}%
  \BibitemOpen
  \bibfield  {author} {\bibinfo {author} {\bibfnamefont {I.}~\bibnamefont
  {Bray}}\ and\ \bibinfo {author} {\bibfnamefont {A.~T.}\ \bibnamefont
  {Stelbovics}},\ }\bibfield  {title} {\enquote {\bibinfo {title} {Convergent
  close-coupling calculations of electron-hydrogen scattering},}\ }\href
  {\doibase 10.1103/PhysRevA.46.6995} {\bibfield  {journal} {\bibinfo
  {journal} {Phys. Rev. A}\ }\textbf {\bibinfo {volume} {46}},\ \bibinfo
  {pages} {6995--7011} (\bibinfo {year} {1992}{\natexlab{b}})}\BibitemShut
  {NoStop}%
\bibitem [{\citenamefont {Bray}(1994)}]{BrayHlike}%
  \BibitemOpen
  \bibfield  {author} {\bibinfo {author} {\bibfnamefont {I.}~\bibnamefont
  {Bray}},\ }\bibfield  {title} {\enquote {\bibinfo {title} {Convergent
  close-coupling method for the calculation of electron scattering on
  hydrogenlike targets},}\ }\href {\doibase 10.1103/PhysRevA.49.1066}
  {\bibfield  {journal} {\bibinfo  {journal} {Phys. Rev. A}\ }\textbf {\bibinfo
  {volume} {49}},\ \bibinfo {pages} {1066--1082} (\bibinfo {year}
  {1994})}\BibitemShut {NoStop}%
\bibitem [{\citenamefont {Fursa}\ and\ \citenamefont
  {Bray}(1997{\natexlab{b}})}]{fursaHe}%
  \BibitemOpen
  \bibfield  {author} {\bibinfo {author} {\bibfnamefont {D.~V.}\ \bibnamefont
  {Fursa}}\ and\ \bibinfo {author} {\bibfnamefont {I.}~\bibnamefont {Bray}},\
  }\bibfield  {title} {\enquote {\bibinfo {title} {Convergent close-coupling
  calculations of electron-helium scattering},}\ }\href
  {http://stacks.iop.org/0953-4075/30/i=4/a=003} {\bibfield  {journal}
  {\bibinfo  {journal} {J. Phys. B}\ }\textbf {\bibinfo {volume} {30}},\
  \bibinfo {pages} {757} (\bibinfo {year} {1997}{\natexlab{b}})}\BibitemShut
  {NoStop}%
\bibitem [{\citenamefont {Rawlins}\ \emph {et~al.}(2016)\citenamefont
  {Rawlins}, \citenamefont {Kadyrov}, \citenamefont {Stelbovics}, \citenamefont
  {Bray},\ and\ \citenamefont {Charlton}}]{rawlinsPos}%
  \BibitemOpen
  \bibfield  {author} {\bibinfo {author} {\bibfnamefont {C.~M.}\ \bibnamefont
  {Rawlins}}, \bibinfo {author} {\bibfnamefont {A.~S.}\ \bibnamefont
  {Kadyrov}}, \bibinfo {author} {\bibfnamefont {A.~T.}\ \bibnamefont
  {Stelbovics}}, \bibinfo {author} {\bibfnamefont {I.}~\bibnamefont {Bray}}, \
  and\ \bibinfo {author} {\bibfnamefont {M.}~\bibnamefont {Charlton}},\
  }\bibfield  {title} {\enquote {\bibinfo {title} {Calculation of antihydrogen
  formation via antiproton scattering with excited positronium},}\ }\href
  {\doibase 10.1103/PhysRevA.93.012709} {\bibfield  {journal} {\bibinfo
  {journal} {Phys. Rev. A}\ }\textbf {\bibinfo {volume} {93}},\ \bibinfo
  {pages} {012709} (\bibinfo {year} {2016})}\BibitemShut {NoStop}%
\bibitem [{\citenamefont {Fursa}\ and\ \citenamefont
  {Bray}(2012)}]{fursaPosNob}%
  \BibitemOpen
  \bibfield  {author} {\bibinfo {author} {\bibfnamefont {D.~V.}\ \bibnamefont
  {Fursa}}\ and\ \bibinfo {author} {\bibfnamefont {I.}~\bibnamefont {Bray}},\
  }\bibfield  {title} {\enquote {\bibinfo {title} {Convergent close-coupling
  method for positron scattering from noble gases},}\ }\href
  {http://stacks.iop.org/1367-2630/14/i=3/a=035002} {\bibfield  {journal}
  {\bibinfo  {journal} {New J. Phys.}\ }\textbf {\bibinfo {volume} {14}},\
  \bibinfo {pages} {035002} (\bibinfo {year} {2012})}\BibitemShut {NoStop}%
\bibitem [{\citenamefont {Kadyrov}\ \emph {et~al.}(2014)\citenamefont
  {Kadyrov}, \citenamefont {Bailey}, \citenamefont {Bray},\ and\ \citenamefont
  {Stelbovics}}]{kadyrovPos}%
  \BibitemOpen
  \bibfield  {author} {\bibinfo {author} {\bibfnamefont {A.~S.}\ \bibnamefont
  {Kadyrov}}, \bibinfo {author} {\bibfnamefont {J.~J.}\ \bibnamefont {Bailey}},
  \bibinfo {author} {\bibfnamefont {I.}~\bibnamefont {Bray}}, \ and\ \bibinfo
  {author} {\bibfnamefont {A.~T.}\ \bibnamefont {Stelbovics}},\ }\bibfield
  {title} {\enquote {\bibinfo {title} {Two-center approach to fully
  differential positron-impact ionization of hydrogen},}\ }\href {\doibase
  10.1103/PhysRevA.89.012706} {\bibfield  {journal} {\bibinfo  {journal} {Phys.
  Rev. A}\ }\textbf {\bibinfo {volume} {89}},\ \bibinfo {pages} {012706}
  (\bibinfo {year} {2014})}\BibitemShut {NoStop}%
\bibitem [{\citenamefont {Utamuratov}\ \emph {et~al.}(2010)\citenamefont
  {Utamuratov}, \citenamefont {Kadyrov}, \citenamefont {Fursa},\ and\
  \citenamefont {Bray}}]{utamuratovMultiHe}%
  \BibitemOpen
  \bibfield  {author} {\bibinfo {author} {\bibfnamefont {R.}~\bibnamefont
  {Utamuratov}}, \bibinfo {author} {\bibfnamefont {A.~S.}\ \bibnamefont
  {Kadyrov}}, \bibinfo {author} {\bibfnamefont {D.~V.}\ \bibnamefont {Fursa}},
  \ and\ \bibinfo {author} {\bibfnamefont {I.}~\bibnamefont {Bray}},\
  }\bibfield  {title} {\enquote {\bibinfo {title} {A two-centre convergent
  close-coupling approach to positron--helium collisions},}\ }\href
  {http://stacks.iop.org/0953-4075/43/i=3/a=031001} {\bibfield  {journal}
  {\bibinfo  {journal} {J. Phys. B}\ }\textbf {\bibinfo {volume} {43}},\
  \bibinfo {pages} {031001} (\bibinfo {year} {2010})}\BibitemShut {NoStop}%
\bibitem [{\citenamefont {Bray}\ \emph {et~al.}(2002)\citenamefont {Bray},
  \citenamefont {Fursa}, \citenamefont {Kheifets},\ and\ \citenamefont
  {Stelbovics}}]{brayelecandphoto}%
  \BibitemOpen
  \bibfield  {author} {\bibinfo {author} {\bibfnamefont {I.}~\bibnamefont
  {Bray}}, \bibinfo {author} {\bibfnamefont {D.~V.}\ \bibnamefont {Fursa}},
  \bibinfo {author} {\bibfnamefont {A.~S.}\ \bibnamefont {Kheifets}}, \ and\
  \bibinfo {author} {\bibfnamefont {A.~T.}\ \bibnamefont {Stelbovics}},\
  }\bibfield  {title} {\enquote {\bibinfo {title} {Electrons and photons
  colliding with atoms: development and application of the convergent
  close-coupling method},}\ }\href
  {http://stacks.iop.org/0953-4075/35/i=15/a=201} {\bibfield  {journal}
  {\bibinfo  {journal} {J. Phys. B}\ }\textbf {\bibinfo {volume} {35}},\
  \bibinfo {pages} {R117} (\bibinfo {year} {2002})}\BibitemShut {NoStop}%
\bibitem [{\citenamefont {Kheifets}\ and\ \citenamefont
  {Bray}(2001)}]{PhysRevA.65.012710}%
  \BibitemOpen
  \bibfield  {author} {\bibinfo {author} {\bibfnamefont {A.~S.}\ \bibnamefont
  {Kheifets}}\ and\ \bibinfo {author} {\bibfnamefont {I.}~\bibnamefont
  {Bray}},\ }\bibfield  {title} {\enquote {\bibinfo {title} {Frozen-core model
  of the double photoionization of beryllium},}\ }\href {\doibase
  10.1103/PhysRevA.65.012710} {\bibfield  {journal} {\bibinfo  {journal} {Phys.
  Rev. A}\ }\textbf {\bibinfo {volume} {65}},\ \bibinfo {pages} {012710}
  (\bibinfo {year} {2001})}\BibitemShut {NoStop}%
\bibitem [{\citenamefont {Cvejanovic}\ \emph {et~al.}(2000)\citenamefont
  {Cvejanovic}, \citenamefont {Wightman}, \citenamefont {Reddish},
  \citenamefont {Maulbetsch}, \citenamefont {MacDonald}, \citenamefont
  {Kheifets},\ and\ \citenamefont {Bray}}]{0953-4075-33-2-311}%
  \BibitemOpen
  \bibfield  {author} {\bibinfo {author} {\bibfnamefont {S.}~\bibnamefont
  {Cvejanovic}}, \bibinfo {author} {\bibfnamefont {J.~P.}\ \bibnamefont
  {Wightman}}, \bibinfo {author} {\bibfnamefont {T.~J.}\ \bibnamefont
  {Reddish}}, \bibinfo {author} {\bibfnamefont {F.}~\bibnamefont {Maulbetsch}},
  \bibinfo {author} {\bibfnamefont {M.~A.}\ \bibnamefont {MacDonald}}, \bibinfo
  {author} {\bibfnamefont {A.~S.}\ \bibnamefont {Kheifets}}, \ and\ \bibinfo
  {author} {\bibfnamefont {I.}~\bibnamefont {Bray}},\ }\bibfield  {title}
  {\enquote {\bibinfo {title} {Photodouble ionization of helium at an excess
  energy of 40 {eV}},}\ }\href {http://stacks.iop.org/0953-4075/33/i=2/a=311}
  {\bibfield  {journal} {\bibinfo  {journal} {J. Phys. B}\ }\textbf {\bibinfo
  {volume} {33}},\ \bibinfo {pages} {265} (\bibinfo {year} {2000})}\BibitemShut
  {NoStop}%
\bibitem [{\citenamefont {Hoszowska}\ \emph {et~al.}(2009)\citenamefont
  {Hoszowska}, \citenamefont {Kheifets}, \citenamefont {Dousse}, \citenamefont
  {Berset}, \citenamefont {Bray}, \citenamefont {Cao}, \citenamefont {Fennane},
  \citenamefont {Kayser}, \citenamefont {Kav\ifmmode \check{c}\else
  \v{c}\fi{}i\ifmmode~\check{c}\else \v{c}\fi{}}, \citenamefont {Szlachetko},\
  and\ \citenamefont {Szlachetko}}]{PhysRevLett.102.073006}%
  \BibitemOpen
  \bibfield  {author} {\bibinfo {author} {\bibfnamefont {J.}~\bibnamefont
  {Hoszowska}}, \bibinfo {author} {\bibfnamefont {A.~K.}\ \bibnamefont
  {Kheifets}}, \bibinfo {author} {\bibfnamefont {J.-C.}\ \bibnamefont
  {Dousse}}, \bibinfo {author} {\bibfnamefont {M.}~\bibnamefont {Berset}},
  \bibinfo {author} {\bibfnamefont {I.}~\bibnamefont {Bray}}, \bibinfo {author}
  {\bibfnamefont {W.}~\bibnamefont {Cao}}, \bibinfo {author} {\bibfnamefont
  {K.}~\bibnamefont {Fennane}}, \bibinfo {author} {\bibfnamefont
  {Y.}~\bibnamefont {Kayser}}, \bibinfo {author} {\bibfnamefont
  {M.}~\bibnamefont {Kav\ifmmode \check{c}\else
  \v{c}\fi{}i\ifmmode~\check{c}\else \v{c}\fi{}}}, \bibinfo {author}
  {\bibfnamefont {J.}~\bibnamefont {Szlachetko}}, \ and\ \bibinfo {author}
  {\bibfnamefont {M.}~\bibnamefont {Szlachetko}},\ }\bibfield  {title}
  {\enquote {\bibinfo {title} {Physical mechanisms and scaling laws of
  $k$-shell double photoionization},}\ }\href {\doibase
  10.1103/PhysRevLett.102.073006} {\bibfield  {journal} {\bibinfo  {journal}
  {Phys. Rev. Lett.}\ }\textbf {\bibinfo {volume} {102}},\ \bibinfo {pages}
  {073006} (\bibinfo {year} {2009})}\BibitemShut {NoStop}%
\bibitem [{\citenamefont {Kheifets}\ and\ \citenamefont
  {Bray}(1998{\natexlab{a}})}]{kheifetsPhoto}%
  \BibitemOpen
  \bibfield  {author} {\bibinfo {author} {\bibfnamefont {A.~S.}\ \bibnamefont
  {Kheifets}}\ and\ \bibinfo {author} {\bibfnamefont {I.}~\bibnamefont
  {Bray}},\ }\bibfield  {title} {\enquote {\bibinfo {title} {Application of the
  {CCC} method to the calculation of helium double-photoionization triply
  differential cross sections},}\ }\href
  {http://stacks.iop.org/0953-4075/31/i=10/a=002} {\bibfield  {journal}
  {\bibinfo  {journal} {J. Phys. B}\ }\textbf {\bibinfo {volume} {31}},\
  \bibinfo {pages} {L447} (\bibinfo {year} {1998}{\natexlab{a}})}\BibitemShut
  {NoStop}%
\bibitem [{\citenamefont {Fursa}\ and\ \citenamefont {Bray}(1995)}]{fursaHeLS}%
  \BibitemOpen
  \bibfield  {author} {\bibinfo {author} {\bibfnamefont {D.~V.}\ \bibnamefont
  {Fursa}}\ and\ \bibinfo {author} {\bibfnamefont {I.}~\bibnamefont {Bray}},\
  }\bibfield  {title} {\enquote {\bibinfo {title} {Calculation of
  electron-helium scattering},}\ }\href {\doibase 10.1103/PhysRevA.52.1279}
  {\bibfield  {journal} {\bibinfo  {journal} {Phys. Rev. A}\ }\textbf {\bibinfo
  {volume} {52}},\ \bibinfo {pages} {1279--1297} (\bibinfo {year}
  {1995})}\BibitemShut {NoStop}%
\bibitem [{\citenamefont {Bostock}\ \emph {et~al.}(2010)\citenamefont
  {Bostock}, \citenamefont {Fursa},\ and\ \citenamefont {Bray}}]{bostokRel}%
  \BibitemOpen
  \bibfield  {author} {\bibinfo {author} {\bibfnamefont {C.~J.}\ \bibnamefont
  {Bostock}}, \bibinfo {author} {\bibfnamefont {D.~V.}\ \bibnamefont {Fursa}},
  \ and\ \bibinfo {author} {\bibfnamefont {I.}~\bibnamefont {Bray}},\
  }\bibfield  {title} {\enquote {\bibinfo {title} {Relativistic convergent
  close-coupling method applied to electron scattering from mercury},}\ }\href
  {\doibase 10.1103/PhysRevA.82.022713} {\bibfield  {journal} {\bibinfo
  {journal} {Phys. Rev. A}\ }\textbf {\bibinfo {volume} {82}},\ \bibinfo
  {pages} {022713} (\bibinfo {year} {2010})}\BibitemShut {NoStop}%
\bibitem [{\citenamefont {Bray}\ \emph {et~al.}(2012)\citenamefont {Bray},
  \citenamefont {Fursa}, \citenamefont {Kadyrov}, \citenamefont {Stelbovics},
  \citenamefont {Kheifets},\ and\ \citenamefont {Mukhamedzhanov}}]{BrayRep}%
  \BibitemOpen
  \bibfield  {author} {\bibinfo {author} {\bibfnamefont {I.}~\bibnamefont
  {Bray}}, \bibinfo {author} {\bibfnamefont {D.~V.}\ \bibnamefont {Fursa}},
  \bibinfo {author} {\bibfnamefont {A.~S.}\ \bibnamefont {Kadyrov}}, \bibinfo
  {author} {\bibfnamefont {A.~T.}\ \bibnamefont {Stelbovics}}, \bibinfo
  {author} {\bibfnamefont {A.~S.}\ \bibnamefont {Kheifets}}, \ and\ \bibinfo
  {author} {\bibfnamefont {A.~M.}\ \bibnamefont {Mukhamedzhanov}},\ }\bibfield
  {title} {\enquote {\bibinfo {title} {Electron- and photon-impact atomic
  ionisation},}\ }\href {\doibase
  http://dx.doi.org/10.1016/j.physrep.2012.07.002} {\bibfield  {journal}
  {\bibinfo  {journal} {Phys. Rep.}\ }\textbf {\bibinfo {volume} {520}},\
  \bibinfo {pages} {135 -- 174} (\bibinfo {year} {2012})},\ \bibinfo {note}
  {electron- and photon-impact atomic ionisation}\BibitemShut {NoStop}%
\bibitem [{\citenamefont {Bray}\ \emph {et~al.}(2015)\citenamefont {Bray},
  \citenamefont {Abdurakhmanov}, \citenamefont {Kadyrov}, \citenamefont
  {Fursa},\ and\ \citenamefont {Bray}}]{bray2015solving}%
  \BibitemOpen
  \bibfield  {author} {\bibinfo {author} {\bibfnamefont {A.~W.}\ \bibnamefont
  {Bray}}, \bibinfo {author} {\bibfnamefont {I.~B.}\ \bibnamefont
  {Abdurakhmanov}}, \bibinfo {author} {\bibfnamefont {A.~S.}\ \bibnamefont
  {Kadyrov}}, \bibinfo {author} {\bibfnamefont {D.~V.}\ \bibnamefont {Fursa}},
  \ and\ \bibinfo {author} {\bibfnamefont {I.}~\bibnamefont {Bray}},\
  }\bibfield  {title} {\enquote {\bibinfo {title} {Solving close-coupling
  equations in momentum space without singularities},}\ }\href {\doibase
  http://dx.doi.org/10.1016/j.cpc.2015.06.015} {\bibfield  {journal} {\bibinfo
  {journal} {Comp. Phys. Commun.}\ }\textbf {\bibinfo {volume} {196}},\
  \bibinfo {pages} {276 -- 279} (\bibinfo {year} {2015})}\BibitemShut {NoStop}%
\bibitem [{\citenamefont {Bray}\ \emph {et~al.}(2016)\citenamefont {Bray},
  \citenamefont {Abdurakhmanov}, \citenamefont {Kadyrov}, \citenamefont
  {Fursa},\ and\ \citenamefont {Bray}}]{bray2016solving}%
  \BibitemOpen
  \bibfield  {author} {\bibinfo {author} {\bibfnamefont {A.~W.}\ \bibnamefont
  {Bray}}, \bibinfo {author} {\bibfnamefont {I.~B.}\ \bibnamefont
  {Abdurakhmanov}}, \bibinfo {author} {\bibfnamefont {A.~S.}\ \bibnamefont
  {Kadyrov}}, \bibinfo {author} {\bibfnamefont {D.~V.}\ \bibnamefont {Fursa}},
  \ and\ \bibinfo {author} {\bibfnamefont {I.}~\bibnamefont {Bray}},\
  }\bibfield  {title} {\enquote {\bibinfo {title} {Solving close-coupling
  equations in momentum space without singularities {II}},}\ }\href {\doibase
  http://dx.doi.org/10.1016/j.cpc.2016.02.028} {\bibfield  {journal} {\bibinfo
  {journal} {Comp. Phys. Commun.}\ }\textbf {\bibinfo {volume} {203}},\
  \bibinfo {pages} {147 -- 151} (\bibinfo {year} {2016})}\BibitemShut {NoStop}%
\bibitem [{\citenamefont {McCarthy}\ and\ \citenamefont
  {Stelbovics}(1983)}]{McCarthyStelbovics}%
  \BibitemOpen
  \bibfield  {author} {\bibinfo {author} {\bibfnamefont {I.~E.}\ \bibnamefont
  {McCarthy}}\ and\ \bibinfo {author} {\bibfnamefont {A.~T.}\ \bibnamefont
  {Stelbovics}},\ }\bibfield  {title} {\enquote {\bibinfo {title}
  {Momentum-space coupled-channels optical method for electron-atom
  scattering},}\ }\href {\doibase 10.1103/PhysRevA.28.2693} {\bibfield
  {journal} {\bibinfo  {journal} {Phys. Rev. A}\ }\textbf {\bibinfo {volume}
  {28}},\ \bibinfo {pages} {2693--2707} (\bibinfo {year} {1983})}\BibitemShut
  {NoStop}%
\bibitem [{\citenamefont {Cohen}\ and\ \citenamefont
  {McEachran}(1967)}]{cohenFrozen}%
  \BibitemOpen
  \bibfield  {author} {\bibinfo {author} {\bibfnamefont {M.}~\bibnamefont
  {Cohen}}\ and\ \bibinfo {author} {\bibfnamefont {R.~P.}\ \bibnamefont
  {McEachran}},\ }\bibfield  {title} {\enquote {\bibinfo {title} {The triplet
  {S} states of the helium isoelectronic sequence},}\ }\href
  {http://stacks.iop.org/0370-1328/92/i=3/a=304} {\bibfield  {journal}
  {\bibinfo  {journal} {Proc. Phys. Soc.}\ }\textbf {\bibinfo {volume} {92}},\
  \bibinfo {pages} {539} (\bibinfo {year} {1967})}\BibitemShut {NoStop}%
\bibitem [{\citenamefont {Mason}(1993)}]{laserass}%
  \BibitemOpen
  \bibfield  {author} {\bibinfo {author} {\bibfnamefont {N.~J.}\ \bibnamefont
  {Mason}},\ }\bibfield  {title} {\enquote {\bibinfo {title} {Laser-assisted
  electron-atom collisions},}\ }\href
  {http://stacks.iop.org/0034-4885/56/i=10/a=002} {\bibfield  {journal}
  {\bibinfo  {journal} {Rep. Prog. Phys.}\ }\textbf {\bibinfo {volume} {56}},\
  \bibinfo {pages} {1275} (\bibinfo {year} {1993})}\BibitemShut {NoStop}%
\bibitem [{\citenamefont {Eisberg}\ and\ \citenamefont
  {Resnick}(1985)}]{eisberg1985quantum}%
  \BibitemOpen
  \bibfield  {author} {\bibinfo {author} {\bibfnamefont {R.~M.}\ \bibnamefont
  {Eisberg}}\ and\ \bibinfo {author} {\bibfnamefont {R.}~\bibnamefont
  {Resnick}},\ }\href {https://books.google.com.au/books?id=1v1QAAAAMAAJ}
  {\emph {\bibinfo {title} {Quantum physics of atoms, molecules, solids,
  nuclei, and particles}}},\ Quantum Physics of Atoms, Molecules, Solids,
  Nuclei and Particles\ (\bibinfo  {publisher} {Wiley},\ \bibinfo {year}
  {1985})\BibitemShut {NoStop}%
\bibitem [{{DLMF}()}]{NIST:DLMF}%
  \BibitemOpen
  {DLMF},\ \href {http://dlmf.nist.gov/} {\enquote {\bibinfo {title}
  {{\textit{NIST Digital Library of Mathematical Functions}}},}\ }\bibinfo
  {howpublished} {Release 1.0.13} (\bibinfo {year} {2016}),\ \bibinfo {note}
  {f.~W.~J. Olver, A.~B. {Olde Daalhuis}, D.~W. Lozier, B.~I. Schneider, R.~F.
  Boisvert, C.~W. Clark, B.~R. Miller and B.~V. Saunders, eds.}\BibitemShut
  {Stop}%
\bibitem [{\citenamefont {Jung}(1979)}]{Jung1979}%
  \BibitemOpen
  \bibfield  {author} {\bibinfo {author} {\bibfnamefont {C.}~\bibnamefont
  {Jung}},\ }\bibfield  {title} {\enquote {\bibinfo {title} {Analysis of the
  differential sum rule for laser-induced free-free transitions},}\ }\href
  {\doibase 10.1103/PhysRevA.20.1585} {\bibfield  {journal} {\bibinfo
  {journal} {Phys. Rev. A}\ }\textbf {\bibinfo {volume} {20}},\ \bibinfo
  {pages} {1585--1589} (\bibinfo {year} {1979})}\BibitemShut {NoStop}%
\bibitem [{\citenamefont {Kheifets}\ and\ \citenamefont
  {Bray}(1998{\natexlab{b}})}]{KB98d}%
  \BibitemOpen
  \bibfield  {author} {\bibinfo {author} {\bibfnamefont {A.~S.}\ \bibnamefont
  {Kheifets}}\ and\ \bibinfo {author} {\bibfnamefont {I.}~\bibnamefont
  {Bray}},\ }\bibfield  {title} {\enquote {\bibinfo {title} {Photoionization
  with excitation and double photoionization of helium isoelectronic
  sequence},}\ }\href {http://prola.aps.org/abstract/PRA/v58/i6/p4501_1}
  {\bibfield  {journal} {\bibinfo  {journal} {Phys.~Rev.~A}\ }\textbf {\bibinfo
  {volume} {58}},\ \bibinfo {pages} {4501--4511} (\bibinfo {year}
  {1998}{\natexlab{b}})}\BibitemShut {NoStop}%
\bibitem [{\citenamefont {Kheifets}\ \emph {et~al.}(2000)\citenamefont
  {Kheifets}, \citenamefont {Ipatov}, \citenamefont {Arifin},\ and\
  \citenamefont {Bray}}]{KIAB00}%
  \BibitemOpen
  \bibfield  {author} {\bibinfo {author} {\bibfnamefont {A.~S.}\ \bibnamefont
  {Kheifets}}, \bibinfo {author} {\bibfnamefont {A.}~\bibnamefont {Ipatov}},
  \bibinfo {author} {\bibfnamefont {M.}~\bibnamefont {Arifin}}, \ and\ \bibinfo
  {author} {\bibfnamefont {I.}~\bibnamefont {Bray}},\ }\bibfield  {title}
  {\enquote {\bibinfo {title} {Double-photoionization calculations of the
  helium metastable $2{}^{1,3}{S}$ states},}\ }\href {\doibase
  10.1103/PhysRevA.62.052724} {\bibfield  {journal} {\bibinfo  {journal} {Phys.
  Rev. A}\ }\textbf {\bibinfo {volume} {62}},\ \bibinfo {pages} {052724}
  (\bibinfo {year} {2000})}\BibitemShut {NoStop}%
\bibitem [{\citenamefont {Wigner}(1955{\natexlab{b}})}]{wigner1955lower}%
  \BibitemOpen
  \bibfield  {author} {\bibinfo {author} {\bibfnamefont {E.~P.}\ \bibnamefont
  {Wigner}},\ }\bibfield  {title} {\enquote {\bibinfo {title} {Lower limit for
  the energy derivative of the scattering phase shift},}\ }\href {\doibase
  10.1103/PhysRev.98.145} {\bibfield  {journal} {\bibinfo  {journal} {Phys.
  Rev.}\ }\textbf {\bibinfo {volume} {98}},\ \bibinfo {pages} {145} (\bibinfo
  {year} {1955}{\natexlab{b}})}\BibitemShut {NoStop}%
\bibitem [{\citenamefont {Kheifets}\ and\ \citenamefont
  {Ivanov}(2010{\natexlab{b}})}]{kheifets2010delay}%
  \BibitemOpen
  \bibfield  {author} {\bibinfo {author} {\bibfnamefont {A.~S.}\ \bibnamefont
  {Kheifets}}\ and\ \bibinfo {author} {\bibfnamefont {I.~A.}\ \bibnamefont
  {Ivanov}},\ }\bibfield  {title} {\enquote {\bibinfo {title} {Delay in atomic
  photoionization},}\ }\href {\doibase 10.1103/PhysRevLett.105.233002}
  {\bibfield  {journal} {\bibinfo  {journal} {Phys. Rev. Lett.}\ }\textbf
  {\bibinfo {volume} {105}},\ \bibinfo {pages} {233002} (\bibinfo {year}
  {2010}{\natexlab{b}})}\BibitemShut {NoStop}%
\bibitem [{\citenamefont {de~Carvalho}\ and\ \citenamefont
  {Nussenzveig}(2002)}]{de2002time}%
  \BibitemOpen
  \bibfield  {author} {\bibinfo {author} {\bibfnamefont {C.~A.~A.}\
  \bibnamefont {de~Carvalho}}\ and\ \bibinfo {author} {\bibfnamefont {H.~M.}\
  \bibnamefont {Nussenzveig}},\ }\bibfield  {title} {\enquote {\bibinfo {title}
  {Time delay},}\ }\href {\doibase
  http://dx.doi.org/10.1016/S0370-1573(01)00092-8} {\bibfield  {journal}
  {\bibinfo  {journal} {Phys. Rep.}\ }\textbf {\bibinfo {volume} {364}},\
  \bibinfo {pages} {83--174} (\bibinfo {year} {2002})}\BibitemShut {NoStop}%
\bibitem [{\citenamefont {D{\"u}rr}\ \emph {et~al.}(2008)\citenamefont
  {D{\"u}rr}, \citenamefont {Dimopoulou}, \citenamefont {Najjari},
  \citenamefont {Dorn}, \citenamefont {Bartschat}, \citenamefont {Bray},
  \citenamefont {Fursa}, \citenamefont {Chen}, \citenamefont {Madison},\ and\
  \citenamefont {Ullrich}}]{durr2008higher}%
  \BibitemOpen
  \bibfield  {author} {\bibinfo {author} {\bibfnamefont {M.}~\bibnamefont
  {D{\"u}rr}}, \bibinfo {author} {\bibfnamefont {C.}~\bibnamefont
  {Dimopoulou}}, \bibinfo {author} {\bibfnamefont {B.}~\bibnamefont {Najjari}},
  \bibinfo {author} {\bibfnamefont {A.}~\bibnamefont {Dorn}}, \bibinfo {author}
  {\bibfnamefont {K.}~\bibnamefont {Bartschat}}, \bibinfo {author}
  {\bibfnamefont {I.}~\bibnamefont {Bray}}, \bibinfo {author} {\bibfnamefont
  {D.~V.}\ \bibnamefont {Fursa}}, \bibinfo {author} {\bibfnamefont
  {Z.}~\bibnamefont {Chen}}, \bibinfo {author} {\bibfnamefont {D.~H.}\
  \bibnamefont {Madison}}, \ and\ \bibinfo {author} {\bibfnamefont
  {J.}~\bibnamefont {Ullrich}},\ }\bibfield  {title} {\enquote {\bibinfo
  {title} {Higher-order contributions observed in three-dimensional (e, 2 e)
  cross-section measurements at 1-k{eV} impact energy},}\ }\href {\doibase
  10.1103/PhysRevA.77.032717} {\bibfield  {journal} {\bibinfo  {journal} {Phys.
  Rev. A}\ }\textbf {\bibinfo {volume} {77}},\ \bibinfo {pages} {032717}
  (\bibinfo {year} {2008})}\BibitemShut {NoStop}%
\bibitem [{\citenamefont {Venuti}\ and\ \citenamefont
  {Decleva}(1997)}]{0953-4075-30-21-020}%
  \BibitemOpen
  \bibfield  {author} {\bibinfo {author} {\bibfnamefont {M.}~\bibnamefont
  {Venuti}}\ and\ \bibinfo {author} {\bibfnamefont {P.}~\bibnamefont
  {Decleva}},\ }\bibfield  {title} {\enquote {\bibinfo {title} {Convergent
  multichannel continuum states by a general configuration interaction
  expansion in a {$B$}-spline basis: application to {H}$^-$ photodetachment},}\
  }\href {http://stacks.iop.org/0953-4075/30/i=21/a=020} {\bibfield  {journal}
  {\bibinfo  {journal} {J. Phys. B}\ }\textbf {\bibinfo {volume} {30}},\
  \bibinfo {pages} {4839} (\bibinfo {year} {1997})}\BibitemShut {NoStop}%
\bibitem [{\citenamefont {Popp}\ and\ \citenamefont
  {Kruse}(1976)}]{POPP1976683}%
  \BibitemOpen
  \bibfield  {author} {\bibinfo {author} {\bibfnamefont {H.~P.}\ \bibnamefont
  {Popp}}\ and\ \bibinfo {author} {\bibfnamefont {S.}~\bibnamefont {Kruse}},\
  }\bibfield  {title} {\enquote {\bibinfo {title} {Negative hydrogen ion
  detachment cross section from radiation measurements on a {PLTE}-arc},}\
  }\href {http://www.sciencedirect.com/science/article/pii/0022407376900601}
  {\bibfield  {journal} {\bibinfo  {journal} {J. Quant. Spectr. Rad. Trans.}\
  }\textbf {\bibinfo {volume} {16}},\ \bibinfo {pages} {683 -- 688} (\bibinfo
  {year} {1976})}\BibitemShut {NoStop}%
\bibitem [{\citenamefont {G\'en\'evriez}\ and\ \citenamefont
  {Urbain}(2015)}]{PhysRevA.91.033403}%
  \BibitemOpen
  \bibfield  {author} {\bibinfo {author} {\bibfnamefont {M.}~\bibnamefont
  {G\'en\'evriez}}\ and\ \bibinfo {author} {\bibfnamefont {X.}~\bibnamefont
  {Urbain}},\ }\bibfield  {title} {\enquote {\bibinfo {title} {Animated-beam
  measurement of the photodetachment cross section of {H}$^-$},}\ }\href
  {http://link.aps.org/doi/10.1103/PhysRevA.91.033403} {\bibfield  {journal}
  {\bibinfo  {journal} {Phys. Rev. A}\ }\textbf {\bibinfo {volume} {91}},\
  \bibinfo {pages} {033403} (\bibinfo {year} {2015})}\BibitemShut {NoStop}%
\bibitem [{\citenamefont {Register}\ and\ \citenamefont
  {Poe}(1975)}]{REGISTER1975431}%
  \BibitemOpen
  \bibfield  {author} {\bibinfo {author} {\bibfnamefont {D.}~\bibnamefont
  {Register}}\ and\ \bibinfo {author} {\bibfnamefont {R.~T.}\ \bibnamefont
  {Poe}},\ }\bibfield  {title} {\enquote {\bibinfo {title} {Algebraic
  variational method - a quantitative assessment in e$^\pm$-{H} scattering},}\
  }\href {\doibase http://dx.doi.org/10.1016/0375-9601(75)90759-8} {\bibfield
  {journal} {\bibinfo  {journal} {Phys. Lett. A}\ }\textbf {\bibinfo {volume}
  {51}},\ \bibinfo {pages} {431--433} (\bibinfo {year} {1975})}\BibitemShut
  {NoStop}%
\bibitem [{\citenamefont {Chen}\ \emph {et~al.}(1997)\citenamefont {Chen},
  \citenamefont {Lin},\ and\ \citenamefont {Tang}}]{PhysRevA.56.2435}%
  \BibitemOpen
  \bibfield  {author} {\bibinfo {author} {\bibfnamefont {M.}~\bibnamefont
  {Chen}}, \bibinfo {author} {\bibfnamefont {C.~D.}\ \bibnamefont {Lin}}, \
  and\ \bibinfo {author} {\bibfnamefont {J.~Z.}\ \bibnamefont {Tang}},\
  }\bibfield  {title} {\enquote {\bibinfo {title} {Hyperspherical
  close-coupling calculation of electron-hydrogen scattering cross sections},}\
  }\href {http://link.aps.org/doi/10.1103/PhysRevA.56.2435} {\bibfield
  {journal} {\bibinfo  {journal} {Phys. Rev. A}\ }\textbf {\bibinfo {volume}
  {56}},\ \bibinfo {pages} {2435--2438} (\bibinfo {year} {1997})}\BibitemShut
  {NoStop}%
\bibitem [{\citenamefont {Chernysheva}\ \emph {et~al.}(1979)\citenamefont
  {Chernysheva}, \citenamefont {Cherepkov},\ and\ \citenamefont
  {Radojevic}}]{CCR79}%
  \BibitemOpen
  \bibfield  {author} {\bibinfo {author} {\bibfnamefont {L.~V.}\ \bibnamefont
  {Chernysheva}}, \bibinfo {author} {\bibfnamefont {N.~A.}\ \bibnamefont
  {Cherepkov}}, \ and\ \bibinfo {author} {\bibfnamefont {V.}~\bibnamefont
  {Radojevic}},\ }\bibfield  {title} {\enquote {\bibinfo {title} {Frozen core
  hartree-fock program for atomic discrete and continuous states},}\ }\href
  {http://www.sciencedirect.com/science/article/pii/0010465579900262}
  {\bibfield  {journal} {\bibinfo  {journal} {Comp. Phys. Comm.}\ }\textbf
  {\bibinfo {volume} {18}},\ \bibinfo {pages} {87--100} (\bibinfo {year}
  {1979})}\BibitemShut {NoStop}%
\bibitem [{\citenamefont {Sadeghpour}\ \emph {et~al.}(2000)\citenamefont
  {Sadeghpour}, \citenamefont {Bohn}, \citenamefont {Cavagnero}, \citenamefont
  {Esry}, \citenamefont {Fabrikant}, \citenamefont {Macek},\ and\ \citenamefont
  {Rau}}]{0953-4075-33-5-201}%
  \BibitemOpen
  \bibfield  {author} {\bibinfo {author} {\bibfnamefont {H.~R.}\ \bibnamefont
  {Sadeghpour}}, \bibinfo {author} {\bibfnamefont {J.~L.}\ \bibnamefont
  {Bohn}}, \bibinfo {author} {\bibfnamefont {M.~J.}\ \bibnamefont {Cavagnero}},
  \bibinfo {author} {\bibfnamefont {B.~D.}\ \bibnamefont {Esry}}, \bibinfo
  {author} {\bibfnamefont {I.~I.}\ \bibnamefont {Fabrikant}}, \bibinfo {author}
  {\bibfnamefont {J.~H.}\ \bibnamefont {Macek}}, \ and\ \bibinfo {author}
  {\bibfnamefont {A.~R.~P.}\ \bibnamefont {Rau}},\ }\bibfield  {title}
  {\enquote {\bibinfo {title} {Collisions near threshold in atomic and
  molecular physics},}\ }\href {http://stacks.iop.org/0953-4075/33/i=5/a=201}
  {\bibfield  {journal} {\bibinfo  {journal} {J. Phys. B}\ }\textbf {\bibinfo
  {volume} {33}},\ \bibinfo {pages} {R93} (\bibinfo {year} {2000})}\BibitemShut
  {NoStop}%
\bibitem [{\citenamefont {Samson}\ \emph {et~al.}(1994)\citenamefont {Samson},
  \citenamefont {He}, \citenamefont {Yin},\ and\ \citenamefont
  {Haddad}}]{SHYH94}%
  \BibitemOpen
  \bibfield  {author} {\bibinfo {author} {\bibfnamefont {J.~A.~R.}\
  \bibnamefont {Samson}}, \bibinfo {author} {\bibfnamefont {Z.~X.}\
  \bibnamefont {He}}, \bibinfo {author} {\bibfnamefont {L.}~\bibnamefont
  {Yin}}, \ and\ \bibinfo {author} {\bibfnamefont {G.~N.}\ \bibnamefont
  {Haddad}},\ }\bibfield  {title} {\enquote {\bibinfo {title} {Precision
  measurements of the absolute photoionization cross sections of {He}},}\
  }\href {http://stacks.iop.org/0953-4075/27/i=5/a=008} {\bibfield  {journal}
  {\bibinfo  {journal} {J.~Phys.~B}\ }\textbf {\bibinfo {volume} {27}},\
  \bibinfo {pages} {887--898} (\bibinfo {year} {1994})}\BibitemShut {NoStop}%
\bibitem [{\citenamefont {Broad}\ and\ \citenamefont {Reinhardt}(1976)}]{BR76}%
  \BibitemOpen
  \bibfield  {author} {\bibinfo {author} {\bibfnamefont {J.~T.}\ \bibnamefont
  {Broad}}\ and\ \bibinfo {author} {\bibfnamefont {W.~P.}\ \bibnamefont
  {Reinhardt}},\ }\bibfield  {title} {\enquote {\bibinfo {title} {One- and
  two-electron photoejectrion from {H$^-$}: A multichannel {$J$} matrix
  calculation},}\ }\href {\doibase 10.1103/PhysRevA.14.2159} {\bibfield
  {journal} {\bibinfo  {journal} {Phys.~Rev.~A}\ }\textbf {\bibinfo {volume}
  {14}},\ \bibinfo {pages} {2159} (\bibinfo {year} {1976})}\BibitemShut
  {NoStop}%
\bibitem [{\citenamefont {Patchkovskii}\ and\ \citenamefont
  {Muller}(2016)}]{Patchkovskii2016153}%
  \BibitemOpen
  \bibfield  {author} {\bibinfo {author} {\bibfnamefont {S.}~\bibnamefont
  {Patchkovskii}}\ and\ \bibinfo {author} {\bibfnamefont {H.~G.}\ \bibnamefont
  {Muller}},\ }\bibfield  {title} {\enquote {\bibinfo {title} {Simple,
  accurate, and efficient implementation of 1-electron atomic time-dependent
  {Schr{\"o}dinger} equation in spherical coordinates},}\ }\href {\doibase
  http://dx.doi.org/10.1016/j.cpc.2015.10.014} {\bibfield  {journal} {\bibinfo
  {journal} {Comput. Phys. Commun.}\ }\textbf {\bibinfo {volume} {199}},\
  \bibinfo {pages} {153 -- 169} (\bibinfo {year} {2016})}\BibitemShut {NoStop}%
\bibitem [{\citenamefont {Dellwo}\ \emph {et~al.}(1992)\citenamefont {Dellwo},
  \citenamefont {Liu}, \citenamefont {Pegg},\ and\ \citenamefont
  {Alton}}]{PhysRevA.45.1544}%
  \BibitemOpen
  \bibfield  {author} {\bibinfo {author} {\bibfnamefont {J.}~\bibnamefont
  {Dellwo}}, \bibinfo {author} {\bibfnamefont {Y.}~\bibnamefont {Liu}},
  \bibinfo {author} {\bibfnamefont {D.~J.}\ \bibnamefont {Pegg}}, \ and\
  \bibinfo {author} {\bibfnamefont {G.~D.}\ \bibnamefont {Alton}},\ }\bibfield
  {title} {\enquote {\bibinfo {title} {Near-threshold photodetachment of the
  {Li}$^-$ ion},}\ }\href {http://link.aps.org/doi/10.1103/PhysRevA.45.1544}
  {\bibfield  {journal} {\bibinfo  {journal} {Phys. Rev. A}\ }\textbf {\bibinfo
  {volume} {45}},\ \bibinfo {pages} {1544--1547} (\bibinfo {year}
  {1992})}\BibitemShut {NoStop}%
\bibitem [{\citenamefont {Ganesan}\ \emph {et~al.}(2016)\citenamefont
  {Ganesan}, \citenamefont {Saha}, \citenamefont {Decshmukh}, \citenamefont
  {Manson},\ and\ \citenamefont {Kheifets}}]{DAMOP2016}%
  \BibitemOpen
  \bibfield  {author} {\bibinfo {author} {\bibfnamefont {A.}~\bibnamefont
  {Ganesan}}, \bibinfo {author} {\bibfnamefont {S.}~\bibnamefont {Saha}},
  \bibinfo {author} {\bibfnamefont {P.~C.}\ \bibnamefont {Decshmukh}}, \bibinfo
  {author} {\bibfnamefont {S.~T.}\ \bibnamefont {Manson}}, \ and\ \bibinfo
  {author} {\bibfnamefont {A.~S.}\ \bibnamefont {Kheifets}},\ }\bibfield
  {title} {\enquote {\bibinfo {title} {Electron correlation effects on
  photoionization time delay in atomic {Ar} and {Xe}},}\ }in\ \href
  {http://meetings.aps.org/Meeting/DAMOP16/Session/K1.34} {\emph {\bibinfo
  {booktitle} {Proceedings of DAMOP}}},\ Vol.~\bibinfo {volume} {61}\ (\bibinfo
  {year} {2016})\BibitemShut {NoStop}%
\bibitem [{\citenamefont {Dahlstr\"om}\ \emph {et~al.}(2012)\citenamefont
  {Dahlstr\"om}, \citenamefont {Gu\'enot}, \citenamefont {Kl\"under},
  \citenamefont {Gisselbrecht}, \citenamefont {Mauritsson}, \citenamefont
  {Huillier}, \citenamefont {Maquet},\ and\ \citenamefont
  {Ta\"ieb}}]{Dahlstrom2012}%
  \BibitemOpen
  \bibfield  {author} {\bibinfo {author} {\bibfnamefont {J.}~\bibnamefont
  {Dahlstr\"om}}, \bibinfo {author} {\bibfnamefont {D.}~\bibnamefont
  {Gu\'enot}}, \bibinfo {author} {\bibfnamefont {K.}~\bibnamefont {Kl\"under}},
  \bibinfo {author} {\bibfnamefont {M.}~\bibnamefont {Gisselbrecht}}, \bibinfo
  {author} {\bibfnamefont {J.}~\bibnamefont {Mauritsson}}, \bibinfo {author}
  {\bibfnamefont {A.~L.}\ \bibnamefont {Huillier}}, \bibinfo {author}
  {\bibfnamefont {A.}~\bibnamefont {Maquet}}, \ and\ \bibinfo {author}
  {\bibfnamefont {R.}~\bibnamefont {Ta\"ieb}},\ }\bibfield  {title} {\enquote
  {\bibinfo {title} {Theory of attosecond delays in laser-assisted
  photoionization},}\ }\href {\doibase 10.1016/j.chemphys.2012.01.017}
  {\bibfield  {journal} {\bibinfo  {journal} {Chem. Phys.}\ }\textbf {\bibinfo
  {volume} {414}},\ \bibinfo {pages} {53--64} (\bibinfo {year}
  {2012})}\BibitemShut {NoStop}%
\bibitem [{\citenamefont {Pazourek}\ \emph {et~al.}(2013)\citenamefont
  {Pazourek}, \citenamefont {Nagele},\ and\ \citenamefont
  {Burgdorfer}}]{C3FD00004D}%
  \BibitemOpen
  \bibfield  {author} {\bibinfo {author} {\bibfnamefont {R.}~\bibnamefont
  {Pazourek}}, \bibinfo {author} {\bibfnamefont {S.}~\bibnamefont {Nagele}}, \
  and\ \bibinfo {author} {\bibfnamefont {J.}~\bibnamefont {Burgdorfer}},\
  }\bibfield  {title} {\enquote {\bibinfo {title} {Time-resolved photoemission
  on the attosecond scale: opportunities and challenges},}\ }\href
  {http://dx.doi.org/10.1039/C3FD00004D} {\bibfield  {journal} {\bibinfo
  {journal} {Faraday Discuss.}\ }\textbf {\bibinfo {volume} {163}},\ \bibinfo
  {pages} {353--376} (\bibinfo {year} {2013})}\BibitemShut {NoStop}%
\bibitem [{\citenamefont {Zielinski}\ \emph {et~al.}(2016)\citenamefont
  {Zielinski}, \citenamefont {Majety},\ and\ \citenamefont
  {Scrinzi}}]{PhysRevA.93.023406}%
  \BibitemOpen
  \bibfield  {author} {\bibinfo {author} {\bibfnamefont {A.}~\bibnamefont
  {Zielinski}}, \bibinfo {author} {\bibfnamefont {V.~P.}\ \bibnamefont
  {Majety}}, \ and\ \bibinfo {author} {\bibfnamefont {A.}~\bibnamefont
  {Scrinzi}},\ }\bibfield  {title} {\enquote {\bibinfo {title} {Double
  photoelectron momentum spectra of helium at infrared wavelength},}\ }\href
  {\doibase 10.1103/PhysRevA.93.023406} {\bibfield  {journal} {\bibinfo
  {journal} {Phys. Rev. A}\ }\textbf {\bibinfo {volume} {93}},\ \bibinfo
  {pages} {023406} (\bibinfo {year} {2016})}\BibitemShut {NoStop}%
\bibitem [{\citenamefont {Pindzola}\ and\ \citenamefont
  {Robicheaux}(1998)}]{PhysRevA.57.318}%
  \BibitemOpen
  \bibfield  {author} {\bibinfo {author} {\bibfnamefont {M.~S.}\ \bibnamefont
  {Pindzola}}\ and\ \bibinfo {author} {\bibfnamefont {F.}~\bibnamefont
  {Robicheaux}},\ }\bibfield  {title} {\enquote {\bibinfo {title}
  {Time-dependent close-coupling calculations of correlated photoionization
  processes in helium},}\ }\href {\doibase 10.1103/PhysRevA.57.318} {\bibfield
  {journal} {\bibinfo  {journal} {Phys. Rev. A}\ }\textbf {\bibinfo {volume}
  {57}},\ \bibinfo {pages} {318--324} (\bibinfo {year} {1998})}\BibitemShut
  {NoStop}%
\bibitem [{\citenamefont {Heinrich}\ \emph {et~al.}(2006)\citenamefont
  {Heinrich}, \citenamefont {Kornelis}, \citenamefont {Anscombe}, \citenamefont
  {Hauri}, \citenamefont {Schlup}, \citenamefont {Biegert},\ and\ \citenamefont
  {Keller}}]{0953-4075-39-13-S03}%
  \BibitemOpen
  \bibfield  {author} {\bibinfo {author} {\bibfnamefont {A.}~\bibnamefont
  {Heinrich}}, \bibinfo {author} {\bibfnamefont {W.}~\bibnamefont {Kornelis}},
  \bibinfo {author} {\bibfnamefont {M.~P.}\ \bibnamefont {Anscombe}}, \bibinfo
  {author} {\bibfnamefont {C.~P.}\ \bibnamefont {Hauri}}, \bibinfo {author}
  {\bibfnamefont {P.}~\bibnamefont {Schlup}}, \bibinfo {author} {\bibfnamefont
  {J.}~\bibnamefont {Biegert}}, \ and\ \bibinfo {author} {\bibfnamefont
  {U.}~\bibnamefont {Keller}},\ }\bibfield  {title} {\enquote {\bibinfo {title}
  {Enhanced {VUV}-assisted high harmonic generation},}\ }\href
  {http://stacks.iop.org/0953-4075/39/i=13/a=S03} {\bibfield  {journal}
  {\bibinfo  {journal} {J. Phys. B}\ }\textbf {\bibinfo {volume} {39}},\
  \bibinfo {pages} {S275} (\bibinfo {year} {2006})}\BibitemShut {NoStop}%
\end{thebibliography}%
%

\end{document}